\documentclass[a4paper,fleqn]{cas-dc}

\usepackage[authoryear]{natbib}
\usepackage{amssymb}
\usepackage{caption}
\usepackage{multirow}
\usepackage{longtable}
\usepackage{soul}
\usepackage[normalem]{ulem}

\def\tsc#1{\csdef{#1}{\textsc{\lowercase{#1}}\xspace}}
\tsc{WGM}
\tsc{QE}
\tsc{EP}
\tsc{PMS}
\tsc{BEC}
\tsc{DE}
 \newcommand{\rf}[1]{} 

\def \astrima {ASTRI Mini-Array}

\def \astrima {ASTRI Mini-Array}
\def \hess{H.E.S.S.}
\def \glast {{\it Fermi}}
\def \degmark{^\circ}

\def \arcmin {\hbox{$^\prime$}}
\def \arcsec {\hbox{$^{\prime\prime}$}}

\def \gray {$\gamma$-ray}

\def \apj {ApJ}
\def \apjl {ApJL}
\def \apjs {ApJS}
\def \aap {A\&A}

\def \baas {BAAS}

\def \jcap {J. Cosmology Astropart. Phys.}
\def \mnras {MNRAS}

\def \pasp {PASP}
\def \prd {PRD}
\def \ssr {SSRv}

\def \nat {Nature}
\def \na {Nature}
\def \nar {New Astronomy Reviews}
\def \procspie {SPIE Conf. Ser.}
\def \prl {PRL}
\def \aapr {AApR}
\def \araa {ARAA}

\begin{document}
\let\WriteBookmarks\relax
\def\floatpagepagefraction{1}
\def\textpagefraction{.001}
\shorttitle{ASTRI Mini-Array Core Science}

\shortauthors{S. Vercellone et~al.}
\title [mode = title]{ASTRI Mini-Array Core Science at the {\em Observatorio del Teide}}                      
\newcounter{ctr_auts}
\refstepcounter{ctr_auts}\label{oab}    
\author[\getrefnumber{oab} ]{S. Vercellone}[orcid=0000-0003-1163-1396]
\cormark[\getrefnumber{oab}]
\ead{stefano.vercellone@inaf.it}
\address[\getrefnumber{oab}]{INAF, Osservatorio Astronomico di Brera, Sede di Merate, Via Emilio Bianchi 46, I-23807 Merate (LC), Italy}


\refstepcounter{ctr_auts}\label{oar}    
\author[\getrefnumber{oar}]{C. Bigongiari}[orcid=0000-0003-3293-8522] 
\address[\getrefnumber{oar}]{INAF, Osservatorio Astronomico di Roma, Via Frascati 33, I-00078 Monte Porzio Catone (Roma), Italy}

\refstepcounter{ctr_auts}\label{oaa}    
\author[\getrefnumber{oaa}]{A. Burtovoi}[orcid=0000-0002-8734-808X] 
\address[\getrefnumber{oaa}]{INAF,  Osservatorio Astrofisico di Arcetri, Largo E. Fermi 5, I-50125, Firenze, Italy}

\refstepcounter{ctr_auts}\label{iaps}    
\author[\getrefnumber{iaps}]{M. Cardillo}[orcid=0000-0001-8877-3996] 
\address[\getrefnumber{iaps}]{INAF, Istituto di Astrofisica e Planetologia Spaziale, Via Fosso del Cavaliere 100, I-00133 Roma, Italy}

\refstepcounter{ctr_auts}\label{ifc}    
\author[\getrefnumber{ifc}]{O. Catalano}[orcid=0000-0002-9554-4128 ] 
\address[\getrefnumber{ifc}]{INAF, Istituto di Astrofisica Spaziale e Fisica Cosmica, Via Ugo la Malfa 153, I-90146 Palermo, Italy}

\refstepcounter{ctr_auts}\label{unipd}    
\author[\getrefnumber{unipd}]{A. Franceschini}[orcid=0000-0002-9979-3383] 
\address[\getrefnumber{unipd}]{UNIPD, Dipartimento di Fisica ed Astronomia ``Galileo Galilei'', Vicolo dell'Osservatorio 3, I-35122 Padova, Italy}

\refstepcounter{ctr_auts}\label{ssdc}    
\author[\getrefnumber{oar},\getrefnumber{ssdc}]{S. Lombardi}[orcid=0000-0002-6336-865X] 
\address[\getrefnumber{ssdc}]{ASI, Space Science Data Center, Via del Politecnico s.n.c., I-00133 Roma, Italy}

\author[\getrefnumber{oab}]{L. Nava}[orcid=0000-0001-5960-0455] 

\author[\getrefnumber{ifc}]{F. Pintore}[orcid=0000-0002-3869-2925] 

\author[\getrefnumber{oar}]{A. Stamerra}[orcid=0000-0002-9430-5264] 

\author[\getrefnumber{oab}]{F. Tavecchio}[orcid=0000-0003-0256-0995] 

\refstepcounter{ctr_auts}\label{oapd}    
\address[\getrefnumber{oapd}]{INAF, Osservatorio Astronomico di Padova, Vicolo dell'Osservatorio 5, I-35122 Padova, Italy}
\author[\getrefnumber{oapd}]{L. Zampieri}[orcid=0000-0002-6516-1329]

\refstepcounter{ctr_auts}\label{uamcsic}    
\author[\getrefnumber{uamcsic}]{R. {Alves Batista}}[orcid=0000-0003-2656-064X]
\address[\getrefnumber{uamcsic}]{Instituto de F\'isica Te\'orica UAM-CSIC, C/ Nicol\'as Cabrera 13-15, 28049 Madrid, Spain}

\refstepcounter{ctr_auts}\label{unifi}    
\author[\getrefnumber{oaa},\getrefnumber{unifi}]{E. Amato}[orcid=0000-0002-9881-8112] 
\address[\getrefnumber{unifi}]{Dipartimento di Fisica e Astronomia, Universit\`a degli Studi di Firenze, Via Sansone 1, I-50019 Sesto Fiorentino (FI), Italy}

\author[\getrefnumber{oar},\getrefnumber{ssdc}]{L.~A. Antonelli}[orcid= 0000-0002-5037-9034] 

\refstepcounter{ctr_auts}\label{nwu}    
\author[\getrefnumber{oapd},\getrefnumber{nwu}]{C. Arcaro}[orcid= 0000-0002-1998-9707 ] 
\address[\getrefnumber{nwu}]{Centre for Space Research, North-West University, 2520 Potchefstroom, South Africa}

\refstepcounter{ctr_auts}\label{iac}    
\refstepcounter{ctr_auts}\label{laguna}    
\author[\getrefnumber{iac},\getrefnumber{laguna}]{J. {Becerra Gonz\'alez}}[orcid=0000-0002-6729-9022] 
\address[\getrefnumber{iac}]{Instituto de Astrofísica de Canarias, E-38205 La Laguna, Tenerife,Spain}
\address[\getrefnumber{laguna}]{Universidad de La Laguna, Dpto. Astrof\'isica, E-38206 La Laguna, Tenerife, Spain}

\author[\getrefnumber{oab}]{G. Bonnoli}[orcid=0000-0003-2464-9077]

\author[\getrefnumber{nwu}]{M. B\"ottcher}[orcid=0000-0002-8434-5692] 

\refstepcounter{ctr_auts}\label{ira}    
\author[\getrefnumber{ira}]{G. Brunetti}[orcid=0000-0003-4195-8613 ] 
\address[\getrefnumber{ira}]{INAF, Istituto di Radioastronomia, Via P. Gobetti 101, I-40129 Bologna, Italy}

\author[\getrefnumber{ifc}]{A.~A. Compagnino}[orcid=0000-0003-4727-9136] 

\refstepcounter{ctr_auts}\label{iasfmi}    
\refstepcounter{ctr_auts}\label{insubria}    
\author[\getrefnumber{iasfmi},\getrefnumber{insubria}]{S. Crestan}[orcid=0000-0002-8368-0616] 
\address[\getrefnumber{iasfmi}]{INAF, Istituto di Astrofisica Spaziale e Fisica Cosmica, Via Alfonso Corti 12, I-20133 Milano, Italy}
\address[\getrefnumber{insubria}]{Università degli Studi dell'Insubria, Via Valleggio 11, 22100 Como, Italy}

\author[\getrefnumber{ifc}]{A. D'A\`i}[orcid=0000-0002-5042-1036] 

\author[\getrefnumber{oapd},\getrefnumber{unipd}]{M. Fiori}[orcid=0000-0002-7352-6818]

\author[\getrefnumber{iasfmi}]{G. Galanti}[orcid=0000-0001-7254-3029]

\author[\getrefnumber{iasfmi}]{A. Giuliani}[orcid=0000-0002-4315-1699]

\refstepcounter{ctr_auts}\label{iagca}    
\author[\getrefnumber{iagca}]{E.~M. {de Gouveia Dal Pino}}[orcid=0000-0001-8058-4752]
\address[\getrefnumber{iagca}]{Instituto de Astronomia, Geofisica e Ci\^encias Atmosf\'ericas, Universidade de S\~ao Paulo, Brazil}

\author[\getrefnumber{oar}]{J.~G. {Green}}[orcid=0000-0002-1130-6692]

\author[\getrefnumber{oar},\getrefnumber{ssdc}]{A. Lamastra}[orcid=0000-0003-2403-913X]

\author[\getrefnumber{oab}]{M. Landoni}[orcid=0000-0001-5570-5081]

\author[\getrefnumber{oar},\getrefnumber{ssdc}]{F. Lucarelli}[orcid=0000-0002-6311-764X]

\author[\getrefnumber{oaa}]{G. Morlino}[orcid=0000-0002-5014-4817]

\refstepcounter{ctr_auts}\label{oapa}    
\author[\getrefnumber{oapa},\getrefnumber{oaa}]{B. Olmi}[orcid=0000-0001-6022-8216]
\address[\getrefnumber{oapa}]{INAF, Osservatorio Astronomico di Palermo, Piazza del Parlamento 1,  I-90146 Palermo, Italy}

\refstepcounter{ctr_auts}\label{bohr}    
\author[\getrefnumber{bohr}]{E. Peretti}[orcid=0000-0003-0543-0467] 
\address[\getrefnumber{bohr}]{Niels Bohr International Academy, Niels Bohr Institute,University of Copenhagen, Blegdamsvej 17, DK-2100 Copenhagen, Denmark.}

\author[\getrefnumber{iaps}]{G. Piano}[orcid=0000-0002-9332-5319] 

\refstepcounter{ctr_auts}\label{mpe}    
\author[\getrefnumber{oab},\getrefnumber{mpe}]{G. Ponti}[orcid=0000-0003-0293-3608]
\address[\getrefnumber{mpe}]{Max-Planck-Institut f\"{u}r extraterrestrische Physik (MPE), Giessenbachstrasse 1, 85748 Garching bei M\"{u}nchen, Germany}

\refstepcounter{ctr_auts}\label{fgg}    
\author[\getrefnumber{oab},\getrefnumber{fgg}]{E. Poretti}[orcid=0000-0003-1200-0473] 
\address[\getrefnumber{fgg}]{INAF, Fundaci\'on  Galileo Galilei, Rambla Jos\'e Ana Fernandez P\'erez 7, 38712 Bre\~{n}a Baja (TF), Spain}

\author[\getrefnumber{oab}]{P. Romano}[orcid=0000-0003-0258-7469]

\author[\getrefnumber{oar},\getrefnumber{ssdc}]{F.~G. Saturni}[orcid=0000-0002-1946-7706]

\author[\getrefnumber{iasfmi}]{S. Scuderi}[orcid=0000-0002-8637-2109] 

\author[\getrefnumber{oar}]{A. Tutone}[orcid=0000-0002-2840-0001]

\refstepcounter{ctr_auts}\label{oact}    
\author[\getrefnumber{oact}]{G. Umana}[orcid=0000-0002-6972-8388]
\address[\getrefnumber{oact}]{INAF, Osservatorio Astrofisico di Catania, Via S.Sofia 78, I-95123 Catania, Italy}


\author[\getrefnumber{iac},\getrefnumber{laguna}]{J.~A. {Acosta-Pulido}}[orcid=0000-0002-0433-9656]

\author[\getrefnumber{iagca}]{P. {Barai}}[orcid=0000-0001-7031-2331]

\author[\getrefnumber{oact}]{A. {Bonanno}}[orcid=0000-0003-3175-9776]

\author[\getrefnumber{oact}]{G. {Bonanno}}[orcid=0000-0002-4792-3983]

\author[\getrefnumber{oact}]{P. {Bruno}}[orcid=0000-0003-3919-9611]

\refstepcounter{ctr_auts}\label{oass}    
\author[\getrefnumber{oass}]{A. {Bulgarelli}}[orcid=0000-0001-6347-0649]
\address[\getrefnumber{oass}]{INAF, Osservatorio di Astrofisica e Scienza dello Spazio, Via Gobetti 93/3, I-40129, Bologna, Italy}

\author[\getrefnumber{oass}]{V. {Conforti}}[orcid=0000-0002-0007-3520]

\author[\getrefnumber{oact}]{A. {Costa}}[orcid=0000-0003-0344-8911]

\author[\getrefnumber{ifc}]{G. {Cusumano}}[orcid=0000-0002-8151-1990]

\author[\getrefnumber{ifc}]{M. {Del Santo}}[orcid=0000-0002-1793-1050]

\author[\getrefnumber{iagca}]{M.~V. {del Valle}}[orcid=0000-0002-5444-0795]

\author[\getrefnumber{oab}]{R. {Della Ceca}}[orcid=0000-0001-7551-2252]

\author[\getrefnumber{iagca}]{D.~A. {Falceta-Gon\c{c}alves}}[orcid=0000-0002-1914-6654]

\author[\getrefnumber{oass}]{V. {Fioretti}}[orcid=0000-0002-6082-5384]

\refstepcounter{ctr_auts}\label{unipg}    
\refstepcounter{ctr_auts}\label{infnpg}    
\author[\getrefnumber{unipg},\getrefnumber{infnpg}]{S. {Germani}}[orcid=0000-0002-2233-6811]
\address[\getrefnumber{unipg}]{Dipartimento di Fisica, Universit\`{a} degli Studi di Perugia, I-06123 Perugia, Italy}
\address[\getrefnumber{infnpg}]{Istituto Nazionale di Fisica Nucleare, Sezione di Perugia, I-06123 Perugia, Italy}

\author[\getrefnumber{iac},\getrefnumber{laguna}]{R.~J. {Garc\'ia-L\'opez}}[orcid=0000-0002-8204-6832]

\author[\getrefnumber{fgg}]{A. Ghedina}[orcid=0000-0003-4702-5152] 

\author[\getrefnumber{oact}]{V. {Giordano}}[orcid=0000-0001-8865-5930]

\author[\getrefnumber{nwu}]{M. {Kreter}}[orcid=0000-0002-3092-3506]

\author[\getrefnumber{oact}]{F. {Incardona}}[orcid=0000-0002-2568-0917]

\author[\getrefnumber{oab}]{S. {Iovenitti}}[orcid=0000-0002-2581-9528]

\author[\getrefnumber{ifc}]{A. {La Barbera}}[orcid=0000-0002-5880-8913]

\author[\getrefnumber{iasfmi}]{N. {La Palombara}}[orcid=0000-0001-7015-6359]

\author[\getrefnumber{ifc}]{V. {La Parola}}[orcid=0000-0002-8087-6488]

\author[\getrefnumber{oact}]{G. {Leto}}[orcid=0000-0002-0040-5011]

\refstepcounter{ctr_auts}\label{infnts}    
\refstepcounter{ctr_auts}\label{units}    
\author[\getrefnumber{infnts},\getrefnumber{units}]{F. {Longo}}[orcid=0000-0003-2501-2270]
\address[\getrefnumber{infnts}]{Istituto Nazionale di Fisica Nucleare, Sezione di Trieste, I-34127 Trieste, Italy}
\address[\getrefnumber{units}]{Dipartimento di Fisica, Università di Trieste, I-34127 Trieste, Italy}

\author[\getrefnumber{iac},\getrefnumber{laguna}]{A. {L\'opez-Oramas}}[orcid=0000-0003-4603-1884]

\author[\getrefnumber{ifc}]{M.~C. {Maccarone}}[orcid=0000-0001-8722-0361]

\author[\getrefnumber{iasfmi}]{S. {Mereghetti}}[orcid=0000-0003-3259-7801] 

\author[\getrefnumber{oab}]{R. {Millul}}[orcid=0000-0003-3760-7861] 

\author[\getrefnumber{unipd}]{G. {Naletto}}[orcid=0000-0003-2007-3138]

\author[\getrefnumber{ifc}]{A. {Pagliaro}}[orcid=0000-0002-6841-1362]

\author[\getrefnumber{oass}]{N. {Parmiggiani}}[orcid=0000-0002-4535-5329]

\author[\getrefnumber{oab}]{C. {Righi}}[orcid= 0000-0002-1218-9555]

\author[\getrefnumber{iagca}]{J.~C. {Rodr\'iguez-Ram\'irez}}[orcid=0000-0001-9980-5973]

\author[\getrefnumber{oact}]{G. {Romeo}}[orcid=0000-0003-3239-6057]

\author[\getrefnumber{ifc}]{P. {Sangiorgi}}[orcid=0000-0001-8138-9289]

\author[\getrefnumber{iagca}]{R. {Santos de Lima}}[orcid=0000-0001-6880-4468]

\author[\getrefnumber{oab}]{G. Tagliaferri}[orcid=0000-0003-0121-0723] 

\author[\getrefnumber{oar}]{V. {Testa}}[orcid=0000-0003-1033-1340]

\author[\getrefnumber{unipg},\getrefnumber{infnpg}]{G. {Tosti}}[orcid=0000-0002-0839-4126]

\author[\getrefnumber{iac},\getrefnumber{laguna}]{M. {V\'azquez Acosta}}[orcid=0000-0002-2409-9792]

\refstepcounter{ctr_auts}\label{lodz}    
\author[\getrefnumber{nwu},\getrefnumber{lodz}]{N. \.{Z}ywucka}[orcid=0000-0003-2644-6441]
\address[\getrefnumber{lodz}]{Department of Astrophysics, The University of \L{}\'{o}d\'{z}, ul. Pomorska 149/153, 90-236 \L{}\'{o}d\'{z}, Poland}

\author[\getrefnumber{iasfmi}]{P.~A. Caraveo}[orcid=0000-0003-2478-8018] 

\author[\getrefnumber{oab}]{G. Pareschi}[orcid=0000-0003-3967-403X] 


\cortext[cor1]{Principal Corresponding author}

\begin{abstract}
 The ASTRI (Astrofisica con Specchi a Tecnologia Replicante Italiana) Project led by the Italian National Institute for Astrophysics (INAF) is developing and will deploy at the {\it Observatorio del Teide} a mini-array (ASTRI Mini-Array) composed of nine telescopes similar to the small-size dual-mirror Schwarzschild-Couder telescope (ASTRI-Horn) currently operating on the slopes of Mt. Etna in Sicily.
 The ASTRI Mini-Array will surpass the current Cherenkov telescope array differential sensitivity above a few tera-electronvolt (TeV), extending the energy band well above hundreds of TeV. This will allow us to explore a new window of the electromagnetic spectrum, by convolving the sensitivity performance with excellent angular and energy resolution figures.
In this paper we describe the Core Science that we will address during the first four years of operation, providing examples of the breakthrough results that we will obtain when dealing with current open questions, such as the acceleration of cosmic rays, cosmology and fundamental physics and the new window, for the TeV energy band, of the time-domain astrophysics.

\end{abstract}

\begin{keywords}
ASTRI \sep Imaging Atmospheric Cherenkov Arrays \sep Very high-energy $\gamma$ ray astrophysics \sep Astroparticle \sep © 2022. This manuscript version is made available under the CC-BY-NC-ND 4.0 license \href{https://creativecommons.org/licenses/by-nc-nd/4.0/}{https://creativecommons.org/licenses/by-nc-nd/4.0/}
\end{keywords}

\newpage

\maketitle

\tableofcontents

\newpage

\section{Introduction}\label{sec:0}
The Universe is populated by extreme particle accelerators, capable of conveying more than $10^{20}$\,eV in a single proton. The \gray{} photons they are able to produce could be used as probes to investigate the laws of Nature at the highest energies. The very high-energy (VHE) portion of the electromagnetic spectrum (above $\approx 100$\,GeV) is currently being investigated by means of both ground-based imaging atmospheric Cherenkov telescopes (IACT) and particle sampling arrays (PSA) \citep[see][for reviews]{2009ARA&A..47..523H,2018EPJP..133..324D, 2019JPhCS1263a2003D}.
The Cherenkov instrumentation already in place, like the H.E.S.S. \citep{2006A&A...457..899A}, MAGIC \citep{2012APh....35..435A}, and VERITAS \citep{2002APh....17..221W} telescope arrays and the future Cherenkov Telescope Array Observatory \citep[CTAO, ][]{2019scta.book.....C}, will allow us to resolve questions related to the origin of both Galactic and extra-galactic cosmic-rays (CRs), the extra-ga\-lac\-tic background light, and to definitively uncover the sources of the most energetic cosmic rays.

Within this science framework, the Italian National Institute for Astrophysics (INAF) is leading  the ``Astrofisica con Specchi a Tecnologia Replicante Italiana'' (ASTRI) Flagship Project~\citep{2013ICRC...33.1151P, 2019SPIE11119E..1EG, 2019EPJWC.20901001S} of the Ministry of Education, University and Research.
Primarily, INAF has designed and developed an end-to-end prototype of the CTAO small-size telescope in a dual-mirror configuration (SST-2M). This prototype is currently taking data at the INAF ``M.C. Fraca\-sto\-ro'' observing station in Serra La Nave (Mount Etna, Sicily). The ASTRI SST-2M prototype was inaugurated during the CTA Consortium Meeting in September 2014. %
On November 2019, the ASTRI prototype was named {\it ASTRI-Horn}, in honor of Guido Horn d’Arturo an Italian astronomer who first proposed in the past century the technology of tessellated mirrors for astronomy.
Since 2014, the ASTRI prototype achieved several milestones, including the first-light optical qualification by means of observation of the Polaris, using a dedicated optical camera~\citep{2017A&A...608A..86G}, and the first detection of very high-energy \gray{} emission from the Crab Nebula by a Cherenkov telescope in dual-mirror Schwarzschild-Coud\'e (SC) configuration~\citep{2020A&A...634A..22L}. 

A remarkable improvement in terms of scientific return will come from the realization of a mini-array of ASTRI telescopes~\citep{2012AIPC.1505..749V}. The ASTRI Mini-Array will be able to study in great detail relatively bright ($\approx 10^{-12}$\,erg\,cm$^{-2}$s$^{-1}$ at 10\,TeV) sources with an angular resolution of  $\sim 3$\arcmin{} and an energy resolution of $\sim 10$\,\% at an energy of about 10\,TeV.
The combination of the array approach and the single-telescope wide field-of-view will make it possible the detection and reconstruction of very high-energy showers with a core located at a distance up to $\sim 1000$\,m.

In this context, the planned ASTRI Mini-Array of imaging atmospheric Cherenkov telescopes will be installed at the Observatorio del Teide (island of Tenerife, Spain) thanks to an agreement between INAF and the {\it Instituto de Astrofisica de Canarias}. The starting activities will be managed by the {\it Fundaci\'on Galileo Galilei-INAF} (FGG-INAF\footnote{\href{http://www.tng.iac.es/}{http://www.tng.iac.es/}}), a Spanish no-profit institution supported by INAF. The FGG-INAF's main aim is to operate the {\it Telescopio Nazionale Galileo} (TNG, located in the island of La Palma), but also to promote INAF activities in the Canary islands. The ASTRI Mini-Array includes national and international partners. On the Italian side, the ASTRI Collaboration encompasses the universities of Perugia, Padova, Catania, Genova and the Milano Polytechnic together with the INFN sections of Roma Tor Vergata and Perugia, while on the international side, apart from the strategic partnership with IAC, the ASTRI Collaboration includes the University of S\~ao Paulo with FAPESP in Brasil and the North Western University in South Africa. 
The ASTRI Mini-Array will provide a fully functional complement of MAGIC and CTAO North. In particular, the ASTRI Mini-Array is expected to improve the MAGIC (and VERITAS) sensitivity in the North for E > few TeV and, at the same time, to operate for a few years before the full completion of CTAO North. Therefore, the ASTRI Mini-Array will have a vast discovery space in the field of extreme gamma-rays, up to 100\,TeV and beyond on a short time-frame.

This paper is organized as follows. Section~\ref{sec:01} discusses the ASTRI Mini-Array performance as derived from detailed Monte-Carlo simulations; Section~\ref{Sec:multi_wave} compares the ASTRI Mini-Array performance with respect to the current IACTs and PSAs ones; Section~\ref{sec:3} introduces the concept of the Science Pillars and describes the science simulation environment; Sections~\ref{sec:4} and \ref{sec:5} discuss the results we expect on the Core Science; Section~\ref{sec:6} shows the expectations for transients events; Sections~\ref{sec:7.2} and \ref{sec:7.1} discuss how the ASTRI Mini-Array can be exploited in non \gray{} science; Section~\ref{sec:8} puts the ASTRI Mini-Array in a multi-wavelength framework; Section~\ref{legacy} briefly provides what the ASTRI Mini-Array legacy will be and draws some conclusions.

We remark that this is Paper-II of a series of four papers devoted to the comprehensive description of the ASTRI Mini Array project from a technological, managerial and scientific point of view: \citet[Paper-I,][]{2021JHEAp...S..XXXS}, \citet[Paper-III,][]{2021JHEAp...A..XXXD} and \citet[Paper-IV,][]{2021JHEAp...F..XXXS}. In the following, we shall focus on the potential science outcome.

\section{ASTRI Mini-Array Expected Performance}\label{sec:01}
%
%

\subsection{Monte-Carlo Simulations}
%
\label{sec:1}
The expected performance of the ASTRI Mini-Array, as well as the instrument response functions (IRFs) needed for the high-level scientific analyses presented in this work, were obtained from dedicated Monte Carlo (MC) simulations.  

\begin{figure}
\center
\includegraphics[width=0.48\textwidth]{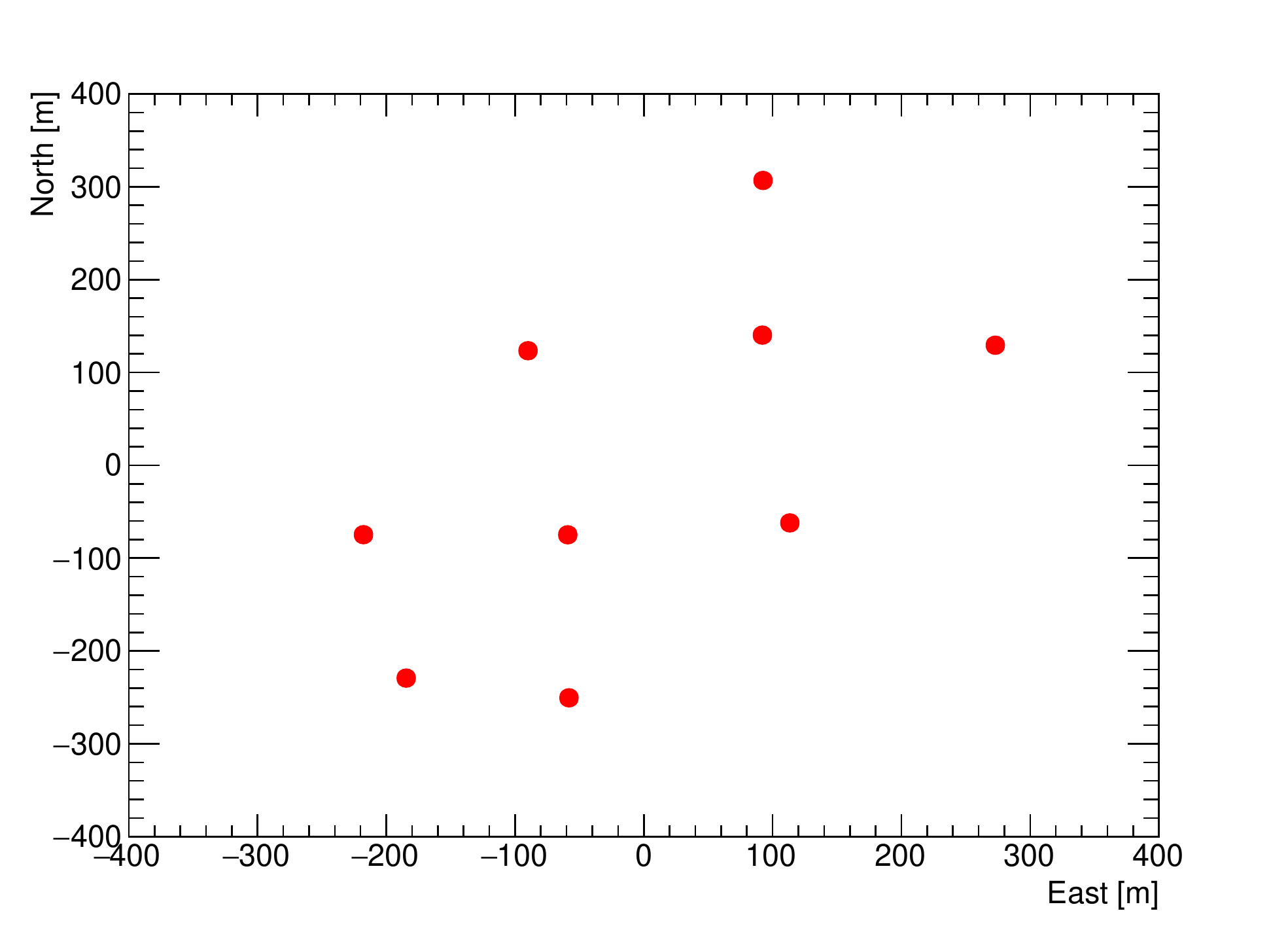}
\caption{Layout of the ASTRI Mini-Array considered in this work. The positions of the 9 ASTRI telescopes (red circles) were chosen according to the most up-to-date (as for June 2020) telescope positions foreseen at the Observatorio del Teide.}
\label{FIG:Chap2_fig1}
\end{figure}

Air showers initiated by \gray{}s, cosmic-ray nuclei, and electrons were simulated using the {\tt CORSIKA} package~\citep{1998cmcc.book.....H}, version 6.99. This publicly available, open-source code is presently used by all the current major IACT arrays and represents a standard tool in the wider astroparticle physics community.
The telescope response was simulated using the {\tt sim\_telarray} package~\citep{2008APh....30..149B}, version 2018-11-07, which propagates photons hitting the primary mirror through the telescope optical system to the camera, and simulates the photon detection, the trigger logic and the readout system.  
The simulation of the peculiar readout system of the ASTRI telescopes has been purposely implemented in the {\tt sim\_telarray} code and cross-checked against a custom code.
The Mini-Array layout chosen for these simulations is shown in Fig.~\ref{FIG:Chap2_fig1} and includes the most up-to-date (as for June 2020) telescope positions foreseen at the Observatorio del Teide site (Lat. 28.30\textdegree{}~North; Lon. 16.51\textdegree{}~West, 2390~m a.s.l.).

Showers produced by the primaries were simulated as coming from a fictitious point-like source at 20\textdegree{} zenith angle and 180\textdegree{} azimuth angle (which corresponds to a direction close \rf{to} the geomagnetic North). The incoming directions of background events (protons and electrons) and diffuse \gray{} events were randomized within a cone with 10\textdegree{}~radius centered on the position of the fictitious point-like source. Such a big diffusion angle is necessary to correctly take into account the contribution of events which can trigger the telescope data-acquisition system even if they are far away from the telescopes full FoV. 
To increase the available number of events, while introducing a negligible statistical bias, each atmospheric shower was used several times (10 times for \gray{}s from point-like sources, and 20 times for all the diffuse primaries), randomizing its impact point on the observational level according to a uniform distribution within a circle with radius equal to 2000~m for \gray{}s from point-like sources and 2400~m for all other primary particles. 
The energies of simulated primary particles were distributed according to a power law of spectral index -1.5 to evenly distribute the CPU time over the entire energy range, between 0.1~TeV and 330~TeV for \gray{}s and electron primaries, and between 0.1~TeV and 600~TeV for protons. The reason for the higher maximum energy for the simulation of proton primaries is that a certain fraction of the primary energy goes into the hadronic component of the atmospheric showers and does not contribute to the emission of Cherenkov light. 
Primary spectra were properly reweighed in the analysis step to match measured spectra of cosmic protons, electrons and \gray{}s from the Crab Nebula (see Sec.~\ref{sec:2} and~\ref{sec:3}).

The most relevant simulation parameters, as well as the available number of simulated events for each primary particle, are summarized in Table~\ref{TAB:Chap2_tab1}.

\begin{table*}[width=1.9\linewidth,cols=8,pos=h!]
\centering
\caption{Parameters describing the MC air shower simulations, used in this work to estimate the expected performance of the ASTRI Mini-Array and to generate the Instrument Response Functions.}
\def\arraystretch{1.1}%
\label{TAB:Chap2_tab1}
\footnotesize
\begin{tabular*}{\tblwidth}{@{} CCCCCCCC@{} }
 \toprule
  Particle & spectral & energy & view cone & scatter & azimuth & zenith & number of\\
type & index & range & radius & radius & direction & angle & simulated\\
 & & [TeV] & [deg] & [m] & [deg] & [deg] & showers\\
\midrule
gamma (point-like) & -1.5 & 0.1--330 & 0 & 2000 & 180 & 20 & 10$^{7}$\\
gamma (diffuse) & -1.5 & 0.1--330 & 10 & 2400 & 180 & 20 & 10$^{8}$\\
electron & -1.5 & 0.1--330 & 10 & 2400 & 180 & 20 & 10$^{8}$\\
proton & -1.5 & 0.1--600 & 10 & 2400 & 180 & 20 & 10$^{9}$\\
\bottomrule
\end{tabular*}
\end{table*}

\subsection{Scientific Software}
%
\label{sec:2}

The MC simulations described in Sec.~\ref{sec:1} were analyzed with {\it A-SciSoft}~\citep{2016SPIE.9913E..15L, 2018SPIE10707E..0RL, 2020A&A...634A..22L}, the data reduction and scientific analysis software of the ASTRI Project. The software package is designed to handle both real and MC data from the raw level up to the generation of scientific products. It comprises a set of independent modules, efficiently wrapped in pipelines, that implement every algorithm to perform the complete data reduction and analysis chain. The scientific products are obtained by means of either specifically developed science tools (ASTRI Science Tools, included in the software package) or external ones currently in use in the CTA Consortium, i.e. {\it ctools} \citep{2016A&A...593A...1K} and {\it Gammapy} \citep{2017ICRC...35..766D}. The software has been extensively checked on a MC basis~\citep{2016SPIE.9913E..15L, 2017ICRC...35..804L, 2018SPIE10707E..0RL}, in single-telescope as well as array mode, and applied to real data acquired with the ASTRI-Horn telescope~\citep{2018SPIE10707E..0RL, 2021A&C...XXXY...XL}. In particular, {\it A-SciSoft} was exploited for the data analysis of the Crab Nebula observations performed with the ASTRI-Horn telescope in December 2018, which led to the first detection of the source at TeV energies with a dual-mirror Cherenkov telescope~\citep{2020A&A...634A..22L}.

In the present work, the primary aim of the MC data reduction was the assessment of the ASTRI Mini-Array performance and the generation of the IRFs. The performance of a given array of IACTs is typically provided in terms of the energy and angular resolution, and differential sensitivity of the system as a function of the energy. These quantities are generally provided for both on-axis and off-axis source observations. The IRFs, instead, contain fundamental quantities representing the system performance and are needed to simulate the observations and to perform the high-level scientific analysis of the simulated sources.

In order to obtain the above mentioned products, the following analysis steps were performed. The raw MC data (of all particle species), containing the full information available per camera pixel (integrated signal amplitude in ADC-counts of the Cherenkov light emitted by the showers), were calibrated separately for each telescope. In this step, the pixel signal is extracted and converted into physically meaningful units (photo-electrons, pe), by means of suitable calibration coefficients. The calibrated data of each telescope underwent, then, an image cleaning procedure aimed at removing pixels which most likely did not belong to a given Cherenkov shower image. The default cleaning method implemented in {\it A-SciSoft} is a two-threshold two-pass cleaning~\citep{2020A&A...634A..22L}, with customizable thresholds. The two cleaning levels for the present analysis were set in order to be 3 and 1.5 times the average RMS of the pixel pedestal, respectively. After this cleaning procedure, the resulting images were parameterized. The image parameters are mainly based on the moments up to the third order of the light distribution on the camera~\citep{1985ICRC....3..445H}. Successively, the data coming from the different telescopes were merged and a set of array-wise shower parameters were computed. Among them, the arrival direction of each incoming shower was estimated from the intersection of the major axes of the images from different telescopes. Once the array-wise parameters were computed, a sample of (diffuse) \gray{} and proton data was used as train sample to compute array-wise look-up-tables (LUTs) for gamma/hadron separation and energy reconstruction, by means of the Random Forest method~\citep{2001MachL..45....5B}. In this step, both telescope-wise and array-wise pieces of information are used. Finally, the LUTs were applied to the remaining (independent) sample of MC data (of all particle species) to get the fully array-wise reconstructed data. At this level of the analysis, the parameters for the arrival direction estimation, energy reconstruction, and gamma/hadron separation are available for each event. The fully array-wise reconstructed data were then used to compute the performance quantities and to generate the IRFs. 

The computation of the performance quantities was carried out by means of a dedicated routine included in the {\it A-SciSoft} software package~\citep{2020A&A...634A..22L}. The background and \gray{} events were reweighed in order to match the experimental fluxes of the proton background (as measured by the BESS Collaboration~\citealp{2000ApJ...545.1135S}), electron background (as measured by the Fermi-LAT~\citealp{2010PhRvD..82i2004A} and H.E.S.S.~\citealp{2008PhRvL.101z1104A} telescopes), and Crab Nebula (as measured by the HEGRA Collaboration~\citealp{2004ApJ...614..897A}). This reweighing procedure is commonly adopted in other IACT MC analyses~\citep{2019APh...111...35A, 2013APh....43..171B} and allows us to derive the performance quantities under the same assumptions. The sensitivity is computed by optimizing, in each considered energy bin and off-axis bin, the cuts on the shower arrival direction and background rejection efficiency. Then, five standard deviations (5$\sigma$, with $\sigma$ defined as in Equation 17 of~\citealp{1983ApJ...272..317L}) are required for a detection in each energy bin, considering an observation time of 50 hours. In addition, the signal excess is required to be larger than 10 and at least five times the expected systematic uncertainty in the background estimation (assumed to be $\sim 1$\%). Finally, a ratio of the off-source to on-source exposure equal to 5 is considered.

The IRFs were generated by means of the default {\it A-SciSoft} executable modules~\citep{2016SPIE.9913E..15L, 2018SPIE10707E..0RL}. In this analysis step, only cuts in the gamma/hadron separation parameter are applied to the MC data. The cuts are dependent on the energy and off-axis angle and are chosen so as to optimize the sensitivity in each estimated-energy bin (21 logarithmic bins between 10$^{-1.9}$$\simeq$~0.01\,TeV and 10$^{2.3}$$\simeq$~200\,TeV) and in each off-axis bin (5 linear bins between 0\textdegree{} and 5\textdegree{}). The definition of the energy bins is compliant with the usual prescription adopted, e.g., in the CTA Consortium~\citep{2019APh...111...35A}. However, it should be noted that the response functions of the ASTRI Mini-Array are meaningful above $\sim$0.3~TeV, because of the energy threshold of the system (on the order of 1\,TeV). The final IRFs include effective collection area, angular resolution, energy dispersion, and residual background rate\footnote{For the computation of the residual background rate, the background events were reweighed in order to match the experimental fluxes as measured by the BESS Collaboration \citep{2000ApJ...545.1135S} for protons and by the Fermi-LAT \citep{2010PhRvD..82i2004A} and H.E.S.S. \citep{2008PhRvL.101z1104A} Collaborations for electrons, respectively.}, as a function of the energy and off-axis angle.

\subsection{Performance}\label{sec:Performance}

\subsubsection{On-axis performance}
\label{sec:3.1}

\begin{figure}
\center
\includegraphics[width=0.5\textwidth]{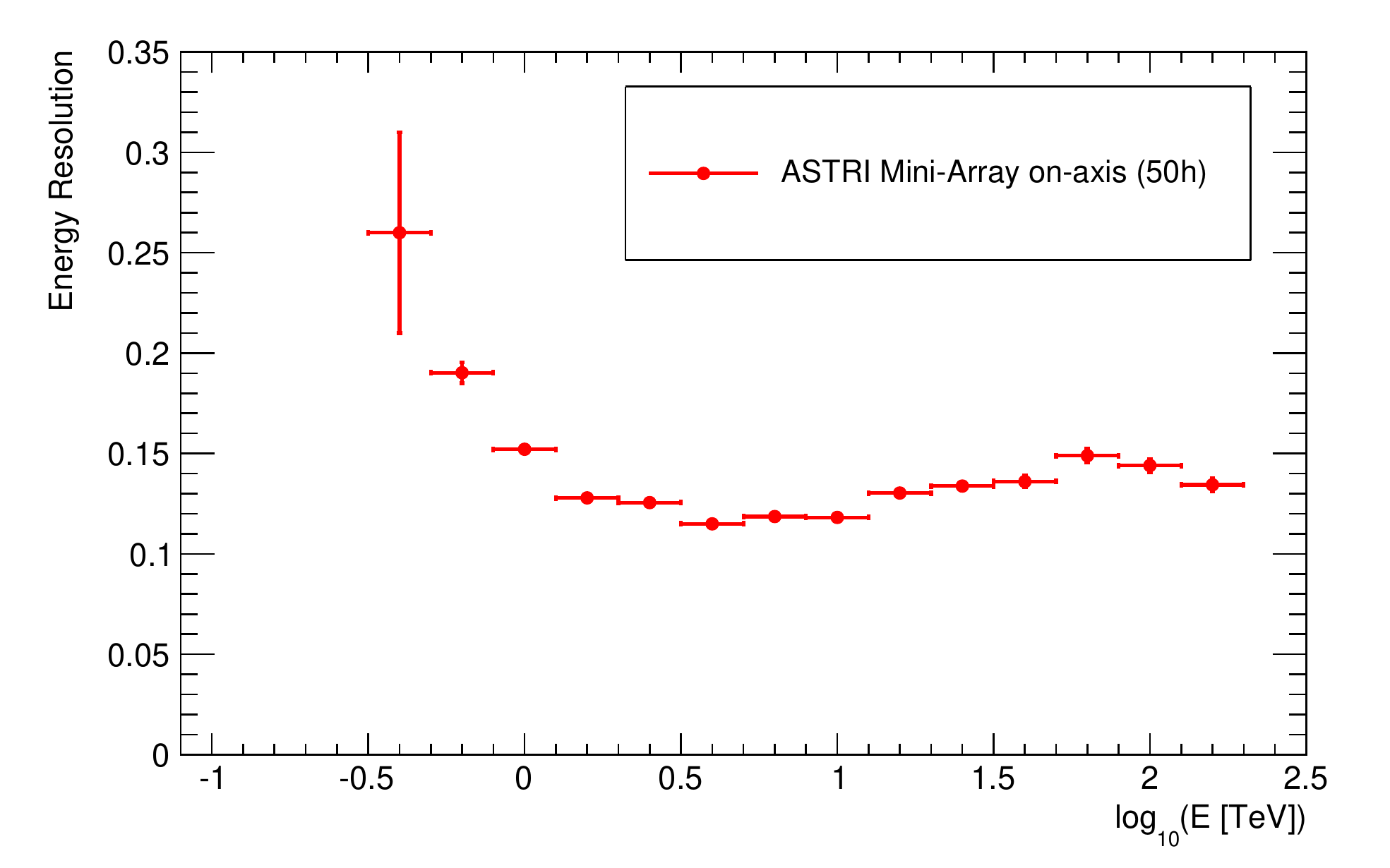}
\caption{On-axis energy resolution of the ASTRI Mini-Array as a function of the energy between $\simeq$~0.3~TeV and $\simeq$~200~TeV.}
\label{FIG:Chap2_fig2}
\end{figure}

\begin{figure}
\center
\includegraphics[width=0.5\textwidth]{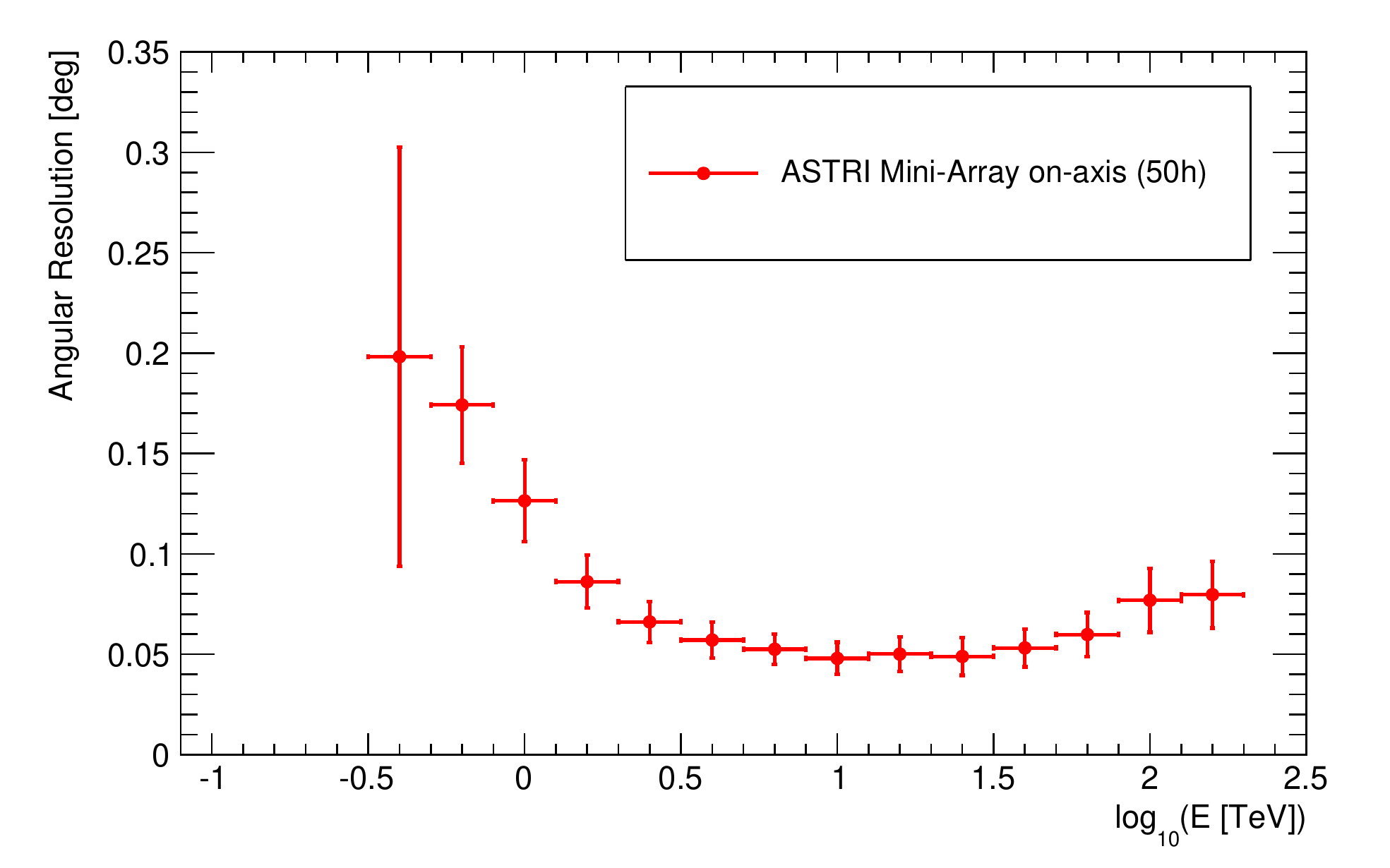}
\caption{On-axis angular resolution of the ASTRI Mini-Array as a function of the energy between $\simeq$~0.3~TeV and $\simeq$~200~TeV.}
\label{FIG:Chap2_fig3}
\end{figure}

\begin{figure}
\center
\includegraphics[width=0.5\textwidth]{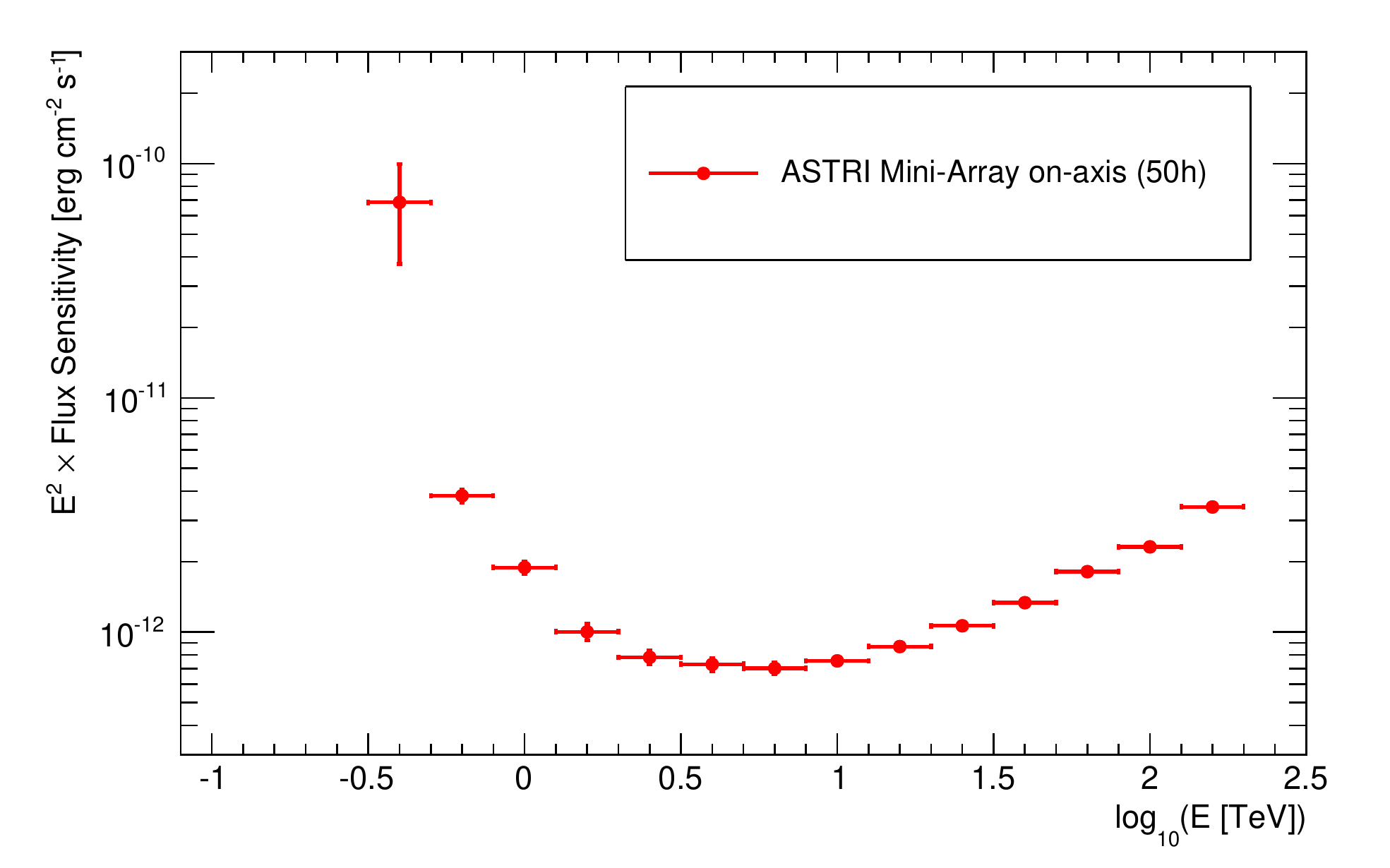}
\caption{On-axis differential sensitivity (multiplied by energy squared) of the ASTRI Mini-Array for 50~hours of observations as a function of the energy between $\simeq$~0.3~TeV and $\simeq$~200~TeV.}
\label{FIG:Chap2_fig4}
\end{figure}

\begin{figure}
\center
\includegraphics[width=0.5\textwidth]{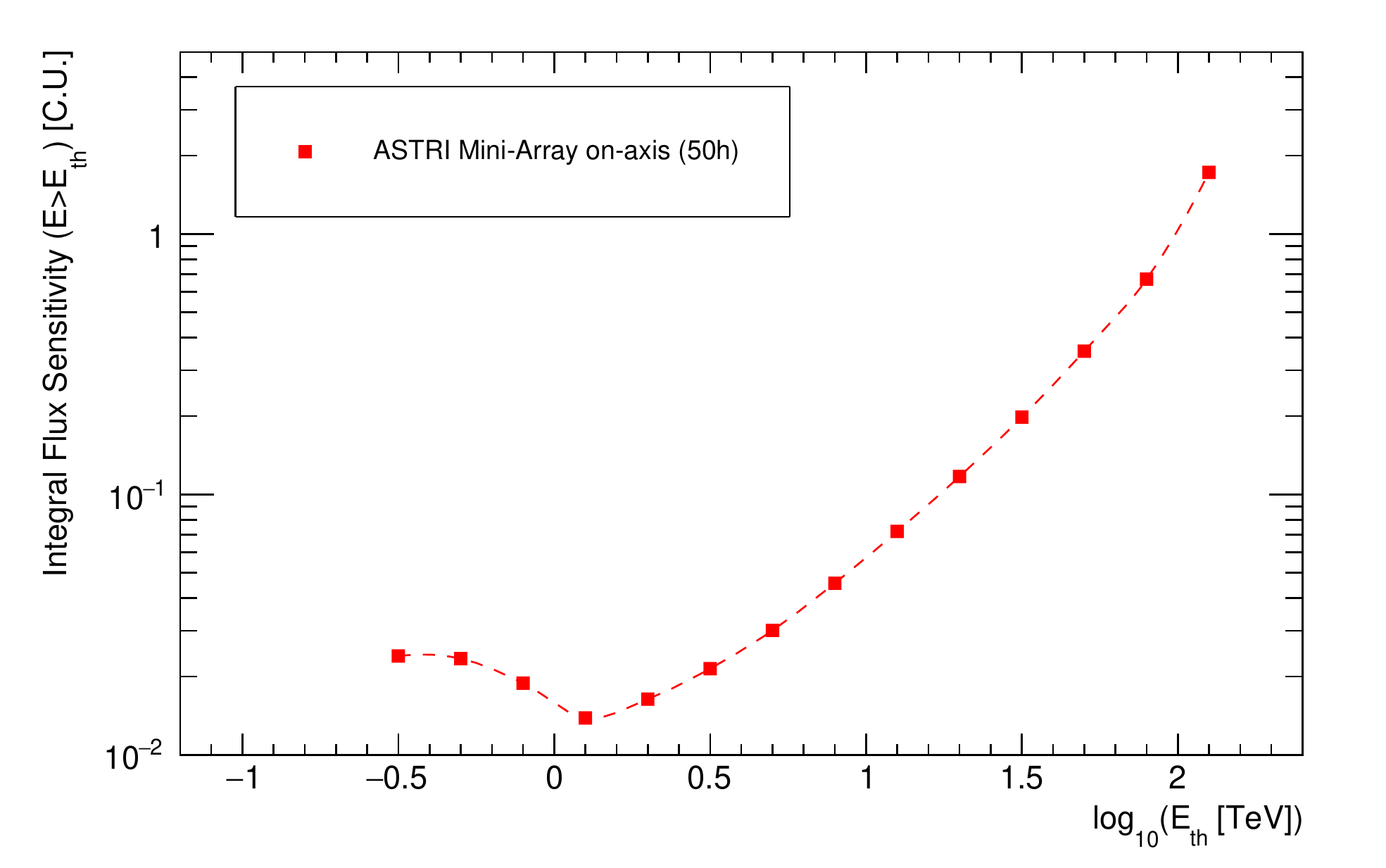}
\caption{On-axis integral sensitivity of the ASTRI Mini-Array for sources with a Crab Nebula-like spectrum for 50~hours of observations as a function of the energy threshold in the range between $\simeq$~0.3~TeV and $\simeq$~150~TeV.}
\label{FIG:Chap2_fig5}
\end{figure}

In order to obtain the on-axis performance, the MC point-like \gray{} and diffuse background samples (see Table~\ref{TAB:Chap2_tab1}) were used. In Fig.~\ref{FIG:Chap2_fig2},~\ref{FIG:Chap2_fig3}, and~\ref{FIG:Chap2_fig4}, the main on-axis performance quantities are displayed, in the energy range between $10^{-0.5}$$\simeq$~0.3\,TeV and 10$^{2.3}$$\simeq$~200\,TeV, considering five logarithmic energy bins per decade. Above a few TeV, the energy resolution is of the order of 10-15\%, while the angular resolution is better than $\sim$4~arcmin, reaching a minimum value of 3~arcmin (0.05\textdegree{}) at $\sim$10~TeV. The differential sensitivity in 50~hours of observations surpasses the ones achieved by the present IACTs (H.E.S.S., MAGIC, and VERITAS) for energies above a few TeV~\citep{2019APh...111...35A} (see also Fig.~\ref{FIG:Chap8_multiplot_IACT_50h}).
In Fig.~\ref{FIG:Chap2_fig5}, the integral sensitivity (expressed in Crab Nebula Units, C.U., as provided in~\citealp{2004ApJ...614..897A}) for sources with a Crab-Nebula-like spectrum above a given energy threshold (5 per decade) for 50~hours of observations is depicted. The best integral sensitivity of the system is on the order of 1.5\% of the Crab Nebula flux above an energy threshold of $\sim$1-2~TeV.
The current estimate of differential sensitivity is based on a conservative data analysis approach. On one hand, the adopted methods for event reconstruction are standard; on the other hand, a number of conservative selection cuts have been applied to the data. More efficient event reconstruction, e.g. by exploiting the temporal information of acquired events and making use of more sophisticated gamma/hadron separation methods, are under investigation and will be implemented in the analysis chain. Future Monte-Carlo productions, based on actual data from the first batch of three telescopes deployed during the commissioning and science verification phase, will allow us to validate new analysis methods, fine-tune selection cuts and possibly obtain improved performance figures.


\subsubsection{Off-axis performance}
\label{sec:3.2}

\begin{figure}
\center
\includegraphics[width=0.5\textwidth]{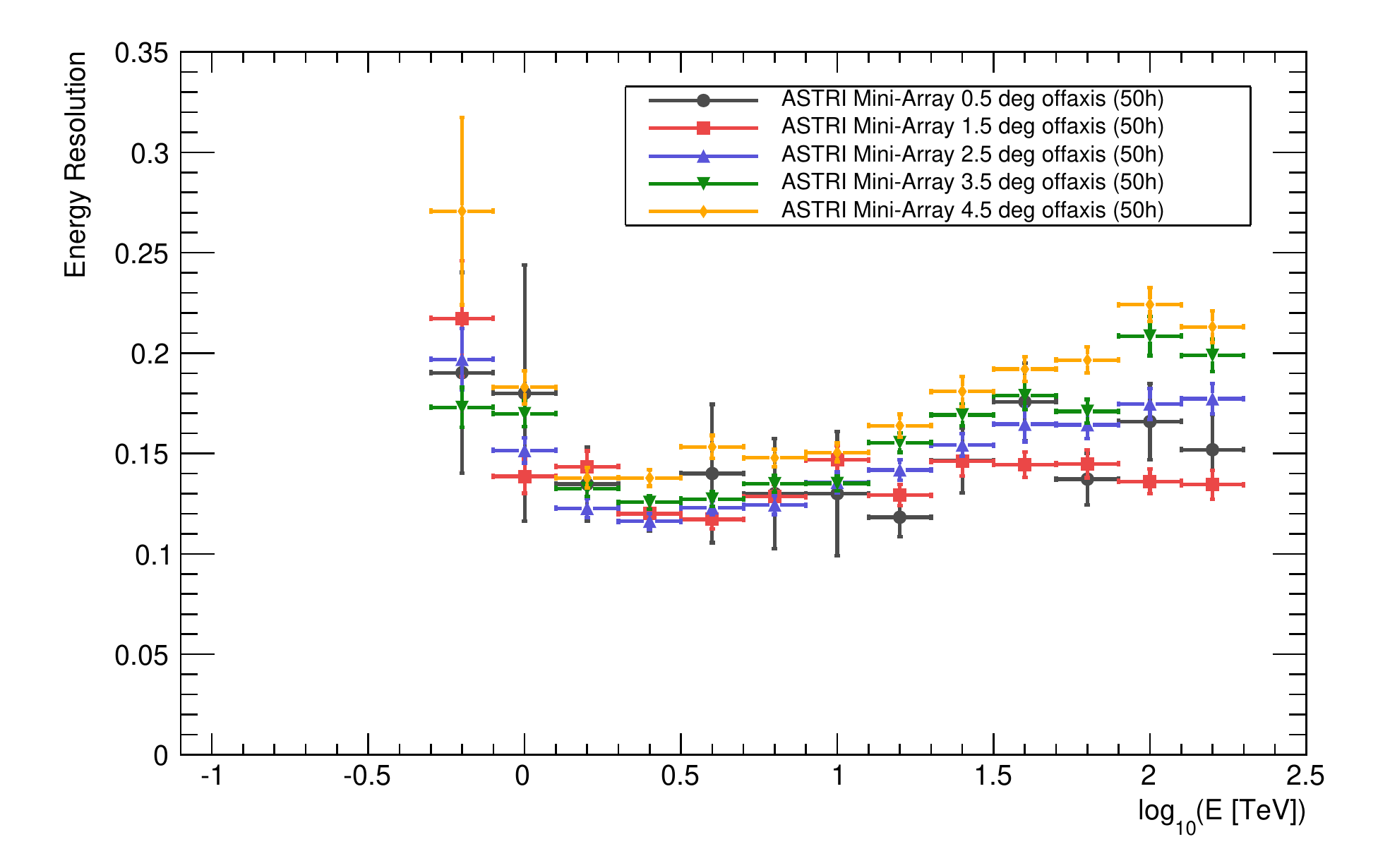}
\includegraphics[width=0.5\textwidth]{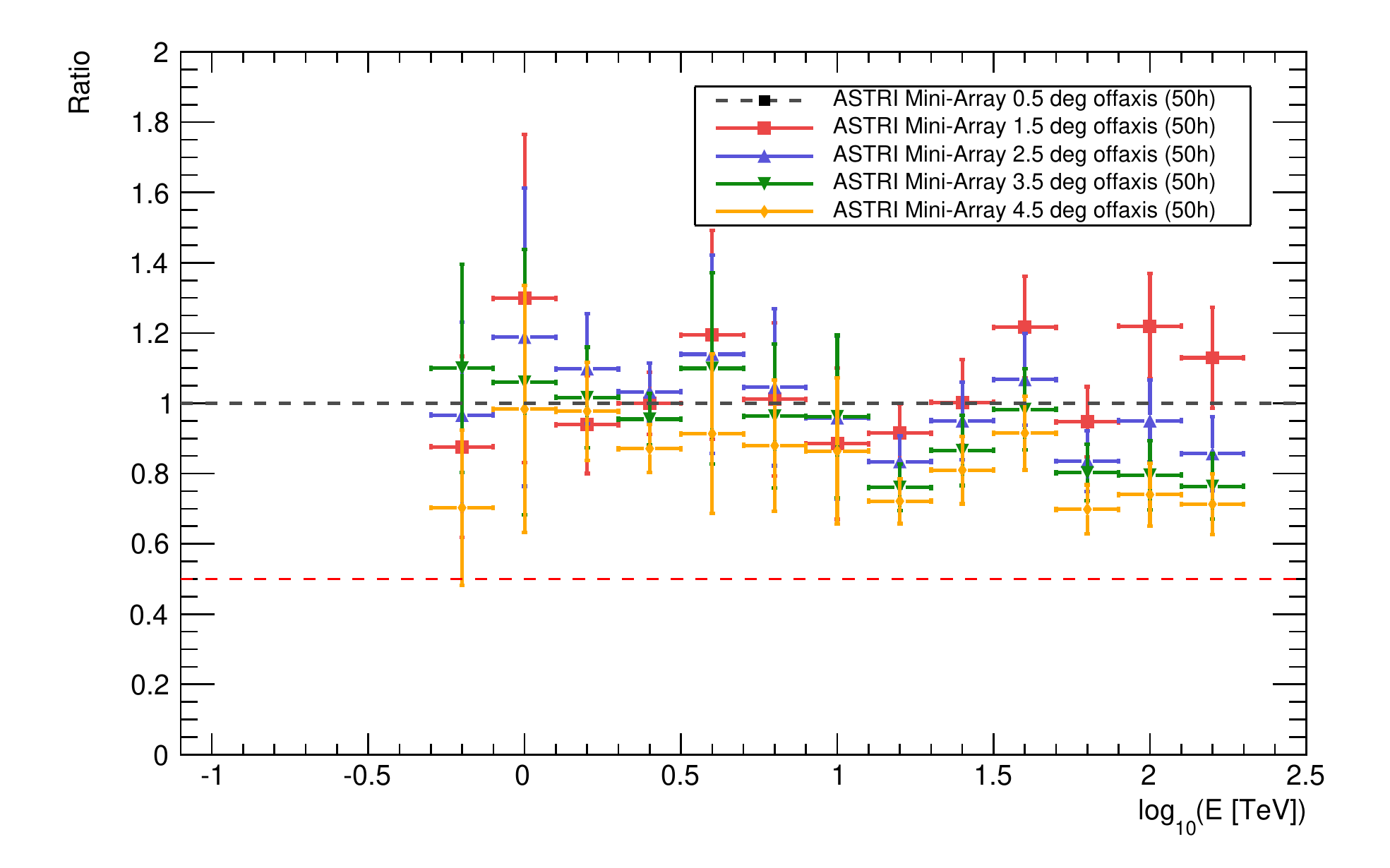}
\caption{{\it Top}: Off-axis energy resolution of the ASTRI Mini-Array as a function of the energy between \rf{ 10$^{-0.3}$$\simeq$~0.5~TeV} 
 and 10$^{2.3}$$\simeq$~200~TeV for 5 source off-axis bins between 0\textdegree{} and 5\textdegree{}. {\it Bottom}: Energy resolution ratios with respect to the energy resolution achieved in the first considered off-axis bin (from 0\textdegree{} to 1\textdegree{}). The ratio is calculated so that higher values correspond to better performance. The dashed, thin red line represents a performance drop of a factor 2.}
\label{FIG:Chap2_fig6}
\end{figure}

\begin{figure}
\center
\includegraphics[width=0.5\textwidth]{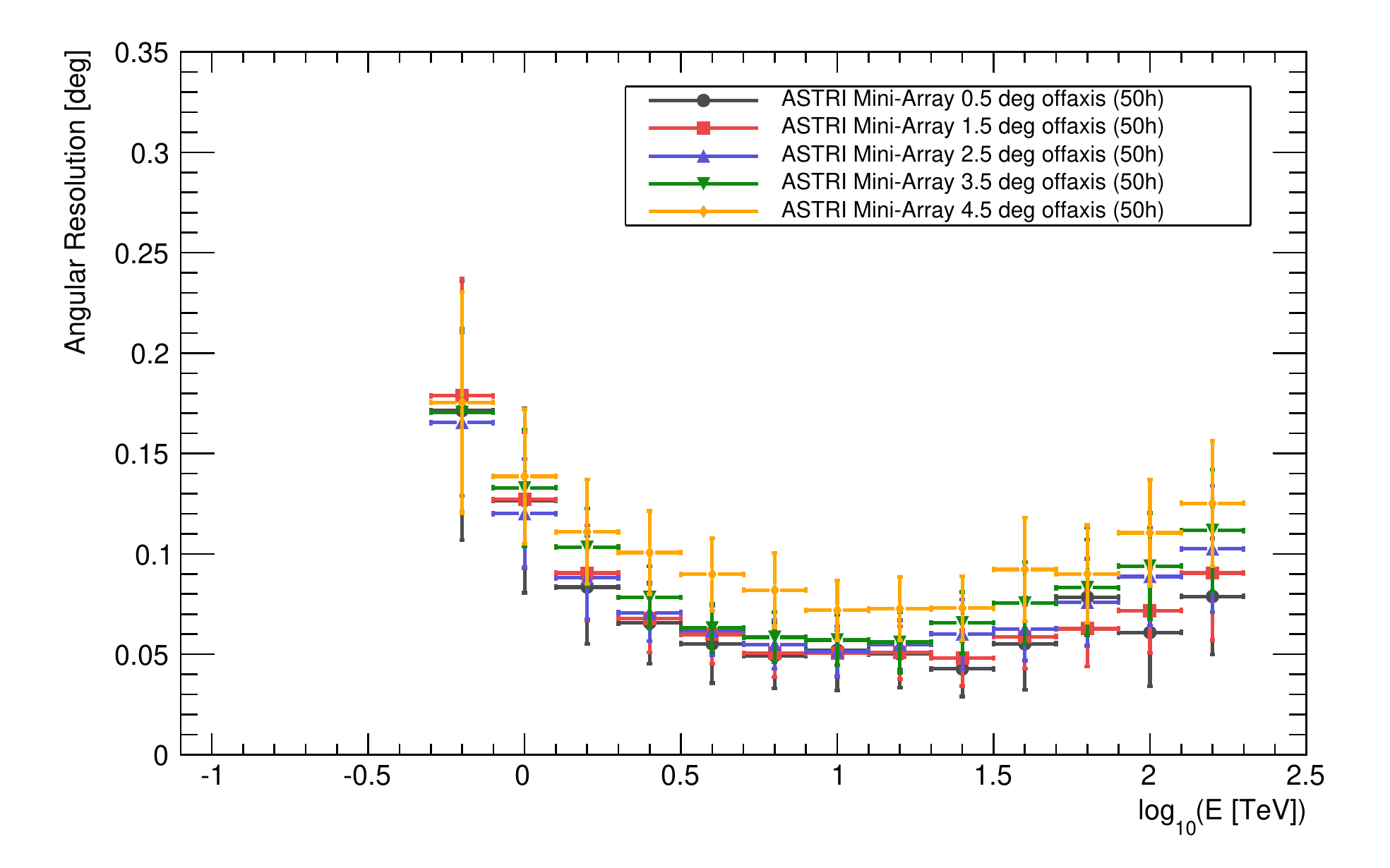}
\includegraphics[width=0.5\textwidth]{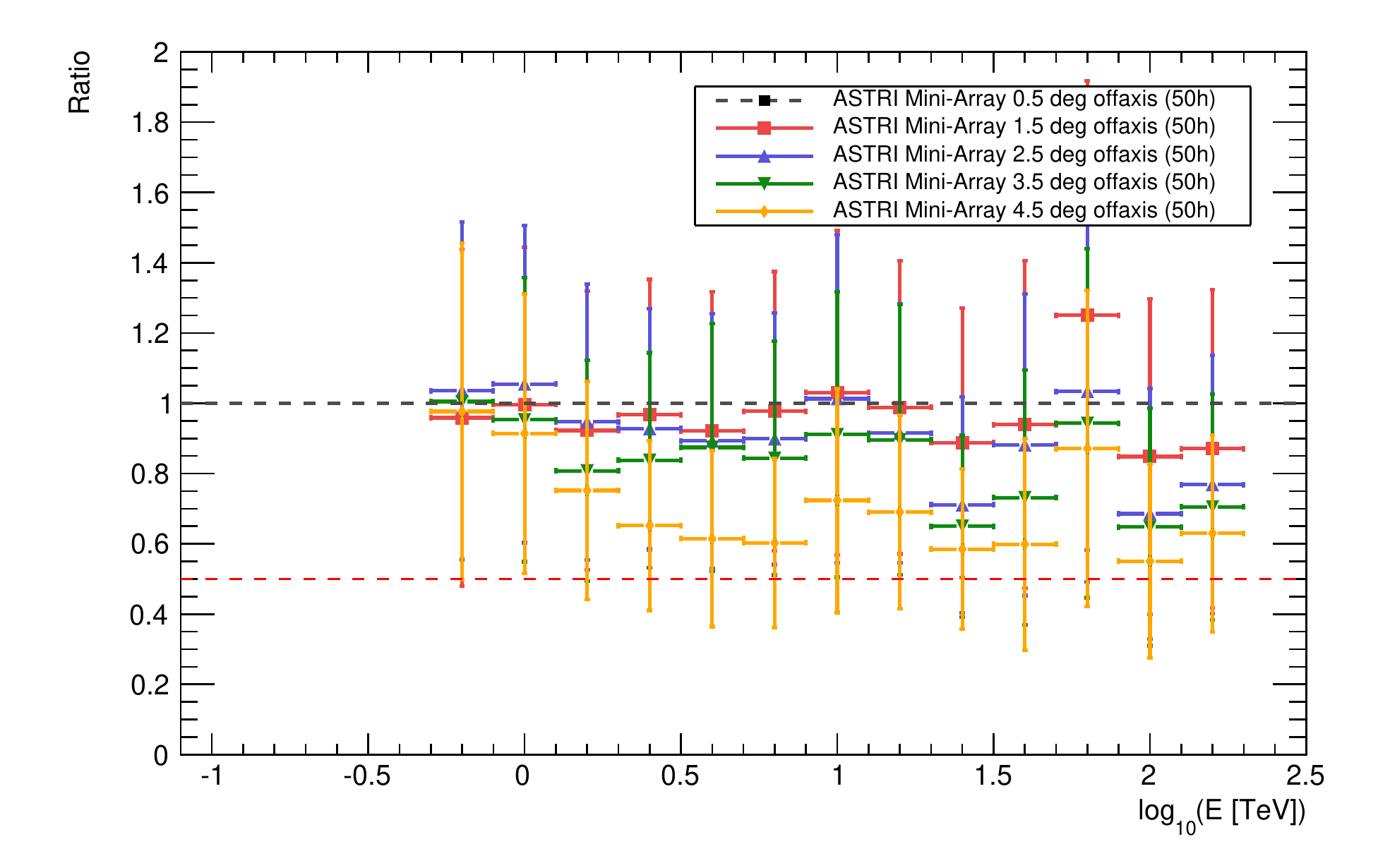}
\caption{{\it Top}: Off-axis angular resolution of the ASTRI Mini-Array as a function of the energy between \rf{ 10$^{-0.3}$$\simeq$~0.5~TeV}  
 and 10$^{2.3}$$\simeq$~200~TeV for 5 source off-axis bins between 0\textdegree{} and 5\textdegree{}. {\it Bottom}: Angular resolution ratios with respect to the angular resolution achieved in the first considered off-axis bin (from 0\textdegree{} to 1\textdegree{}). The ratio is calculated so that higher values correspond to better performance. The dashed, thin red line represents a performance drop of a factor 2.}
\label{FIG:Chap2_fig7}
\end{figure}

\begin{figure}
\center
\includegraphics[width=0.5\textwidth]{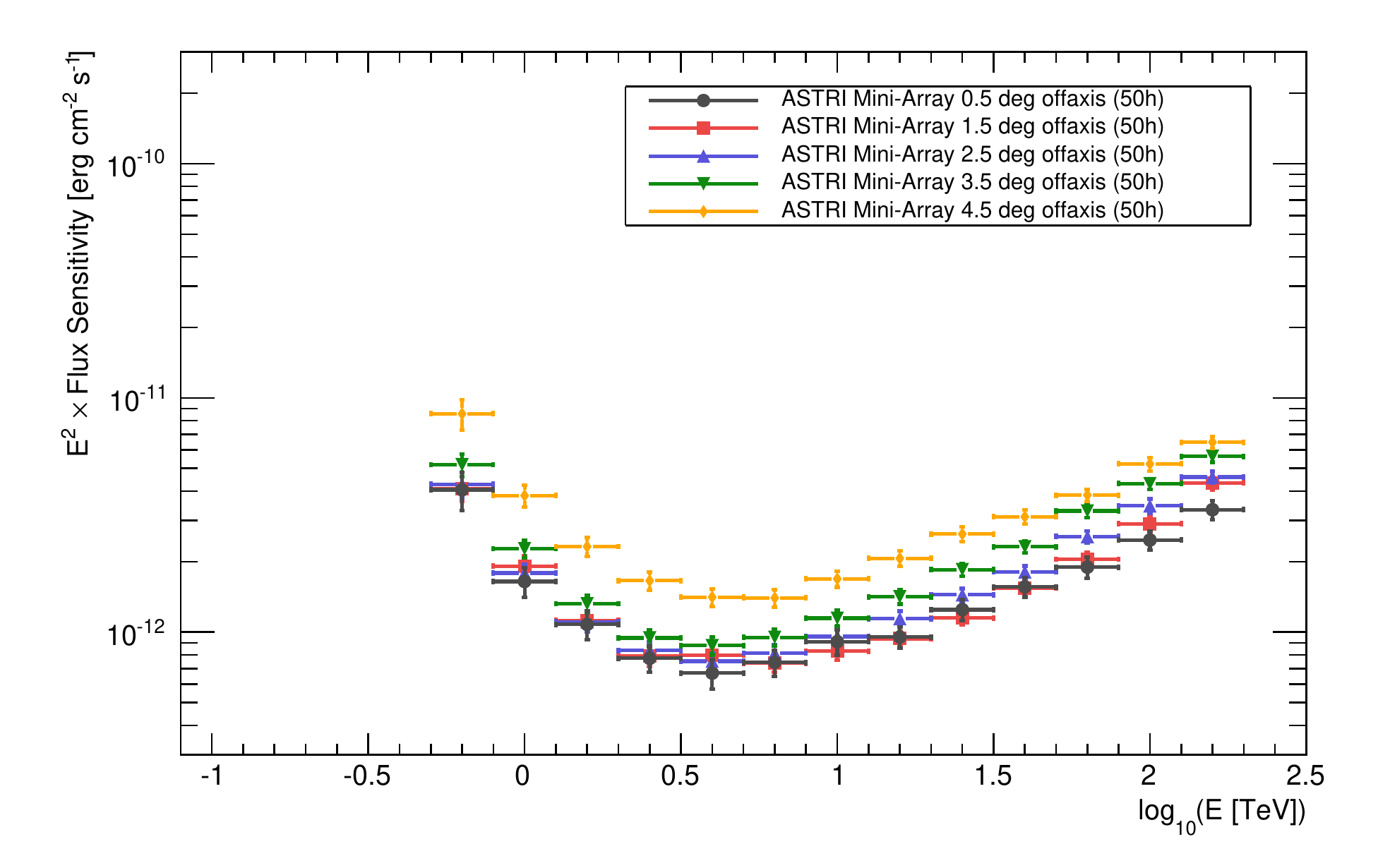}
\includegraphics[width=0.5\textwidth]{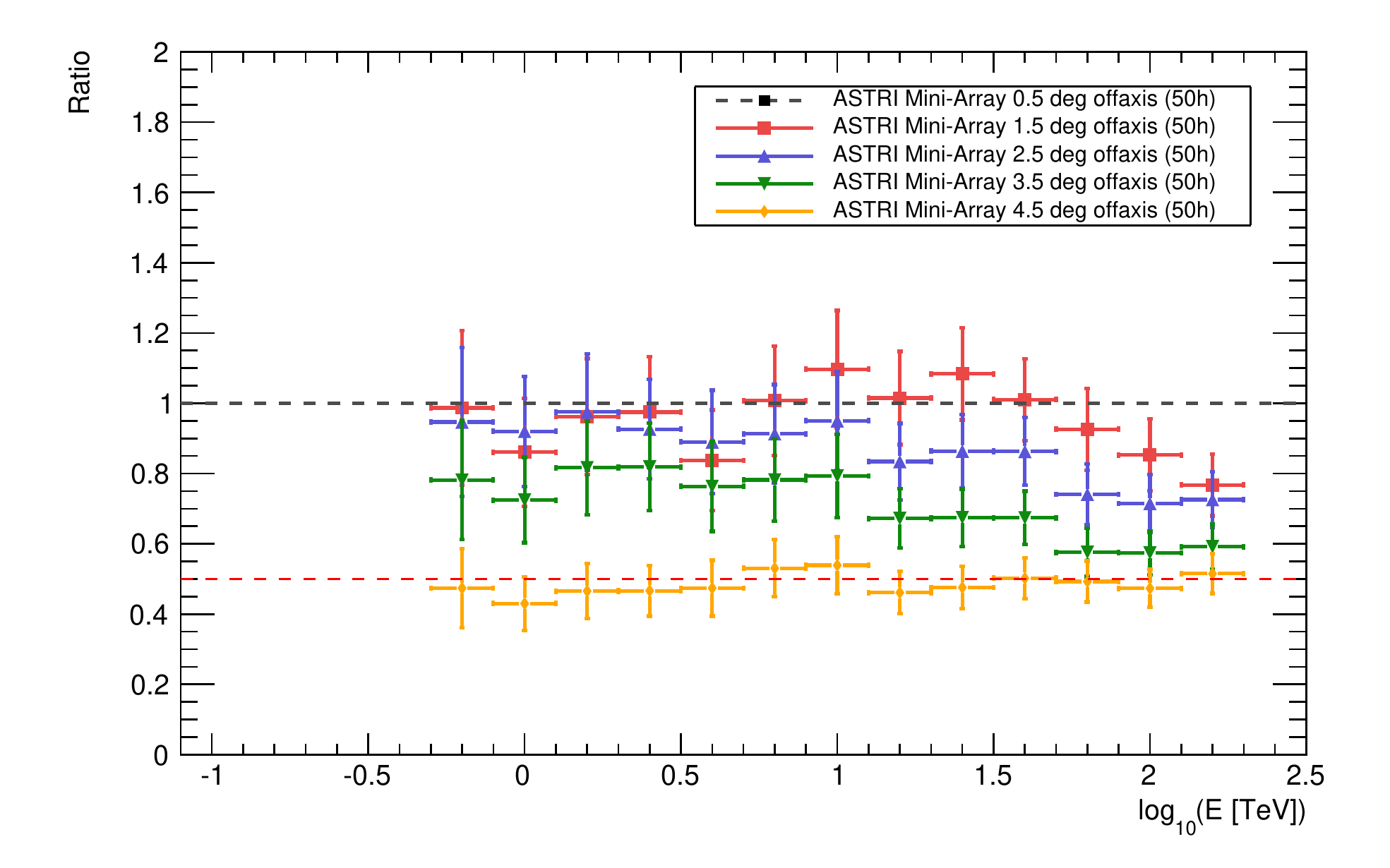}
\caption{{\it Top}: Off-axis differential sensitivity (multiplied by energy squared) of the ASTRI Mini-Array as a function of the energy between \rf{ 10$^{-0.3}$$\simeq$~0.5~TeV}  
and 10$^{2.3}$$\simeq$~200~TeV for 5 source off-axis bins between 0\textdegree{} and 5\textdegree{}. 50 hours of observations are considered. {\it Bottom}: Differential sensitivity ratios with respect to the differential sensitivity achieved in the first considered off-axis bin (from 0\textdegree{} to 1\textdegree{}). The ratio is calculated so that higher values correspond to better performance. The dashed, thin red line represents a performance drop of a factor 2.}
\label{FIG:Chap2_fig8}
\end{figure}

The off-axis performance quantities were derived considering the MC diffuse \gray{} sample, along with the diffuse background samples (see Table~\ref{TAB:Chap2_tab1}). All samples were divided in 5 source off-axis bins between 0\textdegree{} and 5\textdegree{}. In Fig.~\ref{FIG:Chap2_fig6},~\ref{FIG:Chap2_fig7}, and~\ref{FIG:Chap2_fig8}, the main off-axis performance quantities are displayed for each off-axis bins (top plots), in the energy range between \rf{10$^{-0.3}$$\simeq$~0.5~TeV}  
and 10$^{2.3}$$\simeq$~200~TeV, considering five logarithmic energy bins per decade. In the same figures (bottom plots), the ratio between the off-axis performance quantities with respect \rf{to} the one achieved in the first considered off-axis bin are also shown. 
The off-axis performance remains in the entire energy range within a factor of $\sim$1\,($\sim$2) of the on-axis performance up to $\sim$3\textdegree{}($\sim$5\textdegree{}),
allowing the system to preserve a performance close to the best one over a wide field of view of several squared degrees. This feature represents a key factor of the system, particularly important for observations of extended sources, large sky-surveys, and possible serendipitous discoveries.


\section{High- and Very High-Energy Observatories}\label{Sec:multi_wave}
%
%
Table~\ref{TAB:tab_Canarias_comparison_IACTs} shows the performance of the current main imaging atmospheric Cherenkov telescope arrays (H.E.S.S, MA\-GIC and VERITAS) compared with the ASTRI Mini-Array ones.

\begin{table*}[width=2.\linewidth,cols=5,pos=htp!]
\centering
\caption{Summary of the performance of the current main imaging atmospheric Cherenkov telescope arrays compared with those of the ASTRI Mini-Array.
{\bf References.} ASTRI Mini-array: this work. MAGIC:~\cite{2016APh....72...76A}. VERITAS: \cite{2006APh....25..391H} and \href{https://veritas.sao.arizona.edu}{https://veritas.sao.arizona.edu}. H.E.S.S.: \cite{2006A&A...457..899A}. {\bf Notes.} $^{(a)}$: considering the contribution of H.E.S.S.-II telescope unit \citep{2017EPJWC.13603001D}.
}
\def\arraystretch{1.5}%
\label{TAB:tab_Canarias_comparison_IACTs}
 \begin{tabular*}{\tblwidth}{@{}CCCcC@{} }
\toprule
                & {\bf ASTRI Mini-Array} 
                & {\bf MAGIC} 
                & {\bf VERITAS} 
                & {\bf H.E.S.S.} \\
\midrule
{\bf Location}  & 28\textdegree{} 18\arcmin{} 04\arcsec{} N  
                & 28\textdegree{} 45\arcmin{} 22\arcsec{} N
                & 31\textdegree{} 40\arcmin{} 30\arcsec{} N 
                & 23\textdegree{} 16\arcmin{} 18\arcsec{} S \\
                & 16\textdegree{} 30\arcmin{} 38\arcsec{} W  
                & 17\textdegree{} 53\arcmin{} 30\arcsec{} W 
                & 110\textdegree{} 57\arcmin{} 7.8\arcsec{} W
                & 16\textdegree{} 30\arcmin{} 00\arcsec{} E \\
{\bf Altitude} [m]  & 2,390 
                    & 2,396
                    & 1,268
                    & 1,800 \\
{\bf FoV}       & $\sim 10$\textdegree{} 
                & $\sim 3.5$\textdegree{}
                & $\sim 3.5$\textdegree{}
                & $\sim 5$\textdegree{} \\
{\bf Angular Res.}  &  $0.05$\textdegree{} (30\,TeV)
                          & 0.07\textdegree{} (1\,TeV)
                          & 0.07\textdegree{} (1\,TeV)
                          & 0.06\textdegree{} (1\,TeV) \\
{\bf Energy Res.}   & 12\% (10\,TeV)
                          & 16\% (1\,TeV)
                          & 17\% (1\,TeV)
                          & 15\% (1\,TeV) \\
{\bf Energy Range}        & (0.3-200)\,TeV 
                          & (0.05-20)\,TeV
                          & (0.08-30)\,TeV
                          & (0.02-30)\,TeV$^{(a)}$ \\
\bottomrule
\end{tabular*}
\end{table*}

Table~\ref{TAB:tab_Canarias_comparison_PSAs} shows the performance of the current main particle sampling arrays (HAWC, LHAASO and Tibet AS$\gamma$) compared with the ASTRI Mini-Array ones.
\begin{table*}[width=2.\linewidth,cols=5,pos=htp!]
\centering
\caption{Summary of the performance of the current main particle sampling arrays compared with those of the ASTRI Mini-Array.
{\bf References.} ASTRI Mini-array: this work. HAWC: \cite{2017ApJ...843...39A, 2017ApJ...843...40A}. LHAASO: \cite{2010ChPhC..34..249C}. Tibet AS$\gamma$: \cite{2017ExA....44....1K,2019PhRvL.123e1101A}
{\bf Notes.} $^{(a)}$: (0.15--1)\textdegree{} as a function of the event size. $^{(b)}$: angular resolution is (0.70--0.94)\textdegree{} at 10\,TeV; (0.24--0.32)\textdegree{} at 100\,TeV; 0.15\textdegree{} at 1000\,TeV. Energy resolution is (30--45)\% at 10\,TeV; (13--36)\% at 100\,TeV; (8--20)\% at 1000\,TeV;  \cite{2021ChPhC..45b5002A}. $^{(c)}$: angular resolution is $\sim0.5$\textdegree at 10\,TeV and $\sim0.2$\textdegree at 10\,TeV at 50\% containment radius \citep{2019PhRvL.123e1101A}. Energy resolution is $\sim40$\% at 10\,TeV and $\sim20$\% at 100\,TeV \citep{2017ExA....44....1K}. \rf{The different values of the LHAASO angular and energy resolution performance at a given energy have been computed at different Zenith angle, $0<\theta<20$, $20<\theta<35$, and $35<\theta<50$ degrees, respectively. At lower Zenith angles, the performance is better.}
}
\def\arraystretch{1.5}
\label{TAB:tab_Canarias_comparison_PSAs}
 \begin{tabular*}{\tblwidth}{@{}CCCCC@{} }
\toprule
                & {\bf ASTRI Mini-Array} 
                & {\bf HAWC}
                & {\bf LHAASO}
                & {\bf Tibet AS$\gamma$}\\
\midrule
{\bf Location}  & 28\textdegree{} 18\arcmin{} 04\arcsec{} N  
                & 18\textdegree{} 59\arcmin{} 41\arcsec{} N
                & 29\textdegree{} 21\arcmin{} 31\arcsec{} N
                & 30\textdegree{} 05\arcmin{} 00\arcsec{} N \\

                & 16\textdegree{} 30\arcmin{} 38\arcsec{} W  
                & 97\textdegree{} 18\arcmin{} 27\arcsec{} W 
                & 100\textdegree{} 08\arcmin{} 15\arcsec{} E
                & 90\textdegree{} 33\arcmin{} 00\arcsec{} E\\
                
{\bf Altitude} [m]  & 2,390 
                    & 4,100
                    & 4,410
                    & 4,300\\
{\bf FoV}       & $\sim 10$\textdegree{} 
               & 2\,sr
                & 2\,sr
                & 2\,sr\\
{\bf Angular Res.}  &  $0.05$\textdegree{} (30\,TeV)
                          & 0.15\textdegree{}$^{(a)}$ (10\,TeV)
                          & (0.24--0.32)\textdegree{}$^{(b)}$ (100\,TeV)
                          & 0.2\textdegree{}$^{(c)}$ (100\,TeV) \\
{\bf Energy Res.}   & 12\% (10\,TeV)
                          & 30\% (10\,TeV)
                          & (13--36)\% (100\,TeV)$^{(b)}$
                          & 20\%$^{(c)}$ (100\,TeV)\\
{\bf Energy Range}        & (0.3-200)\,TeV 
                          & (0.1-1000)\,TeV
                          & (0.1-1000)\,TeV
                          & (0.1-1,000)\,TeV \\
\bottomrule
\end{tabular*}
\end{table*}

\begin{figure}
	\centering
	\includegraphics[width=0.4\textwidth, angle=90]{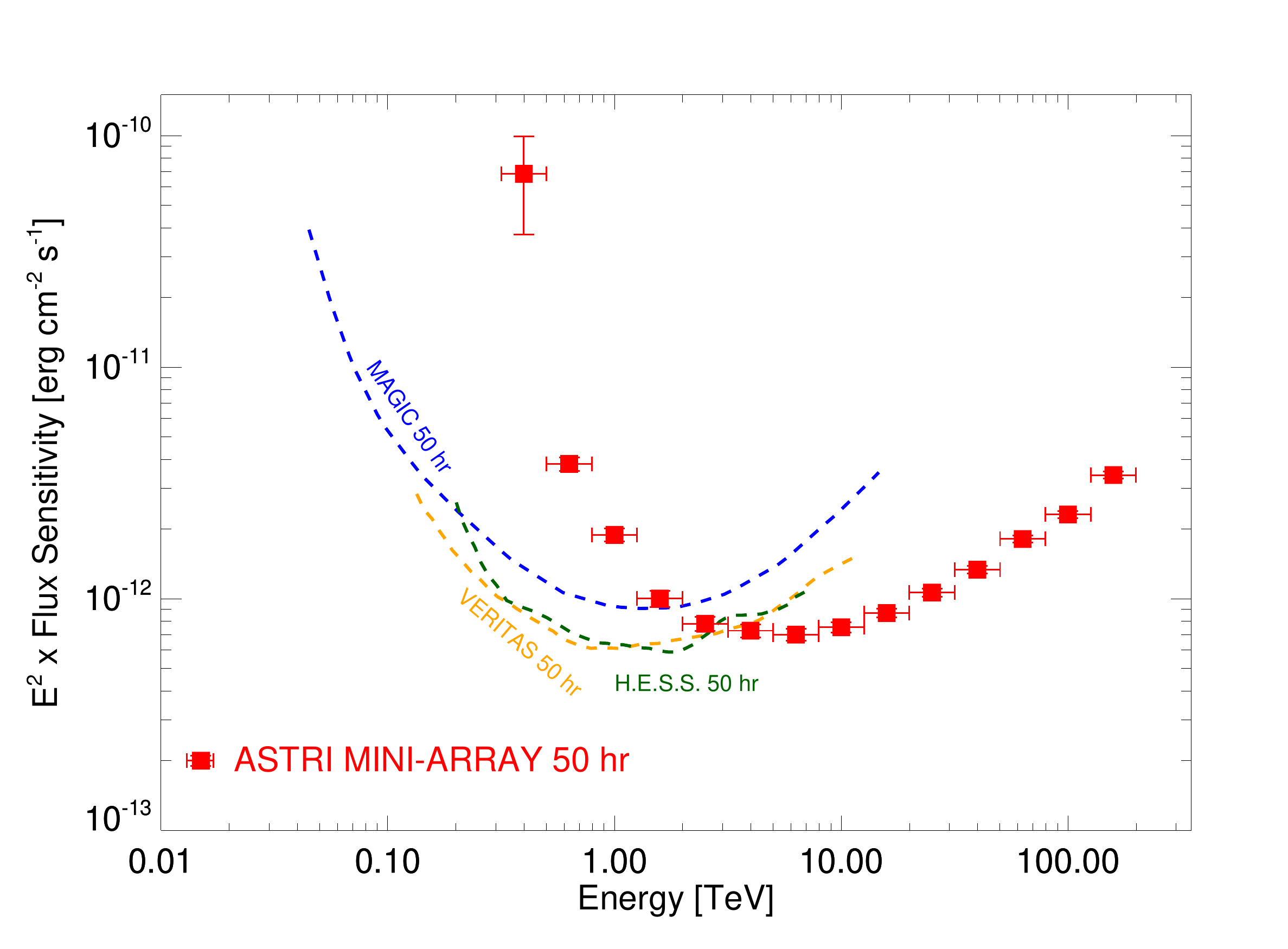}
	\caption{ASTRI Mini-Array differential sensitivity compared with those of current imaging atmospheric Cherenkov telescope arrays. See text for details.}
	\label{FIG:Chap8_multiplot_IACT_50h}
\end{figure}

Figure~\ref{FIG:Chap8_multiplot_IACT_50h} shows the ASTRI Mini-Array differential sensitivity compared with those of current very high-energy imaging atmospheric Cherenkov telescope arrays. The integration time is 50\,hr. The differential sensitivity curves come from \cite{2016APh....72...76A} (MAGIC), the VERITAS official website\footnote{\href{https://veritas.sao.arizona.edu}{https://veritas.sao.arizona.edu}}, and \cite{2015ICRC...34..980H} (sensitivity curve for H.E.S.S.--I, stereo reconstruction).

\begin{figure}
	\centering
	\includegraphics[width=0.4\textwidth, angle=90]{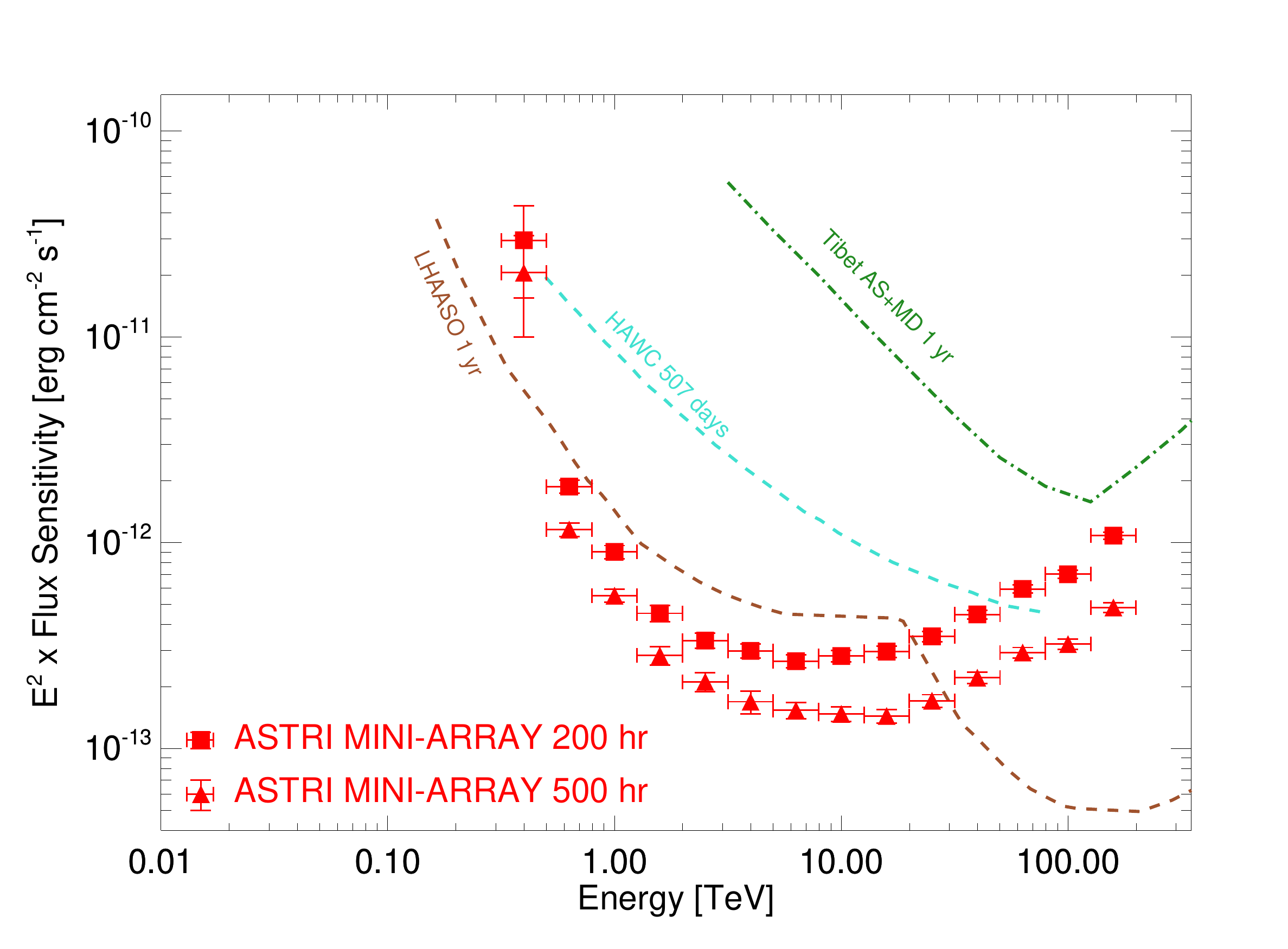}
	\caption{ASTRI Mini-Array differential sensitivity curves for two different typical long-term deep pointings compared with those of current particle sampling arrays. See text for details.}
	\label{FIG:Chap8_multiplot_WCD_200h_500h}
\end{figure}

Figure~\ref{FIG:Chap8_multiplot_WCD_200h_500h} shows the ASTRI Mini-Array differential sensitivity compared with those of current very high-energy PSAs in the northern hemisphere. The integration times are 200\,hr and 500\,hr for the ASTRI Mini-Array and about 1\,yr for PSAs, respectively. The differential sensitivity curves come from \cite{2017ApJ...843..116A} (HAWC), \cite{2016NPPP..279..166D} (LHAASO), \rf{and Takita M. (private communication) based on~\cite{2019PhRvL.123e1101A} (Tibet AS$+$MD).} We note that the 507-day HAWC differential sensitivity curve corresponds to about 3000\,hr of acquisition on a source at declination of 22\textdegree{} within its field of view~\citep{2017ApJ...843..116A}.
Given the very small number of pointings that are planned for the ASTRI Mini-Array, the two different sensitivity curves correspond to a deep observation on a specific sky region at the end of the first year of operations (200\,hr) and to the typical observing time accumulated on a particular target of interest at the completion of the ``Pillar'' observational time-frame (3--4 years, 500\,hr), prior to the ``Observatory'' phase.

\subsection{Beyond the current IACTs}

\begin{figure}
	\centering
	\includegraphics[width=0.49\textwidth]{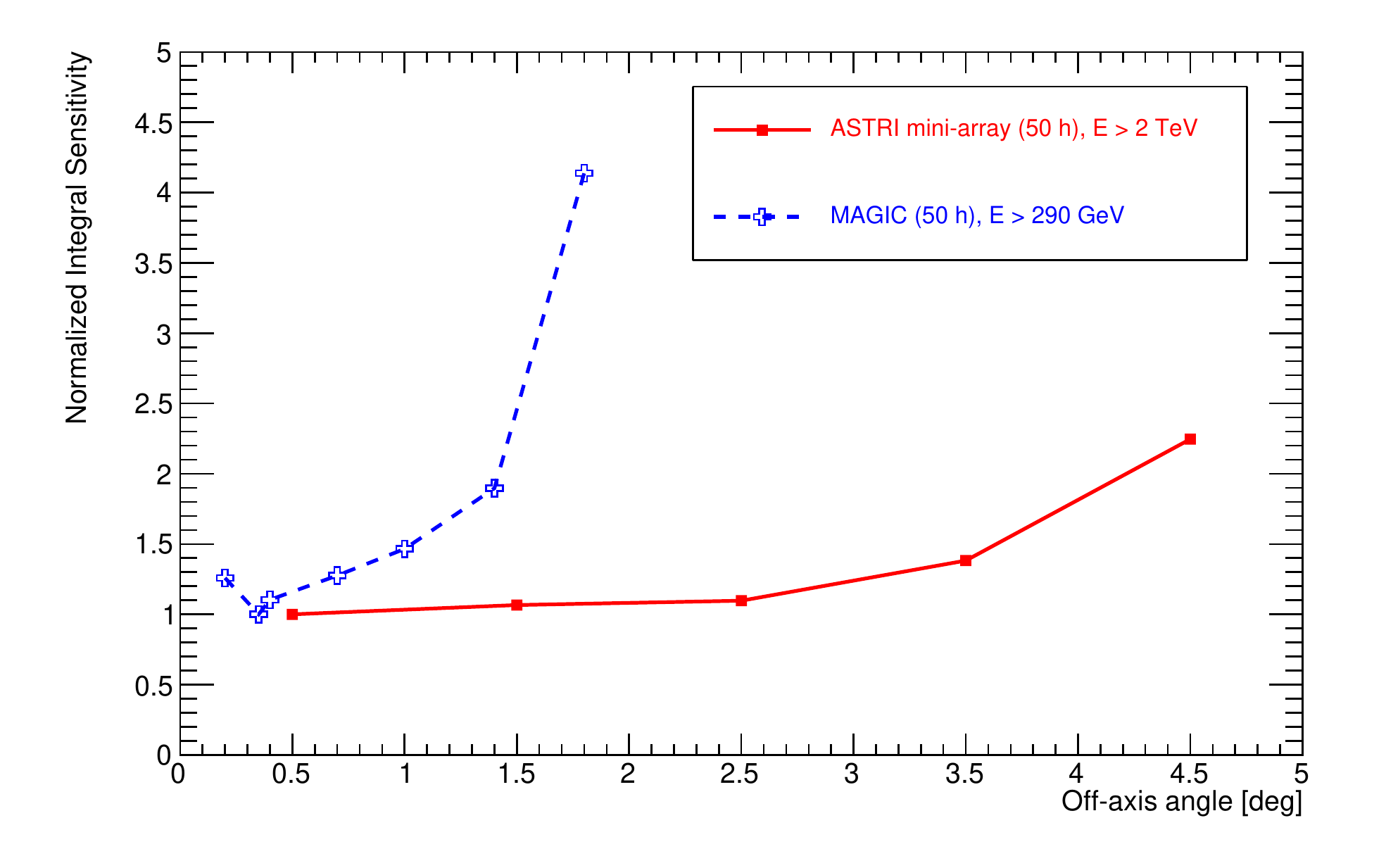}
	\caption{MAGIC and ASTRI Mini-Array normalized integral sensitivity above 290~GeV and 2~TeV, respectively, as a function of the source off-axis angle (see text for details).}
	\label{FIG:Chap8_IntSens_offaxis}
\end{figure}

H.E.S.S., MAGIC and VERITAS allowed the scientific community to access the VHE sky in a systematic fashion. Highlight results include the H.E.S.S. survey of a large fraction of the Galactic plane, detecting both known and still unidentified sources \citep[][]{2018A&A...612A...1H}, the VERITAS discovery of M~82, the first starburst galaxy emitting at VHE \citep[][]{2009Natur.462..770V}, and the flourishing of the transient and multi-messenger era with the detection by MAGIC of the first extra-galactic counterpart of a neutrino event~\citep{2018Sci...361.1378I} and of a gamma-ray burst at TeV energies~\citep{MAGIC2019a}.
The ASTRI Mini-Array will significantly provide a breakthrough step-up both in performance and in science. 
\paragraph{Sensitivity above tens of TeV}--
We extend the differential sensitivity up to several tens of TeV and beyond, an energy range barely accessible to current IACTs. This will allow us to investigate possible spectral features at VHE, such as the presence of spectral cut-offs or the detection of emission at few tens of TeV expected from galactic PeVatrons (see Section~\ref{Sec:Pevatrons})
\paragraph{Field of view}--
Because of the rather flat performance response over a wide FoV of several squared degrees, we will have a better sensitivity at E$>$10\,TeV for extended sources, investigating the VHE emission and spectral properties in different regions of the source. In order to provide a comparison of the typical off-axis performance between the currently operating northern IACTs and the ASTRI Mini-Array, in Figure~\ref{FIG:Chap8_IntSens_offaxis} we show the integral sensitivity for the MAGIC Telescopes~\citep{2016APh....72...76A} and the ASTRI Mini-Array as a function of the source off-axis and above an energy threshold of 290~GeV and 2~TeV, respectively. These energy thresholds typically provide the best integral sensitivity of the two considered systems. Note that, for each system separately, the integral sensitivity values shown in the plot are normalized to the best achieved one.
It is worth mentioning that a decrease in performance similar to the MAGIC one affects H.E.S.S. too. In this case, the relative acceptance for gamma-rays is roughly uniform for the innermost 2\textdegree{} of its 5\textdegree{} FoV, and drops toward the edges to 40\% of the peak value at an off-axis angle of about 2\textdegree{} \citep{2006A&A...457..899A}.
\paragraph{Selected observing fields}--
The ASTRI Mini-Array is most sensitive in an energy regime where very low gamma-ray fluxes are generally expected, because of the intrinsic energy dependence of the astrophysical spectra or because of the possible presence of spectral cut-offs, and eventually because of the severe extra-galactic background light (EBL) absorption for distant extra-galactic sources. These reasons make long exposures ($>>$50 hrs, as described in Sections \ref{sec:4} and \ref{sec:5}) necessary to get the statistics needed to achieve source detection and valuable scientific results. Moreover, for the first three-to-four years, the ASTRI Mini-Array will be operated as an experiment and not as an open observatory, allowing us to focus on a few selected sky regions.

\subsection{HAWC, LHAASO and Tibet AS$\gamma$}
The HAWC array \citep{2013APh....50...26A} has been inaugurated on 2015 March 20, on the flanks of the Sierra Negra volcano near Puebla, Mexico. To record the particles created in cosmic-ray and gamma-ray air showers, the HAWC detector uses an array of water Cherenkov detectors. In this technique, the detector is used to sample air-shower particles at ground level by recording the Cherenkov light produced when particles pass through tanks full of purified water. HAWC is located in the northern hemisphere, its performance in terms of the highest achievable energy range make it an excellent reference for the ASTRI Mini-Array.

The LHAASO array \citep[located in the Daochen site, Sichuan province, P.R. China, ][]{2019arXiv190502773C} covers an area larger than one square kilometer. It is a hybrid particle sampling array, equipped with muon detectors, water Cherenkov detectors and an array of wide field-of-view Cherenkov telescopes.

The Tibet AS$\gamma$ array is operating at Yangbajing in Tibet, 4300\,m above sea level \citep{1999ApJ...525L..93A}. Currently, after several upgrades, it has an effective area of about 65,700\,$m^{2}$ and about 600 detectors.

HAWC, LHAASO and Tibet AS$\gamma$ operate in an energy range similar to the ASTRI Mini-Array one, easily monitoring the sky at multi-TeV energies. There is a strong complementarity between ASTRI Mini-Array and PSAs, and clearly several differences that allow us to explore the same process in a different way.
\paragraph{Angular and energy resolution}--
As shown in Figure~\ref{FIG:Chap8_multiplot_WCD_200h_500h}, the energy range of the ASTRI Mini-Array (from a few hundreds of GeV to 100\,TeV and beyond) has a wide overlap with those of HAWC and LHAASO, allowing us a direct comparison of scientific data (spectra, light-curves, integral fluxes) of those sources which could be simultaneously observed. The different angular and energy resolution figures at the same energy will be extremely important to investigate possible energy-dependent regions in extended sources, such as the southern PWN HESS~J1825$-$137 \citep{2006A&A...460..365A}.
\paragraph{Exposure timescale}--
 At the time of the ASTRI Mini-Array operation, HAWC and LHAASO will have performed a few years of operation, accumulating a sensitivity that, on selected sources, could be reached by the ASTRI Mini-Array in some months of pointings.
\paragraph{Field of view}--
 The region near the Galactic Center will be accessible by all the facilities. Thanks to the wide field of view of the ASTRI Mini-Array (about 10\textdegree{} in diameter) a large portion of the sky will be investigated simultaneously, so that the ASTRI Mini-Array can study, by means of deep observations, sky ``hot-spots'' detected by HAWC and LHAASO, similarly to the ones identified by the MILAGRO \citep{2003ApJ...595..803A} experiment.
\paragraph{Transients}--
 HAWC demonstrated to be able to detect intense flares from a large portion of the sky \citep{2017ApJ...843..116A}. The distribution of these alerts will allow us to promptly re-point the ASTRI Mini-Array to any observable flaring source detected by current PSAs.

\subsection{Performance in a context}\label{Sec:PerfContext}
\begin{figure}[]
	\centering
	\includegraphics[width=0.49\textwidth, angle=0]{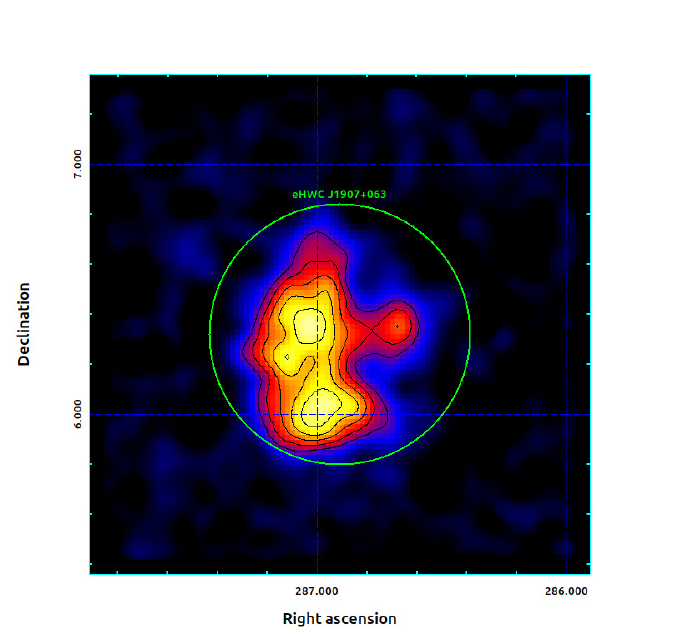}
	\caption{ASTRI Mini-Array 200\,hr simulation of the region of the Galactic source 2HWC~J1908$+$063 for energy up to about 200\,TeV. The light green circle marks the $\sim 0.52$\textdegree{} HAWC error-box for $E>56$\,TeV (see text for details).}
	\label{FIG:ASTRI_HAWC_J1907p063}
\end{figure}

%
The performance discussed in previous Sections can be appreciated when comparing some actual observations of current IACTs and/or PSAs with ASTRI Mini-Array simulations.
Figure~\ref{FIG:ASTRI_HAWC_J1907p063} highlights the importance of the ASTRI Mini-Array angular resolution. The image shows the ASTRI Mini-Array 200\,hr simulation (for energy up to 200\,TeV) of the region around the Galactic source 2HWC~J1908$+$063 (eHWC~1907$+$063/VER~J1907$+$062)~\citep[see also][for morfological details]{2014ApJ...787..166A}. The light green circle marks the $\sim 0.52$\textdegree{} HAWC error-box (for $E>56$\,TeV)~\citep{PhysRevLett.124.021102}. The details of the ASTRI Mini-Array simulations are reported in Section~\ref{Subsec:candidatePeV}.
Although the ASTRI Mini-Array and HAWC have a similar energy range (see Figure~\ref{FIG:Chap8_multiplot_WCD_200h_500h}), their angular resolution is remarkably different, as shown in Table~\ref{TAB:tab_Canarias_comparison_PSAs}. The ASTRI Mini-Array will easily resolve Galactic sources emitting at VHE within the HAWC error-box.

Recently, the LHAASO Collaboration published the detection of a dozen of Galactic sources emitting at energies similar to, or even greater than 1\,PeV~\citep{2021Natur.594...33C}. This discovery is extremely important for the ASTRI Mini-Array science, as discussed in Section~\ref{Sec:Pevatrons}, especially because of its angular resolution which, at energies of about 100\,TeV, is a factor 3--4 times better in radius than the LHAASO one (0.08\textdegree{} vs. 0.24\textdegree{} -- 0.32\textdegree{}). At energies of 10--30\,TeV the difference in angular resolution between the ASTRI Mini-Array and LHAASO is even larger (0.05\textdegree{} vs. 0.70\textdegree{} -- 0.94\textdegree{}), enabling us to accurately investigate the VHE morphology of extended sources.
\begin{figure}[]
	\centering
	\includegraphics[width=0.49\textwidth, angle=0]{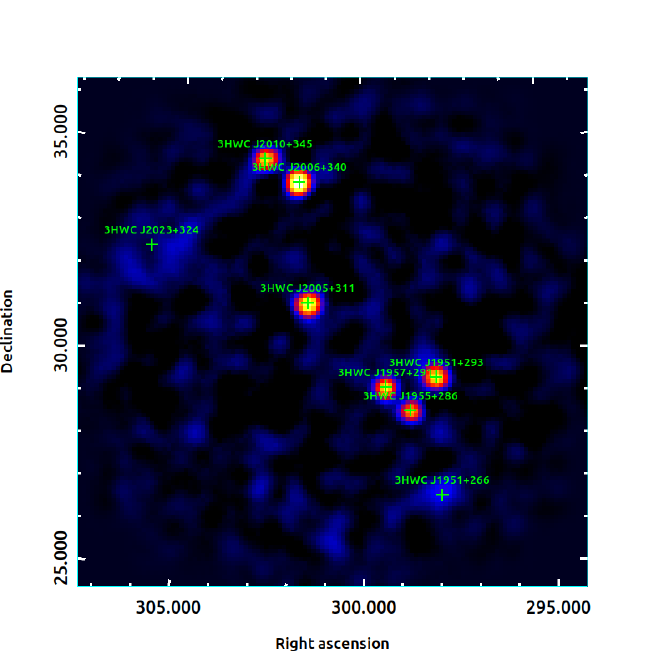}
	\caption{ASTRI Mini-Array 200\,hr simulation of the Cygnus Region. Green crosses mark the positions of the 3HWC sources in a 10\textdegree{} $\times$ 10\textdegree{} field of view (see text for details).}
	\label{FIG:ASTRI_10deg_FoV}
\end{figure}

Figure~\ref{FIG:ASTRI_10deg_FoV} shows the importance of a wide field of view when pointing crowded regions. The panel reports the 200\,hr simulation of a deep pointing towards the Cygnus Region. Several sources of the {\it Third HAWC Catalog of Very-high-energy Gamma-Ray Sources}~\citep[3HWC,][]{2020ApJ...905...76A} can be observed and possibly detected in a single pointing by the ASTRI Mini-Array. This will also allow the ASTRI Mini-Array to effectively monitor variable sources.

\section{ASTRI Mini-Array Core Science and Simulation Setup}\label{sec:3}
%
\subsection{Core science: the Pillar concept}
The ASTRI collaboration will start deploying the first telescopes of the Mini Array on-site starting at the beginning of 2022, commissioning and preliminary observations will begin in 2023 with a first array of three telescopes, while the whole ASTRI Mini-Array lifetime will be of eight years. We plan to devote the first four years to specific science topics, with the aim to provide robust answers to a few well-determined open questions. 
In particular, we will take advantage of the wide field of view ($\sim 10^{\circ}$) to simultaneously investigate more than one source during the same pointing and to study complex Galactic regions, such as the Cygnus region or the Galactic center, where the diffuse emission and the high number of sources need careful data analysis.
We identified the following high-level topics.
\paragraph{The origin of cosmic rays}

We will study how particle are accelerated in both Galactic and extra-galactic sources. In particular, we will tackle the long-standing quest of sources which could accelerate hadrons up to peta-electronvolt energies, by means of the capability of reaching energy boundaries above 100\,TeV in conjunction with an excellent angular resolution.
\paragraph{The EBL and the study of fundamental physics}

The ASTRI Mini-Array, with its excellent sensitivity in the energy range 10--30\,TeV, is perfectly suited to investigate the extra-galactic background light in the far infra-red domain, only accessible at this energies. Moreover, we will address some open issues in fundamental physics, such as the study of the axion-like particle.
\paragraph{Time-domain and multi-messenger astrophysics}

Gamma-ray bursts, gravitational waves and neutrinos of extra-galactic origin recently became topics of paramount importance. Thanks to the excellent performance at energies $E>1$\,TeV, fast reaction, wide field of view, the ASTRI Mini-Array will play an important role in the time-domain and multi-messenger astrophysics.
\paragraph{Ultra high-energy cosmic rays}

The ASTRI Mini-Array will also be able to make use of the tremendous amount of hadrons impinging on the focal plane of its nine camera in order to direct measure muons multiplicity on a statistics base to perform direct measurements of CRs composition. 
\paragraph{Stellar intensity interferometry}

On the other side, thanks to the high number of telescopes, the ASTRI Mini-Array will also be able to study bright stars in the visible light waveband at very high angular resolution using a technique known as stellar intensity interferometry.

\subsection{Scientific simulations setup}\label{Sec:SciSimSet}

%
In the following sub-sections and in Appendix~\ref{Sec:Appendix1} we describe how the simulations were performed with the different tools which are commonly used by the CTAO Consortium and that are freely available.
All the simulations were based on the instrument response functions developed in the framework of the ASTRI Mini-Array project, as described in the previous Sections. 

\subsubsection{Ctools simulations}\label{Sec:ctoolssims}

The simulations were performed with the {\tt ctools}  
\citep[][v.\ 1.6.3]{2016A&A...593A...1K}\footnote{\href{http://cta.irap.omp.eu/ctools/}{http://cta.irap.omp.eu/ctools/}.} analysis package. 
We followed the procedures detailed in \citet{2018MNRAS.481.5046R,2020MNRAS.494..411R} 
and used in several CTA publications  \citep{2019A&A...631A.177C,2019MNRAS.483.1802T,2019ApJS..240...32L,2019APh...112...16L}. 
The spectra of the sources were calculated in several energy bins logarithmically spaced between 0.8 and 199.5 TeV and with exposure times ranging between 100 and 500 hours, only considering the instrumental background included in the IRFs. The IRF energy dispersion was not included in the simulations, since the effect is only prominent below a few hundred GeV while its inclusion would have increased the run time by a factor of 5--10 without producing significanly better results given the ASTRI Mini-Array energy range and our choice of energy binning. 

In each bin, we first used the task {\tt ctobssim} to create event lists based on our input models, and then  {\tt ctlike} to fit a power-law model by using  a unbinned maximum likelihood model fitting. To reduce the impact of variations between individual realisations, we performed sets of $N=100$ statistically independent realisations \citep[see, e.g.][]{2016A&A...593A...1K}. 
In order to reconstruct the flux in each energy bin, the normalisation and photon index parameters of the power-law $\gamma$-ray spectrum were free to vary while the pivot energy was set to the geometric mean of the boundaries of the energy bin. {\tt ctlike} also calculates the test statistics \citep[TS,][]{1979ApJ...228..939C} of the maximum likelihood model fitting, which we used to assess the goodness of the detection in each bin. We considered TS$\geq 9$ as the threshold for a detection. 

\subsubsection{Gammapy simulations}\label{Sec:gammapysims}

We also used {\it Gammapy} v0.17 \citep[e.g.][]{2017ICRC...35..766D} to perform data simulations and analyses of some sources. We adopted a  simulation process that follows the prescriptions described in \citet{2018MNRAS.481.5046R,2020MNRAS.494..411R}.

By means of the {\it Gammapy} simulator (contained in the {\sc MapDatasetEventSampler} class), we generated source and (instrumental) background events in the 0.8--199.5 TeV energy range, adopting for each source a given sky model. We did not include the IRF energy dispersion for the simulations in order to avoid a further degree of complexity in the analyses. To reduce the impact of random variations between individual realisations, we performed a set of $N=100$ statistically independent simulations. 

We then fitted each of the 100 simulated data with a maximum likelihood analysis ({\sc Fit} function in {\it Gammapy}), adopting the same template model of the simulation. From each best-fit model, we calculate the source flux in several energy bins in the 0.8 - 199.5 TeV energy range (adopting the {\sc FluxPointEstimator} function in {\it Gammapy}). For each energy bin, we also estimated the corresponding test statistics \citep[TS,][]{1979ApJ...228..939C}, which defines the source significance. We calculated the distribution of TSs and we deemed a flux point significant when the mean of the TS in that energy bin was greater than 9. In such cases, we estimated the source flux and its uncertainty for each energy bin by determining the mean flux and standard deviation of the distribution from the 100 simulations. Instead, when the TS distribution was not significant in a given energy bin, we calculated a 95\,\% confidence level upper limit on flux from the distribution of the simulated fluxes. Finally, from the distributions of the best-fit spectral parameters obtained in each of the 100 simulations, we estimate the mean and the corresponding standard deviation of each model parameter.
If not explicitly mentioned, this is the general approach adopted along this paper.

\subsubsection{The Naima Package}

Naima \citep[]{2015ICRC...34..922Z} is a python package for evaluating the non-thermal emission of relativistic electrons and protons through some radiative processes as Bremsstrahlung, inverse Compton, synchrotron and neutral pion decay. Naima can also find the spectral energy distribution that best matches a set of data, varying the parameters of the parent-particle distribution for a given astrophysical ambient (e.g. gas density, magnetic field, interstellar radiation field...). In other words, it finds a best-fit of a set of data based on a model that describes the physical features of the source. With Naima it is also possible to fit simultaneously a few sets of data (for example, from different facilities), making it very useful when a multiwavelength  spectrum of an astrophysical object is required. \\

\smallskip
\noindent
In the next two Sections we introduce the concept of {\it Pillar science topics}. They are science fields in which the ASTRI Mini-Array will contribute solid pieces of evidence to significantly improve our understanding of the above key science questions. We present them discussing the minimum requested integration time necessary to fulfill our science goal. Nevertheless, thanks to both our capability to observe also during moonlight periods and to our large field of view, we will add more integration time to each science topic. We discuss their impact in terms of the extension of the observing time.
\section{Pillar--1: Origin of Cosmic Rays}\label{sec:4}
Cosmic rays (CRs) are charged particles detected at the Earth, mainly consisting of protons, with a fraction of about 12\% of helium nuclei and smaller abundances of heavier elements, electrons and anti-particles \citep[see][for recent reviews about the main open problems in CR physics]{2013A&ARv..21...70B,2014IJMPD..2330013A,2018NPPP..297....6D,2019IJMPD..2830022G}. In first approximation, the all-particle spectrum is a power law in energy that spans from \rf{a} few MeV up to $\sim 10^{20}$\,eV. In spite of such an incredibly large energy range, only three clear features are present: the {\it knee}, namely a break with a change in slope from $E^{-2.7}$ to $E^{-3.1}$ at $E_{\rm knee}\simeq 3\times 10^{15}$ eV; the {\it ankle}, namely a hardening, with a change from $E^{-3.1}$ to $E^{-2.6}$ at $E_{\rm ankle} \simeq 3\times 10^{18}$ eV; and a sharp decrease above $\sim 10^{21}$ eV, often called Greisen–Zatsepin–Kuzmin (GZK) cut-off, {due to CR proton interaction with background photons.} 
Energetic arguments and the observed isotropy of their incoming directions, suggest that CRs up to an energy around $10^{17}$ eV originate in our own Galaxy, requiring that protons reach at least $\sim$ PeV energies to explain the break at the knee.
Particles with energy beyond $E_{\rm ankle}$, usually referred to as ultra-high-energy cosmic rays (UHECRs), likely have an extra-galactic origin because they cannot be efficiently confined in our Galaxy ~\citep{2019JCAP...01..002A}, as suggested by observations by the Pierre Auger Observatory ~\citep{2017Sci...357.1266P} and the Telescope Array ~\citep{2020arXiv200507312T}.

Despite the enormous efforts done in very recent years, both theoretically and experimentally, the basic three questions about the CR origin remain without clear answers: what are the main sources? How are these particles accelerated? How do they propagate to us?
The theoretical framework which offers the most convincing scenario, at least for the Galactic CRs, is the so-called {\it Diffusive Shock Acceleration} (DSA) born from Fermi's original idea that particles can gain energy by scattering off magnetic disturbances \cite[see, e.g.][]{1983SSRv...36...57D}. 
DSA applied to Supernova Remnant (SNR) shocks has acquired a large consensus in the community to explain the origin of Galactic CRs. However, even though there is a large amount of circumstantial evidence, we still lack a direct proof that the acceleration occurs efficiently enough and up to the required maximum energies. 
At energies beyond the {\it knee}, DSA encounters increasing difficulties as the main acceleration mechanism and the picture is far from clear~\cite[see e.g.][for a recent review]{2019FrASS...6...23B}.
An alternative mechanism that has been also proved to be very efficient in accelerating particles, especially in magnetized regions of compact sources like pulsar wind nebulae, micro-quasars and relativistic jets, relies on fast magnetic reconnection (\citealt{2005A&A...441..845D, 2012ApJ...746..148C}, see also \citealt{2015ASSL..407..373D, 2020NewAR..8901543M} for recent reviews).
 
Gamma-ray astronomy plays a fundamental role in the search for direct evidence of CR acceleration. Indeed, CR protons interact with target protons producing mainly neutral pions that, in turn, decay into two \gray{} photons with $\sim10$\% of the parent proton energy. The produced photons directly point to the source, making a direct identification of the CR sources possible. Unfortunately, relativistic electrons emit in the same energy band through Bremsstrahlung and inverse Compton scattering, making it difficult to firmly disentangle hadronic and leptonic spectral contributions. Such a degeneracy can be broken looking either at low or at very high energies ($>10$ TeV). In fact, radiation from $\pi^0$ decay shows a characteristic peak at half of the $\pi^0$ mass, 67.5\,MeV, the so called ``pion bump'', which is absent in leptonic processes. At higher energies, instead, a leptonic origin of the emission is generically disfavored due to severe energy losses and to the Klein-Nishina suppression of the Compton cross section.  Consequently, a detection of \gray{} photons with energies above 100 TeV is expected to indicate that their source is accelerating hadronic CRs at PeV energies. The only caveat to this conclusion comes from system hosting very energetic pulsars, with spin-down power $\dot E>10^{37}$ erg/s. In fact, one noticeable exception is the Crab PWN, whose \gray{} emission up to $> 100$\,TeV is likely primarily due to inverse Compton scattering of electrons accelerated up to the maximum pulsar potential drop (\citealt{2019MNRAS.489.2403L}, see Sec.~\ref{Subsec:Crab} for a more detailed discussion). 

In the MeV--GeV band, AGILE and {\it Fermi}-LAT detected a curvature of the \gray{} spectrum, compatible with the pion bump, in some middle-aged SNRs, like W44 \citep{2011ApJ...742L..30G,2014A&A...565A..74C, 2013Sci...339..807A}, IC443 \citep{2013Sci...339..807A} and W51c \citep{2016ApJ...816..100J}.  
However, at higher energies, the \gray{} spectrum of those SNRs is rather steep, suggesting that the acceleration becomes ineffective. Hence, these objects cannot be the main contributors of Galactic CRs (at least at the present stage). In addition, when a SNR enters the radiative phase, the \gray{} emission could result from the re-acceleration and compression of pre-existing CRs \citep{2010ApJ...723L.122U,2015ApJ...806...71L,2016A&A...595A..58C} rather than from freshly accelerated particles.
 
At the highest energies, PeVatrons are predicted to be quite rare and only recently LHAASO reported on the detection of 12 Galactic sources with emission well above 100 TeV, and in one case extending up to 1.4 PeV \citep{2021Natur.594...33C}. LHAASO also observed the Crab Nebula up to an energy of 1.1 PeV~\citep{2021Science373..425C}. Although the PeVatron accelerator was not firmly identified (apart from the Crab Nebula), these sources represent the best examples of PeVatron candidates. A possibility to enlarge the sample of PeVatron sources is to look for \gray{} emission produced by escaping particles which collide with dense molecular clouds in the source surroundings. The feasibility of this approach depends on the diffusion properties of the interstellar medium: a small diffusion coefficient leads to the confinement of particles for a longer time, increasing the chance to detect them.
From the theoretical point of view, quite extreme conditions are required to accelerate particles up to $\sim$ PeV energies, hence one should also explore alternative candidates like massive stellar clusters \citep{2020SSRv..216...42B,2021MNRAS.504.6096M} and supermassive black holes \citep{2016Natur.531..476H}.

Among Galactic sources, PWNe represent the most numerous \gray{} sources and the only known (leptonic) PeVatrons. The \gray{} spectrum of the Crab shows that electron-positron pairs are accelerated up to the maximum potential drop, challenging our current understanding of particle acceleration mechanisms.
In the context of CRs, PWNe are primary sources of $e^{\pm}$ and can be responsible for the rising positron fraction observed in the CR spectrum \citep{2009Natur.458..607A, 2013PhRvL.110n1102A,2017SSRv..207..235B}. 
Efficient escape of multi-TeV leptons has been recently detected from old PWNe, especially thanks to the discovery of the extended \gray{} halo surrounding the Geminga and {PSR~0656$+$14} pulsars \citep{2017Sci...358..911A}.
In addition, the possibility that accelerated hadrons could also be present in PWNe is still open, and observations above $\sim 100$\,TeV could provide the most stringent constraints (see Sec.~\ref{Subsec:Crab}).

The extra-galactic component of CRs is even more puzzling than the Galactic one: neither the sources nor the acceleration mechanism have been identified. Possible sources include the most powerful objects of our Universe, namely gamma-ray bursts (GRBs), active galactic nuclei (AGNs) and star-burst galaxies (SBGs), the latters being the most attractive candidates \citep{2019FrASS...6...23B}.
Interestingly, IceCube recently detected a very-high-energy neutrino in spatial coincidence with a \gray-emitting blazar during an active phase \citep{2018Sci...361.1378I}, which suggests that blazars may be a source of high-energy neutrinos and, as a consequence, of high-energy hadrons.
On the other hand, the incoming direction of UHECRs above 38~EeV suggests a possible correlation with the spatial distribution of SBGs \citep{2018ApJ...853L..29A} (see, however, \citealt{2018ApJ...867L..27A}).
Note that while such correlations may genuinely point to the UHECRs sources, they may also be sporadic, since Galactic and extra-galactic magnetic fields affect the arrival directions.

\smallskip
The aim of next sections is to study a few selected topics in CR physics that can be approached by analyzing the very high energy \gray{} emission above $\sim 10$ TeV, taking advantage of the performance of the ASTRI Mini-Array. In particular we will focus on: 1) the search for Galactic PeVatrons (\S~\ref{Sec:Pevatrons}), 2) high-energy particle escape and propagation around their sources (\S~\ref{Sec:particleProp}), 3) high-energy emission from PWNe (\S~\ref{Sec:PWNe}) and 4) SBGs as possible sources of UHECRs (\S~\ref{Sec:UHECRs}). The selected targets, summarized in Table~\ref{tab:sources-Pillar1}, have been chosen in order to optimize the scientific results.
This work started well before the recent discovery by PSAs of sources emitting above several hundreds of TeV. Some of them had already been selected as possible very high-energy targets by the ASTRI Mini-Array Collaboration and taken into account in our analysis. Interestingly, the recent results reported by the LHAASO Collaboration are in good agreement with our simulations performed prior to the recent publications.

\begin{table*}[width=2.\linewidth,cols=9,pos=htp!]
\caption{List of selected \gray{} sources relevant for the study of CR origin, observable from the Observatorio del Teide and studied with ASTRI Mini-Array simulations}
\centering
\def\arraystretch{1.5}
\begin{tabular}{L@{\hskip 0.05in}C@{\hskip 0.05in}C@{\hskip 0.05in}C@{\hskip 0.05in}C@{\hskip 0.05in}C@{\hskip 0.05in}C@{\hskip 0.05in}C@{\hskip 0.05in} C}
\toprule
Name & RA & Dec & Type &  Zenith Angle\textsuperscript{1}  & Visibility\textsuperscript{2}  & Flux\textsuperscript{3}(1 TeV)  & Index  & Section \\
 & (deg) & (deg) &  &  (deg)  & (hr/yr)  & ($10^{-13}$ TeV$^{-1}$cm$^{-2}$s$^{-1}$) &  &  \\
\midrule
Tycho & 6.36 & 64.13 & SNR & 35.8 & 410+340 & 1.71 & 2.28 & \ref{Subsec:Tycho}\\

Galactic Center & 266.40 & -28.94 & Diffuse & 57.2 & 0+180 & 36& 2.32  & \ref{Subsec:GC} \\ 

VER J1907$+$062 & 286.91 & 6.32 & SNR+PWN & 22 & 400+170 & 0.85 (7 TeV)& 2.33  & \ref{Subsec:candidatePeV}\\

SNR G106.3+2.7 & 337.00 & 60.88  & SNR & 32.6 & 460+300 & 1.15 (3 TeV)& 2.29& \ref{Subsec:candidatePeV} \\
\midrule
$\gamma$-Cygni & 305.02 & 40.76 & SNR & 12.5 & 460+160& 20 (whole SNR) & 2.37 & \ref{Subsec:gammaCyg} \\
 &  &  &  &  &  & 
12 (hot-spot) & &   \\
W28/HESS\,J1800-240B & 270.11 & -24.04  & SNR/MC & 51.6 & 0+300 & 7.5& 2.4 - 2.55& \ref{Subsec:W28} \\
\midrule
Crab  & 83.63 & 22.01 & PWN & 6.3 & 470+170& * & * & \ref{Subsec:Crab} \\
Geminga  & 98.48 & 	17.77 & PWN & 10.5 &460+170 & * & * & \ref{Subsec:Geminga} \\
\midrule
M82 & 148.97 & 69.68 & Starburst & 41.4 & 310+470 & 
2.74 & 2.2 & \ref{Sec:UHECRs} \\
\bottomrule
\end{tabular}

{\raggedright \textsuperscript{1}Culmination angle reachable at Teide from the source.\\
\textsuperscript{2}Maximum available hours of visibility, in moonless conditions, calculated for one year of observations and for two zenith angle intervals [0-45°]+[45°-60°].\\
\textsuperscript{3} Flux and index are the ones of the input model used in the simulation. See the text for the references. \\
\textsuperscript{*} For these sources, we adopted an input model not previously reported in literature. See text for more details. \par}
\label{tab:sources-Pillar1}
\end{table*}
%
%
\subsection{The Quest for PeVatrons}
\label{Sec:Pevatrons}
\paragraph{Scientific Case}-- 
In the standard scenario for the origin of CRs, Galactic sources should be able to accelerate the light component of CRs (p and He) at least up to the knee energy \citep[see][for recent reviews]{2013A&ARv..21...70B,2014IJMPD..2330013A,2019IJMPD..2830022G,2019NCimR..42..549B,Amato2021}. Sources able to accelerate protons up to $\sim 10^{15}$~eV will be referred to as ``PeVatrons''. The steepening of the all-particle spectrum above $E_{\rm knee}$ is usually interpreted as due to the superposition of cut-offs of heavier components whose maximum energy is $\propto Z E_{\rm knee}$ (see \S~\ref{Subsec:knee} for uncertainties concerning the chemical composition at the knee). It is worth noting that alternative scenarios, where the knee is explained as a change in the propagation regime of particles \citep{2014PhRvD..90d1302G}, require even larger maximum energies to be achieved by Galactic sources.

Among the known SNRs, no PeVatrons have been clearly identified up to now. Typical remnants show a power law spectrum with a cut-off energy at $E \lesssim 10$\,TeV. Possible exception are Tycho's SNR and a few composite remnants associated with pulsars. 
HAWC \citep{PhysRevLett.124.021102} {and LHAASO \citep{2021Natur.594...33C}} have recently detected several sources with \gray{} emission above 100 TeV. The majority of them could be PWNe like the Crab nebula, but some sources are also associated with SNRs, probably in collision with dense molecular clouds. The hadronic or leptonic nature of their emission must be clearly disentangled to assess whether these sources are hadronic PeVatrons.

Indeed, theories of particle acceleration at SNRs begin to encounter problems at a few hundred TeV \citep{2016arXiv161007638G}, and PeV energies seem to be reachable only in quite extreme conditions \citep{2013MNRAS.431..415B,2015APh....69....1C}. {On the other hand, LHAASO detected several SNRs \citep{2021Natur.594...33C} emitting well above hundreds of TeV, challenging our understanding of particle acceleration. In addition, there could be other possible PeVatron candidates:} in the last few years, the \hess{} array detected \gray{} emission from the region around the Galactic center (GC) with a power law photon spectrum up to 10--30\,TeV without a clear cut-off \citep{2016Natur.531..476H,2018A&A...612A...9H}. Recently,  \citet{2020A&A...642A.190M} also reported new observations that confirmed the diffuse emission up to $\sim$\,50 TeV from the GC region.
A deeper analysis of the spatial and spectral characteristics of this emission found strong similarities with \gray{} emission detected in some stellar clusters as Westerlund 1 or Cygnus OB2 \citep{2019NatAs...3..561A} suggesting that massive stellar clusters in the GC may be responsible for the \gray{} emission. {The recent results published by LHAASO collaboration confirm this hypothesis with detection of LHAASO~J2032$+$4102 a\-bo\-ve 1~PeV \citep{2021Natur.594...33C}}. Such a finding could be pointing towards a paradigm shift, where other sources may substantially contribute to the Galactic CRs in addition to SNRs \citep{Amato2021}. The last published spectrum of the Cygnus Cocoon detected above 100 TeV by HAWC \citep{2021NatAs...5..465A} also points in this direction.
An alternative  scenario for the production of hadronic \gray{} emission from the GC assumes that CRs are accelerated by turbulent magnetic reconnection in the accretion flow around Sgr\,A$^{\star}$. Such a mechanism will result into a VHE emission flux at a few 10 to 100 TeV \citep{2019ApJ...879....6R}, consistent with the H.E.S.S. upper limits for the GC and potentially detectable by the ASTRI Mini-Array.

In this context, the ASTRI Mini-Array will provide a fundamental contribution with its unparalleled sensitivity and spatial resolution at $E > 10$\,TeV and its wide FoV, helping us to unveil whether SNRs or other classes of sources are the long sought PeVatrons. In the present work, a few most likely hadronic sources will be analyzed: the Tycho SNR, the region around the GC and two composite sources from the HAWC catalogue, VER~J1907$+$062 and VER~J2227$+$608.

\subsubsection{Supernova Remnants: Tycho}  \label{Subsec:Tycho}

\paragraph{Immediate Objective}-- 
Tycho is one of the youngest and best studied SNRs. Assessing the shape of its \gray{} spectrum is of the outermost importance because combining it with information from other wavelengths can strongly constrain the shock acceleration mechanism. The presence of synchrotron X-ray filaments at the shock location implies a strong magnetic field of the order of hundreds of $\mu$G \citep{2006AdSpR..37.1902B}, a necessary condition to reach very high energies.
Its \gray{} spectrum is $\propto E^{-2.3}$ and multi-wavelength studies clearly point towards a hadronic origin of this emission \citep{2012A&A...538A..81M}. VERITAS data \citep{2015ICRC...34..769P, 2017ApJ...836...23A} suggest a cut-off energy of $\sim 10$ TeV but a larger value cannot be excluded due to the large error bars.
Only an effective area and a sensitivity better than the currently available, as the ones of the ASTRI Mini-Array, can better constrain the spectrum at TeV energies, and hence confirm or disprove the PeVatron nature of Tycho.

\paragraph{Observing Time, Pointing Strategy, Visibility and Simulation Setup}-- 
The Tycho SNR is a very faint source in the \gray\ band and it is observable from Teide for about 400 hr per year between zenith angle $0^\circ$-$45^\circ$ and about 350 hr per year at angles $>45^\circ$~above horizon, in moonless conditions. We investigated the spectrum of this important but very weak source for the representative exposure times of 100\,hr, 200\,hr and 500\,hr.
We modeled the SNR following \cite{2017ApJ...836...23A}, which describe the source spectrum as a simple power law with an index of about 2.3, without a cut-off. Given its small size of $\sim 8'$, we simulated the Tycho SNR as a point-like source but we stress that ASTRI Mini-Array, thanks to its resolution of 3', could resolve it. We used {\tt Gammapy} v0.17 \citep[e.g.][]{2017ICRC...35..766D} for data simulations and analysis {(see Sec.~\ref{Sec:gammapysims})}. {We generated source and (instrumental) background events in the 0.5--199.5 TeV energy range, adopting for Tycho a sky model with the spectral and morphological properties described above.}

\begin{figure*}
	\center
		\includegraphics[scale=.55]{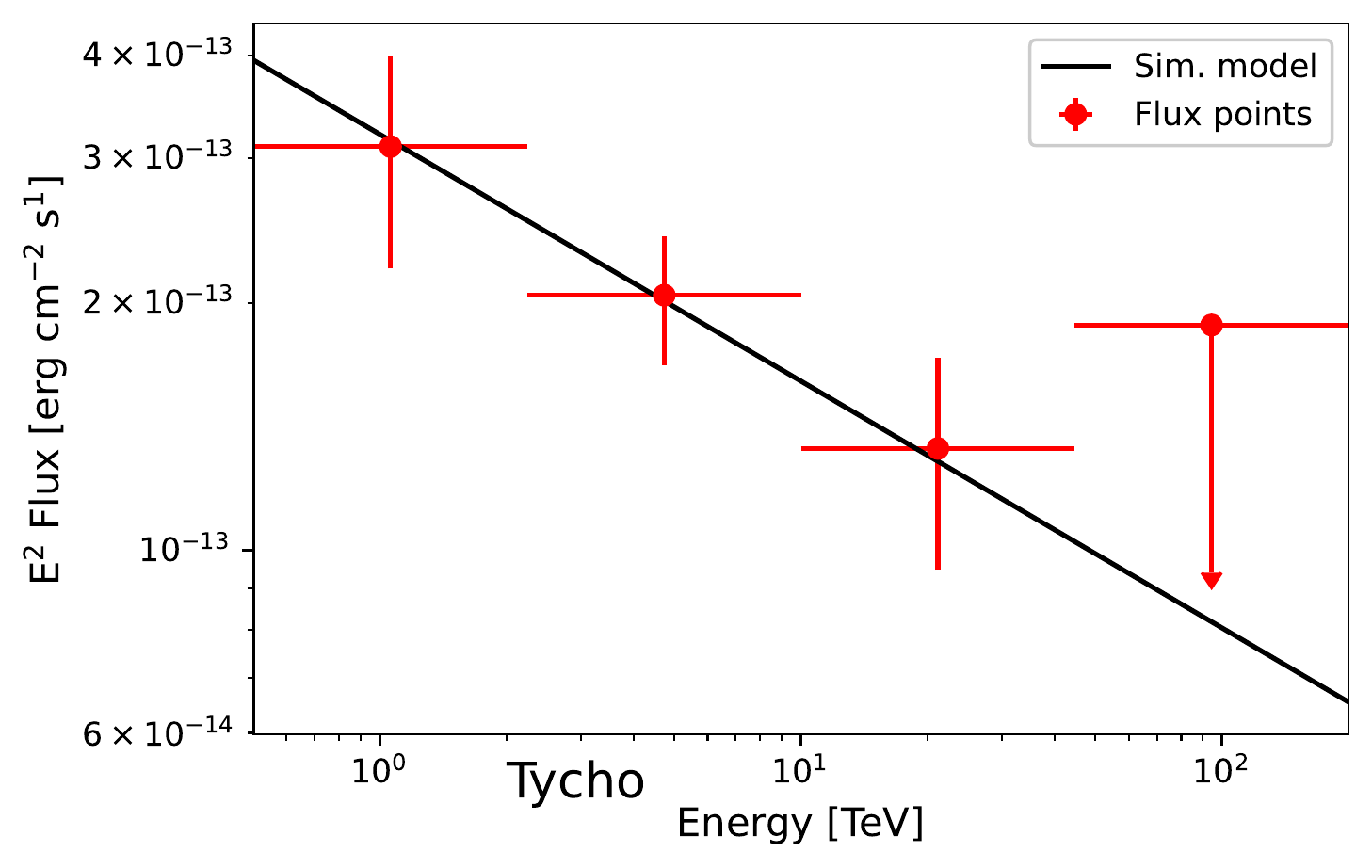}
		\includegraphics[scale=.55]{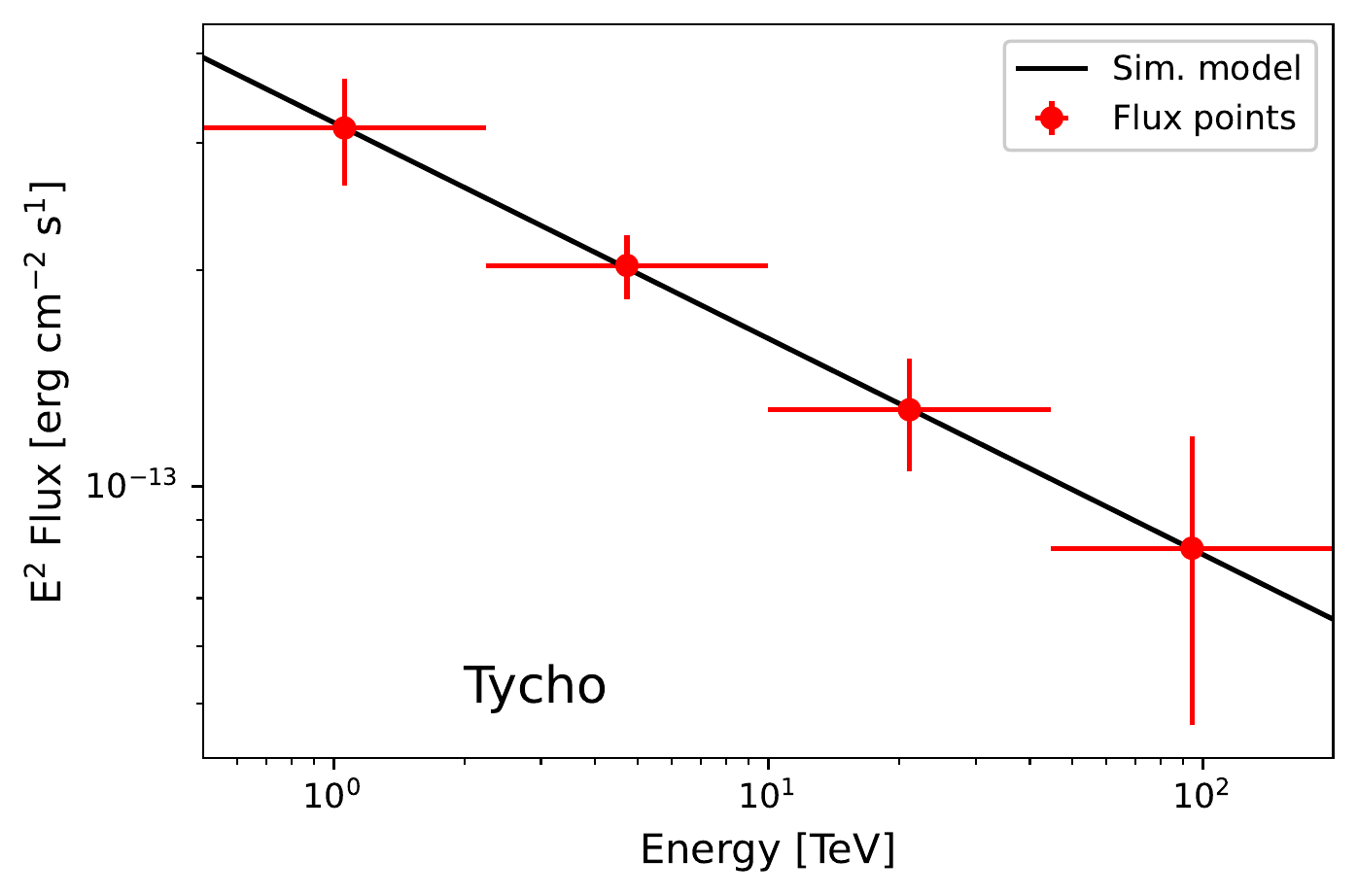}
	\caption{Tycho SNR: ASTRI-MA spectrum (red points) and input model (black line) obtained with 200 hr (left) and 500 hr (right) of observations. The spectral points and the error bars are obtained from the distribution of 100 simulations. See text for more details.}
	\label{FIG:Tycho_spectrum}
\end{figure*}

\paragraph{Analysis Method}-- 
From the distributions of the best-fit spectral parameters obtained in each of the 100 simulations, we estimated: a differential flux of
 $\rf{N}_{\text{1TeV}}=(2.1 \pm 0.7)\times10^{-13}$ TeV$^{-1}$ cm$^{-2}$ s$^{-1}$ and a spectral index $\Gamma=(2.3\pm0.2)$, for 100\,hr of exposure, $\rf{N}_{\text{1TeV}}=(2.1 \pm 0.6)\times10^{-13}$ TeV$^{-1}$ cm$^{-2}$ s$^{-1}$ and a spectral index $\Gamma=(2.3\pm0.1)$, for 200\,hr of exposure and $\rf{N}_{\text{1TeV}}=(2.0 \pm 0.5)\times10^{-13}$ TeV$^{-1}$ cm$^{-2}$ s$^{-1}$ and a spectral index $\Gamma=(2.3\pm0.1)$, for 500\,hr of exposure.
In Fig.~\ref{FIG:Tycho_spectrum}, we show both the 200\,hr and 500\,hr average spectra obtained from the 100 realizations. If the Tycho SNR is a PeVatron, the ASTRI Mini-Array will be able to detect its \gray\ emission at 100 TeV.

\begin{figure*}
	\centering
		\includegraphics[scale=.45]{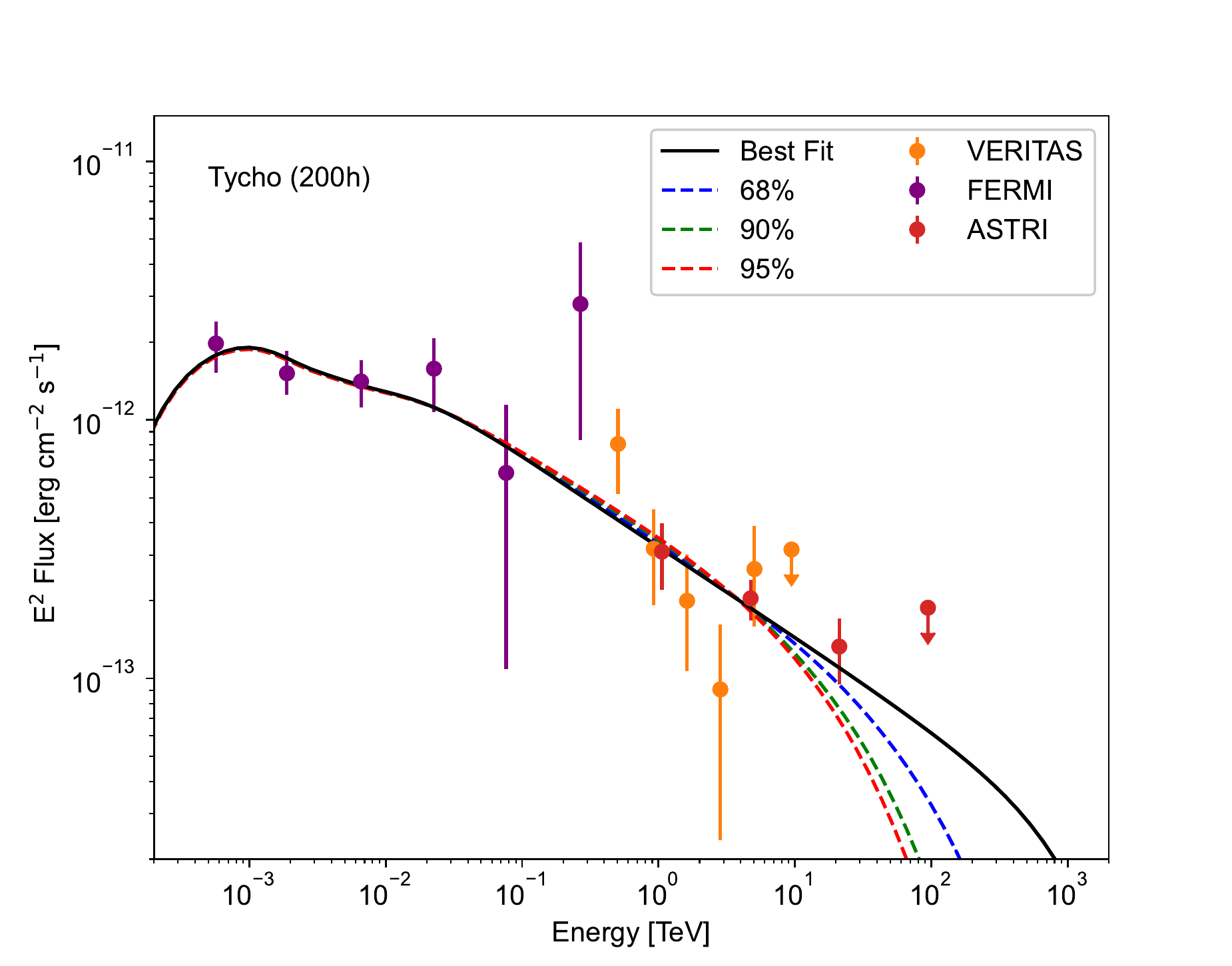}
		\includegraphics[scale=.45]{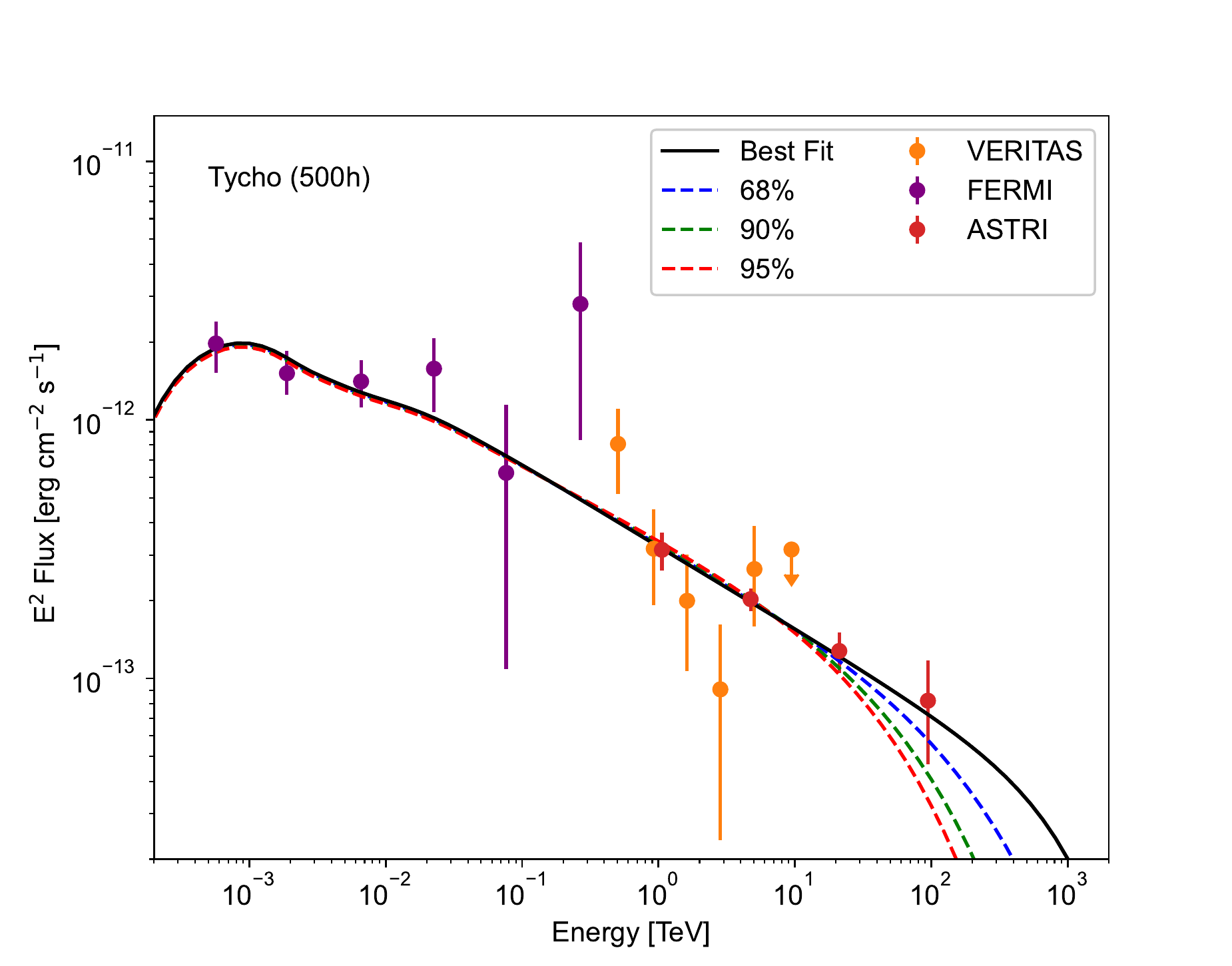}
	\caption{Tycho SNR: \gray{} data from Fermi-LAT (purple, 84 months)and VERITAS (orange, 147 hr) \citep{2017ApJ...836...23A} together with the ASTRI Mini-Array simulations (red). Left: 200 hr of observations. The dashed lines show the PL fit with cut-off energies of 0.29, 0.41, 1.27\,PeV (corresponding to 95, 90 and 68\% of confidence levels). Right: 500 hr of observations. The dashed lines show the PL fit with cut-off energies of 0.85, 1.36, 3.96\,PeV (corresponding to 95, 90 and 68\% of confidence levels)}
	\label{FIG:Tycho_NAIMA}
\end{figure*}

As a next step, we tried to constrain the \gray{} spectrum of the source taking into account the ASTRI Mini-Array data in combination with lower energy ones collected by Fermi-LAT and VERITAS with 84 months and 147 hours of observations, respectively \citep{2017ApJ...836...23A}. This allows us to cover Tycho's \gray{} emission over six orders of magnitude (from $\sim 100$ MeV up to $\sim 100$ TeV). We used the {\sc naima} package \citep{ 2015ICRC...34..922Z} to fit simultaneously Fermi-LAT and VERITAS observations with the ASTRI Mini-Array simulated data (see Fig.~\ref{FIG:Tycho_NAIMA}).

We modeled the \gray{} emission in a purely hadronic scenario, produced by a proton population following a power law distribution with a high energy cut-off. Considering the flux normalization, power law index and cut-off energy as free parameters, we obtained as best fit a value of the energy cut-off well beyond 1~PeV in the proton spectrum.
We evaluated the cut-off lower limit as a function of the confidence level (see Fig. \ref{FIG:Tycho_NAIMA}), following the approach used in \cite{2016Natur.531..476H}. In the case of an exposure of 200 hr, we estimated that ASTRI Mini-Array data plus Fermi-LAT and VERITAS data will allow us to exclude a cut-off below 1.27 PeV, 0.41 PeV and 0.29 PeV at 68$\%$, $90\%$ and $95\%$ confidence level, respectively (see Fig.\ref{FIG:Tycho_NAIMA}, left).
These limits can be further improved increasing the exposure time up to 500 hr. In such a case, we can exclude a cut-off below 4 PeV at 68\% of confidence level, and below 0.9 PeV at 95\% (see Fig.\ref{FIG:Tycho_NAIMA}, right). Without the ASTRI Mini-Array data points, instead, the 68\% lower limit is only 4 TeV. Such a  preliminary analysis clearly shows that the ASTRI Mini-Array will provide a fundamental contribution to constraining the particle acceleration process even in a SNR as faint as Tycho in $\gamma$-rays.
%

\subsubsection{Galactic Center}  
\label{Subsec:GC}

\paragraph{Immediate Objective}-- 
The other strong PeVatron candidate is the Galactic center, which comprises approximately a region of 1.5\textdegree\ in Galactic longitude and $\sim$\,0.2\textdegree\ in Galactic latitude.
This volume hosts at its center the super-massive black-hole Sgr~A$^{\star}$, many star-forming regions, young and recycled pulsars, heated and shocked gas from past and recent supernova explosions, and many other astrophysical sites of potential particle acceleration. All these objects could contribute, in a way which is not yet fully understood, to the TeV \gray{} excess detected by H.E.S.S. \citep{2016Natur.531..476H, 2018A&A...612A...9H}, VERITAS \citep{2016ApJ...821..129A} and
MAGIC \citep{2017A&A...601A..33A, 2020A&A...642A.190M}.
This emission, spatially associated with the giant gas clouds, shows a hard spectrum without evidence of a cut-off at least up to 40 TeV, and lack of variability on the year-long time-scale. These findings strongly suggest a hadronic origin of the \gray{} emission.
Extending the spectral measurements from the present upper threshold of 40 TeV up to 100 TeV, and possibly beyond, could firmly confirm the likely PeVatron nature of the source and constrain the parent hadronic population up to PeV energies \citep[see e.g.,][]{2019ApJ...879....6R}. In addition, the excellent angular resolution of the \astrima{} could help to identify the HE source among several candidates. For example, the projected distances of the most powerful stellar clusters in the GC region, Arches and Quintuplet, from Sgr A$^*$ is $\sim 30$~pc, corresponding to an angular separation of $0.20^\circ$, well above the \astrima{} angular resolution of $0.05^\circ$. Moreover, thanks to the very large FoV, the ASTRI Mini-Array will be able to map the whole GC region in a single observation.

\paragraph{Observing Time, Pointing Strategy, Visibility and Simulation Setup}-- 
The GC is observable from the Teide site for about 180 hr per year in moonless conditions, at a maximum culmination angle of $\sim$57\textdegree. 
We investigated the spectral constraints achievable in 100 hr, 260 hr and 500 hr of observation (the second exposure time is the same used by HESS in \citealt{2018A&A...612A...9H}). 
For the aims of the present analysis, we use the spatial and spectral characterization of the inner Galactic ridge emission obtained from $\sim$\,260 hr of H.E.S.S. observations \citep{ 2018A&A...612A...9H}.
The template model describes the GC diffuse emission as a combination of 
three spatial components taken as Gaussians of different widths and normalizations and all centered at Galactic coordinates l\,=\,0\textdegree\ and b\,=\,0\textdegree. 
The brightest component, accounting for about half of the total emission, 
is associated with the dense gas environment, mostly present along the inner Galactic ridge ($\sim$\,150 pc extension); the second 
component is much more compact ($\sim$\,15 pc width); the third, fainter, component shows different widths in longitude
(around 140 pc) and latitude (30 pc). 
The components have all the same spectral shape: 
a power-law with photon index of 2.28 and no cut-off.
Over-imposed to the diffuse emission, two bright point sources, HESS J1745$-$245 and 
HESS J1746$-$285, were also taken into account and simulated for self-consistency
\citep{2018A&A...612A...9H}. 
We used {\tt Gammapy} v0.17 for data simulation and analysis, following the same approach discussed for Tycho. We made 100 simulations from the sky-model described above, in the energy range 0.5--199 TeV, adopting the IRF background model.

\begin{figure}
	\begin{center}
	\includegraphics[scale=.7]{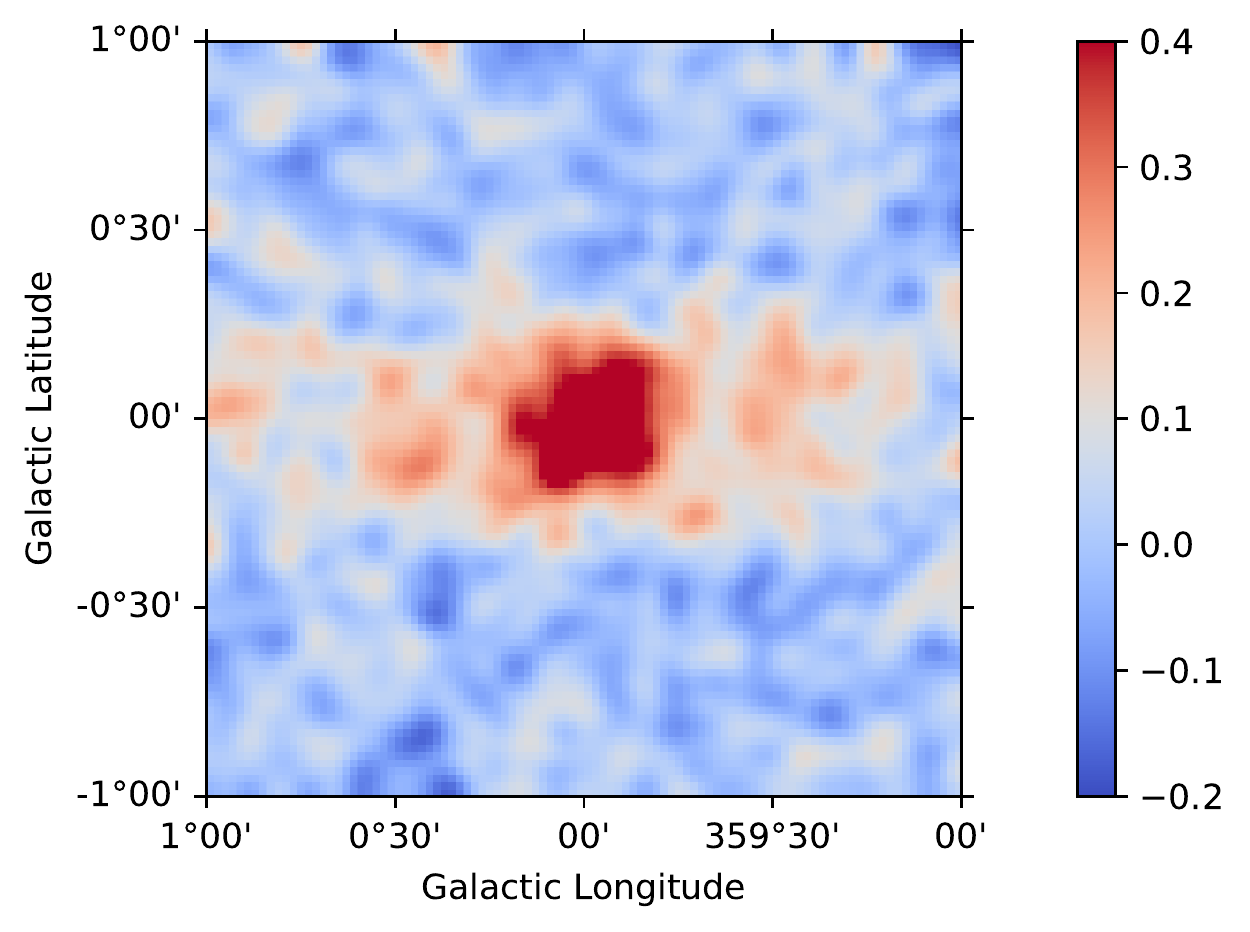}
	\caption{Residual map of the GC above 3 TeV and for 260hr of exposure. The color bar represents the residual per pixel in terms of (data-model)/model units. {The model is given by the best-fit background model, in order to show only the net residuals from the GC.} }
	\label{FIG:GC_residuals}
	\end{center}
\end{figure}

\paragraph{Analysis Method}-- 
We fitted each dataset with the template sky-model, leaving only the spectral parameters of the GC diffuse emission free to vary. The distribution of the 100 best-fit parameters allowed us to estimate the power-law spectral parameters as photon index and differential flux: $\Gamma=(2.27\pm0.04)$ and $N_{\text{1TeV}}=(3.5 \pm0.3)\times10^{-12}$ TeV$^{-1}$ cm$^{-2}$ s$^{-1}$, $\Gamma=(2.27\pm0.03)$ and $N_{\text{1TeV}}=(3.5\pm 0.2)\times10^{-12}$ TeV$^{-1}$ cm$^{-2}$ s$^{-1}$, and $\Gamma=(2.28\pm0.03)$ and $N_{\text{1TeV}}=(3.5\pm0.2)\times10^{-12}$ TeV$^{-1}$ cm$^{-2}$ s$^{-1}$, for 100\,hr, 260\,hr and 500\,hr, respectively. In Fig.~\ref{FIG:GC_residuals} we show a residual map {defined as (data-model)/model, where the model is given only by the best-fit background}, obtained for the 260 hr exposure time and selecting only events above 3 TeV. As expected, the source and its morphology is clearly {detectable} in 260\,hr of exposure.

\begin{figure}
	\centering
		\includegraphics[scale=.5]{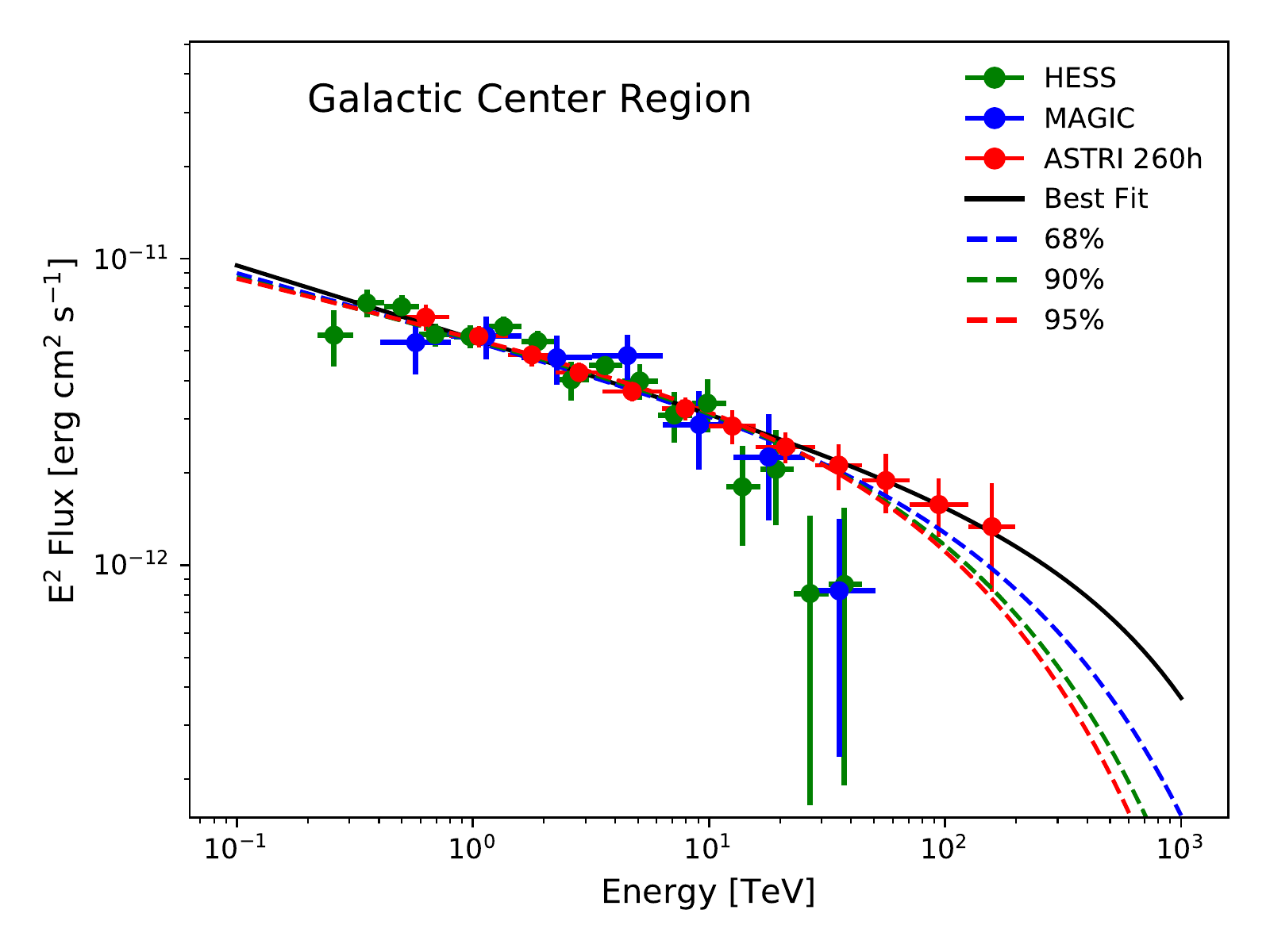}
	\caption{Galactic Center region: HESS (green points), MAGIC (blue points) and ASTRI Mini-Array (260 hr; red points) spectra fitted with a proton population with a best fit cut-off at 120 PeV (black solid line). The blue, green and red solid lines indicates the 68$\%$ (3.50 PeV), 90$\%$ (2.0 PeV) and 95$\%$ (1.7 PeV) confidence level for the cut-off, respectively. The ASTRI Mini-Array spectral points and their error bars are obtained from the distribution of 100 simulations. See text for more details.}
	\label{FIG:GC_bestfit}
\end{figure}

Also in this case, we estimated the cut-off energy of the hadronic population assuming a pure hadronic model as origin of the \gray{} emission. 
We did this by fitting simultaneously the HESS, MAGIC \citep{2020A&A...642A.190M} and ASTRI Mini-Array simulated data with the {\sc naima} package, as we did for Tycho. The best-fit model for 100 hr of ASTRI Mini-Array  estimates an upper limit for the proton spectrum cut-off above $1$ PeV, with a lower limit of 0.40 PeV at 95\% confidence level. The constraint can be improved in 260 hr of ASTRI Mini-Array exposure: 95\% lower limit for a cut-off at 2 PeV. The latter results are shown in Fig.~\ref{FIG:GC_bestfit} and they indicate that the best-fit is mainly guided by the ASTRI Mini-Array data points because of their small error bars, especially at energies above 10 TeV; the same is true also in the remaining spectra discussed in the next sections.
The MAGIC telescope location is very similar to the ASTRI Mini-Array one; consequently, the MAGIC result indicates that high-zenith angle observations are particularly rewarding in this context. They allow larger effective area in the highest energy range, with respect to low zenith angle observations, confirming that even from the Teide site ASTRI Mini-Array will play a first-class role for the understanding of the GC region.
\rf{We note that the use of IRF developed for a fixed zenith angle of 20\textdegree when observing sources at moderate and high zenith-angles ($40^{\circ} \le ZA \le 60^{\circ}$) would not be fully appropriate. The ASTRI Collaboration is developing ad-hoc Monte Carlo productions to be use for moderate and high zenith-angle observations that will be ready in the next months. In the meantime, by considering the public Prod-5 CTA IRFs for the southern array (where SSTs will be deployed) and comparing the performance at different zenith angles, we could estimate that, for energies greater than a few tens of TeV, the use of the IRF developed for ZA=20\textdegree should not be considered optimistic for both the differential sensitivity and the energy resolution. For the angular resolution, instead, a worsening by a factor on the order of 2 is expected, although this should not substantially affect the capabilities of the Mini-Array to provide a significant morphological characterisation of sources observed at high zenith angles. Therefore, the impact on our studies can be considered minimal at this stage.}

\subsubsection{Other Possible PeVatron Candidates}
 \label{Subsec:candidatePeV}

\paragraph{Immediate Objective}--
Among other PeVatron candidates we present here the most promising TeV sources: \\
VER J190\rf{7}$+$062 (MGRO J1908$+$06/eHWC 1907$+$063) and VER J2227$+$608 (SNR G106.3$+$2).

VER~J1907$+$062 was discovered by the MILAGRO collaboration \citep{2007ApJ...664L..91A} which reported VHE emission corresponding to $\sim80\%$ of the Crab Nebula flux at 20\, TeV and an upper limit on the intrinsic source extension of 2.6\textdegree. The VERITAS observations (about 62\,hr of useful exposure) show strong diffuse TeV emission, whose origin is not firmly established because of the complex morphology and the limited angular resolution of the current instrument.
VER J1907$+$062 has been detected also by HAWC (2HWC J1908$+$063) up to $\sim 100$\,TeV with a relatively flat spectrum and no evidence of a cut-off \citep{PhysRevLett.124.021102}. {This source was recently detected by  LHAASO~\citep{2021Natur.594...33C} up to $\sim 500$\, TeV}.
According to \cite{2014ApJ...787..166A}, the emission in the northern region of VER J1907$+$062 has probably a hadronic origin connected to the SNR G40.5$-$0.5.
In this scenario, protons accelerated at the shock front collide with target protons of the surrounding ISM, producing TeV photons via neutral pion decay. From the analysis of the spatial distribution of the $^{12}$CO in the vicinity of the SNR G40.5$-$0.5, \cite{2020MNRAS.491.5732D} found molecular clouds that match the eastern, southern, and western borders of the remnant and partially overlap with peaks of the TeV emission from VER J1907$+$062. Other possible counterparts, such as the PSR J1907+0602 cannot be excluded with the current data \rf{\citep[see also ][]{2021ApJ...913L..33L}}.
Observations at very high energy, with increased spatial resolution with respect to HAWC, can firmly constrain the origin of the emission from the northern region of VER J1907$+$062, thus assessing its PeVatron nature.

VER J2227$+$608~\citep{2009ApJ...703L...6A} is associated with the SNR G106.3$+$2.7 and is a potential target of great interest for two main reasons: it is a SNR  with one of the highest TeV fluxes (5\% in Crab units) and shows a hard spectrum (photon index $\Gamma=2.29$) with no detected cut-off up to 10 TeV.
The remnant G106.3$+$2.7 is extended and shows two compact and close-by regions:  the “head” formed by the SNR shock, which also contains the bright pulsar PSR J2229$+$6114, powering a PWN, and an elongated, fainter, “tail” region, which contains VER J2227$+$608.
It is worth noting that MILAGRO~\citep{2007ApJ...664L..91A,2009ApJ...700L.127A} and, very recently, HAWC~\citep{2020ApJ...896L..29A}, Tibet AS $\gamma$~\citep{2021NatAs.tmp...41T} {and LHAASO~\citep{2021Natur.594...33C}} detected \gray s from the remnant's region (up to $\sim 100$\,TeV in the case of HAWC, Tibet AS~$\gamma$ and LHAASO) although no clear association with a specific region of SNR G106.3$+$2.7 is possible, due to their low angular resolution.
The VHE morphology can be well superimposed on the molecular gas images  as traced by $^{12}$CO radio maps, thus suggesting a potential hadronic origin, even if a leptonic origin is also possible~\citep{2022NewA...9001669Y, 2020ApJ...897L..34L}. 
We show here the main constraints on the spectral shape (close to 100 TeV) which would be obtained with an exposure of $\sim$\,500 hours with the ASTRI Mini-Array.
These observations would firmly constrain the hadronic origin of the VER J2227$+$608 emission. 
\newline
\newline
\noindent{\bf VER~J1907$+$062}
\paragraph{Observing Time, Pointing Strategy, Visibility and Simulation Setup}--
VER~J1907$+$062 is observable from the Teide site for about 400 hours per year with a zenith angle between 0\textdegree  and 45\textdegree and 170 hr per year with a zenith angle between 45\textdegree-60\textdegree, in moonless condition. To better understand what spectral constraints we can achieve within the available range of observing times, we simulate the source for 100, 200, 500 hours of exposure time.
We simulated VER J1907+062 as a diffuse source, with the morphology taken from \cite{2014ApJ...787..166A}. In terms of the spectrum, instead of adopting the VERITAS results, we made the conservative choice of describing it as a power-law with an index of 2.33, reflecting the steepening seen by HAWC above $\sim$ 5 TeV  according to \cite{2017ApJ...843...40A}. 
This spectrum is based on the model of VER~J1907$+$062 described in \cite{2021MNRAS.505.2309C}, made before the LHAASO detection of this source  \citep{2021Natur.594...33C}. 
LHAASO reported a spectrum up to 500\,TeV, with a flux value at 100\,TeV ($\sim 2\times 10^{-12}$ \, erg cm$^{-2}$ \, s) well compatible with our predictions.
The simulations were performed according to the methods described in Sect.~\ref{Sec:ctoolssims}. 

\paragraph{Analysis Method}--

\begin{figure}
	\centering
		\includegraphics[scale=.5]{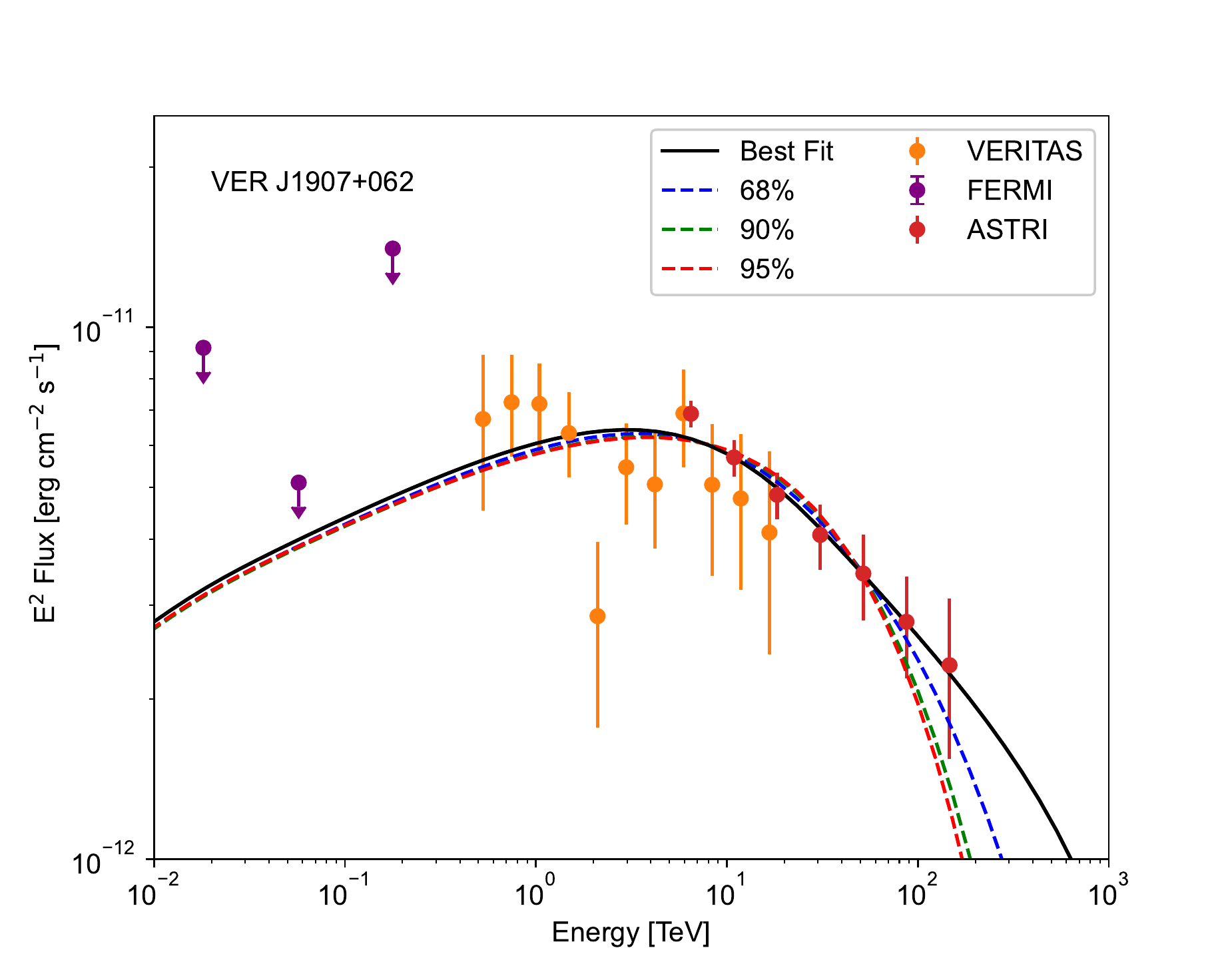}
	\caption{\textbf{VER J1907+062}: \gray\ data from \glast~\citep{2013ApJ...773...77A} (purple dots), VERITAS~\citep{2014ApJ...787..166A} (orange dots) and  ASTRI Mini-Array simulations for 100h of observations (red dots).
	The blue, green and red lines show the broken power law fit with a cut-off energy of 1.67, 0.54 and 0.4\,PeV, corresponding to 68\%, 90\% and 95\% of confidence level respectively. The ASTRI Mini-Array spectral points and their error bars are obtained from the distribution of 100 simulations. See text for more details.}
	\label{FIG:1907_100h}
\end{figure}

We detect the source up to an energy of above 100\,TeV. We then combine the ASTRI Mini-Array simulated data with VERITAS ~\citep{2014ApJ...787..166A} and Fermi-LAT~\citep{2013ApJ...773...77A} observations but we do not include the HAWC data because HAWC tends to measure higher fluxes (a difference of about a factor of two) and larger source extents than the IACTs \citep{PhysRevLett.124.021102}.
We computed the \gray{} spectrum by means of the {\sc naima} package, assuming a proton distribution described by a broken power law with a cut-off. We fixed only the cut-off energy, leaving the other parameters as free.
We compute the profile log-likelihood curve as a function of the cut-off energies. From the curve, we estimated a lower limit on the cut-off energy of the proton population, at different confidence levels, with and without the ASTRI Mini-Array data.
With 100 hours of exposure, the ASTRI Mini-Array data allow us to find lower limits of 1.67, 0.54 and 0.40\,PeV at 68$\%$, 90$\%$ and 95$\%$ confidence levels, respectively. The results are shown in Fig.~\ref{FIG:1907_100h}. 
Increasing the exposure time, we can fix a cut-off at about 1 PeV within the 95\% confidence level. With 200 hour of exposure we obtain 3.95, 1.54 and 0.96\,PeV at 68\%, 90\% and 95\% confidence level, respectively and with 500 hour of exposure similar lower limits of 8.52, 4.31 and 3.58\,PeV at 68\%, 90\% and 95\% confidence level, respectively.

Without the ASTRI Mini-Array data instead, the fit value estimates an UL for the cut-off below 1\,PeV, with a lower limit of $\sim50$ TeV at $95\%$ confidence level. This preliminary analysis shows that the ASTRI Mini-Array will make a substantial contribution to the study of the maximum acceleration of protons in this source.
The ASTRI Mini-Array will also investigate the nature of this strong diffuse emission. Indeed, thanks to its good angular resolution it will be able to resolve the possible counterparts present in the \gray{} error box ($\sim 0.5$\textdegree) and to distinguish among the different contributions. 
\newline
\newline
\noindent{\bf VER J2227$+$608/SNR G106.3$+$2.7}

\paragraph{Observing Time, Pointing Strategy, Visibility and Simulation Setup}--
The source SNR G106.3$+$2.7 is observable at a zenith angle $<45\degmark$ for about 450 hours per year, in moonless conditions. Following the same simulation procedure used for VER~1907$+$062, we simulated SNR G106.3$+$2.7 as an extended elliptical source with an angular extent of 0.27$\degmark$ along the major axis, 0.18$\degmark$ along the minor axis, an orientation angle of 22$\degmark$ East to North. Following~\citet{2009ApJ...703L...6A}, we described the spectrum with a power law with an index of 2.29. We simulated event files with an exposure of 100, 200 and 500 hours. For each exposure, we obtained the spectral energy distribution using ten logarithmically spaced energy bins between 3\,TeV and 200\,TeV.

\begin{figure}
	\centering
		\includegraphics[scale=.5]{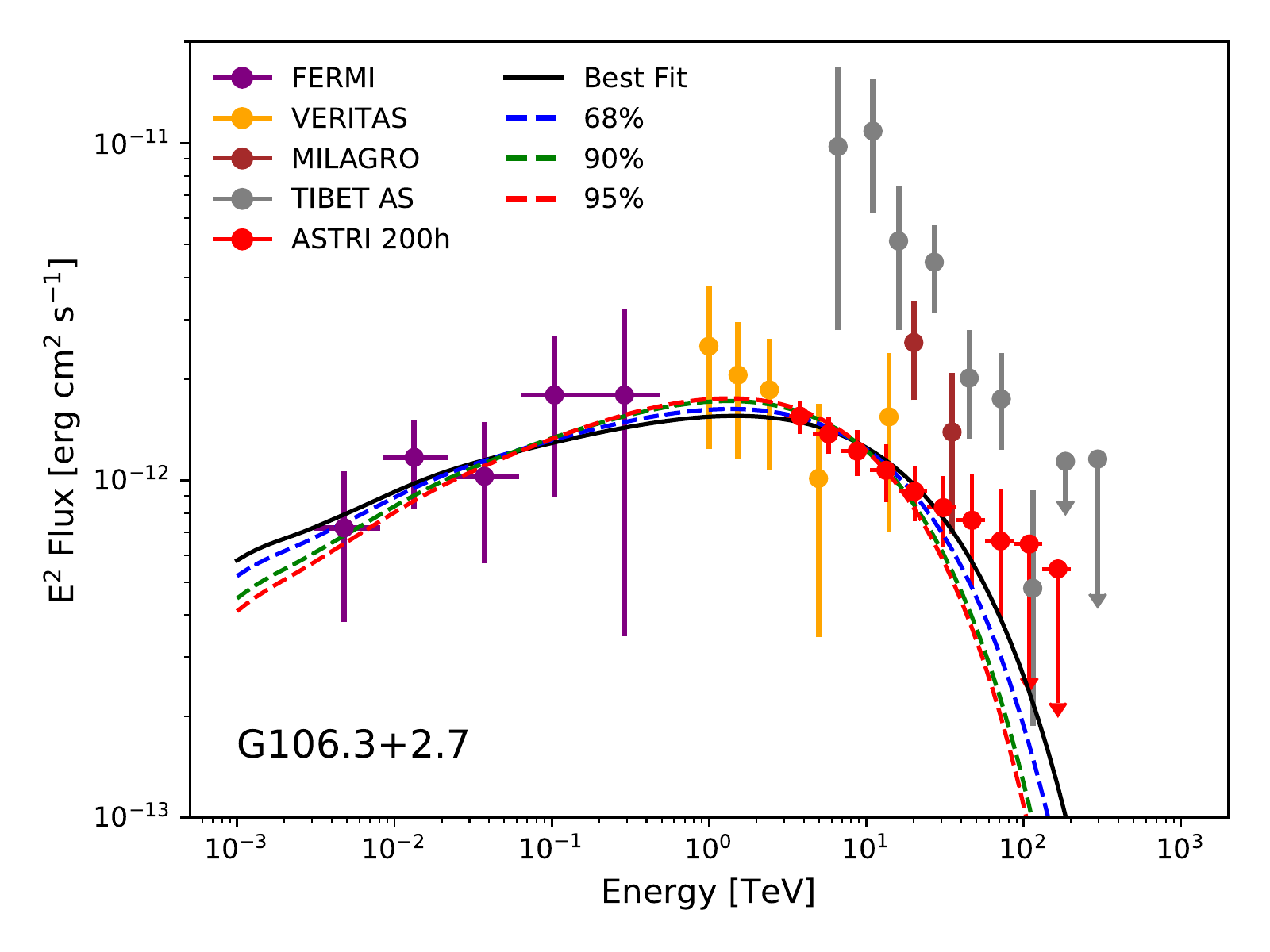}
	\caption{G106.3+2.7: \gray\ data from \glast~\citep{2019ApJ...885..162X} (purple dots), VERITAS~\citep{2009ApJ...703L...6A} (orange dots) and ASTRI Mini-Array (200 hours exposure time) (red dots). Milagro and Tibet AS $\gamma$ data points from~\citet{2007ApJ...664L..91A,2009ApJ...700L.127A,2021NatAs.tmp...41T} are shown for reference only.
	The solid lines show the emission from a proton population with a best-fit cut-off energy of 350 TeV (black line) and lower-limit energy of 250 (blue line), 180 (green line) and 160 (red line) TeV, (corresponding to 68, 90 and 95\% of confidence levels). The ASTRI Mini-Array spectral points and their error bars are obtained from the distribution of 100 simulations. See text for more details.}
	\label{FIG:Chap4_G106}
\end{figure}

\paragraph{Analysis Method}-- 
\rf{The distribution of the 100 best-fit parameters allowed us to estimate the power-law spectral parameters as photon index and differential flux: $\Gamma=(2.37\pm0.10)$ and $N_{\text{3TeV}}=(1.2 \pm 0.2)\times10^{-13}$ TeV$^{-1}$ cm$^{-2}$ s$^{-1}$, $\Gamma=(2.34\pm0.06)$ and $N_{\text{3TeV}}=(1.2\pm 0.1)\times10^{-13}$ TeV$^{-1}$ cm$^{-2}$ s$^{-1}$, and $\Gamma=(2.32\pm0.04)$ and $N_{\text{3TeV}}=(1.2\pm0.1)\times10^{-13}$ TeV$^{-1}$ cm$^{-2}$ s$^{-1}$, for 100\,hr, 200\,hr and 500\,hr, respectively.}
We found that we can detect the SNR up to energies of $\sim$100 TeV with at least 500 hr of exposure. 
To obtain better constraints on the \gray\ emission, we combined both the 500 hours and 200 hours ASTRI Mini-Array data points with published data obtained by \glast~\citep{2019ApJ...885..162X} and VERITAS Collaborations~\citep{2009ApJ...703L...6A}.
In our analysis, we did not consider the MILAGRO, HAWC, Tibet AS$\gamma$ and LHAASO~\citep{2007ApJ...664L..91A,2009ApJ...700L.127A,2020ApJ...896L..29A,2021NatAs.tmp...41T,2021Natur.594...33C} points; in the case of HAWC, MILAGRO and LHAASO because of their unclear spatial association within the SNR/pulsar region and, in the case of Tibet AS$\gamma$, because of the higher observed flux compared to the previous observations.
Using the {\sc naima} package, we computed the expected \gray{} emission assuming a proton population described by a power-law with a cut-off.
We fixed the cut-off energy and considered both the PL index and normalization as free parameters. We then produced a profile log-likelihood curve as a function of different cut-off energies, maximized over the power law index and normalization.
From the curve, we estimated the cut-off lower limit of the proton population, at different confidence levels, with and without the ASTRI Mini-Array data.
With the 200 hours exposure ASTRI Mini-Array data, we obtain a best fit value of the proton cut-off energy of 350\,TeV, with a lower limit of 250, 180, 160\,TeV at 68\%, 90\% and 95\% confidence level respectively (results are showed in Fig.~\ref{FIG:Chap4_G106}). These constraints improve considering an exposure time of 500 hours.  In this case we obtain a best fit value of the proton cut-off energy of 530\,TeV, with a lower limit of 415, 340, 310\,TeV at 68\%, 90\% and 95\% confidence level, respectively. Without the ASTRI Mini-Array data, the fit value of the proton cut-off energy is below 100\,TeV, with a lower limit of $\sim$10 TeV at 95\% confidence level.
As for previous cases our preliminary analysis shows that the ASTRI Mini-Array could provide a fundamental contribution to the study of particle acceleration in this SNR.
The ASTRI Mini-Array will also investigate other potential interesting properties of this source, as its morphology. Thanks to its good angular resolution, the ASTRI Mini-Array will help to firmly distinguish, at energies > 10\,TeV, the emission in the ``head'' region, where the pulsar is located, from the emission in the ``tail'' region, where molecular clouds are located. Another important point to be investigated is a possible energy-dependent morphology for this source. This could explain the differences between VERITAS and Tibet AS$\gamma$ spectra, being the latter steeper than the former, perhaps indicating different morphological characteristics. The ASTRI Mini-Array would help to study the possible implications of such differences.

\subsection{Particle escape and propagation}
\label{Sec:particleProp}
\paragraph{Scientific Case}--
Understanding the escape of accelerated particles from expanding spherical shocks is a key ingredient to establish a connection between SNRs and the origin of Galactic CRs. It is often assumed that the spectrum of particles released into the Galaxy by a single SNR resembles the instantaneous spectrum of particles accelerated at the shock. However, such an assumption is based on several subtleties of the shock acceleration theory, like the magnetic field amplification and the temporal behaviour of the acceleration efficiency, which are still active subjects of research \cite[see, e.g.][]{2011MNRAS.415.1807D,2013A&ARv..21...70B,2014IJMPD..2330013A}.

The escape process is tightly connected with the maximum energy achievable by the shocks. In fact, reaching PeV energies requires a strong amplification of the magnetic field: this is needed to ensure efficient scattering and reduce the acceleration timescale.
One of the most efficient mechanisms to amplify the magnetic field up to hundreds of $\mu$G is the non-resonant streaming instability \citep{ 2004MNRAS.353..550B} which, however, requires a current of escaping particles to be triggered. Hence, a fraction of the particles at the highest energies are required to escape at every moment. As a consequence, looking for signatures of the escape process can shed light on the magnetic field amplification mechanism and, in turn, on the maximum attainable energies.
One distinctive feature of the escape process is the presence of a break in the spectrum of CRs confined inside the SNR, with a steepening just above the current maximum energy  \citep{2019MNRAS.490.4317C,2020A&A...634A..59B}. The slope above the break is related to two main factors: the time dependence of magnetic field amplification and the diffusion coefficient immediately around the SNR. The peculiar shape of the spectrum of the accelerated particles is directly reflected into a break in the \gray{} spectrum produced by hadronic collisions. There are already indications of the presence of such a break in several SNRs \cite[see][for a summary]{ 2019ApJ...874...50Z}, but it is often difficult to discriminate between a break and a cut-off, the latter being rather connected to the lack of ongoing escape and/or to a fast diffusion out of the source. The discrimination can be achieved only through the observation of the \gray{} spectrum at energies much larger than that of the break.

According to the theoretical models invoking the non-resonant streaming instability \citep{2014MNRAS.437.2802S,2015APh....69....1C,2020APh...12302492C}, $\sim$ PeV energies can be achieved only at very early times ($\lesssim 100$ yr) during the evolution of a SNR resulting from a core-collapse event, namely when the SN blast wave is propagating inside the dense wind of the progenitor with a very large speed ($\sim 10^4$ km s$^{-1}$). Hence, in the majority of known SNRs, the highest energy particles are expected to have left the source already.
However, if the diffusion coefficient immediately around the source is small enough to confine the high energy particles for a long time, the chance to detect their presence is still open by looking at molecular clouds nearby young SNRs. 
The detection of high energy \gray{} emission from those clouds allows us to put lower limits on the maximum energies achieved in the past and also to put constraints on the diffusion coefficient around the SNR \citep{2009MNRAS.396.1629G, 2011MNRAS.410.1577O, 2018ApJ...868...36X, 2011ApJ...731...35Y}

The aim of this study is to constrain the time dependence of the escape process, as well as the diffusion coefficient around SNRs, taking advantage of the high sensitivity of the ASTRI Mini-Array at energies above 10 TeV. We have chosen two promising SNRs to perform complementary studies, namely $\gamma$-Cygni and W28. Both sources are middle-aged SNRs that show signatures of the escape process.

%
\subsubsection{$\gamma$-Cygni}
 \label{Subsec:gammaCyg}
\paragraph{Immediate Objective}--
%
The $\gamma$-Cygni SNR (G78.2+2.1) is a middle-aged SNR located in the Cygnus region. Its age has been estimated to be between 5 kyr and 10 kyr while its distance from Earth between 1.7 kpc and 2.6 kpc \citep{2013MNRAS.436..968L}.
This SNR has been detected both in the GeV and the TeV energy range. The GeV emission has two components: an extended ‘disk’ coincident with the radio shell plus a hot-spot in the north-western quadrant of the remnant \cite[see e.g.][]{2016ApJ...826...31F}. In the VHE regime, VERITAS reported an extended emission from the north-western hot-spot \citep{2013ApJ...770...93A} with a size of $\sim 0.25^{\circ}$ and a flux $\sim 3.7\%$ of the Crab Nebula one in the same energy range. 
Also HAWC observed this source, reporting a significant detection for the ‘disk’ component up to $\sim 10$\,TeV, but without a preference for including the hot-spot. However, the low HAWC angular resolution does not allow one to drive firm conclusions on the spatial structure.

Combining GeV and TeV emission, the whole spectrum shows a break at $\sim 300$ GeV  changing from $\sim E^{-2}$, at lower energies, to $\sim E^{-2.4}$ at higher ones. 
The emission from the hot-spot could be due to  particles interacting with a dense cloud or a shell swept up by the SN progenitor's wind and located just in front of the forward shock.
Given the low estimated shock speed ($\sim 700$ km s$^{-1}$) any hadronic TeV \gray{} emission would have to arise from particles accelerated during an earlier epoch that escaped the SNR and are only now interacting with the shell.
This picture is strengthened by the MAGIC detection \citep{2017ICRC...35..685S}, where, in addition to the disk and the hot-spot, a third region has been revealed and located just outside of the disk with an arc-shaped geometry in the western direction~\citep[see also ][for a morphological study of $\gamma$-Cygni with {\it Fermi}-LAT and MAGIC]{2020arXiv201015854M}. The \gray{} emission from this region could also be due to escaping particles interacting with a lower density medium with respect to the hot-spot. 

In the escape scenario, the shape of the spectrum above the break mainly depends on three parameters: the time dependence of the maximum energy, its highest value reached in the past and the diffusion coefficient in the circum-stellar medium.
Detection of \gray{} emission at energies above $\sim 10$ TeV can constrain these parameters.

\paragraph{Observing Time, Pointing Strategy, Visibility and Simulation Setup}--
$\gamma$-Cygni is observable from the Teide site for about 460 hr per year in the zenit angle interval $[0^\circ-45^{\circ}]$ plus additional 160\,hr for $[45^{\circ}-60^{\circ}]$, in moonless conditions.

Having in mind the escape scenario outlined in the previous section, we assumed that the \gray{} emission is of purely hadronic origin, and we modelled $\gamma$-Cygni using the framework  developed by \cite{2019MNRAS.490.4317C} which accounts for particle acceleration and escape during the Sedov-Taylor phase. In this model the instantaneous accelerated spectrum is
\begin{equation}
\label{EQ:Chap5_Eq_dNdE}
\text{dN/dE} \propto E^{-2} e^{-E/E_{\max}(t)},
\end{equation}
where the maximum energy decreases with time as 
\begin{equation}
\label{EQ:Chap5_Eq_Emax(t)}
E_{\max}(t)= E_{\max,0} (t/t_{\rm Sed})^{-\delta},
\end{equation}
$E_{\max,0}$ being the maximum energy reached by protons at the beginning of the Sedov-Taylor phase, $t_{\rm Sed}$. We assume that the SNR evolves into a uniform medium with density $0.2$ cm$^{-3}$ and has age and distance equal to 7\,kyr and $1.7$\,kpc, respectively.
In addition, the diffusion coefficient in the circumstellar medium is taken proportional to the average Galactic one, i.e. $D(E)= \xi D_{\rm gal}(E)$, where $D_{\rm gal}(E) = 3.6 \times 10^{28} E_{\rm GeV}^{1/3}\, \rm cm^2\, s^{-1}$.
Concerning the extension of the emitting region, in order to be consistent with the VERITAS detection, we only consider the hot-spot, simulated as a spherical structure with uniform surface brightness and size of $0.25^{\circ}$. To fit the observed flux, we assume that this region has a density of $\sim 10\, \rm cm^{-3}$, hence 50 times denser than the average circum-stellar region.

In order to show the discrimination power of the ASTRI Mini-Array, we consider two different models, both compatible with the existing data. For model A, we fix $E_{\max}= 500$\,TeV, $\delta= 4.2$ and $\xi=0.02$, while model B has $E_{\max}= 50$\,TeV, $\delta=2.7$ and $\xi=0.013$. We have used {\tt Gammapy} v0.17 for data simulation and analysis averaging 100 realizations of observations with exposure time of 200 hr each.

\begin{figure}
	\centering
	    {\includegraphics[scale=.55]{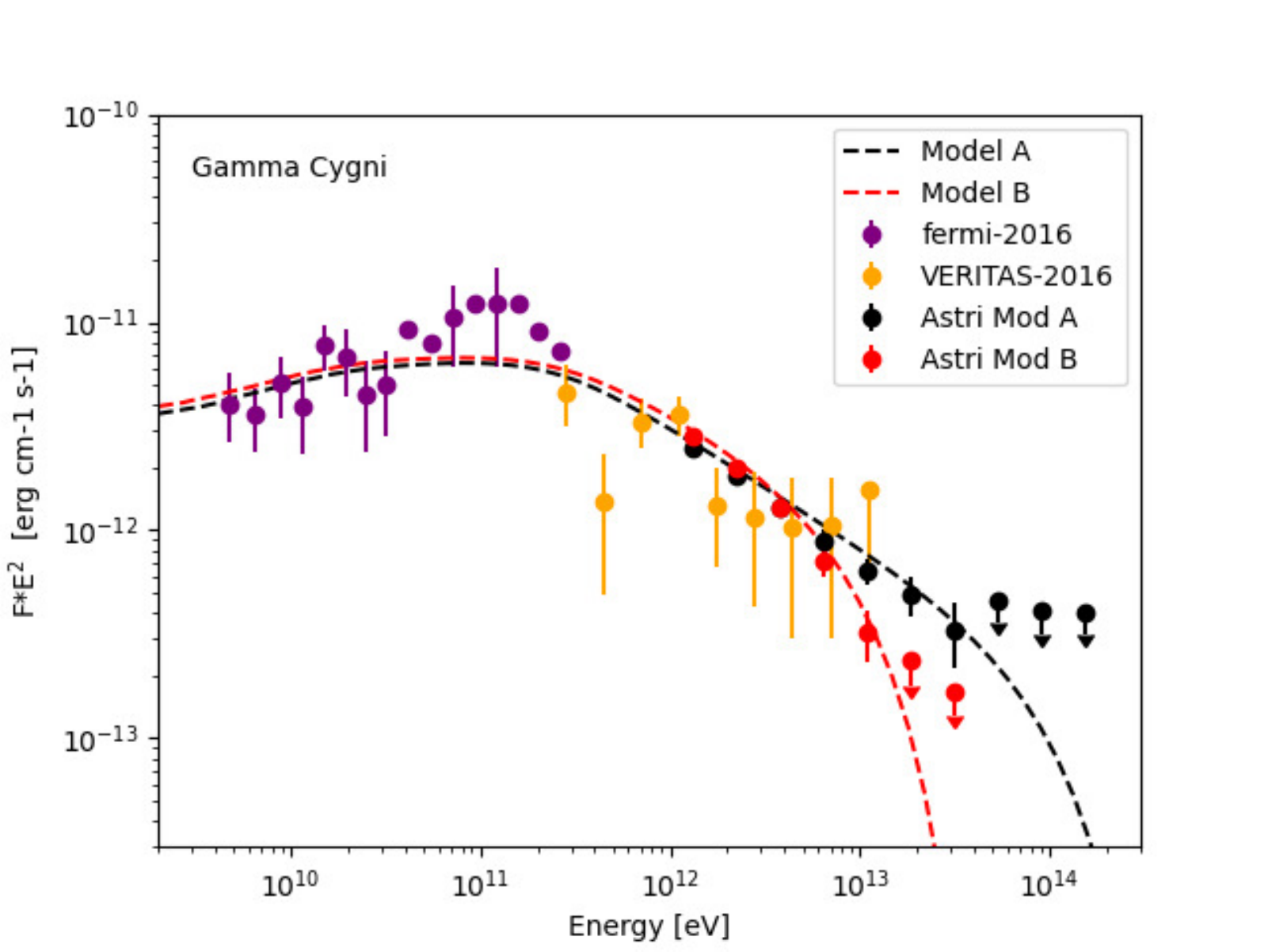}}
	\caption{\gray{} spectrum of $\gamma$-Cygni. Data are from Fermi-LAT and VERITAS while theoretical models A and B (described in the text) are showed with dashed curves. Black and red dots show the ASTRI Mini-Array simulations for model A and B, respectively, for 200 hr of exposure.}
	\label{FIG:gamma-Cygni_spectrum}
\end{figure}

\paragraph{Analysis Method}--
In Figure~\ref{FIG:gamma-Cygni_spectrum}, we compare the two theoretical models described above with existing Fermi-LAT and VERITAS data taken from \cite{2013ApJ...770...93A} plus the simulated ASTRI Mini-Array data for both models. 
The detection has a significance of $\sim 15 \sigma$ for model A and $\sim 12 \sigma$ for model B. For model A the measured flux above 1 TeV is $1.11 \times 10^{-12}$   cm$^{-2}$ s$^{-1}$. 
Figure~\ref{FIG:gamma-Cygni_spectrum} shows that above $\sim 10$ TeV the detection can discriminate between the two models: $E_{\max,0}$ can be well constrained in Model B while only a lower limit can be derived for model A. The latter can be better constrained increasing the exposure time up to 500 hr which is, hovewer, not necessary to estimate $\delta$ and $\xi$.
We notice that existing data already point towards $\xi < 1$, underlining that local diffusion coefficient should be suppressed with respect to $D_{\rm gal}$. 
We repeated the analysis also using {\tt ctools}, obtaining similar results.

\subsubsection{W28}
 \label{Subsec:W28}
%
\paragraph{Immediate Objective}--
%
W28 is a SNR surrounded by at least three molecular clouds, located at a projected distance of 10-20\,pc from the SNR shell and detected by H.E.S.S up to several TeV \citep{2008A&A...481..401A}.
The \gray{} emission from these clouds has been successfully interpreted as produced by those CRs that have escaped the SNR shell at some point of its evolution, and have reached the cloud after diffusing (with an energy-dependent diffusion coefficient) in the surrounding medium.
Studies performed within this scenario have pointed out the need to invoke a relatively low value of the CR diffusion coefficient $D(E)$ as compared to the average Galactic value $D_{\rm gal}(E)$ \citep{2009MNRAS.396.1629G,2011MNRAS.410.1577O}. A suppression of the diffusion may indicate that a self-confinement mechanism, such as streaming instability, is at work \citep{1970MNRAS.147....1S,2008AdSpR..42..486P,2016MNRAS.461.3552N,2016PhRvD..94h3003D,2019MNRAS.484.2684N}. 
The level of suppression $\xi\equiv D(E)/D_{\rm gal}(E)$ inferred from observations is in the range $\xi = 0.05-0.1$ when CRs are assumed to diffuse isotropically around the SNR \citep{2009MNRAS.396.1629G}, while a larger value is inferred for an anisotropic geometry \citep{2013MNRAS.429.1643N}.
The energy dependencies of the escape process, diffusion, and possible self-confinement mechanisms are expected to produce cut-offs and/or breaks in the \gray{} spectrum, and affect the spectral slope.
Surrounded by MCs, likely located at different distances, W28 represents a perfect candidate for the study of particle escape, diffusion around SNRs, and particle-wave interactions.

Here we study how observations from the ASTRI Mini-Array can be used to shed a light on the properties of the diffusion and then on mechanisms regulating the diffusion itself in the environment surrounding SNRs.
In spite of the non-optimal observing conditions, the results show that observations with the ASTRI Mini-Array can extend the knowledge of this source to much higher energies as compared to currently available observations and constrain the high-energy part of the spectrum.

\paragraph{Observing Time, Pointing Strategy, Visibility and Simulation Setup}--
%
The region of W28 is observable by the ASTRI Mini-Array only at zenith angles larger than 45$^\circ$, for about 300\,hr per year in moonless conditions. Similarly to the case of $\gamma$-Cygni, we assume that the maximum energy decreases with time as $E_{\max}= E_{\max,0} (t/t_{\rm Sed})^{-\delta}$ and that particles (accelerated with a spectrum $\text{dN/dE}\propto E^{-2}$) diffuse in the surrounding medium with an energy dependent diffusion coefficient $D(E)\propto E^{\rm s}$. We adopt the value $\delta=4$ and fit Fermi-LAT and H.E.S.S. observations of the different clouds surrounding W28 by varying the value of the parameter $s$. 
Current observations give large uncertainties in the value of the spectral index of the \gray\ emission, and hence on the energy dependence of the diffusion process, escape mechanism, and  spectrum of the accelerated particles.
To test the capability of ASTRI Mini-Array in reducing the uncertainties by constraining the slope of the \gray\ spectrum we consider the brightest MCs in the vicinity of W28 (HESS~J1800-240B) and predict its \gray{} emission over a broad range of energies (from GeV to $\gtrsim$\,100\,TeV) under different assumptions for the energy dependence of the diffusion coefficient.
In particular, we consider two extreme cases: $s=0.35$ (model A, consistent with predictions from a Kolmogorov-like turbulence spectrum) and $s=0.50$ (model B, consistent with a Kraichnan-like turbulence spectrum).
When tested against currently available Fermi-LAT and H.E.S.S. observations, both models are acceptable, but predict a different flux at energies $>$\,1\,TeV: we perform simulations to test the possibility to discriminate among the two different scenarios with observations by the ASTRI Mini-Array, and constrain whether the background turbulence responsible for the diffusion in the ISM surrounding the remnant has a Kolmogorov-like or a Kraichnan-like spectrum.

\begin{figure}
	\centering
	\includegraphics[scale=.48]{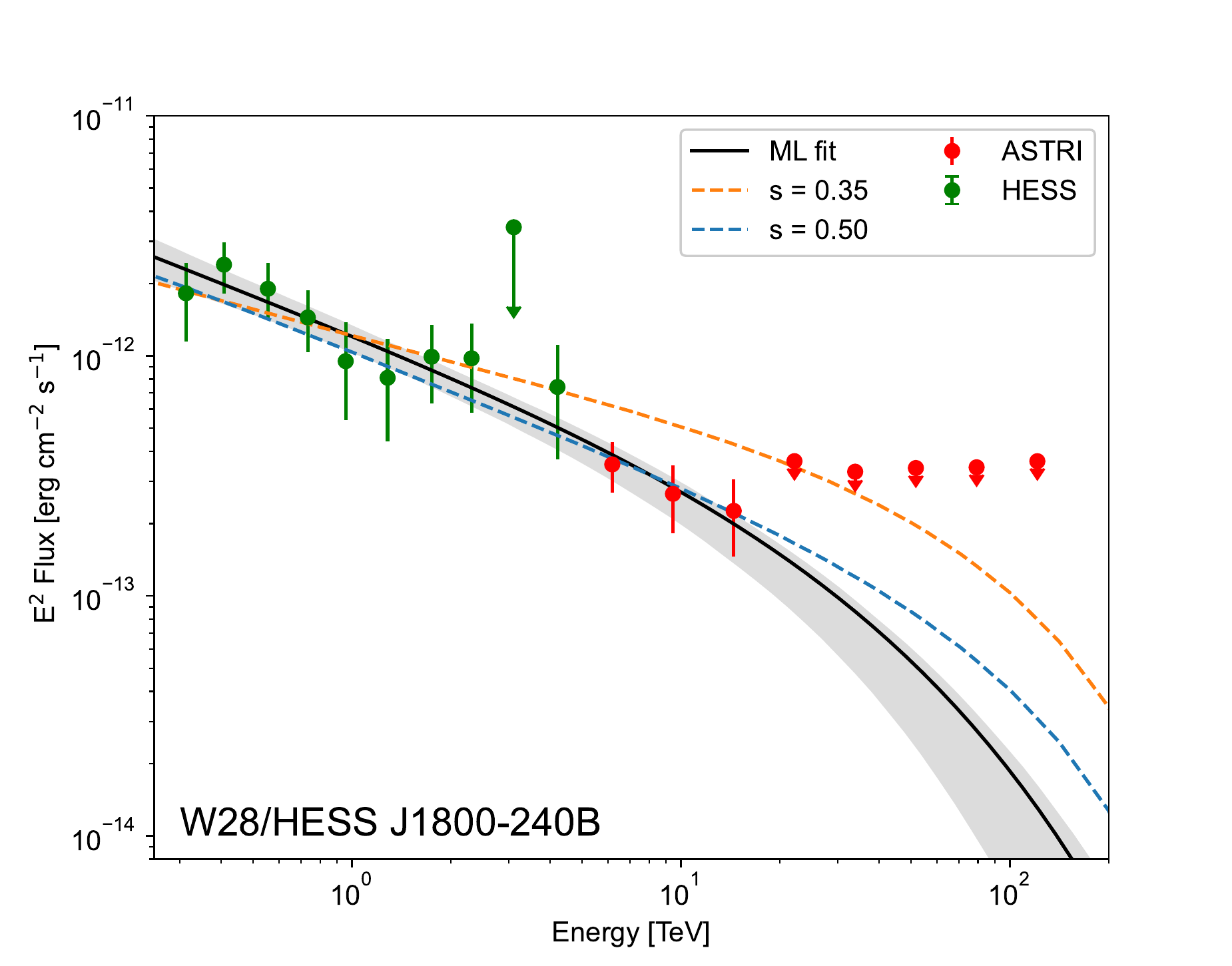}
	\caption{\gray{} spectrum from the source HESS~J1800-240B, coincident with one of the MCs in the vicinity of W28. Data are from H.E.S.S. (green dots) and ASTRI Mini-Array simulations (red dots) for 200hr of exposure for model B ($s=0.5$, blue dashed line). The best fit and its 1$\sigma$ uncertainty are represented by the solid black line and the shaded gray region. The orange dashed line shows the predicted spectrum for the model with $s=0.35$.}
	\label{FIG:W28}
\end{figure}

We simulated and analyzed a dataset of an ASTRI Mini-Array observation toward the W28 region using the {\tt Gammapy} package (v0.17).
We focused our analysis on the source HESS J1800-240B which was modeled as a point source with spectrum described by the model B (obtained with $s=0.5$, dashed blue curve in Fig.~\ref{FIG:W28}), predicting a lower flux in the ASTRI Mini-Array energy range as compared to the model with $s=0.35$ (orange dashed curve). 
Simulations have been repeated 100 times, for an exposure of 200\,hours (red points in Fig.~\ref{FIG:W28}). 

%

\paragraph{Analysis Method}--
%
Using the {\sc naima} package we fitted simultaneously H.E.S.S. observations (green points) and the simulated ASTRI Mini-Array spectrum. We assumed a proton distribution described by power-law with exponential cut-off. All parameters are considered free. The best fit and the 1$\sigma$ contour region are shown with a solid black line and a grey shaded area, respectively. 
A \gray{} spectrum produced by particles diffusing with $s=0.35$ (orange dashed curve) is inconsistent with the simulated data, demonstrating that ASTRI Mini-Array observations may allow to clearly discriminate among different spectral indices and then different energy dependencies of the diffusion coefficient in the vicinity of W28. We also remark that the angular separation between those clouds is $\lesssim 0.5^\circ$ hence a proper analysis can only be performed with an angular resolution much better than this value, a requirements perfectly fulfilled by the \astrima{}.
%
\subsection{High energy emission from Pulsar Wind Nebulae}
\label{Sec:PWNe}
%
%
\paragraph{Scientific Case}--
Pulsar Wind Nebulae (PWNe) are the most numerous class of identified Galactic sources of VHE photons. As such, they will be a primary target of future $\gamma$-ray surveys \citep{2013APh....43..287D}. A second category of sources, closely related to PWNe, is expected to be dominant in the sky: the recently discovered, but quickly growing, class of objects named "TeV halos" \citep{2019PhRvD.100d3016S}. 
These are the two main classes of galactic leptonic emitters that we address in the following. 

PWNe are produced by the interaction of the wind emanating from a fast-spinning magnetized neutron star (the pulsar) with the surroundings (the debris of the supernova explosion or the ISM).
The pulsar wind (PW) is highly relativistic, magnetized and cold, mainly composed by electron-positron pairs produced by electromagnetic cascades in the pulsar magnetosphere. 
It is terminated by a strong shock that dissipates the flow kinetic energy. Efficient particle acceleration, up to very high energies, seems a universal feature of the dissipation process: in the Crab nebula -- the PWNe prototype -- we directly observe PeV pairs and $\sim 20\%$ efficiency of conversion of the pulsar spin-down energy into synchrotron radiation \citep{2008ARA&A..46..127H}. 

The study of their very broad non-thermal spectrum, extending from radio to $\gamma$-rays, is the main diagnostic of PWNe physics. 
Leptons are bound to be the dominant component of the PW by number, while the possible presence of hadrons has been so far only loosely constrained by observations. 
Hadrons, in principle, could be present and even energetically dominant in the wind, in spite of being a minority by number \citep{2012SSRv..173..341A,2019hepr.confE..33A}. This possibility, being one of the fundamental question related to the physics of pulsars and PWNe, might require a complete change in our understanding of the physics of PWNe \citep{2006ApJ...653..325A} and also in the origin of CRs. PW ions would be hadrons with energies in the 100 TeV--PeV range, making PWNe possible contributors of CRs in this energy range, which SNRs strive to explain \citep{2015APh....69....1C, 2020MNRAS.494.2760C}. 
In addition the detection of ions in PWs would lend support to the theories that invoke young magnetars as the main sources of UHECRs \citep{2015JCAP...08..026K}. 
The presence of hadrons could also provide a viable explanation for the still mysterious acceleration mechanism that guarantees such efficient acceleration of the leptonic component of the PW, through resonant cyclotron absorption by the pairs of the cyclotron waves emitted by the ions at the crossing of the termination shock \citep{2006ApJ...653..325A}. Acceleration by fast magnetic reconnection driven by turbulence and instabilities also arises as a potential mechanism in these highly magnetized systems \citep[e.g.,][]{2012ApJ...746..148C}. 

Relativistic hadrons are highly elusive, because they can only be revealed through the by-products of nuclear collisions, namely neutrinos and VHE $\gamma$-rays. 
While neutrino telescopes are still not sensitive enough to put stringent constraints on the hadronic content of the PW \citep{2014IJMPS..2860160A, 2017ApJ...836..159D, 2020ApJ...898..117A}, observations of PWNe in the TeV range are completely dominated by the bright inverse-Compton Scattering (ICS) emission of these objects. 
Observations in the 100 TeV range -- where IC scattering becomes suppressed by Klein-Nishina effects -- open for the first time the possibility of finding direct evidence of hadrons. The availability of high sensitivity data beyond 100 TeV promises to put the most stringent constraints ever on the hadronic content of the PW.

The nature of the PW and its composition are also naturally connected with the fundamental role of PWNe as antimatter factories in the Galaxy, being the most likely astrophysical candidates to explain the positron excess measured in the CR spectrum \citep{2009Natur.458..607A, 2013PhRvL.110n1102A, 2011ASSP...21..624B, 2018AdSpR..62.2731A}. 
Additional constraints on this subject come from the recent detection of extended TeV halos around evolved PWNe. 
First detected around the Geminga PWN \citep{2017Sci...358..911A, 2017PhRvD..96j3016L}, TeV halos extend in a region well beyond the PWN size, indicating that particles responsible for TeV emission have escaped from the nebular boundary and then diffuse in the surroundings. 
Even if \rf{they are} very large compared with the size of the PWN, TeV halos are too small to be compatible with the average Galactic diffusion coefficient, indicating that diffusion is effectively suppressed in the vicinity of the pulsar \citep{2018PhRvD..98f3017E, 2019ApJ...884..124F}. 
Such a suppression implies an enhanced scattering efficiency in the PWN surroundings, either due to a locally reduced coherence scale of the galactic magnetic field \citep{Giacinti2018} or to an enhanced turbulence level, possibly induced by the streaming of accelerated particles away from their source, either the parent SNR or the PWN \citep{Olmi2019}. 
The study of TeV halos, including their spectral properties and spatial profile, is then the best way to understand the efficiency of particle escape from PWNe and the properties of particle transport in their surroundings, with direct implications on their ability to end up in the general CR galactic pool and contribute to the positron excess detected at Earth.

%
\subsubsection{Crab nebula} \label{Subsec:Crab}
\paragraph{Immediate Objective}--
The Crab nebula is the perfect target within the PWNe class to investigate whether ASTRI Mini-Array can help gain insight in the PW composition. 
The Crab is one of the best studied astrophysical objects in the sky: {its flux, morphology and even variability have} been observed in great detail at all wavelengths.
{In the last years water Cherenkov experiments have become able to investigate the Crab spectrum at energies $\gtrsim 150$ TeV} \citep{2019PhRvL.123e1101A,2021ChPhC..45b5002A},{well beyond the operational range of H.E.S.S. and MAGIC.} and approximately the point at which the possible hadronic component is expected to become detectable \citep{2003A&A...402..827A}.
{Recent data from LHAASO \citep{2021Science373..425C} revealed the Crab spectrum at PeV energies for the first time, and seem to suggest the presence of interesting features in the PeV range.}

In {one-zone models of the Crab nebula spectrum \citep{2009ApJ...703.2051G,2011MNRAS.410..381B,2014JHEAp...1...31T},} leptons are usually assumed to carry $\sim95-98\%$ of the spin-down luminosity of the pulsar, with the rest going into magnetic field. The wind luminosity fraction carried by particles can be written as $\chi=(1-\eta)$, with $\eta=L_B(t)/L(t)$ the magnetic fraction, $L_B(t)$ the magnetic power and $L(t)$ the pulsar spin-down luminosity.  The magnetic fraction is then connected with the wind magnetization $\sigma$ as $\eta=\sigma/(1+\sigma)$.
If the wind contains both hadrons and leptons, the particle luminosity fraction can be written as: $\chi=\chi_e + \chi_p$, with $\chi_e$ being the spin-down luminosity fraction carried by leptons and $\chi_p$ that carried by protons. The energy distribution of the hadronic component is modeled as a Dirac $\delta$ centered on the proton energy $E_p= m_p c^2 \gamma$ \citep{2006ApJ...653..325A}, with the Lorentz factor varying from the lowest predicted value for the wind ($\gamma_{\rm min}\approx10^4$) up to the maximum available energy associated with the pulsar potential drop {$\gamma_{\rm max}=e/(2m_p c^2)\sqrt{3\dot{E}/2c}$}. 
%
\paragraph{Observing Time, Pointing Strategy, Visibility and Simulation Setup}--
The Crab nebula is observable at a zenith angle < 45° for about 470 hours per year, in moonless conditions.
It will be largely observed as a calibrator, being expected to be highly stable in the TeV range. Thus a few hundreds hours of observations per year can be easily reached, making it the perfect target for the proposed investigation. 

We test different assumptions on the wind properties by simulating the Crab nebula spectrum expected for different amounts of protons in the wind (i.e. varying $\chi_p$) and different wind Lorentz factors.
The maximum value of $\chi_p$ is fixed by the requirement that the combination of $\chi_e$ and $\eta$ is still such as to reproduce the nebular synchrotron and ICS emission.

The models needed as input for the simulations were computed using the GAMERA software \citep{2015ICRC...34..917H} and as implemented in \cite{2020MNRAS.499.3494F}, with the sole difference of having added a hadronic component to the PW as described before.

For the simulations of the ASTRI Mini-Array data of the Crab nebula we made use of the software \textit{ctools}. 
The source morphology and spectrum for the different scenarios were given as input, together with the expected background contamination from cosmic rays (IRF background).
Our spatial template was, in all cases, a point source model centered on the best-fit position reported by \cite{2020NatAs...4..167H}. 
As spectral templates, we used different models corresponding to different values of $\chi_p$ and $\gamma$, as already discussed.
We simulated the source for 100 hr, 200 hr, and 500 hr to find the minimum observation time required to obtain results of sufficiently good quality above $50$ TeV.
For each spectral model considered, we simulated data for 100 realizations, to ensure the robustness of the results. 
Final spectral points are then computed as the average values.
\begin{figure*}
\centering
	\includegraphics[width=0.99\textwidth]{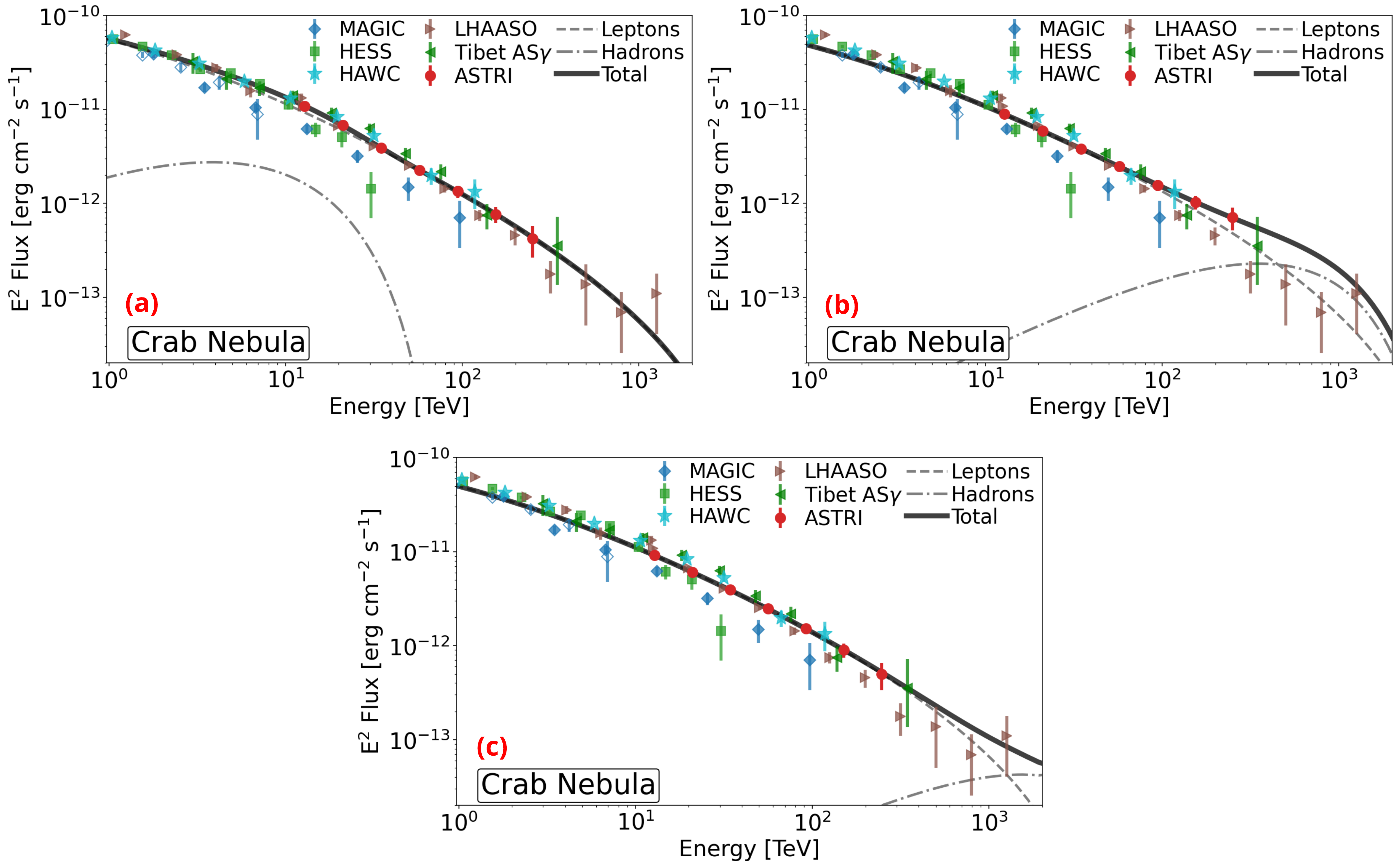}
	\caption{
	\textbf{(a):} 
    spectrum of the Crab Nebula modelled with a hadronic component characterized by $\gamma=10^5$ and $\chi_p=15\%$;
	\textbf{(b)}: the hadronic component here corresponds to $\gamma=5\times 10^6$ and $\chi_p=4\%$; 
	\textbf{(c):} the hadronic component here corresponds to $\gamma=2\times 10^7$ and $\chi_p=4\%$  
	(see text for more details).
	Grey lines indicate the leptonic (dashed) and hadronic (dot-dashed) components, black line shows the total emission. 
	Data from different instruments are shown with various symbols/colors, namely: Blue diamonds for MAGIC (void from \citealt{2008ApJ...674.1037A}, filled from \citealt{2020A&A...635A.158M}); Green squares for HESS \citep{2006A&A...457..899A}; Cyan stars for HAWC  \citep{2019ApJ...881..134A}; Brown triangles for LHAASO \citep{2021ChPhC..45b5002A, 2021Science373..425C}; Pink triangles for Tibet AS$\gamma$ \citep{2019PhRvL.123e1101A}. The simulated ASTRI Mini-Array data (500 hours) are instead shown as red circles.}
\label{fig:pwneC_Crab1}
\end{figure*}
\paragraph{Analysis Method}--
%
As an input for our analyses, we have not considered a simple analytical model (e.g. a power-law) as done for other sources. We performed an un-binned maximum likelihood analysis on the simulated data with \textit{ctools} to obtain the best-fit models and extract the spectral data points. The final spectral data are logarithmically spaced in 7 energy bins.
We made use of a specific IRF extending up to 316.2 TeV, optimized for the particular case of the Crab nebula.
As reported in Table \ref{tab:sources-Pillar1-Emax}, we found that already with 100 hr the source can be detected above $\sim75$ TeV in all the simulated cases, while with 500 hr it can be detected up to the last energy bin, centered at $\sim250$ TeV.

In Fig.~\ref{fig:pwneC_Crab1}, we show {three simulations} of the Crab nebula spectrum at energies in the 10--316.2 TeV range, with different assumptions for the proton injection energy and fraction, and a simulated observation time of 500 hr. 
{Plot (a) shows the spectrum for the case $\gamma=10^5$ and $\chi_p=15\%$: in this case, the hadronic component, peaking below 10 TeV, is basically hidden by the leptonic spectrum within the data error bars. The leptonic component alone can very well reproduce the measurements by HAWC \citep{2019ApJ...881..134A}, Tibet AS$\gamma$ \citep{2019PhRvL.123e1101A} and LHAASO \citep{2021ChPhC..45b5002A} in the 1-400 TeV range.} 
{In plot (b) we show the spectrum for the case with } $\gamma=5\times10^6$, very close to the value originally inferred by 1D magneto-hydro-dynamical models \citep{1984ApJ...283..694K} and consistent with a value of the pulsar multiplicity  -- the amount of pair production in the pulsar magnetosphere --  $\kappa\approx 10^4$, as required by optical/X-ray synchrotron emission modelling \citep{2014MNRAS.438.1518O, 2019hepr.confE..33A}.
{Here the proton fraction is lowered to $\chi_p\simeq4\%$. In this case the overall spectrum is compatible with the highest energy data point by Tibet AS$\gamma$ and LHAASO, while LHAASO measurements in the 0.2-0.9 PeV range are overpredicted.}
{Finally, in plot (c) we increase the protons Lorentz factor to a value very close to the maximum available, namely $\gamma=2\times10^7$. The proton fraction is kept as $\chi_p\simeq4\%$. In this case the model spectrum is compatible with all the available data.
All three plots highlight the excellent performance expected by the \astrima{} (red symbols): the input spectrum is always recovered with very high accuracy with 500 hr of observations.
The \astrima{} will be the only IACT operating in the northern sky able to provide {an energy resolution of 10-15\% and angular resolution of few arcmin up to E$\sim$100\,TeV.} {The panels in Fig.~\ref{fig:pwneC_Crab1} show that this goal will be achieved with a relatively modest time investment, for a whole range of plausible spectral shapes}. Therefore, the possibility of assessing potential systematic differences between particle shower arrays and imaging atmospheric Cherenkov telescopes critically depends on their performance in terms of precision and accuracy. Being the Crab the primary calibration source in the multi-TeV range, verifying that its spectrum is accurately recovered is of fundamental importance. 

Aside from calibration purposes, the accurate recovery with high precision of the Crab spectrum at $\sim 100$ TeV opens the door to exciting scientific investigations. From the plots in Fig.~\ref{fig:pwneC_Crab1}
it is apparent that the highest energy data point from \astrima{} can provide an essential contribution in reducing the uncertainty on the Crab spectrum, so as to clarify whether there is room for a hadronic component, and starting from what energies. The LHAASO data do not require a hadronic contribution, but cannot exclude it either. As one can see from comparison of panel (b) and (c), the \astrima{} measurements in the 100-300 TeV range should definitely be able to provide at least a lower limit on the mean energy and an upper limit on the total energy of the proton component, with fundamental implications on the physics of pulsars and PWNe \citep[e.g.][]{2019hepr.confE..33A}, on particle acceleration in relativistic outflows \citep[e.g.][]{2015SSRv..191..519S}, and  even far reaching consequences on the possible sources of UHECRs \citep{2015JCAP...08..026K,2020A&A...635A.138G}. }
%

\subsubsection{Geminga} \label{Subsec:Geminga}

\paragraph{Immediate Objective}--
The TeV halo surrounding the Ge\-min\-ga pulsar is the perfect target among this emerging class of sources. The wide FoV of the ASTRI Mini-Array will provide a major advancement for the investigation of this very extended source, already with just one pointed observation, and will allow us to perform its spectro-morphological study with an unprecedented resolution: we will compare the new results with previous studies in \gray{}s and X-rays and investigate the source energy--dependent morphology.
We expect to extend its spectrum up to $\sim50$--100 TeV. 
This study will constrain the particle spectrum injected by the Geminga pulsar up to the highest energies together with the spatial and energy dependence of CR propagation in the surrounding region. In this way it allows us to evaluate the impact of Geminga on the measured positron-excess at Earth, and to provide new insights on the physics of this emerging class of TeV emitters.
\paragraph{Observing Time, Pointing Strategy, Visibility and Simulation Setup}--
Geminga is visible from the Teide site from October to March (about 4-5 months per year) with a culmination angle of about $20^\circ$. The source is very extended (more than 10\textdegree) and a single pointing of the ASTRI Mini-Array will not observe the whole TeV halo. Hence a possible observational strategy is to perform a number of tiling observations around the source position. Even with the large FoV of the ASTRI Mini-Array, the size of the source will limit the accuracy of background extraction. To overcome this issue, the background level will be necessarily estimated by performing ON-OFF mode observations. The source will be visible for about 500--600 hr per year in moonless conditions.

We simulated the Geminga TeV halo by creating an energy-dependent morphological model with parameters chosen in such a way as to reproduce HAWC recent results  \citep{2017Sci...358..911A}. This was obtained from the solution of the diffusion-loss equation for electrons and positrons emitted by the pulsar and propagating through the medium surrounding Geminga. The basic assumption is that particles are emitted isotropically (spherical symmetry) by the source, and then propagate with a diffusion coefficient depending on energy as $E^{1/3}$ (Kolmogorov type diffusion), which is compatible with the recent results from AMS-02 for hadronic cosmic rays \citep{2015NPPP..265..245T}. In terms of spatial dependence of the diffusion coefficient, we assumed a one-zone model, with spatially constant diffusion.
The electron/positron population was generated by assuming a continuous injection of particles having a power-law spectrum with index $1.7$ from the pulsar during its entire life ($\sim3.4\times10^5$ years). 

We stored the predicted \gray{} flux of the Geminga TeV halo into a model data-cube, where we assumed a radially symmetric morphology. The cube was then used to perform the simulations. More details on the solution of the diffusion equation and generation of the cube can be found in \cite{BuonoG}.

The simulations were carried out with {\tt Gammapy} v0.17 and we considered 100 hr, 200 hr and 500 hr as possible exposure times for the Geminga TeV halo. We simulated one single observation per exposure, {limiting the FoV} to 3.5\textdegree of radius {to avoid degradation of the sensitivity for large off-axis angles}, and including only the IRF background. We did not simulate an OFF field, which will have to be obtained, however, in a real observation in order to correctly estimate the background (see above). 
Geminga is close to the pulsar PSR B0656+14 (Monogem), but we did not include it in our sky-model because the source falls outside the ASTRI Mini-Array FoV. The Geminga halo and background events were generated in the energy range 1--50 TeV (where the source is significantly detected). For the analysis only, we modelled the source by adopting a radial \rf{Gaussian} morphology with an exponentially cut-off power-law spectrum. The background was instead calculated from the IRF. 
%
%
\begin{figure*}
\centering
	\includegraphics[width=8.6cm]{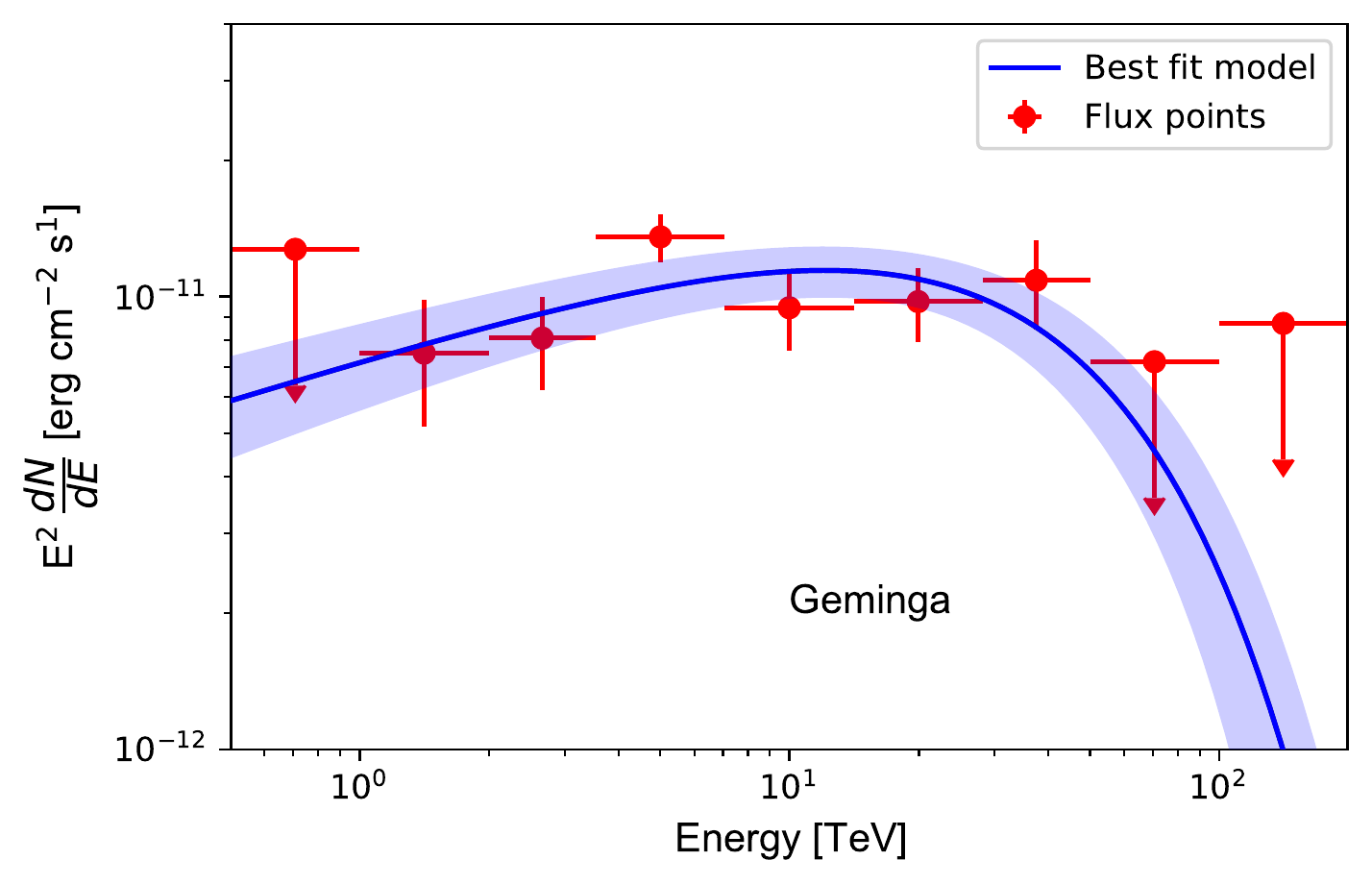} 
	\includegraphics[width=8.2cm]{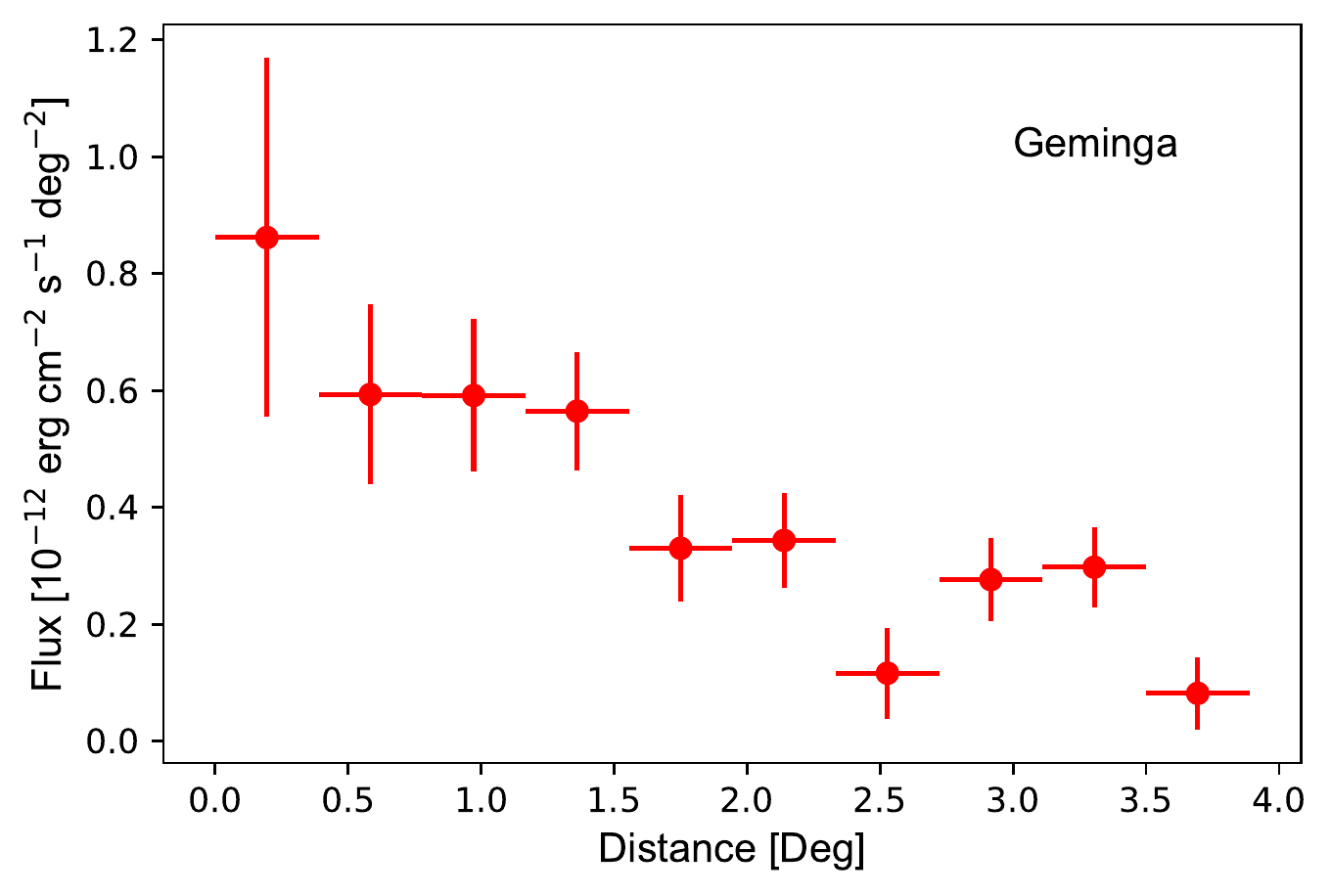} 

	\caption{\textit{Left panel:} Spectrum (red points) and best-fit model (blue line and purple shadow) of the 200 hours of simulation for the Geminga TeV halo. \textit{Right panel:} Geminga TeV halo radial profile obtained with 200 hr of simulation.}
	\label{fig:pwneC_Geminga1}
\end{figure*}
\newline

\paragraph{Analysis Method}--
We carried out the spectral analysis on the simulated data with {\tt Gammapy} v0.17. We found that the Geminga halo was significantly detected ($>10\sigma$) already with 100 hr of exposure. As reference, we report here the best-fit parameters in the case of 200 hr: we obtained a photon index $\Gamma=1.70\pm0.07$, a cut-off energy $E_{\rm cut}=41\pm 10$ TeV, an amplitude N$_{\text{1 TeV}} = (5\pm1)\times10^{12}$ erg cm$^{-2}$ s$^{-1}$ TeV$^{-1}$ and a width of the radial \rf{Gaussian} of $\sigma=1.9\pm0.2$\textdegree. In Fig.~\ref{fig:pwneC_Geminga1}-left, we show the spectrum of Geminga: the source is significantly detected up to $\sim50$ TeV (above 3$\sigma$ uncertainty in each energy bin). Therefore, with the selected amount of time, robust constraints can be obtained on the possible existence of a spectral cut-off. The spectral characterization that will be reached in a few hundred hours by the ASTRI Mini-Array will significantly improve the current HAWC spectrum and it will allow us to constrain the particle spectrum injected by Geminga and to determine if this source can significantly contribute to multi-TeV positrons at Earth.
  
The large FoV of the ASTRI Mini-Array will allow us to observe a large fraction of the Geminga TeV halo and to build a radial profile of the $\gamma$-ray emission. On the 200 hr simulated data, we firstly obtained a best-fit for both the source and the background. 
We subsequently removed the source model from the overall model fitting the data between 1 and 50 TeV (the energy range in which the source is significantly detected): this allowed us to create a residual map (i.e. observed counts minus predicted background counts).
Using this map, we created a radial profile, centered on the Geminga pulsar (RA= 98.48\textdegree, Dec=17.77\textdegree), for a number of annuli extending up to a radius of $3.5$\textdegree (which is about the border of the ASTRI Mini-Array FoV). In Fig.~\ref{fig:pwneC_Geminga1}-right, we show the observed radial profile. The source pops up significantly above the background at all radii, with small uncertainties in the flux estimated in each bin, and with a shape comparable with the HAWC radial profile. We note that our results have been obtained in a single observation, therefore at large distances from the center of the FoV the sensitivity degrades. This reflects in a larger relative error on the flux points at those radii. As mentioned in the previous section, multiple observations will allow us to better track the source morphology.
%

\subsection{Ultra High Energy Cosmic Rays from Starburst Galaxies}
\label{Sec:UHECRs}
\paragraph{Scientific Case}--
Starburst galaxies, together with AGNs (and in particular radio galaxies such as M~87 and Cen~A), represent the main
candidates for the acceleration of UHECRs.
Recently, the Pierre Auger Collaboration reported a $4.0\sigma$-significant correlation between the arrival direction of CRs with energy $E > 38$\,EeV and a model based on starburst galaxies \citep{2018ApJ...853L..29A}. Results by the Telescope Array for $E > 43 \; {\rm EeV}$, however,  cannot make a statistically significant confirmation or confutation of this correlation~\citep{2018ApJ...867L..27A}.
In addition, constraints based on the isotropic \gray{} background at $E \lesssim$ TeV measured by the Fermi-LAT \citep{2016PhRvL.116o1105A} point towards almost local sources of UHECRs (i.e. with a distance $\lesssim 200$\,Mpc) supporting the association of UHECRs and starburst galaxies \citep{2016PhRvD..94d3008L}. 

In the central regions of starburst galaxies, the intense star formation rate (SFR) and extreme properties of the interstellar medium (ISM) provide ideal conditions for CR acceleration and confinement. 
Possible acceleration sites include the \rf{nuclear region around the galactic nucleus}, as well as the termination shock of the powerful wind that originates from the nucleus itself \citep{1999PhRvD..60j3001A,2018A&A...616A..57R,2022MNRAS.511.1336P}.
The interaction of CRs with the surrounding ISM and interstellar radiation field (IRF) produce non-thermal emission in the \gray{} band and neutrinos. Accelerated electrons and positrons predominantly lose energy via inverse Compton and synchrotron processes, whereas accelerated protons predominantly via proton-proton interactions which lead to the production of neutral and charged pions. Neutral pions decay into two \gray{}s while charged pions decay into secondary electrons, positrons and neutrinos.  
Typical energies of the secondaries are about 10\% of the parent proton energy for the \gray{}s from neutral pion decay and about 5\% for the neutrinos.

The hadronuclear neutrino production in starburst galaxies is considered an important source of the energetic neutrinos observed by the IceCube Observatory \citep[e.g.][]{2014JCAP...09..043T,2016PhRvD..94j3006M,2019JCAP...09..004P}.
 However, the estimates of the  source  population contribution  to the observed neutrino background rely on the extrapolation of the characteristic source \gray{} spectrum to the region between the HE band and the IceCube energy scale. 
Analyses that assume a spectral index of $p\simeq$2.2  and cut-off energy $E_{\rm cut}$=10 PeV yield a contribution to the diffuse neutrino background of 30\% at 100 TeV, and 60\% at 1 PeV, and they are found to be consistent with the bounds from the residual non-blazar component of the extragalactic \gray{} background \citep{2017ApJ...836...47B}. Softer spectra, as expected, lead to a smaller flux of neutrinos at the highest energies, and the \gray{} fluxes are also reduced correspondingly.   
The diffuse neutrino flux is also very sensitive to the highest energy reached by accelerated protons. The neutrino flux starts declining rather steeply as the maximum energy of accelerated protons decreases \citep{2020MNRAS.493.5880P}.
In general, \gray{}s set an upper limit for the associated neutrino flux, but this condition can be partially relaxed if gamma-gamma absorption inside the source is efficient. A careful understanding of both the production and absorption of \gray{s} in starburst galaxies is therefore instrumental to constrain the starburst contribution to the diffuse neutrino flux. To reach this goal, a detailed measurement of the \gray{} spectral properties (spectral index, high-energy cut-off) of starburst galaxies is necessary.

The last decade has seen an increase in our knowledge of the \gray{} emission from star-forming galaxies beyond our own\rf{~\citep[e.g. through studies of the correlation between non-thermal radiation and the star formation rate as discussed in][and references therein]{2022A&A...657A..49K}}. At GeV energies, the Fermi/LAT satellite has detected seven  starburst galaxies  \citep{2020ApJS..247...33A}. The starburst galaxies  M~82 and NGC~253 were also detected  by Imaging Atmospheric Cherenkov Telescopes \citep{2009Natur.462..770V,2009Sci...326.1080A} indicating that very-high energy (VHE) photons can be produced in the nuclei of these galaxies. M~82 is also expected to be one of the dominant sources of UHECRs in the full-sky starburst model presented in \cite{2018ApJ...853L..29A}. One of the brightest galaxies in the GeV band is the composite starburst/Seyfert galaxy NGC~1068, {located at 0.35 deg from the ``hottest'' neutrino spot in the 10-year survey data of IceCube}  \citep{2020PhRvL.124e1103A}. 
The exact origin of the  \gray{} emission in NGC~1068 is still undetermined owing to the presence of different particle accelerators, like the starburst nucleus and  AGN-driven  jets and winds \citep{2010A&A...524A..72L,2014ApJ...780..137Y,2016A&A...596A..68L}. On contrast, the \gray{} emission of the other starburst galaxies, including the ultra luminous infrared galaxy Arp~220, can be explained by hadronic interactions of  CR particles accelerated by stellar winds and SN explosions  \citep[e.g.][]{2008A&A...486..143P,2009ApJ...698.1054D,2018MNRAS.474.4073W,2019MNRAS.487..168P}. 
With the ASTRI Mini-Array we will have the chance to extend the \gray{} spectrum towards the highest part of the VHE spectrum {for some of these sources}, allowing us to study the emission from the most energetic particles, and to constrain the maximum energy attained by the accelerated particles.

\paragraph{Immediate Objective}--
The aim of the present study is to determine the \gray{} spectrum above a few TeV of some starburst galaxies in order to address the questions of the maximum energy of the accelerated particles, and the potential absorption of \gray{}s in this class of sources.
To this aim, we performed dedicated simulations of the spectrum predicted by the starburst model for the starburst galaxies M~82, NGC~2146, Arp~299, and for the ULIRG Arp~220 \citep{2019MNRAS.487..168P}, and that predicted by the starburst and AGN wind model for NGC~1068 \citep{2019ApJ...883..135A}. {All these galaxies are in the Northern Hemisphere and represent possible targets for the ASTRI Mini-Array}. The results of these simulations indicated that the most promising target for observations with the ASTRI Mini-Array is the starburst galaxy M~82.
The combined Fermi-LAT and VERITAS spectrum of M~82 is described by a power-law with spectral index p$\sim$2.2 up to $\sim$3.5 TeV. The presence of a high energy cut-off in the VHE spectrum is not well constrained by observations, and models predicting the presence or not of this spectral feature equally fit the current data \citep{2008A&A...486..143P,2009ApJ...698.1054D,2019MNRAS.487..168P}. {Here we investigate the capability of the ASTRI Mini-Array to measure the high-energy cut-off.}
ASTRI Mini-Array observations of the most nearby sources, like M~82, will potentially allow us to constrain the EBL in the still unexplored region of the far-IR band (see \ref{sec:EBL}). 

\begin{figure}
	\centering
	\includegraphics[scale=.42]{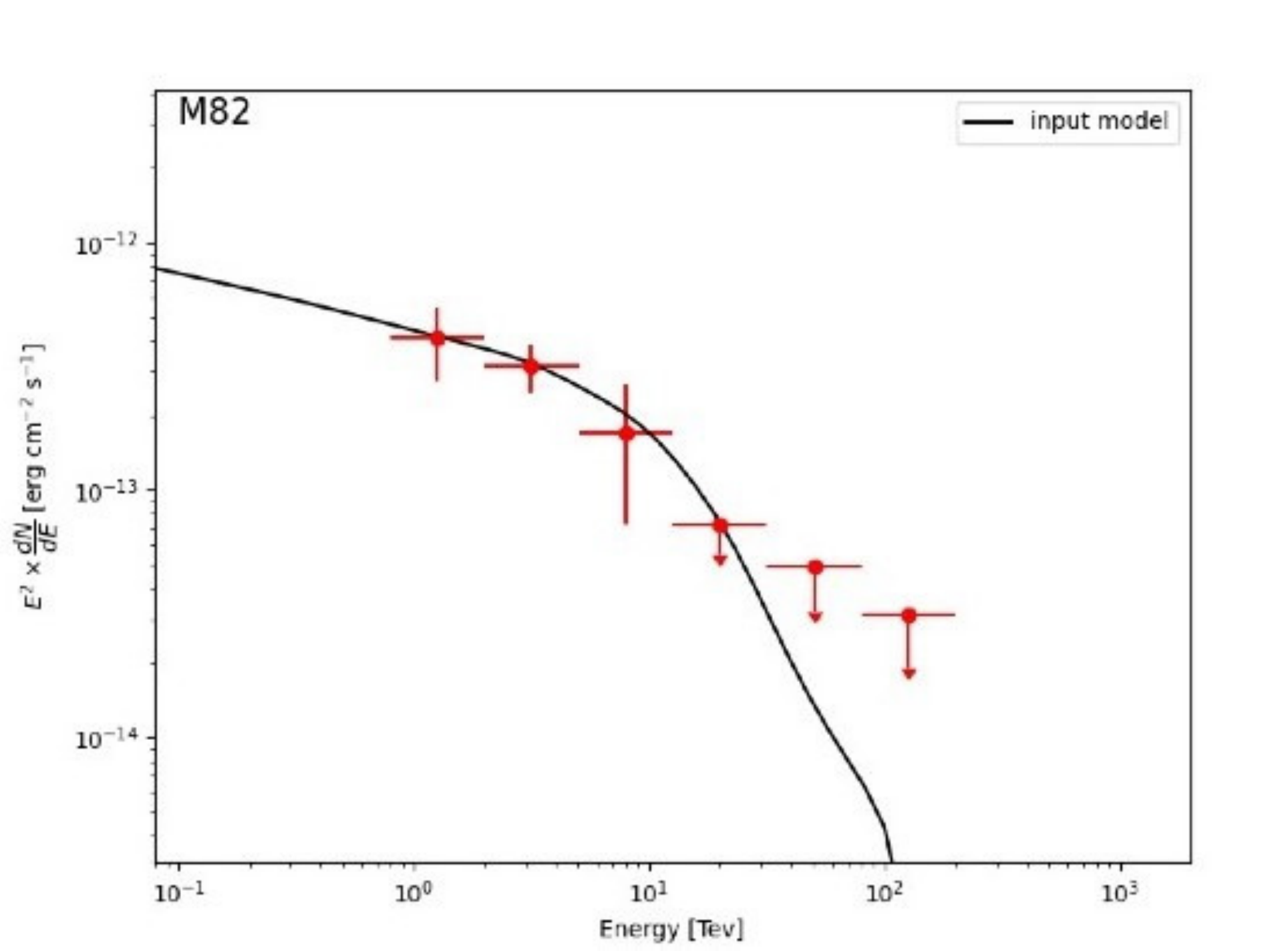}
	\caption{Simulated \gray{} spectrum of M~82. Red points represent the signal that can be observed by ASTRI Mini-Array after 100 hours of observation, and assuming the spectrum predicted by the starburst model \citep[black line,][]{2019MNRAS.487..168P}. The spectral points and their error bars are obtained from the distribution of 100 simulations.}
	\label{UHECR_Fig03}
\end{figure}

\paragraph{Observing Time, Pointing Strategy, Visibility and Simulation Setup}--
The starburst galaxy M~82 is observable from the Teide site for about 310 hr per year at a zenith angle of $<$45 $\deg$, in moonless conditions. 
To give an estimate of the observing time needed to reach our scientific objectives (source detection and extended spectral measurements), we have performed simulations of the expected spectrum using {\tt ctools} (version 1.7.2) coupled to the IRF produced for the Teide site.
We simulated M 82 as a point-like source, located at the known coordinates. 
We  simulated the spectrum predicted by a starburst model that takes into account the absorption of \gray{} due to electron-positron pair production inside the starburst region~\citep{2019MNRAS.487..168P}. This model predicts a \gray{}-spectrum with a high-energy cut-off at energies \rf{above a} few TeV.
In the model definition XML file, this spectral model was introduced as a {\sc FileFunction} type that defines the intensity values at specific energies.
We simulated the source for 100, 200, and 500 hours of exposure time.

\paragraph{Analysis Method}--
Figure~\ref{UHECR_Fig03} shows the simulated spectrum assuming  an exposure of 100 hours and  6 energy bins logarithmically spaced between  0.8  and 199.5 TeV. 
We found that, with an exposure time of  100 hours we will be able to measure the spectrum in the energy range $\sim$1-10 TeV; with  500 hours of exposure, we will extend the spectral measurement up to $\sim$30 TeV.

\begin{figure}
	\centering
	\includegraphics[scale=.55]{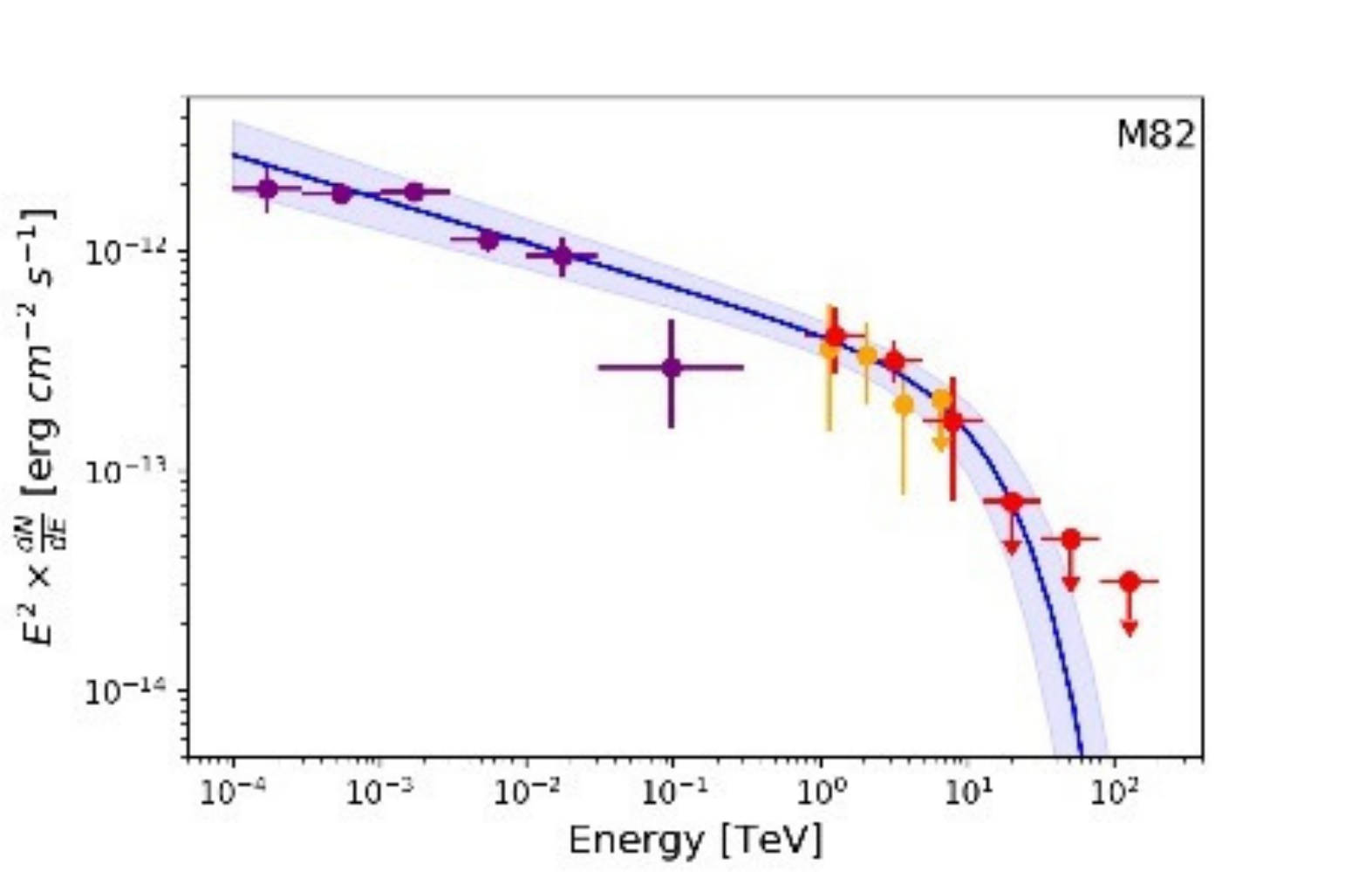}
	\caption{\gray{} spectrum of M82. Data are from Fermi 4FGL-DR2 (purple points, \citealt{2020arXiv200511208B}), from VERITAS (orange points, \citealt{2009Natur.462..770V}), and from ASTRI Mini-Array simulations (100 h, red points). The best fit and its 1$\sigma$ uncertainty are represented by the solid blue line and the shaded region.}
	\label{UHECR_Fig04}
\end{figure}

In order to constrain the  \gray{} spectrum of the source we used the {\sc naima} package to fit simultaneously  the ASTRI Mini-Array simulated data  and the Fermi-LAT and VERITAS  data.
We assumed  an exponential cut-off power-law model with normalization, power law index, and cut-off energy as free parameters. We obtained the best fit values of $F_{\text{1TeV}}=(2.70 \pm 0.43)\times10^{-13}$ TeV$^{-1}$ cm$^{-2}$ s$^{-1}$,  $\Gamma=2.20\pm0.02$, and  $\log(E_{cut}/TeV)=1.22\pm0.19$ for 100 h of exposure time (see Figure~\ref{UHECR_Fig04}). By increasing the exposure time we can improve the measurement of the differential flux and cut-off energy (see Table \ref{tab:sources-Pillar1-Emax}), while the measurement of the spectral index remains almost unchanged.
Performing the same fit without the ASTRI Mini-Array simulated data, we obtained $F_{\text{1TeV}}=(2.64 \pm 0.62)\times10^{-13}$ TeV$^{-1}$ cm$^{-2}$ s$^{-1}$, $\Gamma=2.20\pm0.04$, and  $\log(E_{cut}/TeV)=0.92\pm2.8$. This analysis clearly shows that the ASTRI Mini-Array will give a fundamental contribution in  constraining the spectral parameters, especially the high-energy cut-off.

\begin{table*}[width=2.\linewidth,cols=5,pos=htp!]
\caption{List of cut-off energies established with ASTRI Mini-Array data analysis for the \gray{} sources analysed with different exposure times. The maximum energy indicated is the cut-off energy established with the best-fit models. If this is, instead, the maximum energy of the last significant spectral bin, it is indicated with a ``$^\star$'' symbol. $^\maltese$: values for model A (see text)}
\centering
\def\arraystretch{1.5}
\begin{tabular}{L@{\hskip 0.05in}C@{\hskip 0.05in}C@{\hskip 0.05in}C@{\hskip 0.05in}C@{\hskip 0.05in}C@{\hskip 0.05in}C@{\hskip 0.05in}C@{\hskip 0.05in} C}
\toprule
Name & $E_{M}$(100h) & $E_{M}$(200h) & $E_{M}$(500h) & Section \\
 & (TeV) & (TeV) & (TeV) & \\
\midrule
Tycho & $21.1^\star$ & $21.1^\star$  & $94.4^\star$ & \ref{Subsec:Tycho}\\

Galactic Center & $56.1^\star$ & $158.1^\star$ & $158.1^\star$ & \ref{Subsec:GC} \\ 

VER J1907$+$062 & $146.5^\star$ & $146.5^\star$   & $146.5^\star$ & \ref{Subsec:candidatePeV}\\

SNR G106.3+2.7 & $46.8^\star$ & $71.3^\star$  & $108.4^\star$ & \ref{Subsec:candidatePeV} \\
\midrule
$\gamma$-Cygni$^{ \maltese}$ & $40.8^\star$ & $40.8^\star$ & $69.3^\star$ & \ref{Subsec:gammaCyg} \\

W28/HESS\,J1800-240B & $17.9^\star$ & $17.9^\star$  & $27.4^\star$ &  \ref{Subsec:W28} \\
\midrule
Crab ($\gamma=5\times10^6$, $\chi_p = 4\%$) & $119.7^\star$ & $161.4^\star$ & $250.8^\star$ & \ref{Subsec:Crab} \\
Crab ($\gamma=1\times10^5$, $\chi_p = 15\%$) & $78.7^\star$ & $119.7^\star$ & $250.8^\star$ & \ref{Subsec:Crab} \\
Geminga  & $70\pm40$ & $40\pm9$ & $55\pm9$ & \ref{Subsec:Geminga} \\
\midrule
M82 & 16.6$^{+9.1}_{-5.9}$ & 15.1$^{+7.8}_{-5.1}$ & 14.8$^{+4.7}_{-3.6}$ & \ref{Sec:UHECRs} \\
\bottomrule
\end{tabular}
\label{tab:sources-Pillar1-Emax}
\end{table*}
\section{Pillar--2: Cosmology and Fundamental Physics}\label{sec:5}
%
The propagation of high-energy gamma-rays emitted by cosmic sources can be exploited to probe the properties and the content of the space they traverse and the processes involving ultra-energetic particles. The most direct use of this approach is the determination of level and evolution of the extragalactic background light (EBL), a prime source of absorption for photons above few tens of GeV emitted at cosmological scales, in particular by blazars \citep[e.g.][]{2008A&A...487..837F,2011MNRAS.410.2556D,2012MNRAS.422.3189G}. Similarly, this process also enables studies of intergalactic magnetic fields ~\citep[IGMFs; for a review see][]{2021Univ....7..223A}. 
Furthermore, the powerful beam of VHE photons from blazars and galactic sources recently identified by HAWC can be exploited to perform studies of fundamental physics well beyond the reach of terrestrial labs \citep[e.g.][]{2020MNRAS.491.5268G}. In particular, the focus is on tests able to identify specific signatures related to: 1) the breaking of the Lorentz invariance (Lorentz invariance violation, LIV) close to the Planck scale, expected from several quantization schemes of gravity \citep[e.g.][]{2013CQGra..30m3001L}, and 2) the mixing of photons with the so-called axion like particles (ALP), light pseudoscalar particles predicted by several extension of the standard model of the elementary particles, most notably String Theories \citep[e.g.][]{2010ARNPS..60..405J}. Last, but not least, observations at energies above 10 TeV can also be used to test the exciting idea that the VHE emission from extreme blazars \citep{2020NatAs...4..124B} is the by-product of beams of ultra-high energy cosmic rays, energized and launched by the jets of these sources \citep{2010APh....33...81E}.

In the following, we describe how to use the ASTRI Mini-Array at energies above few TeV to:
1) constrain the EBL in the still unexplored region of the far-IR band; 
2) put the “hadron beam” (HB) model for extreme BL Lacs to the test; 
3) look for anomalies in the spectrum related to photon-ALP mixing;  
4) perform tests of LIV; and
5) probe IGMFs.
The detection of the signatures associated with these effects would have far-reaching consequences for physics and astrophysics. All these studies can be performed through relatively long pointing of few carefully selected targets, also suitable for complementary studies on particle acceleration, emission processes and photon opacity of jets \citep[see also][]{2021JHEAp...F..XXXS}. 

\subsection{TeV observations and constraints on the extra-galactic infrared background light (IR EBL)}\label{sec:EBL}
%
%
\paragraph{Scientific Case}
The EBL not only is an important radiative constituent of the local universe, but also offers critical constraints on all astrophysical and cosmological processes taking place during the formation of cosmic structure.
Unfortunately, the only spectral region where its direct measurement has been possible is the sub-millimeter band ($200 <\lambda < 900$ $\mu$m), where the local foreground emissions are minimal. At all shorter wavelengths, from the mid-IR to the far-IR, where the IR background intensity is maximal, these measurements are prevented by the overwhelming dominance of local emission from both the Galaxy and the Solar System. This is particularly the case over the wide wavelength interval from 3 to 300 $\mu$m, an unfortunate occurrence because deep observations with infrared space and ground-based telescopes have revealed that this spectral region is very rich in astrophysical and cosmological information. Very luminous sources at high redshifts have been detected and identified, which are interpreted as clearly tracing major episodes of the formation of galaxies, AGNs and quasars, when dust, present in the interstellar and circum-nuclear media, absorbs optical-UV light and re-radiate it in the far-IR. Unfortunately, IR telescopes can only observe a few of the most luminous of them at high z, because of confusion and sensitivity limitations, and the bulk of the population cannot be detected.  The recent launch of the James Webb Space Telescope Observatory~\citep[JWST,][]{2006SSRv..123..485G} will allow the scientific community to carry out wide band observations with NIRCam in the 2-5\,$\mu$m and with MIRI in the  6-25\,$\mu$m energy bands, respectively, on small fields, that will fill in crucial gaps in our knowledge of the EBL intensity at these wavelengths. In spite of its expected sensitivity, however, the JWST Observatory will be essentially blind to diffuse emissions, which ASTRI Mini-Array will instead effectively constrain. For these reasons, the information registered in the IR EBL would make a very important contribution to the understanding of the processes of cosmic structure formation, particularly those related to the build-up of stellar populations and metals in galaxies \citep[e.g.][]{2014ARA&A..52..415M,2011A&A...532A..49B,2001A&A...378....1F}.

So far, observations with Cherenkov observatories have helped in constraining the EBL in the spectral range from the UV to the optical and near-infrared, thanks to the extensive monitoring and detection up to a few TeV energies of low- and high-redshift blazars. 
The constraints are obtained through the analysis of GeV to TeV spectra of blazars and the identification of the exponential cutoff due to the cosmic opacity from the interaction of the VHE photons with the EBL photons and the consequent pair production. From educated extrapolations of the HE blazar spectra to the VHE regime and comparison to the observed spectra, the number density of EBL photons along the line-of-sight is inferred \citep[e.g.][]{2008A&A...487..837F}.

The portion of the EBL from the UV to the near-infrared so far investigated includes only half of the total extra-galactic light. It is dominated by star-light from low-z quiescent galaxies, but misses the most important phases of galaxy and stellar formation, when the infrared emission dominates, that are registered in the IR EBL.

\paragraph{Immediate Objectives}--
The goal of the proposed ASTRI Mini-Array observations is to monitor a few local AGNs in the northern hemisphere to characterize their VHE spectra as accurately as possible up to the highest energies. These spectra will manifest the exponential absorption effects of the interactions of the VHE source’s photons and the low-energy IR EBL ones.
We have recently discussed \citep{2019A&A...629A...2F} various extra-galactic source populations, essentially blazars and radio-galaxies, that are best suited to the above purposes. In order to constrain the EBL at the longest wavelengths, we need to observe gamma-rays at the highest energies, from the usual relation:
\begin{equation}
\label{EQ:Chap5_Eq0}
\lambda_{\rm max}\simeq 1.24\times E_{\rm TeV} \; [{\rm \mu m}]
\end{equation}
expressing the maximum of the pair-production cross section.

\begin{figure}
	\centering
		\includegraphics[scale=.4]{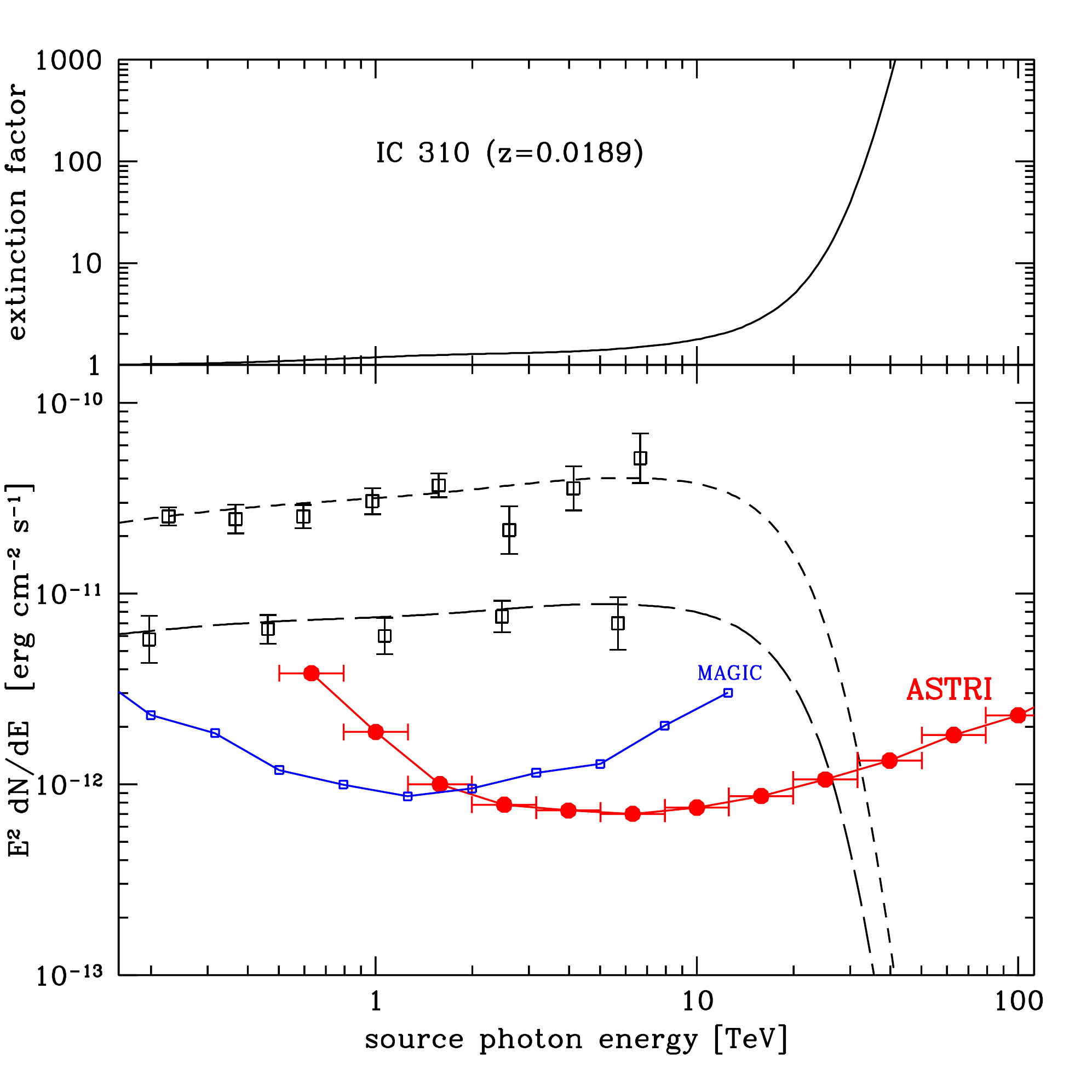}
	\caption{The upper panel reports the $e^{\tau}$ extinction factor for photon-photon interaction on EBL at the IC 310 source distance. Bottom panel reports data-points for two states of the source, a flare in blue and high-state in red, together with best-fitting curves including EBL extinction. Red line marks the ASTRI Mini-Array 50 hours $5\sigma$ limit while the blue line is the MAGIC one.}
\label{FIG:Chap5_Fig03_old}
\end{figure}

\begin{figure}
	\centering
		\includegraphics[scale=.4]{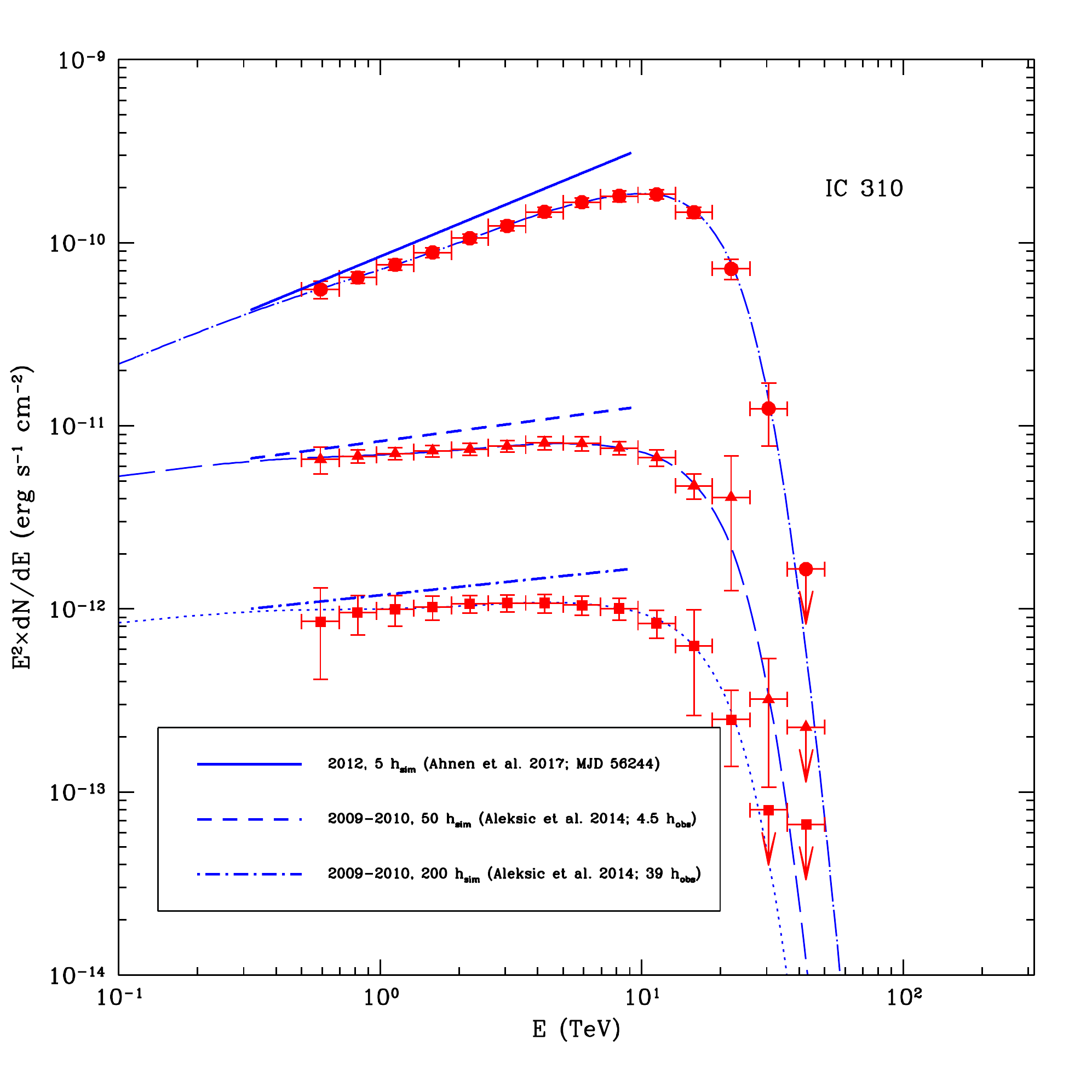}
                \caption{Simulations of different source states of IC~310 as observed by MAGIC between 2009 and 2010 during a low (blue, dotted line; 39\,h of observations) and a high state (blue, long-dashed line; 4.5\,h of observations, \citealt{Aleksic2014}), as well as during a major flare in 2012 (blue, dotted-long-dashed line; MJD 56244.066-56244.082,  \citealt{Ahnen2017}). Depending on the source state, 5, 50 and 200\,h (red points) were simulated respectively, considering the intrinsic source spectra (solid, dashed and dot-dashed lines) and EBL absorption.
                }
	\label{FIG:Chap5_Fig03}
\end{figure}

\begin{figure}
	\centering
		\includegraphics[scale=.4]{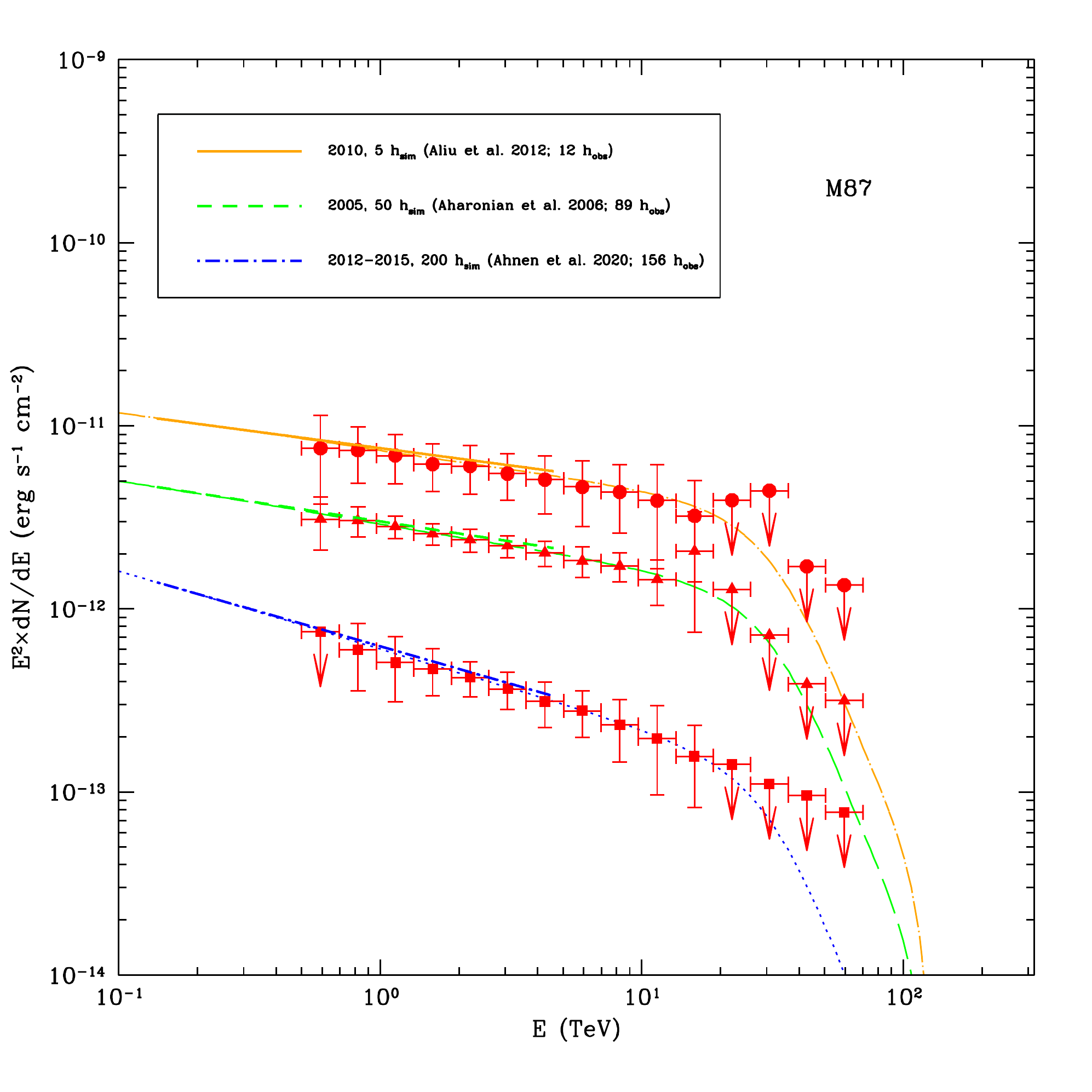}
                \caption{Same as Fig.\ref{FIG:Chap5_Fig03} for M~87 considering the low (blue lines), high (green lines) and flaring (orange lines) states as reported by \citet{Acciari2020,Aharonian2006,Aliu2012}, and simulating integrations of 5, 50 and 200\,h respectively.
                }
	\label{FIG:Chap5_Fig04}
\end{figure}

\begin{figure}
	\centering
		\includegraphics[scale=.4]{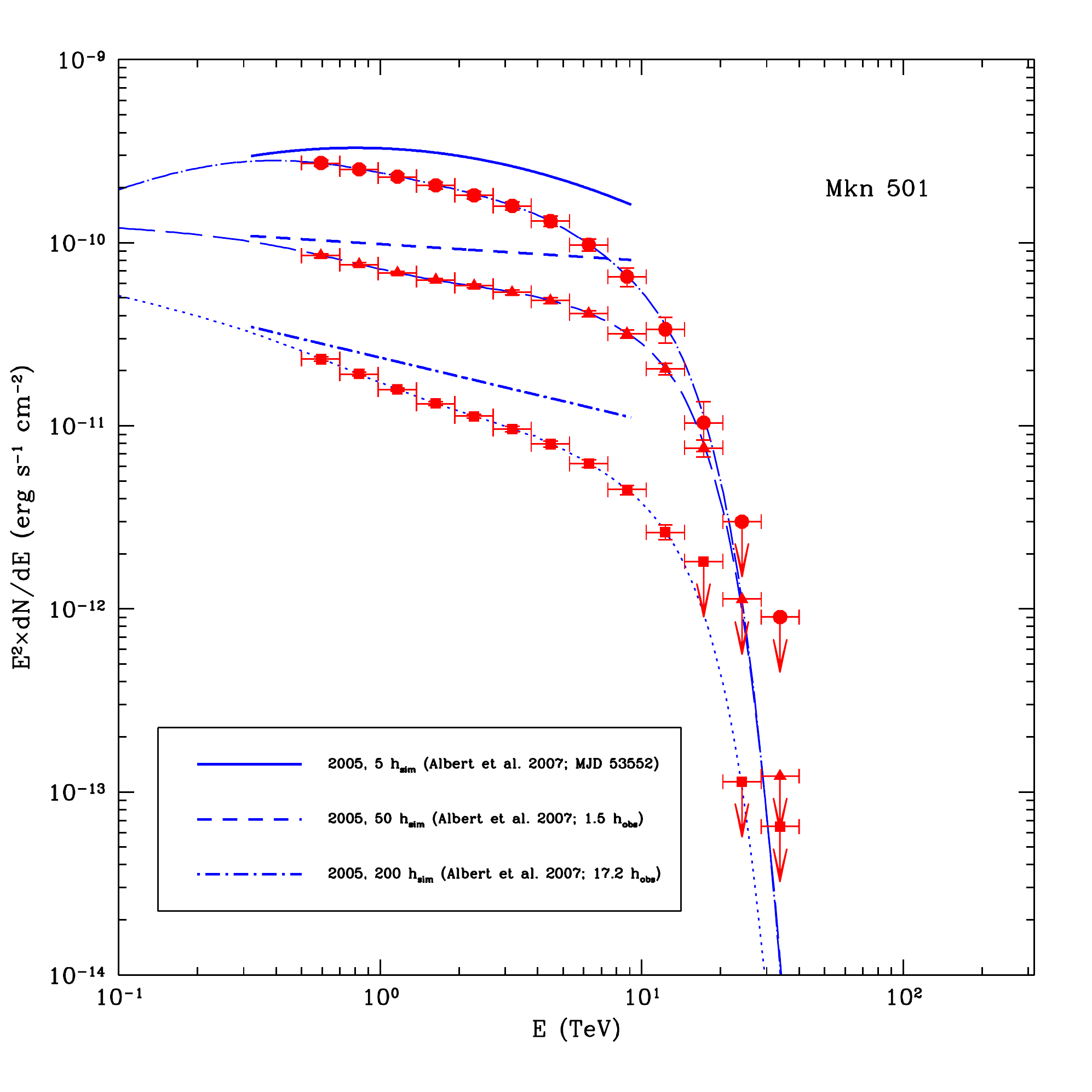}
                \caption{Same as Fig. \ref{FIG:Chap5_Fig03} for Mkn~501 considering the low, high and flaring states reported by \citet{Albert2007}, and simulating integrations of 5, 50 and 200\,h respectively.
                }
	\label{FIG:Chap5_Fig05}
\end{figure}

\begin{table*}[width=2.\linewidth,cols=8,pos=htp!]
\centering
\caption{Summary Table of observations discussed in sect. 6.1.}
\def\arraystretch{1.5}
\label{tab:6.1}
 \begin{tabular*}{\tblwidth}{@{}CCCCCCCC@{} }
\toprule
Target & Class & RA (J2000) &  DEC (J2000) & Obs. time & ZA & Moon & Strategy, analysis, notes \\
IAU Name & &       &       & [hr] & [deg] & [\%] & \\
\midrule
IC 310 & Radio gal. & 03 16 43.0 & +41 19 29 & 50-100 & 45 & 25 & Better suited for ToO observations of high states\\
M87 & Radio gal. & 12 30 47.2 & +12 23 51 & 50-100 & 45 & 25 & Better suited for ToO observations of high states\\
Mkn 501& Blazar & 16 53 52& +39 45 38 & 50-100 & 45 & 25 & Better suited for ToO observations of high states\\
\bottomrule
\end{tabular*}
\end{table*}

At the same time, we should consider that only VHE photons from the most nearby sources could be detected from Earth, because the absorption is a strong function of both energy and distance.
It turned out that, even under the most extremely favorable conditions of huge emission flares, extreme high-energy emitting blazars are not very useful for our purpose because they are much too distant ($>100$ Mpc the nearest ones, Mkn 501 and Mkn 421).
Observations of more local VHE emitting AGNs, like low-redshift radio galaxies (M~87, IC~310, Centaurus~A), or local star-bursting and active galaxies (M~82, NGC~253, NGC~1068) are better suited and will potentially allow us to constrain the EBL up to $\lambda \sim 100\mu$m.
As an illustration, we report in Fig. \ref{FIG:Chap5_Fig03_old} the case for one of the most interesting potential targets. The red line corresponds the sensitivity limits of ASTRI Mini-Array for a 50 hours total integration time at 5$\sigma$. The blue one denotes the limits achievable by MAGIC.

\paragraph{Observing Time, Pointing Strategy, Visibility and Simulation Setup}--
%

Because of the very soft spectra expected above 10 TeV, the feasibility of the proposed observations can not be assessed from the sensitivity curve. Detailed simulations, as reported here, are needed to precisely assess the effectiveness of the observations.

In Figs. \ref{FIG:Chap5_Fig03} and \ref{FIG:Chap5_Fig04} we report results from detailed simulations of the VHE section of the SED of a few proposed targets, in particular IC~310 and M~87, for which we show data-points from current observations.
Simulations were performed according to the scheme described in Section~\ref{Sec:ctoolssims}.

Depending on the source state, 5, 50 and 200\,h were simulated, considering the intrinsic source spectra and EBL absorption (\citealt{2017A&A...603A..34F}). The ctools \citep{2016A&A...593A...1K} were used for the simulations and analysis.
Parameters of the proposed targets are reported in Table \ref{tab:6.1}. We see that all of them have optimal visibility from the planned Teide site.

Fig. \ref{FIG:Chap5_Fig05} reports the observations of the blazar Mkn 501 during the famous 1997 outburst. A similar event taking place during the ASTRI Mini-Array operations would obviously make an interesting target, in spite of the large distance of the source that compromises the detection at the highest energies. Similar considerations might apply to the other low-redshift blazar Mkn 421.

In all cases, the spectra should be detected up to at least ~30 TeV for moderate ASTRI Mini-Array integrations (50 hours) for observations during high-states of the sources. Longer integrations would be needed otherwise.

As it is the case for blazars, also radio galaxies are variable objects. One such case is illustrated for IC 310 in Fig. \ref{FIG:Chap5_Fig03}, where solid, dashed and dotted-dashed lines correspond to a short-lived flare, high-state emission and low state emission, respectively.  We can take this property as an advantage, by selecting to observe them during enhanced emission stages.
This will require to organize the campaigns like ToO’s. It may be considered if the continuous sky monitoring by the Fermi satellite might be sufficiently sensitive to detect enhanced emission phases in these sources, that may likely be the case. In addition, continuous half-sky optical monitoring by dedicated ground-based observatories, like Pann-STARS, the Zwicky Transient Facility, and even the Large Synoptic Survey Telescope (LSST) for part of the targets visible from Chile \citep{2019ApJ...873..111I} could be helpful.
Note finally that other local AGNs of potential interest for IR EBL studies are mentioned e.g. in Sects. \ref{Sec:UHECRs} and \ref{Sec:hadrons}.

\paragraph{Analysis Method}--
The emphasis is on the precise determination of shape and energy of the cut-off, in order to disentangle an intrinsic spectral cut-off and the one expected from EBL, and to determine the shape of the intrinsic continuum. As a consequence, a combination with data from the Fermi and ground-based surveys in the optical will be requested, in order to trigger the ToO. For a proper characterization of the spectra, it will also be ideal (if not required) to coordinate the observations with simultaneous ones by the MAGIC observatory.

\subsection{Probing intergalactic magnetic fields}
%
%
\paragraph{Scientific Case}--
The origin of magnetic fields in the Universe is an open problem with important implications for understanding the formation and evolution of cosmic structures, the propagation of charged particles over cosmological distances, and possibly even the processes that led to the matter--anti-matter asymmetry in the early Universe. For details, the reader is referred to reviews by \citet{durrer2013a} and \citet{2021Univ....7..223A}.

Intergalactic magnetic fields (IGMFs) are a fundamental ingredient to understand the propagation of high-energy gamma rays. At high energies, gamma rays interact with photons from the EBL, producing electron-positron pairs. Being charged, the pairs are deflected by intervening magnetic fields before they up-scatter CMB photons to high energies. The effects of IGMFs on these electromagnetic cascades lead to characteristic signatures in the arrival directions and arrival times of gamma rays, as well as their flux. This avenue has been explored by a number of authors to set limits on IGMF properties~\citep[e.g.,][]{neronov2010a, tavecchio2010a, tavecchio2011a, vovk2012a, finke2015a, alvesbatista2020a}. Furthermore, the maximum intrinsic gamma-ray energy at the sources that can be inferred from observations could, in principle, be degenerate with respect to IGMFs parameters~\citep{dolag2009a, saveliev2021a}, thereby affecting our understanding of particle acceleration in various astrophysical environments, as discussed Sec.~\ref{sec:4}.

IACTs are in general sensitive to relatively strong IGMFs, with strengths $B \sim 10^{-15}$--$10^{-12}$~G. Current Cherenkov telescopes have performed searches for extended emission from blazars, the so-called pair haloes~\citep{magic2010a, hess2014a, veritas2017a}. The absence of such magnetically-broadened features enabled the derivation of some limits on IGMFs. CTA will likely improve these constraints substantially~\citep{cta2021b}. The weaker IGMFs ($B \lesssim 10^{-17} \; \text{G}$) are better probed by Fermi-LAT which, combined with observations from the ASTRI Mini-Array, could enable detailed studies of IGMFs. 

Note that IGMF constraints based on electromagnetic cascades could be compromised due to interactions between the high-energy beam and the intergalactic medium, which could generate plasma instabilities~\citep[see, e.g., ][]{broderick2012a, schlickeiser2012a, vafin2019a}. Nevertheless, the role played by this effect is not clear~\citep{miniati2013a, perry2021a}, and recent works suggest that even if it is indeed relevant, IGMF estimates could still be possible~\citep{alvesbatista2019g, yan2019a}.

\paragraph{Immediate Objectives}--
Observations of 1ES~0229+200 and Mkn~501 with the ASTRI Mini-Array can be used to probe IGMFs at all energies accessible to the instrument. Searches for extended emission around point-like extragalactic sources and precise flux measurements can be used either to infer the strength of IGMFs or to set limits on it.

\paragraph{Observing Time, Pointing Strategy, Visibility and Simulation Setup}--
Suitable sources for this type of study are those at cosmological distances whose spectra extend up to tens of TeV. Optimal targets include extreme blazars~\citep{bonnoli2015a}, as discussed in Sections~\ref{sec:axions} and~\ref{sec:lorentzInvariance}. GRBs could also be used for these studies, although so far they have not been been observed at multi-TeV energies.

To discuss the performance of the ASTRI Mini-Array to probe IGMF, we select 1ES~0229+200 and perform simulations of gamma-ray propagation in the intergalactic space using the CRPropa code~\citep{alvesbatista2016a}. We consider turbulent IGMFs with strengths $10^{-15}$, $10^{-14}$, and $10^{-13}$~G, for a coherence length of 1~Mpc, in addition to a scenario with no IGMFs. The intrinsic spectrum for this object is assumed to be a power law with spectral index $-1.5$ and an exponential cut-off at 7~TeV. We also assume that this source is a steady emitter over time scales of $10^7$~years, with its jet pointing directly at Earth with an opening angle of $5^\circ$. More details on the simulation procedure can be found in \citet{2021Univ....7..223A}.

\begin{figure}[]
    \centering
    \includegraphics[width=\columnwidth]{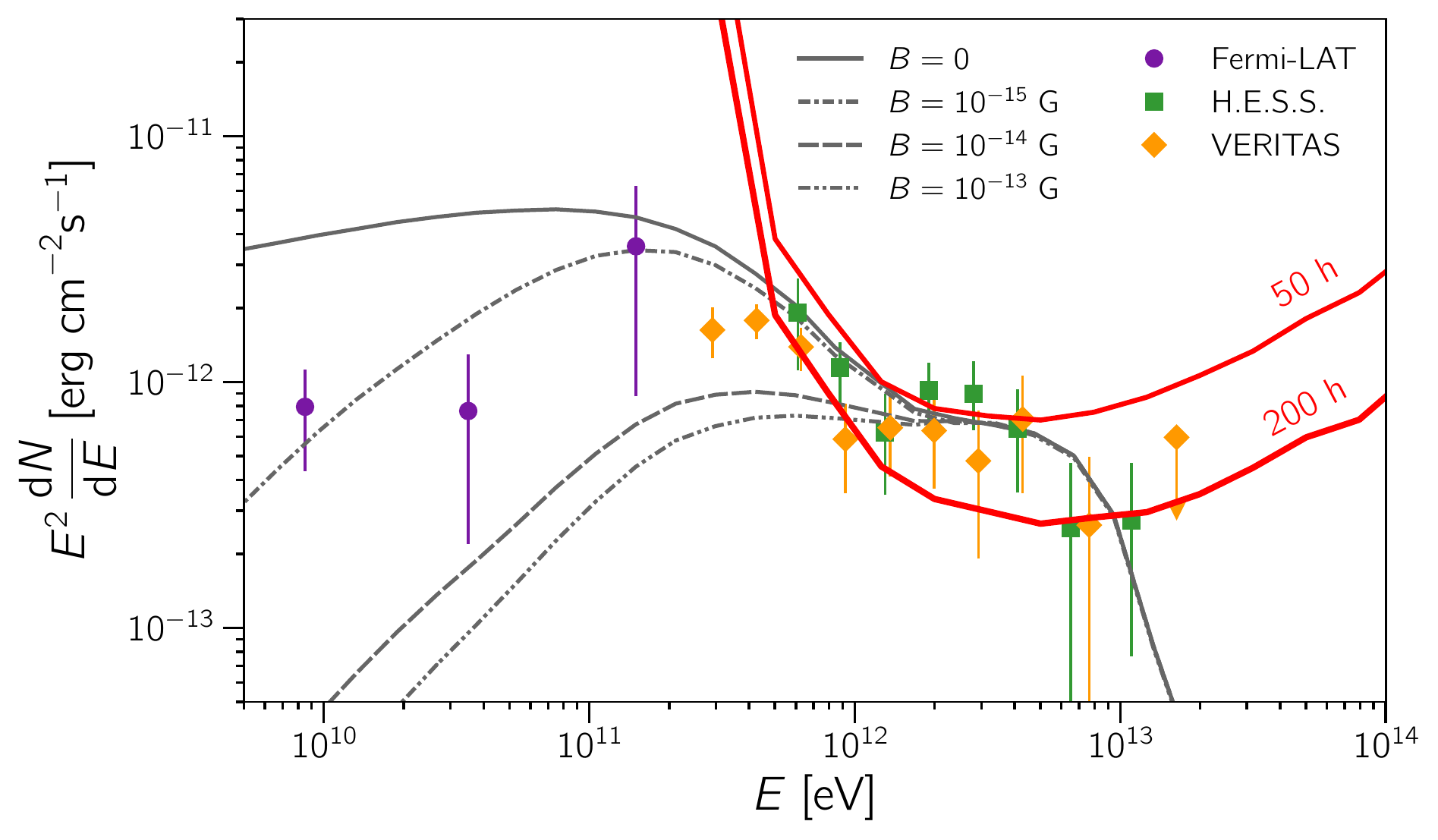}
    \caption{The simulated gamma-ray spectrum for 1ES~0229+200 is shown as dark-grey lines for different magnetic-field strengths, assuming a stochastic IGMF with coherence length 1~Mpc. The intrinsic spectrum of the object is a power law with slope $-1.5$ and an exponential cut-off at 7~TeV. The EBL model by \citet{2012MNRAS.422.3189G}. The thick red lines represent the projected differential sensitivity of the ASTRI Mini-Array for 50~and 200~hours. Measurements by Fermi-LAT~\citep[purple circles;][]{vovk2012a}, H.E.S.S.~\citep[green squares;][]{hess2007a}, and VERITAS~\citep[orange diamonds;][]{veritas2014a} are also shown for reference.}
    \label{FIG:igmf}
\end{figure}

In Fig.~\ref{FIG:igmf} we show the point-like fluxes resulting from these simulations for the angular resolution of the instrument. The differential sensitivity curves for the ASTRI Mini-Array suggests that for $B \gtrsim 10^{-14.5}$ the spectral suppression at energies below $\sim 1 \; \text{TeV}$ could be identified with 200~hours of observations. Note, however, that the cascade is fully isotropized for fields with strength $B \gtrsim 10^{-12} \; \text{G}$, which means that the region of the IGMF parameter space that can effectively be probed is $10^{-14.5} \lesssim B / \text{G} \lesssim 10^{-12}$, for coherence lengths $L_B \gtrsim 1$~Mpc.

We have not studied the prospects for measuring IGMFs with observations from Mkn~501. Nevertheless, this object has been widely used for this purpose~\citep[e.g., ][]{magic2010a, neronov2012a, takahashi2012a}, such that we anticipate it to be suitable for ToOs during high states. Complementary observations of the low state could also enable two-component (low and high states) analyses and IGMF constraints based on its light curve, setting bounds also on the coherence scale of IGMFs~\citep{neronov2013a, alvesbatista2020a}.

\paragraph{Analysis Method}--
It is clear from the sensitivity curve shown in Fig.~\ref{FIG:igmf} that the ASTRI Mini-Array will be able to probe part of the parameter space for IGMFs.
The emphasis lies on the precise determination of the energy spectrum and on the measurement of the angular distribution of the arriving gamma rays around the point-like sources.

\subsection{Blazars as probes for hadron beams}\label{Sec:hadrons}

\paragraph{Scientific Case}--
The identification of the sources responsible for the production of ultra-high energy co\-smic rays (UH\-ECR), with energies exceeding $10^{20}$ eV, is a formidable task, made difficult by the low statistics and by the deviation suffered by \rf{charged} nuclei in the cosmic magnetic web. Among the potential accelerators of UHECR, extragalactic relativistic jets have been widely discussed in the literature (e.g. \citealt{Biermann1998}). 
An interesting scenario connecting UHECR and blazars is the one postulating the existence of the so-called hadron beams ejected by extreme BL Lacs \citep[e.g.][]{2010APh....33...81E,2019MNRAS.483.1802T}. Introduced to explain some of the special features displayed by these peculiar sources \citep{2020NatAs...4..124B}, the hadron beam (HB) scenario assumes that the jets of these BL Lacs produce a collimated beam of high-energy protons/nuclei. 
Another powerful mechanism to produce UHECRs in blazars and specially the often observed variable \rf{VHE gamma-rays} is the acceleration in misaligned magnetic reconnection layers which may be naturally  driven by MHD kink instabilities in the helical fields of the magnetically dominated regions (\cite{2009MNRAS.395L..29G}; \cite{2020IAUS..342...13D};\cite{2020arXiv200908516M}).
In the HB scenario, while travelling towards the Earth, UHECR lose energy through photo-meson and pair production (Bethe–Heitler) reactions, triggering the development of electromagnetic cascades in the intergalactic space, producing cosmogenic photons and neutrinos~\citep{2019JCAP...01..002A}. Because of the reduced distance, high-energy gamma-rays produced by the cascades experience a less severe absorption by the interaction with the EBL and can reach the Earth \citep[e.g.][]{2010APh....33...81E}. Because of the reduced absorption, a distinctive prediction of this model is that the observed gamma-ray spectrum extends at energies much higher than those allowed by the conventional propagation through the EBL. For sources located at low redshift ($z<0.3$), the spectra should be characterized by a hard tail above 10 TeV, whose detection is considered the smoking gun of this model \citep[e.g.][]{2012ApJ...749...63M}. The ASTRI Mini-Array will be the first instrument with a sensitivity above 10 TeV high enough to test this scenario. The detection of even a few events at energies around 20-30 TeV for a source located at z=0.1 would give a strong support to this model, pointing to extreme blazars (and their misaligned counterparts) as UHECR sources.
\paragraph{Immediate Objectives}--
We propose ASTRI Mini-Array observations to test the prediction of the hadron beam model for the prototypical extreme blazar 1ES 0229+220. The relevant signature is a hard-tail extending at energies well above the expected EBL cut-off, detectable with the Mini-Array up to $\sim$20 TeV.
\paragraph{Observing Time, Pointing Strategy, Visibility and Simulation Setup}--
1ES 0229+200 displays an almost quiescent VHE spectrum and, therefore, it is suitable for “fill-in” observations. With 100 h (or, better 200 h) of observations, the sensitivity is expected to reach the level for which a detection is expected. A non-detection will severely constrain or even rule-out the model.

\begin{figure}
	\centering
		\includegraphics[width=0.49\textwidth]{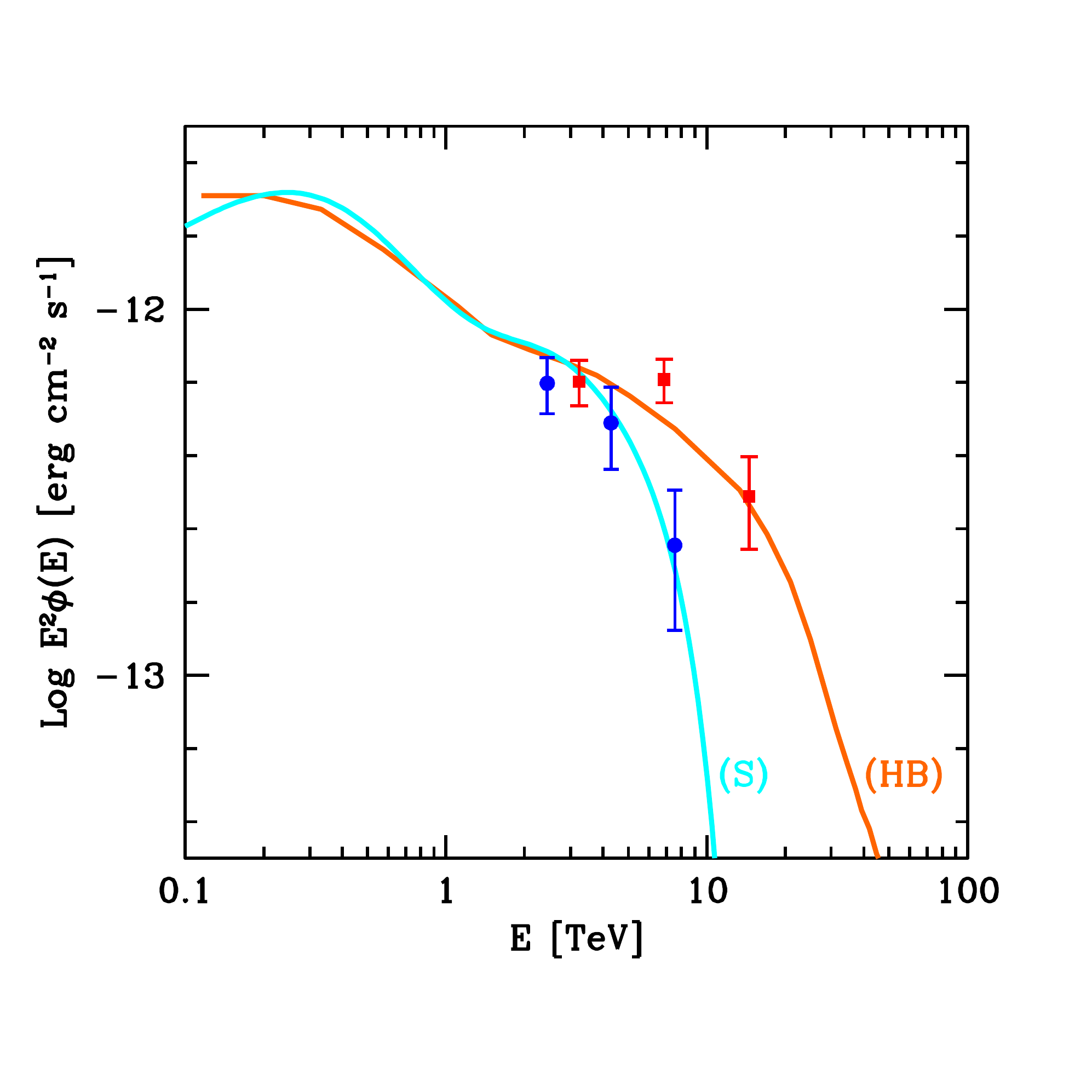}
	\caption{Simulated VHE spectrum of 1ES 0229+220 for the standard (light blue) and hadron beam (red) scenarios (from \citealt{2019MNRAS.483.1802T}). In the latter case the ASTRI Mini-Array would be able to obtain a detection up to 20 TeV, well above the cut-off expected in the standard case.}
	\label{FIG:Chap5_Fig01}
\end{figure}

\paragraph{Analysis Method}--
The feasibility of the proposed observations can be already judged from the sensitivity curve. However, given the peculiarity of the searched signal, specific simulations must be performed to precisely assess the potentiality of the observations. A first attempt performed with the ASTRI Sensitivity Calculator\footnote{\href{https://www.iasf-milano.inaf.it/\~giuliani/sgamati/VT/}{https://www.iasf-milano.inaf.it/\~giuliani/sgamati/VT/}} (v6) is reported in Fig.\ref{FIG:Chap5_Fig01}. The blue points correspond to the model assuming standard EBL absorption (cyan line) with an exposure of 200 hours, the red points with the HB model by \citet{2012ApJ...749...63M} for 250 h of exposure time. In this case, the clear ($\sim 4\sigma$) detection at 15 TeV demonstrates the capability of the ASTRI Mini-Array to confirm or rule out the model.

The analysis should be tailored on the capability to detect hard tails at low fluxes. Recent works remark that the angular distribution of the expected emission (halo) could be observable and useful to further strengthen the detection.

\subsection{Test on the existence of axion-like particles} \label{sec:axions}
Despite the success of the Standard Model (SM) of elementary particles in explaining the subatomic world, the SM is viewed as a the low-energy manifestation of some more fundamental and complete theory of all elementary-particle interactions, including gravity. Among the many attempts to shed light on the “ultimate” unified theory, the most promising ones are represented by four-dimensional ordinary and supersymmetric models, Kaluza-Klein theories, and especially superstring theories (for a review see \citealt{2010ARNPS..60..405J}). Remarkably, all these theories predict the existence of axion-like particles (ALPs). ALPs are spin-zero, neutral and very light pseudo-scalar bosons. They are a generalization of the axion (for a review, see e.g. \citealt{2010RvMP...82..557K}) which was proposed as a natural solution to the strong CP problem. Unlike the original axion, ALPs are supposed to interact primarily only with two photons and they do not present a strict relationship between their mass $m_a$ and their two-photon coupling, $g_{a\gamma\gamma}$. The Lagrangian describing the ALP field $a$ and photon-ALP interaction reads:
\begin{equation}
\label{EQ:Chap5_Eq1}
{\cal L}_{\rm ALP} = \frac{1}{2} \, \partial^{\mu} a \, \partial_{\mu} a - \frac{1}{2} \, m_a^2 \, a^2 + g_{a\gamma\gamma} \, {\bf E} \cdot {\bf B} \, a~,
\end{equation}
where ${\bf E}$ and ${\bf B}$ denote the electric and magnetic fields. In the presence of an external magnetic field ${\bf B}$, photon–ALP oscillations may occur (in the previous equation, E represents the propagating photon field) in a similar way as for neutrino oscillations. As a result, every magnetized medium represents a possible environment for photon-ALP oscillations to take place. Therefore, many attempts in laboratory experiments (such as the Light-shining-through a wall experiment, \citealt{2011ConPh..52..211R}) to identify photons from photon-to-ALP reconversion use strong magnetic fields. However, since photon-ALP interaction is faint, very strong magnetic fields and/or very long distances are necessary in order to produce the effect.
 This is the reason why astrophysical background represents the best candidate in order to identify ALP effects \citep{2019arXiv191109372G}. In particular, ALPs can be produced inside blazar magnetic fields, mitigating the absorption of VHE photons by the IR-optical-UV backgrounds. In a similar way, but in the extragalactic space, photon-ALP oscillations, in the presence of the extragalactic magnetic field, may increase the Universe transparency to VHE photons, mitigating their absorption due to their interaction with the EBL \citep{2018PhRvD..98d3018G,2018JHEAp..20....1G}. In addition, ALPs strongly modify the observable spectra of BL Lacs, inducing a photon excess for energies above $\sim$10 TeV and spectral distortions in the form of a pseudo-oscillatory behavior with respect to the energy \citep{2019MNRAS.487..123G}.

The absence of signals of ALPs produced via the Primakoff scattering in the Sun (ALPs are reconverted back to photons inside the magnetic field of a decommissioned magnet of the LHC) set firm bounds for ALPs: $g_{a\gamma\gamma} < 0.66 \times 10^{-10}$ GeV$^{-1}$ for $m_a < 0.02$ eV \citep{2017JInst..12P1019A}. Comparison of ALP-induced modification of stellar evolution in globular clusters with observations set the same bound: $g_{a\gamma\gamma} < 0.66 \times 10^{-10}$ GeV$^{-1}$ \citep{2014PhRvL.113s1302A}. No preference for ALP-induced spectral irregularities in fitting the spectrum of gamma-rays from the Perseus cluster indicates $g_{a\gamma\gamma} < 5 \times 10^{-12}$ GeV$^{-1}$ for $5 \times 10^{-10} < m_a < 5 \times 10^{-9}$ eV \citep{2016PhRvL.116p1101A}.

With its high sensitivity in the TeV energy band, ASTRI Mini-Array represents the current best observatory in order to detect possible deviations of the BL Lac spectra from the standard physics and shed light on the existence of the ALPs. In particular, through a dedicated observational campaign of Mkn 501 and 1ES 0229+200 we expect to observe a photon excess for energies above ~10 TeV. Such an eventual detection would tell us that standard physics is incomplete; however, we would be unable to discriminate if the responsible process is the hadron beam, the LIV, or the photon-ALP oscillations. In fact, as discussed in \citet{2020MNRAS.491.5268G}, for Mkn 501 both LIV and photon-ALP oscillations produce this excess, while for 1ES 0229+200 hadron beam and photon-ALP oscillations may be invoked to explain the photon surplus. A way out from this conundrum would be to have a very high energy resolution in order to detect, in the observed spectra, energy oscillations, which are an exclusive ALP imprinting. This task appears challenging for Mkn 501 and totally prohibitive for 1ES 0229+200 because of their low fluxes in the energies of interest ($>$ 1 TeV). However, concomitant observations with high energy resolutions at lower energies ($0.2-2$ TeV) from other IACTs may detect/miss energy oscillations, thus disentangling the different physical processes. 
\paragraph{Immediate Objectives}--
We propose to use the ASTRI Mini-Array at energies above 10 TeV to detect the spectral tail at high energies, expected as a result of the photon-ALP mixing.
\paragraph{Observing Time, Pointing Strategy, Visibility and Simulation Setup}--

\begin{table*}[width=2.\linewidth,cols=8,pos=htp!]
\centering
\caption{Summary Table of sources proposed to test of HB, LIV and ALP.}
\def\arraystretch{1.5}
\label{tab:6.2}
 \begin{tabular*}{\tblwidth}{@{}CCCCCCCC@{} }
\toprule
Target & Class & RA (J2000) &  DEC (J2000) & Obs. time & ZA & Moon & Strategy, analysis, notes \\
IAU Name & & & & [hr] & [deg] & [\%] & \\
\midrule
Mkn 501& Blazar & 16 53 52.2 & +39 45 36.6 & 50-100 & 45 & 25 & LIV, ALP. Better suited for\\
 & & & & & & & ToOs in high states.\\
1ES 0229+200 & Blazar & 02 32 48.6 & +20 17 17.5 & 200 & 45 & 25 & HB, LIV, ALP. Almost steady\\
 & & & & & & & source, possible ``fill in" target.\\
\bottomrule
\end{tabular*}
\end{table*}

The source selection is based on the spectral properties of the two sources. In fact, for the detection of effects related to ALP, the ideal target should display a hard spectrum, possibly extending up to tens of TeV.

For Mkn 501, the best opportunities are offered by observations during (relatively frequent) high and hard states. Historical records show that these states can last for several days, allowing to easily accumulate 50 hours of data during a single event. As shown in Fig.\ref{FIG:Chap5_Fig02}, with 50 hours, the ASTRI Mini-Array offers the possibility to detect the ALP-induced tail up to 50 TeV.

1ES 0229+200 displays an almost quiescent VHE spectrum and, therefore, it is also suitable for “fill-in” observations. With 100-200 hours of observations, and depending on the behavior of the intrinsic spectrum, the tail produced by ALP-photon mixing is detectable up to 20 TeV.

The feasibility of the proposed observations can be already judged from the sensitivity curve. However, given the peculiarity of the searched signal, specific simulations (in progress) must be performed to precisely assess the potentiality of the observations. For 1ES 0229+200, the similarity of the spectrum with that expected for the HB model (see Fig.~\ref{FIG:Chap5_Fig01}) suggests that also for ALPs we can expect a result similar to that shown above for the HB model, i.e. a solid detection above 10 TeV.
\paragraph{Analysis Method}--
Given the similarity of the expected signal, as for the HB case also for ALP the emphasis is on the capability to detect hard tails.
\subsection{Lorentz Invariance violation studies} \label{sec:lorentzInvariance}
%
%
\paragraph{Scientific Case}--
Being one of the most fundamental symmetries of Nature, Lorentz invariance lies at the heart of modern physics and shapes the most elementary physical laws. 
In recent times,  
in the context of emergent gravity models  and various quantum theories of gravity,
the idea  that this symmetry is broken at and beyond the Planck scale
\citep{gambini99,carroll01,smolin02,horava09,abonanno13,kharuk,2020PhRvD.102b6007E}
has been proposed, thus implying interesting observable consequences at low energies \citep[e.g.][]{2013CQGra..30m3001L}.
In particular, in a quantum gravitational framework, one expects 
that the full diffeomorphism invariance is broken and only the stability group of the metric 
is unbroken, leading to the presence of a preferred frame \citep{abonanno19}. 

Among the possible effects of LIV, those involving the behavior of highly energetic photons can be effectively probed by astrophysical observations. In fact, the long distances involved in the propagation of photons from cosmic sources allows the tiny effects predicted by LIV schemes to accumulate and become visible. 
From the phenomenological point of view, an effective way to model the expected LIV effects is through modified dispersion relations. For photons, the most commonly assumed modified dispersion relations reads:
\begin{equation}
\label{EQ:Chap5_Eq2}
E^2=p^2c^2 \pm \frac{E^{n+2}}{E_{\rm LIV}^n}
\end{equation}
where $E_{\rm LIV}$ is the energy scale at which LIV should occur (thought to be of the order of the Planck energy) and $n$ is the order of the term. The sign defines the so-called superluminal ($+$) or subluminal ($-$) case. In the subluminal case the modified dispersion relation leads to the modification of the kinematics of the standard pair production reaction which regulates the absorption of gamma-rays by the EBL. In particular, the threshold is strongly modified and, above an energy of the order of few tens of TeV, the resulting cosmic opacity is strongly reduced, allowing the unimpeded propagation of VHE photons from cosmological distances. The powerful gamma-ray beam of blazars, potentially extending up to several tens of TeV, is the natural probe for such effects \citep[e.g.][]{2016A&A...585A..25T,2020NatAs...4..124B}.
\begin{figure}
	\centering
		\includegraphics[width=0.49\textwidth]{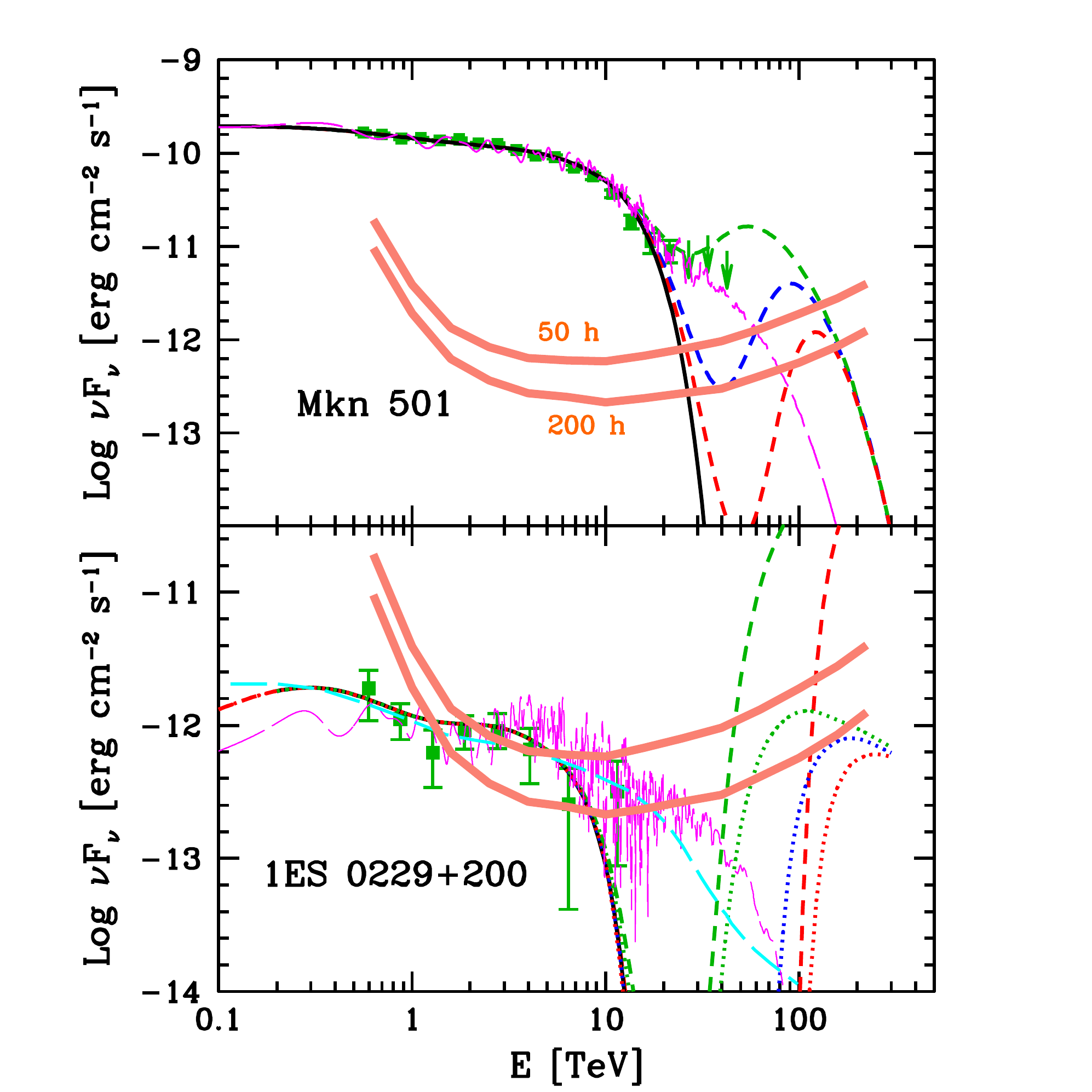}
	\caption{Upper panel: VHE spectrum of Mkn 501 measured by HEGRA during the extreme outburst in 1997 (green triangles). The black solid curve reports an intrinsic cut-off power-law spectrum absorbed by interaction with EBL. The magenta long-dashed line shows the observed spectrum assuming mixing of photons with ALPs (from \citealt{2020MNRAS.491.5268G}). The dashed curves report the observed spectrum assuming an intrinsic cut-off power-law spectrum and LIV occurring at different energy scales (from \citealt{2016A&A...585A..25T}). Lower panel: as above for the case of 1ES 0229+200 (green symbols: data from HESS). For the LIV case we consider the intrinsic spectrum described by an unbroken (short dashed) or a broken (dotted) power law (see \citealt{2016A&A...585A..25T} for details). In both panels, the red thick lines show the expected sensitivity of the ASTRI Mini-Array for 50 hours and 200 hours of exposure.}
	\label{FIG:Chap5_Fig02}
\end{figure}
The detection of galactic sources at energies exceeding 100 TeV offers a complementary powerful diagnostic tool for LIV studies. The method is based on the fact that, if LIV holds, photons above a certain threshold (related to the energy scale of LIV) quickly decay after their emission. Quite remarkably, this kind of test probes a different sector of the LIV parameter space, interesting for the opacity test (decay occurs in the superluminal LIV, opposite to the subluminal case probed by cosmic opacity). The two methods (photon decay and reduction of the cosmic opacity) can thus be used in a complementary way. Due to the huge absorption caused by the interaction with the EBL, only galactic sources are accessible in this energy range. A handful of sources recently detected by HAWC  \citep{2020ApJ...896L..29A} and LHAASO \citep{2021Natur.594...33C} appears particularly interesting. For instance, \citet{2020PhRvL.124m1101A} obtains quite strong constraints on the LIV energy scale, improving by almost two order of magnitude previous results. Even stronger limits can be obtained exploiting the recent sources detected by LHAASO up to 1 PeV.
\paragraph{Immediate Objectives}--
Observations with the ASTRI Mini-Array at energies above 10 TeV can be exploited to perform tests of the Lorentz Invariance Violation (LIV) in the subluminal sector. The main goal of the proposed observations is the identification of deviations in the spectra of Mkn 501 and 1ES 0229+200. In particular, for LIV the expected signal would correspond to an excess of photons at energies around $30-50$ TeV (the exact position depending on the energy scale of LIV), hardly explicable within the standard framework (i.e. spectrum absorbed by EBL). 
\paragraph{Observing Time, Pointing Strategy, Visibility and Simulation Setup}--
The source selection is based on the spectral properties of the sources. In fact, the ideal target should display a hard spectrum, ideally extending up to tens of TeV.

For Mkn 501, the best opportunities are offered by observations during (relatively frequent) high and hard states. Historical records show that these states can last for several days, allowing to easily accumulate 50 hours of data during a single event. As shown in the Fig.\ref{FIG:Chap5_Fig02}, with 50 hours, the ASTRI Mini-Array offers the access to LIV-modified spectra with energy scale $10^{20}$ GeV and to probe the existence of the ALP tail up to 50 TeV. 1ES 0229+200 displays an almost quiescent VHE spectrum and therefore it is also suitable for ''fill-in" observations. With 50 h (or, better, 100-200 h) of observations, and depending on the shape of the intrinsic spectrum, LIV at energy scales above $10^{20}$ GeV can be explored and the existence of a hard tail can be tested.

In the plot we report the VHE section of the SED of our preferred targets, Mkn 501 and 1ES 0229+200. The black line shows the standard spectrum attenuated by the absorption with EBL. The short-dashed lines report the spectrum expected for LIV at different energy scales (green: $3\times10^{19}$ GeV, blue: $10^{20}$ GeV, red: $2\times 10^{20}$ GeV) and, for 1ES 0229+200 only (dotted lines), for different assumptions on the intrinsic spectrum of the source (unbroken, dashed, vs broken, dotted, power law). 

The feasibility of the proposed observations can be already judged from the sensitivity curve. However, given the peculiarity of the searched signal, specific simulations (in progress) must be performed to precisely assess the potentiality of the observations.

\paragraph{Analysis Method}--
The analysis should be tailored on the search of excesses around 30-50 TeV.

\section{GRB \& Time-domain Astrophysics}\label{sec:6}
The detection of VHE counterparts to transient events has proved fundamental to constrain radiative processes, particle acceleration, and the physics of mechanisms responsible for the most extreme astrophysical sources.
Examples of recent, major advances in this field are the discovery of TeV radiation from gamma-ray bursts (GRBs, \citealt{MAGIC2019a}) and the observation of VHE radiation from the blazar TXS~0506+056 in association with the detection of a HE astrophysical neutrino by the IceCube observatory \citep{2018Sci...361.1378I}.

The detection by the MAGIC telescopes of GRB\,190114C \citep[]{MAGIC2019a}, located at $z=0.42$, opened a new window in the exploration of GRB physics. MAGIC observations, complemented by multi-wavelength observations from radio to \gray s \citep[]{MAGIC2019b}, revealed the presence of a new and energetically relevant component in GRBs, likely synchrotron self Compton (SSC) emission, extending into the TeV energy range. 
The production of VHE radiation in GRBs and its detectability by Cherenkov telescopes have been confirmed by the observations of three additional events, two by the H.E.S.S. telescopes hours after the burst (GRB\,190829A at $z=0.078$ and GRB\,180720B at $z=0.65$, \citealt[]{Abdalla:2019qy,2019ATel13052....1D}) and one by the MAGIC telescopes, starting $\sim 1$ minute after the burst (GRB\,201216C at $z=1.1$, \citealt{magic:gcn29075}).
The extension in energy of this new emission component (at least up to 1\,TeV in GRB\,190114C) and its presence in low-luminosity, nearby events (e.g., GRB 190829A) open to the possibility of GRB studies with the ASTRI Mini-Array.
In particular, in the case of nearby GRBs, ASTRI Mini-Array can be unique for the study of the spectral shape and evolution at energies $>1$\,TeV of the VHE component, whose exact nature and implications on the physics of the emission region still need to be fully investigated.
Besides, GRBs detected at these energies can probe the IR component of the EBL (see Sect.~\ref{sec:EBL}) and spot suggested anomalies that may point to dark matter candidates like axions, as described in Sect.~\ref{sec:axions}. We plan to set up a follow-up program for the ASTRI Mini-Array to specifically follow alerts of nearby and particularly bright GRBs that can be followed under favorable observing conditions.

All four GRBs with clear TeV emission detected so far belong to the class of long events. A hint of TeV emission from a short GRB (observed in association with a Kilonova) was found by MAGIC in GRB\,160821B ($z=0.16$) \citep{Inoue2019,magic:160821b}, representing a possible forerunner of the detection of a TeV counterpart to a gravitational wave (GW) event.
Follow-up observations of GW\,170817A by H.E.S.S. resulted only in upper limits on the VHE flux associated with the GW source \citep{hess:170817_obs}, but provided useful constraints on the strength of the magnetic field \citep{hess:170817_Bfield}.

A dedicated follow-up program is required also for alerts provided on high-energy neutrinos. The energetic astrophysical neutrino IC\,170922A, detected by the IceCube observatory, has been associated with the \gray\ emitting blazar TXS~0506+056, detected from the GeV up to TeV range by {\it Fermi}/LAT and MAGIC \citep{2018Sci...361.1378I}. 
The association between HE neutrinos and blazars needs to be confirmed with other sources.
This motivates the search of similar neutrino counterparts by the ASTRI Mini-Array.
Observations of distant blazars by ASTRI Mini-Array, however, are strongly affected by EBL. 
Moreover, a measurable neutrino flux implies a large density of the target radiation field in some emission scenarios, resulting in a strong internal absorption of TeV photons.
A dedicated selection strategy on the neutrino alerts is needed to minimize the observation time and to let the ASTRI Mini-array be competitive with respect to HAWC and to the other IACTs observing in similar and contiguous energy ranges.

\subsection{Alerts from GRBs and GWs}

\paragraph{Immediate Objective}--
We propose to set up an observational program for follow-ups by the ASTRI Mini-Array on alerts provided on GRBs and GWs. 
Facilities observing GRBs (multi-frequencies observatories and space observatories) and GWs (LIGO-Virgo and Kagra) distribute alerts through the GCN network within few seconds to few minutes from the burst detection. 
We aim at observing the VHE transient candidates in a \rf{relatively}  short time (less than one up to few minutes) from the communication of the alert. This might result in the detection of TeV emission from nearby GRBs in the early afterglow phase, up to tens of minutes from the burst. To test this possibility, we perform simulations of TeV emission from GRBs as a function of time for different source distances.

\paragraph{Observing Time, Pointing Strategy, Visibility and Simulation Setup}--\label{trans_strategy}
We plan to build a dedicated procedure to select promising targets according to the parameters and information available at the moment of the alert. 
In particular, target of opportunity selection will be based upon: 
\begin{itemize}
\item visibility and sky position (zenith-angle),
\item time of the burst (visibility of the target/region),
\item uncertainty on the arrival direction,
\item nature of the event (e.g., short/long GRB, 
or NS-NS or NS-BH for GWs),
\item distance (if available, e.g. for GW detections),
\item fluence and counting rate measured by the alerting instrument (for GRBs), or the X-ray/$\gamma$-ray flux (a proxy for the flux of the TeV emission, as shown in \citealt[]{MAGIC2019a}).
\end{itemize}

More specifically, selection of good candidates will be restricted to low-redshift targets (e.g. $z\lesssim 0.4$) when this information is available, or to those events with the highest fluence, 
limited uncertainty in the reconstructed position (comparable or smaller than the FoV of ASTRI, namely few tens of square degrees), and that can be observed in fairly good conditions (zenith angles < 60 deg) immediately after the burst or within few hours. 

The selection criteria limit the observed transient candidates to a few per year.
When a good ToO candidate is selected, a fast reaction is required, and an observation with the ASTRI Mini-Array for a relatively 
limited amount of time, comparable to the visibility of the source within the same night (1-3 hours), has to be performed. 

If preliminary scientific results on the ToO observation will be available within one hour from the start of the observation, they can be evaluated in order to extend the observation for the rest of the night or in the next day.
Late follow-up or prolongation of observations will also be evaluated by means of the successive (minutes/hours) follow-up observations by other multi-frequency observatories (e.g. optical or X-ray).
The aim of the simulations performed (see below) is to understand what are the prospects for ASTRI Mini-Array in terms of redshift, starting time and duration of the observations.

Concerning GW alert, when the uncertainty region is relatively large (yet within few tens of square degrees), a scan of the region will be evaluated, which will trade-off between the limited observing time and probability of covering a good fraction of the confidence region. This can be accomplished using the set of parameters derived from the burst, e.g. the density of galaxies within the 3D GW map, or by means of a uniform coverage of the 90\% or 50\% uncertainty region \citep{Patricelli:2018ec,2017ApJ...846...62S,2019MNRAS.490.3476B}.

\paragraph{Analysis Method}--\label{GRBsim}
We adopt the temporal and spectral properties of the VHE component recently discovered by MAGIC in GRB\,190114C as a template to study the detection prospects with the ASTRI Mini-Array.
GRB\,190114C was detected by MAGIC up to $\gtrsim$1\,TeV, from $\sim$\,60\,seconds to $\sim$\,40\,minutes after the GRB trigger. 
To predict theoretical spectra at energies higher than 1\,TeV, relevant for the ASTRI Mini-Array, and to study detectability at earlier/later times, we have adopted the same theoretical model (and model parameters) used to describe GRB~190114C VHE \citep[SSC radiation from electrons accelerated in the interaction between the jet and the external medium,][]{MAGIC2019b}. 
Besides, we used GRB\,190114C as a template to simulate the emission from GRBs at shorter distances: $z=0.078$ (corresponding to the redshift of GRB~190829A, detected by H.E.S.S.), and the intermediate redshift $z=0.25$. The luminosity in their rest frame has been assumed to be the same of GRB~190114C, but their emission has been rescaled for the different cosmological distances. The proper amount of EBL absorption (from \citealt{2017A&A...603A..34F}) has been included, to compute the observed emission at those redshifts. The resulting theoretical lightcurves and spectra have been used as input to perform simulations of detectability with the ASTRI Mini-Array.

\begin{figure}
	\centering
		\includegraphics[width=0.375\textwidth,angle=-90]{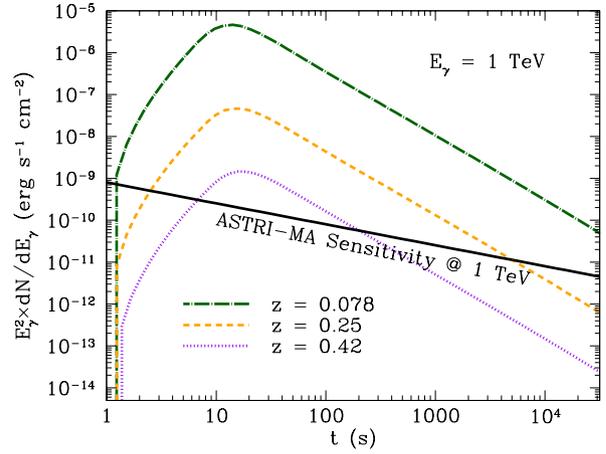}
	\caption{Synthetic light curves at 1\,TeV of the three simulated GRBs, obtained adopting GRB\,190114C ($z=0.42$) as a template and moving the GRB at shorter distances ($z=0.25$ and $z=0.078$), as described in the text. The black line shows the sensitivity of the ASTRI Mini-Array at 1\,TeV, rescaled for the corresponding integration time on the $x$-axis.
	}
	\label{fig:GRBLC}
\end{figure}

\begin{figure}
	\centering
		\includegraphics[width=0.5\textwidth]{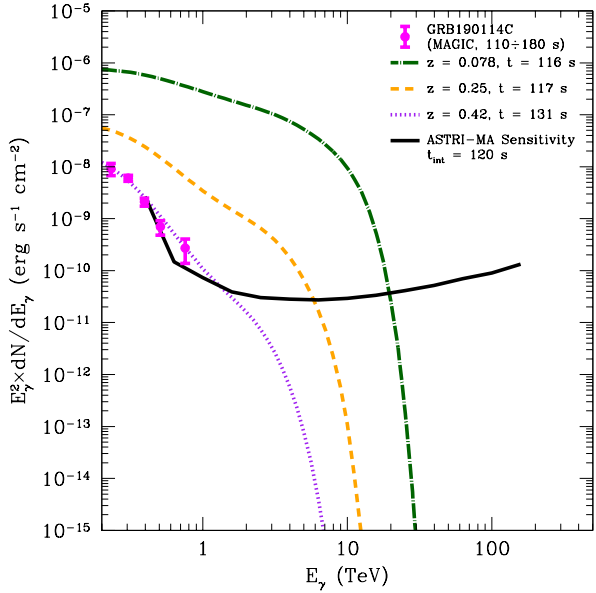}
	\caption{SEDs of the three simulated GRBs, computed at about 2 minutes after the burst. The black line shows the differential
	sensitivity of the ASTRI Mini-Array for an integration time of 2 minutes. The data points superimposed to the SED for $z=0.42$ are the spectral points measured by MAGIC on GRB~190114C, in the time interval 110-180\,s.}
	\label{fig:GRBSED}
\end{figure}

Figure~\ref{fig:GRBLC} shows the synthetic lightcurves at the three different redshifts at 1\,TeV, and the corresponding ASTRI Mini-array sensitivity.
The SED at $E>200$\,GeV are shown in Fig.~\ref{fig:GRBSED}, and are computed at about 2 minutes after the burst. For comparison, the spectral points on GRB\,190114C are shown, as measured by MAGIC in the time interval [110-180\,s] from the burst.
Moving GRB~190114C at smaller redshifts, the observed flux considerably increases, especially above 1\,TeV, as a result of the shorter distance of the source and the smaller attenuation caused by the EBL.

To simulate the response of the ASTRI Mini-Array in the three different cases, we considered an observation starting 200\,seconds after the burst and lasting 600\,s.
The simulations were performed with the  {\tt ctools}  \citep[][v.\ 1.6.3]{2016A&A...593A...1K}\footnote{\href{http://cta.irap.omp.eu/ctools/}{http://cta.irap.omp.eu/ctools/}.} analysis package.
The results are reported in Fig.~\ref{fig:MA-GRB}. 
The simulations clearly show the feasibility of the detection of  TeV emission  by the ASTRI Mini-Array, and allow us to draw the following conclusions:

\begin{figure}
	\centering
		\includegraphics[width=0.5\textwidth]{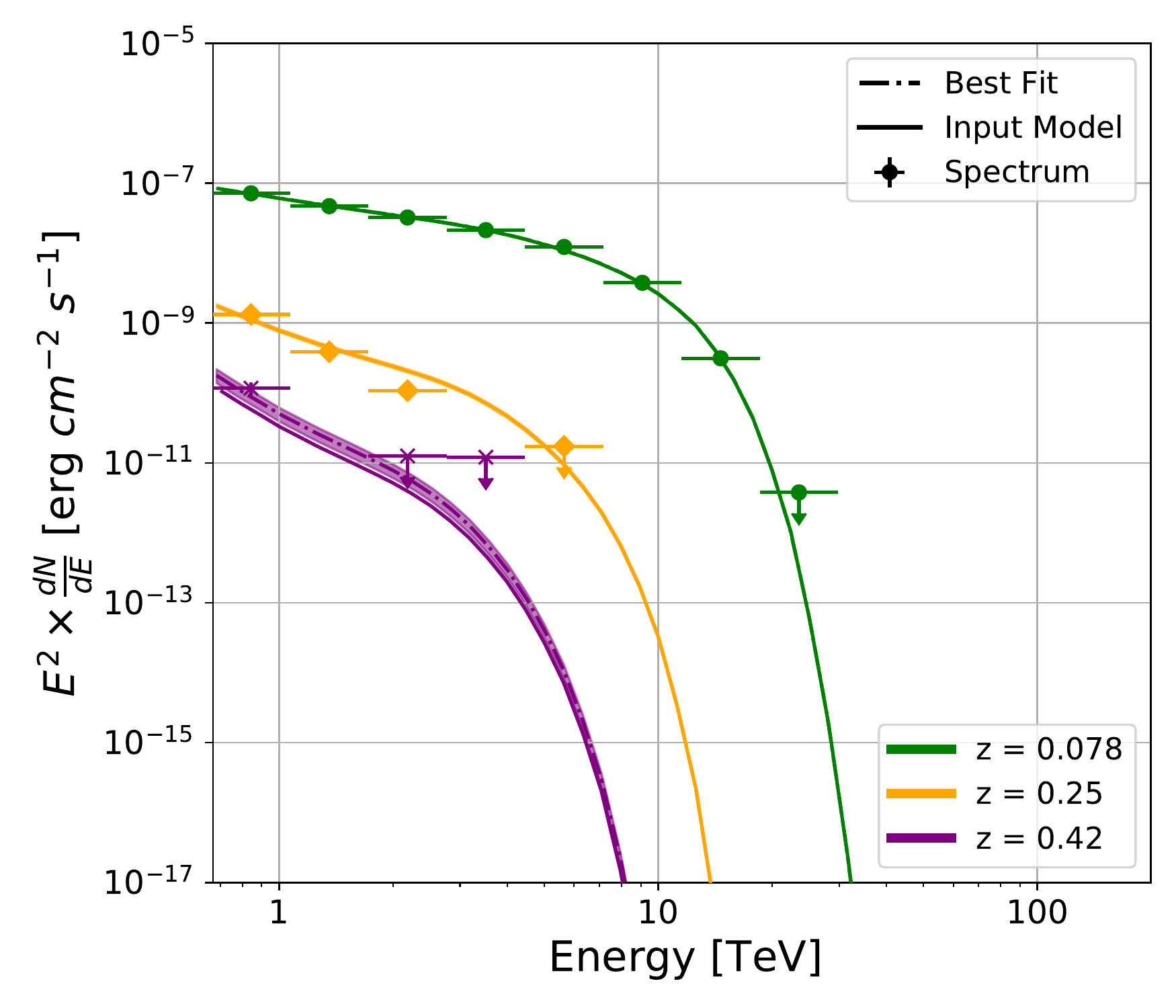}
	\caption{Simulated response of the ASTRI Mini-Array to the emission from three GRB\,190114C-like bursts, at three different redshifts, $z=0.078$, $z=0.25$ and $z=0.42$. The simulation considers an observation started at $t\sim 200\,$s after the initial burst, with flux decaying according to the lightcurves in Fig.~\ref{fig:GRBLC}, integrated for $\sim 600\,$s.
}
	\label{fig:MA-GRB}
\end{figure}

\begin{enumerate}
    \item
the ASTRI Mini-Array might have detected emission from GRB~190114C, as shown in  Fig.~\ref{fig:MA-GRB} (violet points and line); 
\item
the ASTRI Mini-Array is able to confirm afterglow emission at $E>$1\,TeV from close GRBs at redshift smaller than $\sim 0.4$, if observations start within the first tens of seconds, up to few minutes from the onset of the burst (Fig.~\ref{fig:GRBLC});
\item
  in case of detection, the ASTRI Mini-Array can measure the spectral cutoff, either originated by the EBL absorption or intrinsic (if greater than $\sim 1\,$TeV, see Fig.~\ref{fig:GRBSED} and Fig.~\ref{fig:MA-GRB}).
\end{enumerate}

The number of GRBs that are expected to be followed by the ASTRI Mini-Array is low. From the observations performed by existing Cherenkov telescopes, e.g. the MAGIC telescopes located in the same region of the Mini-Array \citep[]{2019ICRC...36..634B}, the expected number of follow-ups on observable GRBs is $\sim 1$ per month. In consideration of the stricter constrains (see previous paragraph), this can be considered as an upper limit on the number of GRBs that can be pointed and observed by the ASTRI Mini-Array, soon after the burst.

On the contrary, GW alerts are expected with a higher rate during the LIGO-Virgo scientific run O4 (starting in 2022). The number of valid GW alerts will be tuned to set the rate similar to GRBs; this can be accomplished tuning the parameters defining the visibility (e.g. area of the uncertainty region) or physical parameter (e.g. distance, or nature of progenitor), to allow us to select the most promising GW alerts. Besides, an ad hoc observing strategy, e.g. selecting regions of the sky with clumps of galaxies and with a smart tiling depending on the integration time \citep{Patricelli:2018ec}, will allow to optimize the observing time to enhance the probability of detection. Despite the amount of observing time devoted by ASTRI Mini-Array to the observation of transients being negligible in the overall time budget of the ASTRI Mini-Array, our preliminary studies show that there are the premises for a significant scientific reward.

\subsection{Alerts from neutrino and associated blazars}

\paragraph{Immediate objective}--
We propose to set up an observational program for follow-ups by ASTRI Mini-Array on alerts provided on high-energy neutrinos. Facilities observing astrophysical neutrinos (like IceCube), distribute alerts through the GCN network within few seconds to few minutes from the burst detection. 
The detection of $\gamma$-rays at energies greater than a few TeVs will have an immediate impact on the description of the emission mechanism and acceleration processes of the associated source. The large flux level expected from these events, when promptly observed, provides a unique test-bench for the propagation of TeV photons in the intragalactic medium, that allow us to test LIV predictions and EBL models (see also Sections \ref{sec:EBL} and \ref{sec:axions}).

\paragraph{Analysis Method}--
For a neutrino event, there is no as a stringent request on the begin of the ToO observation as in the case of a GRB. 
Like for TXS\,0506+056, claimed as the first VHE neutrino source ever detected \citep[]{2018Sci...361.1378I}, the flaring activity observed by MAGIC at GeV-TeV energies from this object was observed a few days after the IC-170922A detection alert \citep[]{Ansoldi:2018qa}. The possible EM counterparts, within the uncertainty region on the IceCube neutrino reconstructed arrival direction, should be carefully evaluated before deciding to schedule the ToO follow-up. Detection of enhanced MeV-GeV \gray{} activity in the proximity of the neutrino trigger time as well as of high hard-X-ray state from known AGN blazars (in particular, those ones of the \rf{high synchrotron peaked} blazar sub-class) positionally compatible with the neutrino position, might be decisive to decide whether to observe the neutrino region or not, even on the days 
after the trigger time.

\subsection{Legacy Products, Multi-wavelength Synergies, Coordinated Observations. }

This observation program is strictly related to the coordination with other observatories, and to the possibility of a coverage in all the frequencies.
A multi-wavelength (MWL) and multimessenger network is already in place, providing key information on the burst and successive follow-ups by other instruments. 
This information is communicated through automatic GCN and circulars. 
The automatic GCN will be received and will be elaborated either in a vetted or in an automatic mode. Also, a dedicated group within the collaboration (burst advocates) will have the task to check the circulars and react promptly for any action requiring human intervention (e.g. retraction due to new information on the burst, or selection of scan regions).
Since a network of many different observatories and facilities is in place, there is no need of a direct coordination for the observations. The corresponding MWL data will be available and can be agreed for successive publication. Instead, it may be useful to have coordinated observation programs with similar facilities in the same region, like the Cherenkov telescopes MAGIC and LST, located on La Palma island. Besides, similar coordination with other optical facilities (e.g. TNG at La Palma) can be investigated. 
\section{Direct measurement of cosmic rays}\label{sec:7.2}
%

\subsection{Scientific Rationale}
Gamma-ray astronomy is undoubtedly the core science of the ASTRI Mini-Array. However, considering that more than 99\% of the observable component is hadronic in nature, it is useful to think about how to use this enormous amount of information contained in the hadronic channel of cosmic rays (CRs), considering that, even today, the origin of CRs, the acceleration mechanism, and the propagation process in the interstellar medium, are being studied. With its 9 telescopes, each with a field of view of about 10 degrees, the ASTRI Mini-Array will be the first array of Cherenkov telescopes with such a wide field of view. The potential outcomes exploiting these unique characteristic, focused to \gray{} astronomy, are widely described in the previous Sections.
The main challenge in detecting \gray{}s is to distinguish them from the much wider background of hadronic CRs. This background, recorded during normal \gray{} observations, could be used to perform direct measurements and detailed studies of some very significant non \gray{} Astrophysics topics.

\subsection{Cosmic ray heavy nuclei}
As proposed by \cite{2001APh....15..287K}, a promising method to measure CR composition in the energy band from few TeV to PeV consists in the detection of Cherenkov light emitted from primary particles, prior to their first interaction in the atmosphere. The more efficient method for heavy nuclei (iron) relies on the identification  of a single high intensity pixel in the camera images  of the detected Extensive Air Shower (EAS). This \rf{pixel lies} between the reconstructed shower direction and the center of gravity of the shower. The charge of the primary particles is   proportional to the intensity of the Cherenkov direct-light (I$_{DC}$), and can be estimated by the relation:
\begin{equation}
    Z^{\star} = d(E,\theta)\sqrt{I_{DC}}
\end{equation}
where $d(E,\theta)$ is a normalization factor that takes into account energy ($E$) and zenith angle ($\theta$) of the primary particles, as deduced by simulations.
\begin{figure}
	\centering
	\includegraphics[width=0.4\textwidth]{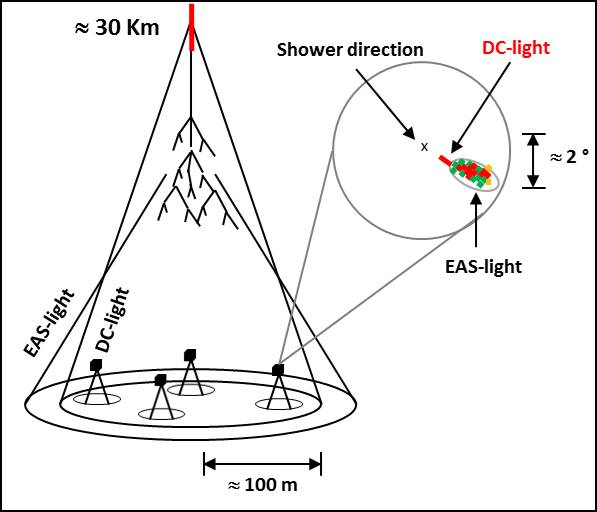}
	\caption{Schematic representation of the Cherenkov emission from a cosmic-ray primary particle and the light distribution on the ground and in the camera plane of an Cherenkov telescope. Direct Cherenkov light is emitted by  the primary particle before the first interaction with atmospheric nuclei. In the image-camera most of the DC light is concentrated in an area of angular size of $0.15^{\circ}$ to $0.3^{\circ}$(in a pixel for ASTRI Mini-Array camera), typical emission angle for DC-light.}
	\label{FIG:Chap7_sec2_f1}
\end{figure}
Figure~\ref{FIG:Chap7_sec2_f1}, adapted from \cite{2001APh....15..287K}, shows schematically this approach. 
To date, no investigation has been carried out on the ASTRI-Horn prototype data mainly because, with a single telescope, it is very difficult to demonstrate that a pixel of the image, that meets the above conditions, is actually the pixel of DC light, even if the topology is similar to the one predicted by the simulations. In the case of many telescopes, and therefore of the same event captured by more than one camera, the analysis could lead to a confirmation with a level of significance increasing with the number of image-cameras involved in the same shower. For illustrative purposes only, Figure~\ref{FIG:Chap7_sec2_f2} shows one of the limited number of events detected with the prototype in a run of March 2019, that could be interpreted as the signature of an heavy nucleus.
\begin{figure}
	\centering
	\includegraphics[width=0.4\textwidth]{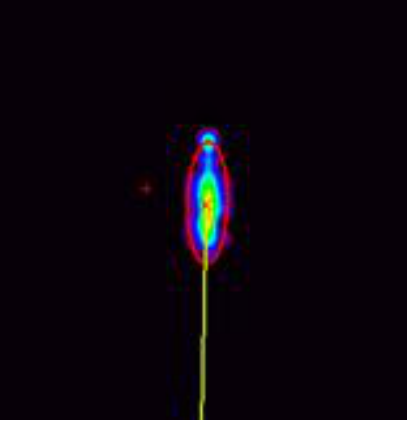}\\
	\includegraphics[width=0.4\textwidth]{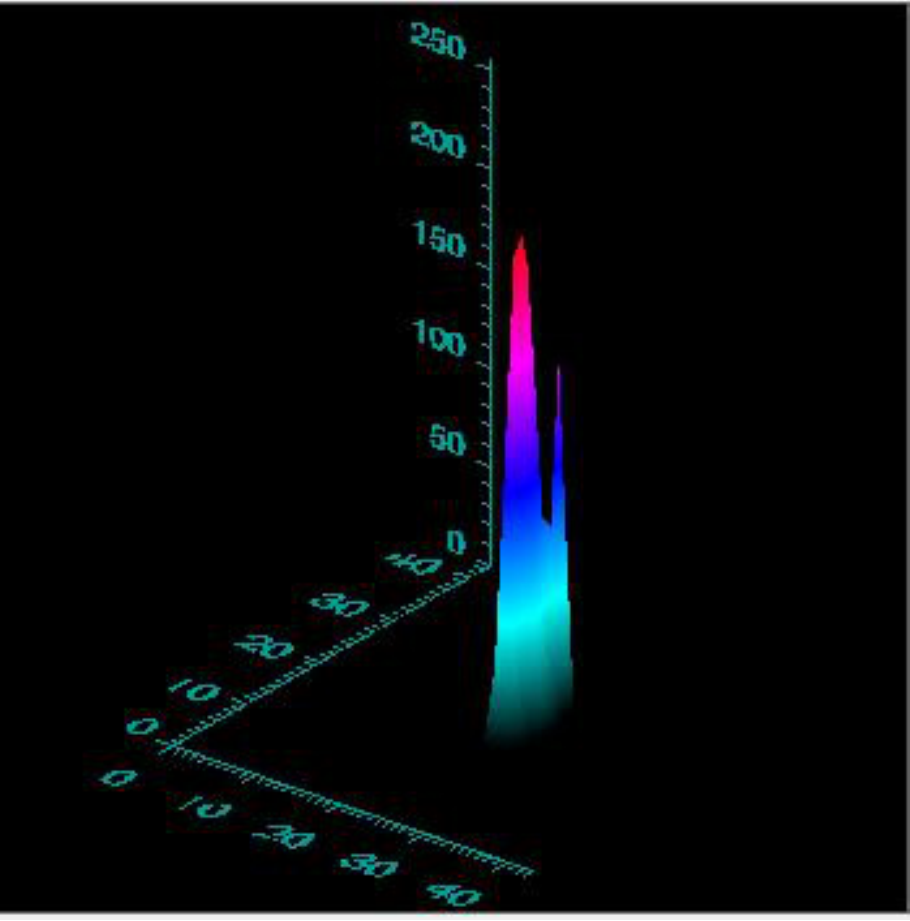}
	\caption{An event  recorded in the ASTRI-Horn camera. As can be seen, both in the 2-D and in the 3-D representations, a single bright pixel is outside the main Cherenkov image.}
	\label{FIG:Chap7_sec2_f2}
\end{figure}
This technique adapts well to the detection sensitivity of the ASTRI Mini-Array, since the estimated minimum energy threshold for the detection of heavy nuclei is above 10 TeV. The ASTRI Mini-Array should be able to measure the iron spectrum from 10\,TeV to 1\,PeV and beyond.

\subsection{Measurement of the {\it Knee}}
\label{Subsec:knee}
The elemental composition of CRs in the {\it knee} region \citep{2013APh....47...54A} is not a completely well settled issue \citep[see e.g.][and references therein]{2015APh....69....1C}. However this information is of paramount importance to constrain the origin of CRs of different energies and assess the transition between Galactic and extra-galactic CRs.
The ASTRI Mini-Array could contribute to this widely discussed topic by exploiting the mesonic channel of the hadronic showers.

\begin{figure}
	\centering
	\includegraphics[width=0.4\textwidth]{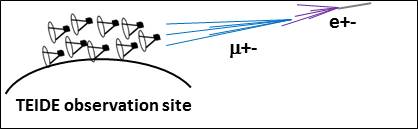}
	\caption{Schematic representation of a very inclined shower.}
	\label{FIG:Chap7_sec2_f3}
\end{figure}

\begin{figure}
	\centering
	\includegraphics[width=0.44\textwidth]{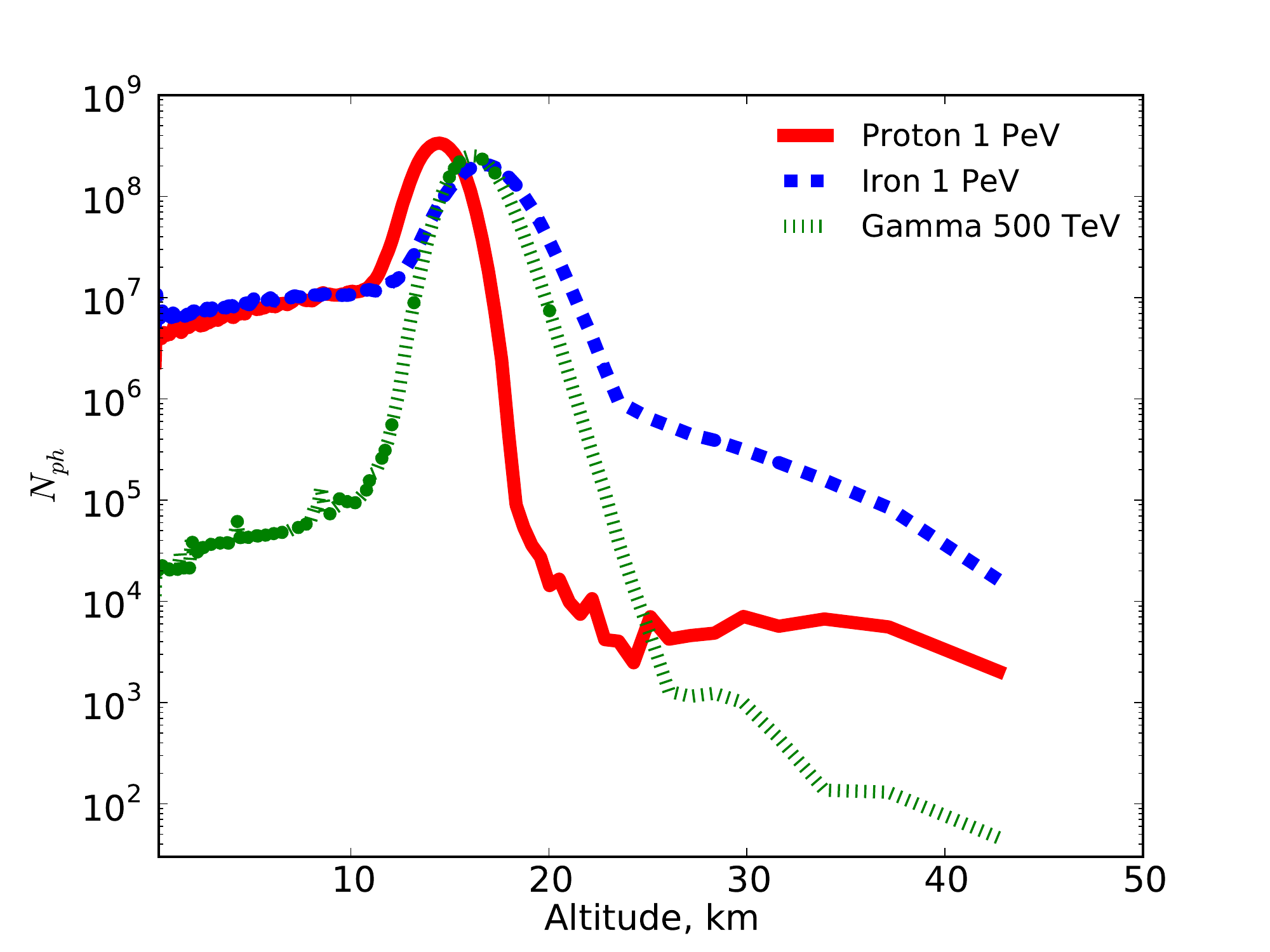}
	\caption{Simulated  longitudinal profile of 1\,PeV proton, 1\,PeV iron and 500\,TeV Gamma-ray at incident angle of $87^{\circ}$. Reproduced from \cite{2016PhRvD..94l3018N}.}
	\label{FIG:Chap7_sec2_f4}
\end{figure}

\begin{figure}
	\centering
	\includegraphics[width=0.44\textwidth]{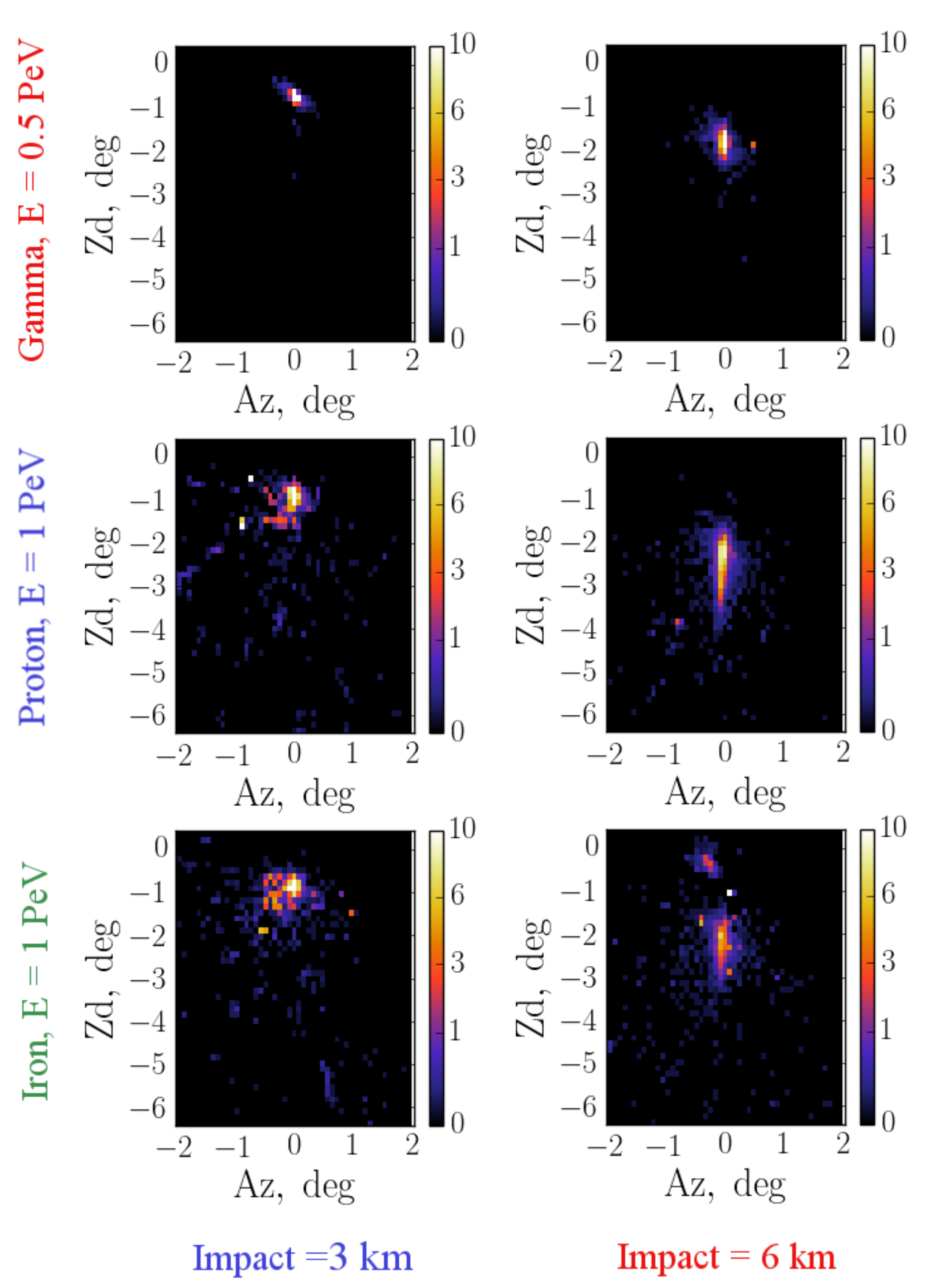}
	\caption{Simulated showers at inclined angle of $87^{\circ}$. Reproduced from \cite{2016PhRvD..94l3018N}.}
	\label{FIG:Chap7_sec2_f5}
\end{figure}

\begin{figure}
	\centering
	\includegraphics[width=0.44\textwidth]{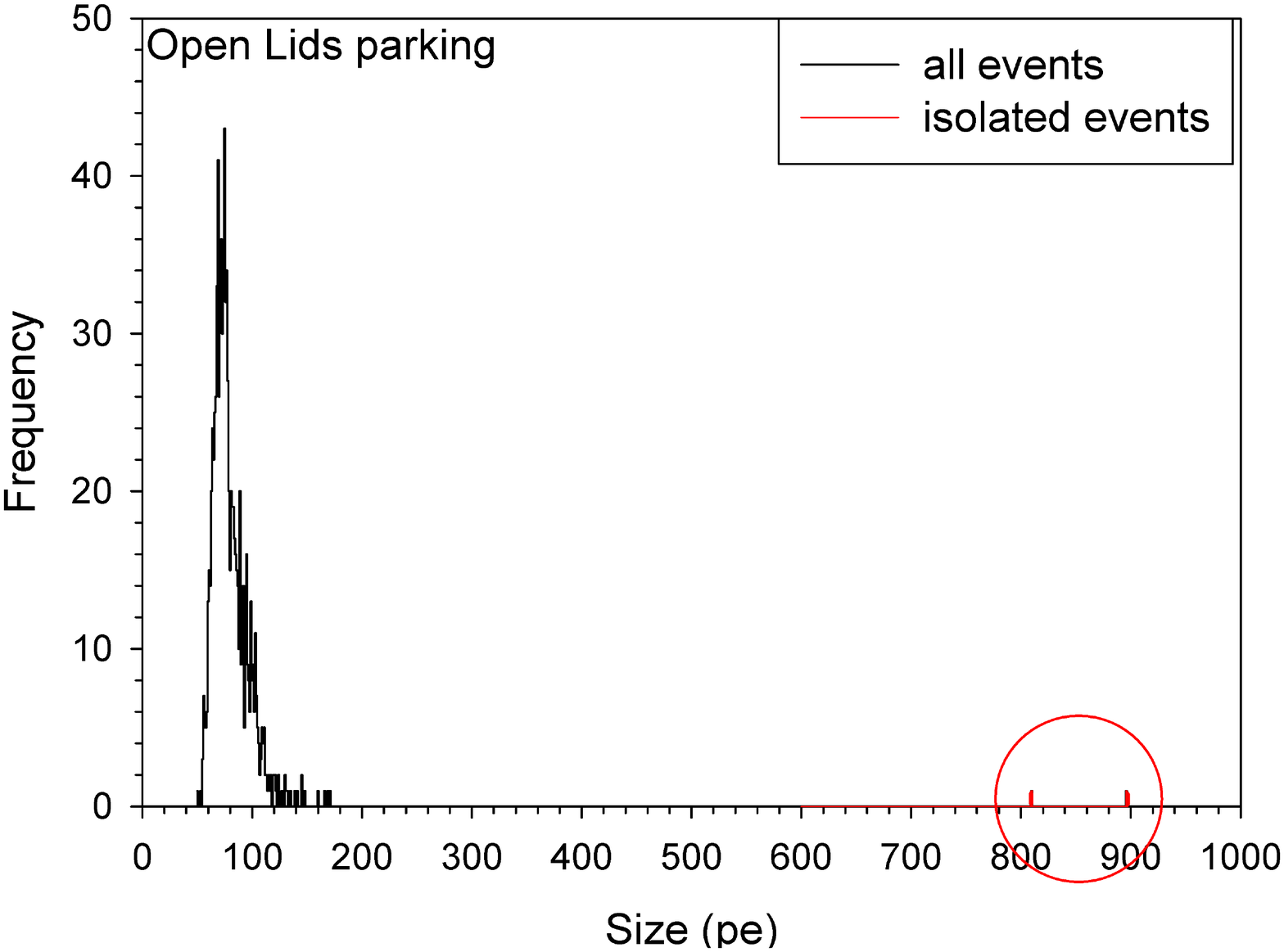}
	\caption{Size distribution of all the events detected in the six minutes run. The size is calculated, event by event, summing the image-pixels greater than 12\,pe. Isolated events are shown in red.}
	\label{FIG:Chap7_sec2_f6}
\end{figure}

\begin{figure*}
	\centering
	\includegraphics[width=0.44\textwidth]{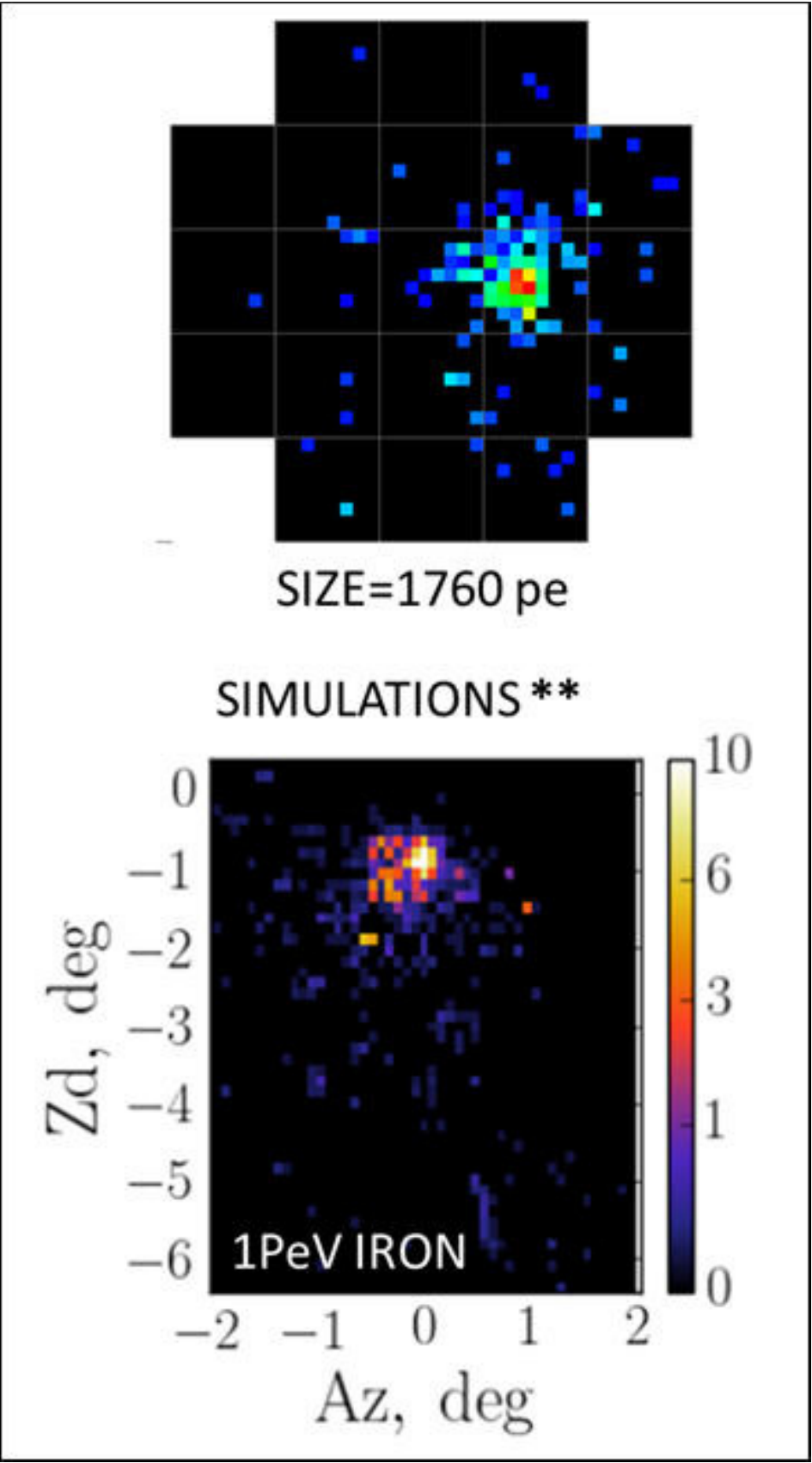}
	\includegraphics[width=0.44\textwidth]{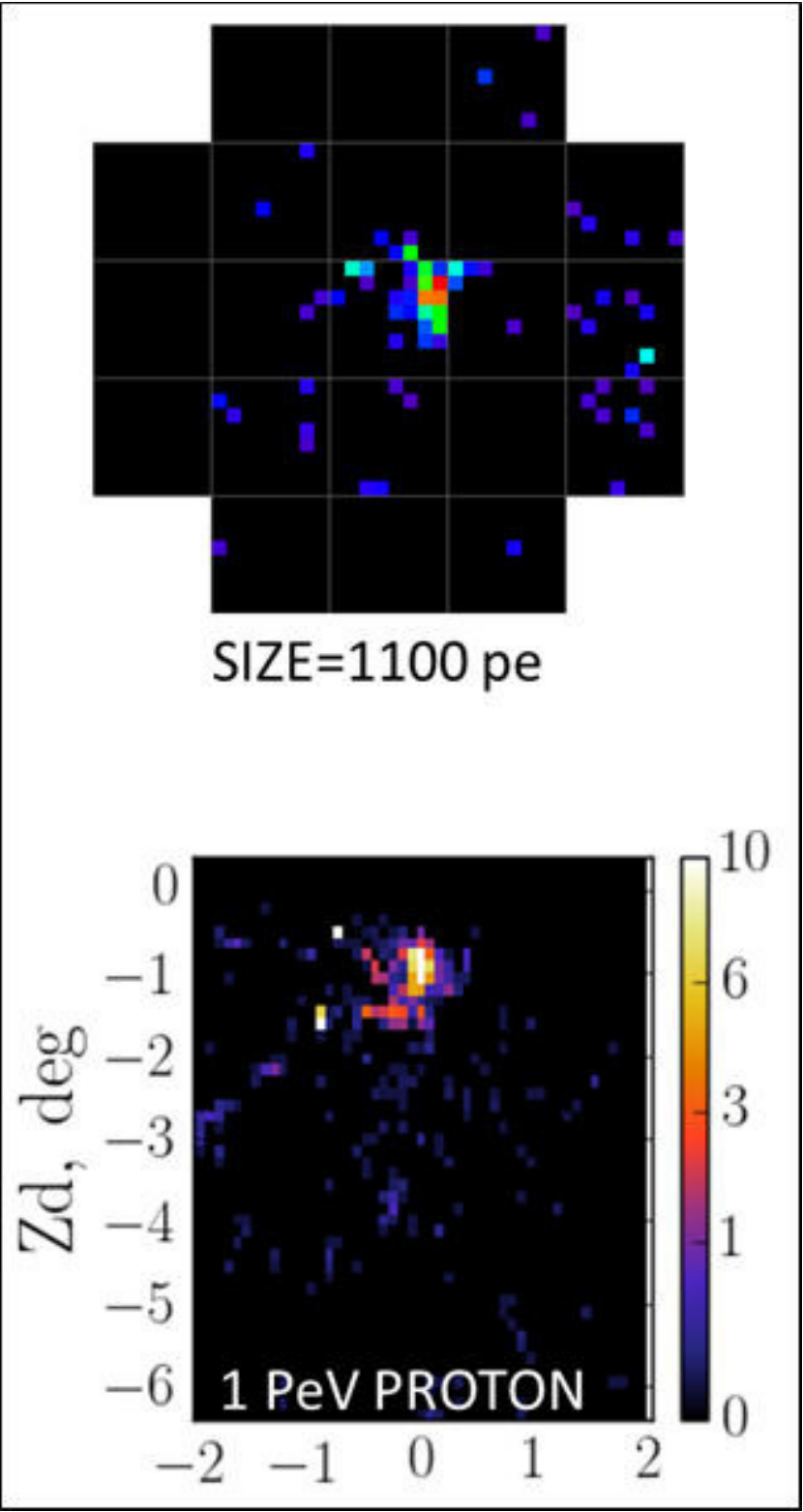}
	\caption{{\bf Top panel:} images of the two detected events by the ASTRI-Horn prototype. The size is here calculated summing image-pixels greater than 6\,pe. {\bf Bottom panel:} images of simulated showers at inclined angle of 87\textdegree{} for 1\,PeV Iron (left) and 1\,PeV proton (right), respectively.
}
	\label{FIG:Chap7_sec2_f6ab}
\end{figure*}

Observations with telescopes pointing to very inclined angles ($>70^{\circ}$), with respect to the Zenith, allow us to detect high energy events. This is expected due to the absorption of the electromagnetic component of the showers in the slant atmosphere, making more clear the separation between electromagnetic and muon component in the shower-image. Showers of energy around 1\,PeV, and above, should be detected through the Cherenkov light emitted by the charged particles in the early shower development and by the Cherenkov light produced by the numerous surviving muons, that is the hard component of the shower. A toy representation of an very inclined shower is shown in Figure~\ref{FIG:Chap7_sec2_f3}.
This interesting scenario, introduced by \cite{2016PhRvD..94l3018N}, leads to a longitudinal profile of the emitted Cherenkov light by the EAS initiated by proton, iron and gamma-ray as the one, as shown in Figure~\ref{FIG:Chap7_sec2_f4} \citep[from ][]{2016PhRvD..94l3018N}.
The difference between the Cherenkov profiles, generated by primary protons, iron nuclei and gamma ray, is due to the change of the particle content of the shower. The bump, evident in the altitude range 10-20 km, is attributed to the Cherenkov emission from electrons, while the flat region from 0 to 10 km is due to the Cherenkov emission from muons. The difference, in the flat region of the profile, of more than two orders of magnitude between the number of Cherenkov photons between the gamma ray and the hadrons is also evident. Cherenkov images are then expected to contain the compact electromagnetic component, together with the more spread muonic component. In the images of the proton and the iron showers, a sort of "halo" can be distinguished around the bulk of the signal. This is due to muons that dominate at great depth in the atmosphere. The image of the gamma ray shower is more compact due to the absence of muons. The simulated images, shown in Figure~\ref{FIG:Chap7_sec2_f5}, assume a total efficiency of the telescope system (product of optical system and photon detection efficiency) of 20\%. The telescope mirror is assumed to have the diameter of 4\,m and a FoV of $10^{\circ}$. This value fit well with the ASTRI Mini-Array telescope characteristics.
Again, without any claim, we report a couple of events that have been detected in a run of 6 minutes with the ASTRI-Horn prototype  telescope in parking position. As shown in Figure~\ref{FIG:Chap7_sec2_f6}, the plot of the size distribution for all the recorded events shows two isolated events with size significantly higher than the average size of all the events. For this short run, the camera worked properly for all the time of data taking and at a constant trigger rate.
Figure~\ref{FIG:Chap7_sec2_f6ab} shows the images of the two detected events. The size is here calculated summing image-pixels greater than 6\,photo-electrons (pe). The comparison of these two events with the ones simulated by \cite{2016PhRvD..94l3018N} is impressive.
If this is confirmed by increasing the statistics in the next months, the ASTRI Mini-Array could provide elemental composition data in the energy range around the {\it knee}. Furthermore, if events without {\it halo} are detected, the gamma-ray astronomy on the PeV energy scale will become a reality.

\subsection{Muons sampling technique}
From the point of view of CRs, the ASTRI Mini-Array can be considered as nine particle sampling units of approximately 0.12\,m$^{2}$ each. The filter window of each camera generates Cherenkov photons when it is crossed by muons. The signal produced (hundreds of photons) is detected by the SiPM and converted in photo-electrons. Because of the proximity of the focal plane detector to the filter window, for geometrical reasons, the typical Cherenkov ring cannot be formed but, instead, the photons cumulate on few nearby pixels, or in a string of pixels if the incident muon is very inclined with respect to the focal plan. With lids closed, during the day and with the telescopes in parking position, for example, a consistent flux of muons is detected by the camera (very high statistics, considering that the telescopes operate only during the night).
\begin{figure}
	\centering
	\includegraphics[width=0.44\textwidth]{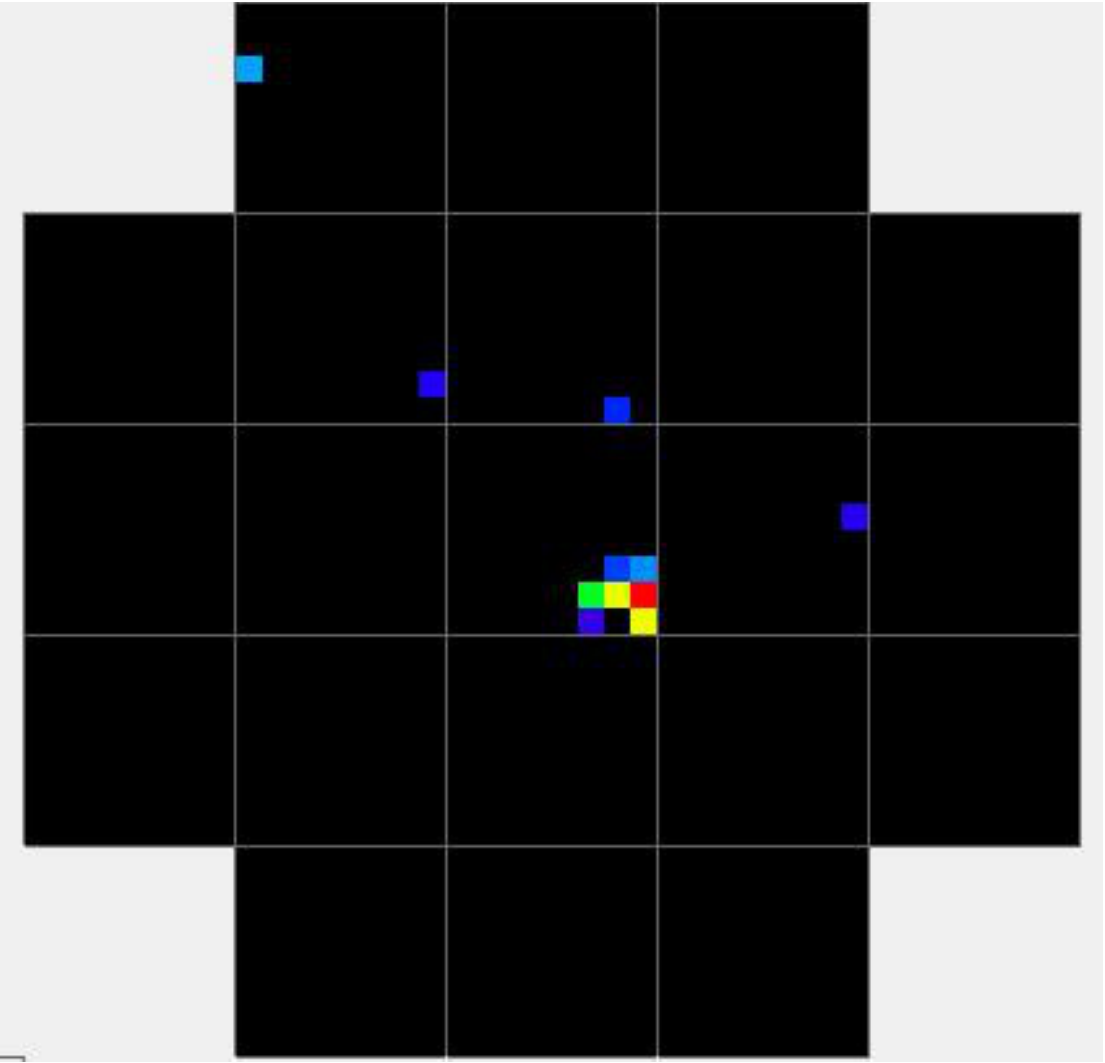}
	\caption{A {\it direct} muon detected in the ASTRI-Horn camera. The light spot is of about 70 photo-electrons.}
	\label{FIG:Chap7_sec2_f7}
\end{figure}
These events are easily reconstructed because the low level noise is only due to the dark current of the SiPM. A typical event of direct muon, as detected by the ASTRI {\it Horn} camera prototype, is shown in Figure~\ref{FIG:Chap7_sec2_f7}, while Figure~\ref{FIG:Chap7_sec2_f8} shows an event triggered by multiple muons.
\begin{figure}
	\centering
	\includegraphics[width=0.44\textwidth]{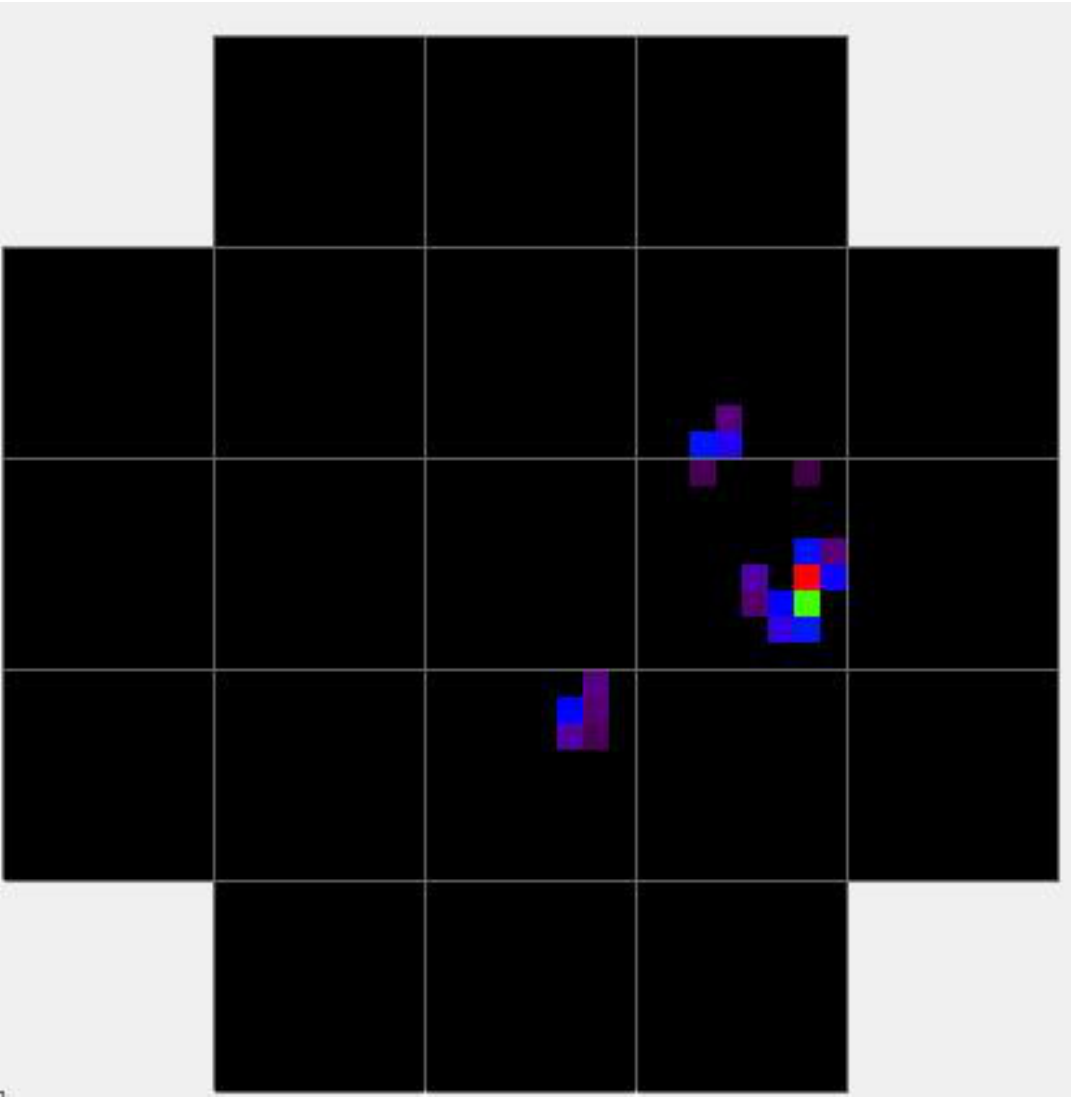}
	\caption{An event with three spots that very likely are muons. }
	\label{FIG:Chap7_sec2_f8}
\end{figure}
One application, among the different possible uses of detection of {\it direct} muons, consists in measuring muons multiplicity on a statistics base to perform direct measurements of CRs composition. The interaction of a CR with the atmosphere  produces multiple muons. The multiplicity of muons increases with the energy of the interacting particles and it is sensitive to the CR chemical composition. It is known that, for the same energy, a heavy nucleus that interacts with an atmospheric nucleus, produces, on average, a greater number of $\pi$ and $K$ mesons than a light nucleus. The mesons produced by a heavy nucleus, however, are on average less energetic than those produced by a light nucleus, since in the first case the energy of the incident nucleus must be divided among a greater number of particles. As a consequence, for the same energy, muons produced by heavy primaries are on average less energetic and with greater multiplicity than those produced by light primaries. The muons propagate through the atmosphere in a narrow cone whose opening angle with respect to the direction of the incident CR is determined by the energy and the transverse momentum of the parent meson, and by the production height. Measurement of the multiplicity in the nine cameras of the ASTRI Mini-Array could lead at interesting results at zero cost.

\subsection{Analysis Method}
Monte-Carlo simulations including camera response are needed. The simulations should be tailored for each of the items reported. For the muon sampling technique is strongly suggested to use GEANT-4~\citep{2003NIMPA.506..250A} and Monte-Carlo EAS simulations.
%
%
%

\section{Stellar Intensity Interferometry}\label{sec:7.1}
%
\paragraph{Scientific Case}--\label{sec:7.1.1}
Imaging a celestial object has always been a primary goal in astronomy, since much of our understanding depends on our ability to resolve it, measure its size, and determine its spatial structure. For the first time, we are in a position to image bright stars in the visible light waveband at very high angular resolution using a technique known as stellar intensity interferometry (SII), which is based on the second order coherence of light \citep{1963PhRv..130.2529G}. Angular resolutions below 100 microarcsec ($\mu$as) are achievable with this technique, using large collecting area telescopes separated by hundreds to thousands of meters baselines. At this level of resolution it turns out to be possible to reveal details on the surface and of the environment surrounding brights stars on the sky, that typically have angular diameters $\lesssim 1-10$ milli-arcsecond (mas) \citep{Kieda2019}.

SII was pionereed by Robert Hanbury Brown and Richard Q. Twiss between the '50s and the '70s (see, e.g., \citealt{Hanbury_Brown1974} and references therein). They built the Narrabri Stellar Intensity Interferometer using twin 6.5 m diameter telescopes movable along a circular track at Narrabri, New South Wales, Australia, and performed the first direct astronomical measurements of stellar radii via SII. After the successfull Narrabri experiment, SII was shelved for about 40 years. The possibility to operate simultaneously an array of large area telescopes and to connect them electronically, with no need to directly combine the photons they detect, has recently renewed interest for SII as a tool for performing imaging observations in the optical band using a detection method similar to long-baseline radio interferometic arrays (e.g. \citealt{Le_Bohec2006,2013APh....43..331D}). Indeed, this possibility is offered by the sparsely distributed arrays of Imaging Air Cherenkov Telescopes (IACTs), such as the ASTRI Mini-Array, which have adequate optical properties, sufficiently large mirror areas, and telescope time available during the full Moon. SII also requires the measurement of photon arrival times with a precision better than one ns at each telescope, over baselines extending to km distances. This accuracy corresponds to a meter light-travel distance, and thus any instrumental or atmospheric delay smaller than a fraction of one meter can be tolerated. New implementations of SII technology to astronomy have then been recently pursued by several groups, either simulating thermal sources in the laboratory (e.g. \citealt{Dravins2015}), or performing pilot experiments or observations with 1-3 meter class telescopes (e.g. \citealt{Zampieri2016, Guerin2017, Matthews2018, Rivet2020}). Eventually, the capability of performing SII measurements with the MAGIC and VERITAS IACTs has been convincingly demonstrated by \cite{2020MNRAS.491.1540A} and \cite{2020NatAs...4.1164A}, respectively.

Since the beginning of 2019 also the INAF ASTRI Collaboration recognizes the scientific value of SII and endorses the development of a SII observing mode. Despite being limited to bright targets because of the limited collecting area, the ASTRI Mini-Array will provide a major improvement compared to present installations thanks to the imaging capabilities achievable through the 9 ASTRI Mini-Array telescopes (36 baselines). This will be rivaled only by the full deployment of the CTA observatory \citep{Zampieri2020_prep}.

\paragraph{Expected goal/Immediate Objective}--\label{sec:7.1.2}
The ASTRI Mini-Array equipped with a SII instrument will provide the first images of bright Galactic stars with sub-mas angular resolution. This capability will open up unprecedented frontiers in some of the major topics in stellar astrophysics.
Measuring the angular shape of a selected number of stars (including main sequence stars) with a resolution of $\sim 100 \mu$as will provide their oblateness and enable direct measurements of the stellar rotation, extending in the visible band the still limited sample of IR stellar images collected with the Center for High Angular Resolution Astronomy (CHARA) interferometer (e.g. \citealt{2011ApJ...732...68C}).
Imaging with this resolution can allow also the detection of dark/bright spots or other surface features \citep{2012MNRAS.424.1006N}.
A low-resolution measurement of this type is provided by the visible light images of the extended red supergiant Betelgeuse, taken with SPHERE using the Very Large Telescope\footnote{Montarg\`es, M., Cannon, E., Kervella, P., Ferreira, B. 2020, ESO Press release ESO2003; \href{https://www.eso.org/public/news/eso2003/}{https://www.eso.org/public/news/eso2003/}} during its recent 2019-2020 pronounced dimming. The images clearly revealed a substantial asymmetry in the surface brightness distribution of the star, possibly caused by a mass ejection event that cooled to form a dust cloud in the southern hemisphere \citep{2020ApJ...899...68D}.
Furthermore, observing stars with circumstellar discs/eruptions will reveal details of the disc structure, density gradients, and scale height, and will show how these systems evolve and dynamically interact.
An astonishing example of the results potentially achievable, and their relevance for understanding stellar astrophysics, is reported in \citet{Kloppenborg2010}. The (infrared) images taken with the CHARA interferometer show clearly that the 18-month long partial eclipse of the star eps Aur is produced by a disc orbiting the companion (see Figure 2 in \citealt{Kloppenborg2010}). The cause of the long eclipse has been a subject of controversy for nearly 200 years.

\begin{figure}
	\centering
	\includegraphics[width=0.49\textwidth]{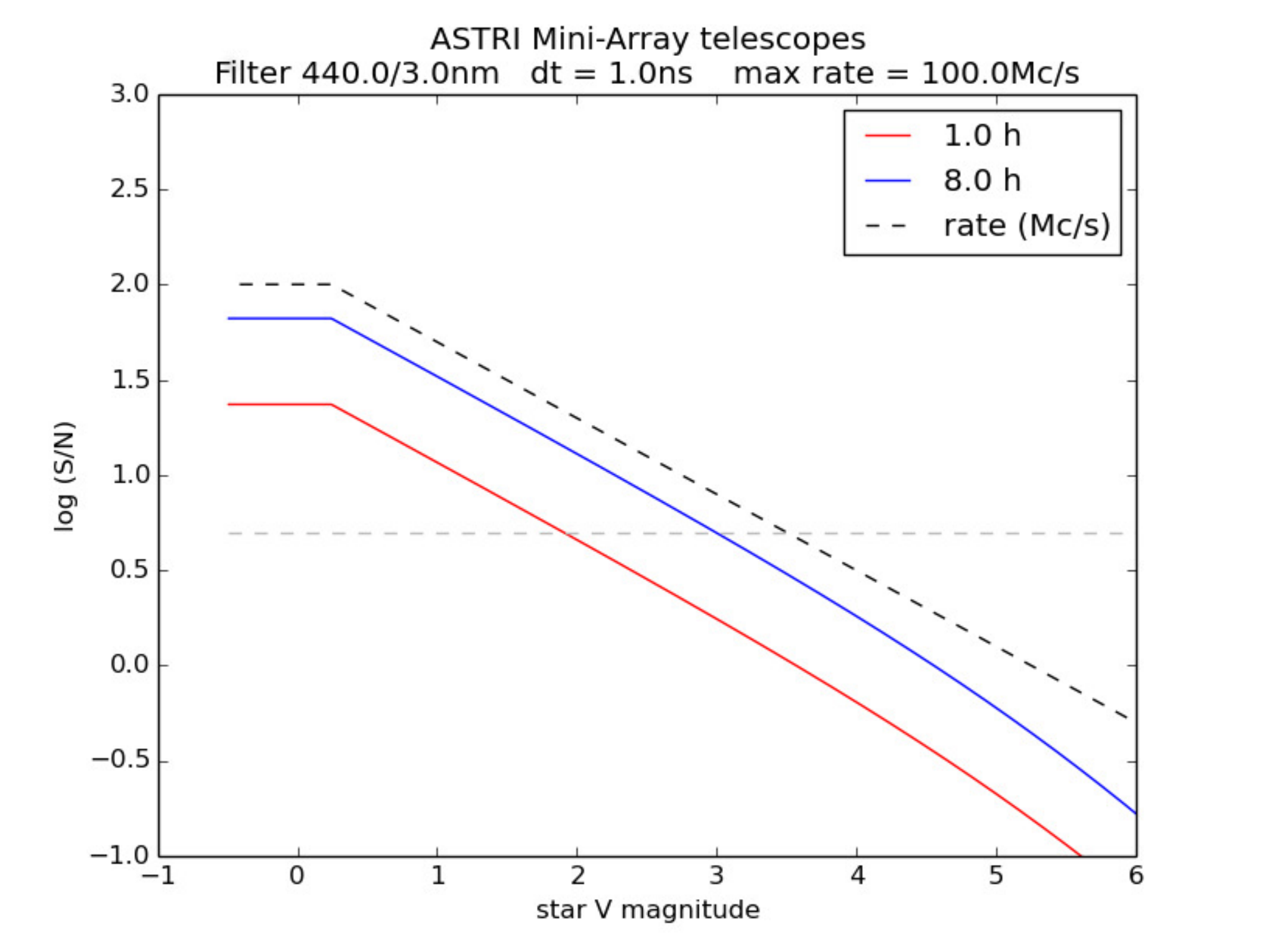} \\
	\caption{S/N ratio for a SII measurement with two ASTRI Mini-Array telescopes as a funtion of stellar magnitude. The source photon flux is limited in order to give a maximum rate of 100 Mcounts/s. The simulation is done using a narrow-band filter centered at 440 nm and with a FWHM of 3 nm (plus a polarizer). The bin time is 1 ns and the observing time is 1 hr (red line) and 8 hrs (blue line). The gray dashed line corresponds to S/N$=5$.}
	\label{FIG:Chap7_sec1}
\end{figure}

\paragraph{Observing Time, Pointing Strategy, Visibility and Simulation Setup}--\label{sec:7.1.3}
The quality of a SII measurement is dictated by the signal-to-noise (S/N) ratio of the main SII observable, the degree of coherence (e.g. \citealt{Zampieri2016}). Figure \ref{FIG:Chap7_sec1} shows the S/N for a measurement with two ASTRI Mini-Array telescopes as a function of stellar magnitude, assuming to correlate the photon arrival times with a bin time of 1 ns \citep{Zampieri2020_prep}.
The instrumental noise is taken into account. 
Stars with magnitude V$<$3 are observable with the ASTRI Mini-Array telescopes with a S/N$>$5, for an exposure time of $\lesssim$8 hours.

Assuming that the time allocated for SII observations is 3 nights/month and that the time lost for unfavourable weather conditions is $\sim 20$\%, the total effective observing time is $\sim 240$ hrs/year. We estimate that, for a bright ($V < 1$) star, 8-24 hrs are needed to perform 100-300 measurements of the correlation using all the baselines of the ASTRI Mini-Array, each with a S/N $>10$. An average ($V \sim 2$) star needs 16 hrs for 36 measurements using all the baselines of the ASTRI Mini-Array, each with a S/N $>10$. For bright stars we expect to be able to perform accurate image reconstruction. For average targets, we will perform image reconstruction, but the number of baselines will allow to obtain also well-constraining high angular resolution measurements of surface features. With 240 hrs/year we then expect to be able to observe 3-8 bright and 14 average stars per year. A detailed list of targets in this brightness interval and with potentially interesting properties for sub-mas optical imaging is included in Table 2 of \citet{CTA_SII_WB2019}. The number of targets and their distribution on the sky ensures that at any time a sizeable fraction of them is visible from Tenerife.

\paragraph{Analysis Method}--\label{sec:7.1.4}
Specific equipment and methods are required for acquiring, reducing and analyzing data taken in the SII observing mode. This mode will make use of a dedicated instrument that is being designed and will be installed on the ASTRI Mini-Array telescopes \citep{Zampieri2020_prep}.

\paragraph{Legacy Products, Multi-wavelength Synergies, Coordinated Observations}--\label{sec:7.1.6}
The ASTRI Mini-array operated in SII mode will leave an extraordinary legacy of \rf{sub-mas} images of the brightest nearby stars and their environments.

\section{The Multi-wavelength Landscape}\label{sec:8}
%
\subsection{Science at the ``Islas Canarias''}\label{sec:Canarias}
%
The very high-energy view from the Islas Canarias is complemented with the MAGIC array \citep{2016APh....72...76A}, \rf{and} the CTA large-size telescope prototype \citep[LST-1, ][]{2019ICRC...36..653C}. At the Observatorio del Roque de los Muchachos (ORM) site in La Palma we have also the INAF Telescopio Nazionale Galileo \citep[TNG,][and references therein]{1998SPIE.3352...91B, 2018SPIE10704E..1AG}. \\
Members of IAC have access to a set of small-size robotic telescopes located at the Observatorio del Teide: Las Cumbres Global Observatory (LCO) \citep{Brown2013}, the 1.2m STELLA Robotic telescopes  \citep{Strassmeier2010} and the 0.4m PIRATE operated by the Open University \footnote{\href{https://www.telescope.org}{https://www.telescope.org}}. All these facilities are operated in robotic mode. 
LCO is composed of 23 telescopes (3 different sizes: 2m, 1m and 0.4 m) at 7 different sites around the world. This feature permits to have a rapid response, and to have continuous monitoring of the sources, which is crucial for transient events. There are two possibilities to access the LCO observing time for IAC members: guaranteed time at the 0.4 m network through a local Time Allocation Committee (TAC); open time access to the whole network through the observatory TAC. Calls for proposal are on a per-semester basis. Rapid reaction observations can be requested. Most telescopes are dedicated to imaging although the 2m telescopes provide spectroscopic facilities.     
The 1.2 m STELLA telescopes combine a wide-field imager (WIFSIP) and a high-resolution spectrograph (SES). The most useful instrument will likely be the telescope equipped with the camera WIFSIP. There is a possibility to request rapid reaction observations, although on a best effort basis. The way to access to observing time is through the Spanish TAC. This committee announces calls for proposals on a per-semester basis.  
Finally, the 0.4 PIRATE telescope can also be requested by IAC members, it is equipped with a wide field camera. The way to access to the observing time is through the mentioned local TAC. 
 In all cases the overhead subscription rates have been relatively low up to now.     
The data are provided in a fully reduced mode by LCO and STELLA but currently this is not the case for PIRATE. 
In addition the 2.5m Liverpool Robotic Telescope located at the Observatorio del Roque de Los Muchachos can also be accessed by the Spanish community through the Spanish TAC. This telescope permits ToOs requests and offers a flexible suite of instruments. Moreover, in case of need of more sensitive instruments, it is possible to use the 10.4m Gran Telescopio de Canarias (GTC), which can be occasionally accessed as Discretionary Director Time (DDT) in addition to the standard proposal calls.

 \subsection{X-ray and $\gamma$-ray facilities}\label{sec:X_Gamma_Facilities}
%
The Neil Gehrels {\it Swift} Observatory \citep[{\it Swift} hereafter,][]{2004ApJ...611.1005G}, AGILE \citep{2009A&A...502..995T} and Fermi \citep{2009ApJ...697.1071A} provide an invaluable complement to the ASTRI Mini-Array observations, thanks to their large field of view and surveying capabilities, joined with a rapid dissemination system of transient events.
Both {\it Swift} and Fermi have been successfully ranked in the last NASA Astrophysics Senior Review of Operating Missions, with the suggestion of prolonging these missions up to 2022, when a new Senior Review will be held\footnote{\href{https://science.nasa.gov/astrophysics/2019-senior-review-operating-missions}{https://science.nasa.gov/astrophysics/2019-senior-review-operating-missions}}. The AGILE satellite has been recently prolongued up to at least the end of May 2022\footnote{\href{http://agile.ssdc.asi.it/}{http://agile.ssdc.asi.it/}}, with possible further extensions.
Since the ASTRI Mini-Array preliminary science operations will start in 2023, all these major wide field-of-view facilities should be still operational. This makes an excellent opportunity both for time-domain studies and for spectral ones. The former ones will benefit of the wide field of view and fast reaction, while the latter ones will benefit of a wide energy coverage from optical-UV up to hundreds of GeV.

Both XMM-Newton~\citep{2001A&A...365L...1J} and Chandra~\citep{2002PASP..114....1W} observatories have been extended and will overlap with the ASTRI Mini-Array observations. We will benefit of both their spectral and imaging capabilities especially for multi-wave\-length studies involving new Galactic extended sources (SNR, PWN) discovered in radio and observed at TeV energies.

In the hard X-ray domain INTEGRAL~\citep{1994ApJS...92..327W} and NuSTAR~\citep{2013ApJ...770..103H} will allow us to complement and extend the spectral performance of both XMM-Newton and Chandra in a domain where the inverse Compton emission process (e.g., in blazars) is dominant.

The launch of eROSITA/SRG~\citep{2021A&A...647A...1P} in 2019 yielded the generation of the first eROSITA/SRG sky survey and in particular the detection of large scale X-ray structures in the Galactic halo~\citep{2020Natur.588..227P}, possibly correlated with the $\gamma$-ray Fermi bubbles~\citep{2010ApJ...724.1044S}. The northern one of these large-scale structures, most likely caused by large energy injection from the Galactic center and whose nature (AGN or starburst) is still debated~\citep{2018Galax...6...27K}, will be an important target for the ASTRI Mini-Array.

Recently, an XMM-Newton {\it Multi-Year Heritage Program}~\citep[PI: G. Ponti, and see also][for further details]{2015MNRAS.453..172P,2019Natur.567..347P} has been granted 3.5\,Ms to survey the Galactic Plane in its central region ($|l|<10$\textdegree{}\,;$|b|<1.5$\textdegree{}). This important program is focussed on the study of the X-ray diffuse emission, in synergy with the eROSITA/SRG {\it Hot Milk ERC Program} (PI: G. Ponti). The XMM-Newton program will allow us to investigate the TeV emission from sources in that region, making it a perfect match with the Galactic Center study described in Section~\ref{Subsec:GC}.

\rf{In the Fall 2021, the Imaging X-ray Polarimetry Explorer \citep[IXPE,][]{2016SPIE.9905E..17W} was successfully laun\-ched and, after completing the in orbit verification phase, is currently collecting data}, allowing, among many other topics, to shed light on the geometry and the emission mechanism of AGNs and to investigate how particles are accelerated in PWNe.

 \subsection{Optical and radio facilities}\label{sec:Opt_Radio_Facilities}
%
In addition to the optical facilities reported in Section~\ref{sec:Canarias}, the ASTRI Mini-Array can count on the GLAST-AGILE Support Program of the Whole-Earth Blazar Telescope (GASP\--WEBT) Collaboration~\citep{2008A&A...481L..79V}, dedicated to the observation of blazars in the radio, millimetre, infrared and optical wavelength, whose contribution is fundamental during multi-wavelength campaigns in order to study the synchrotron portion of the blazar's SED.

Radio observations probe the accelerated electrons population through their synchrotron emission and, when used in synergy with VHE observations, they provide important clues to disentagle different origin of the observed VHE emission. Analysis of radio maps allows a spatial comparison of radio and  VHE emission and to identify radio counterpart/s to a VHE \rf{source/s}. A good angular resolution is necessary for accurate radio flux density estimates. These measurements are used as constraints to the modelling of the broad non-thermal spectrum to  derive  relevant physical parameters of the source.
Multi-frequency radio observations yield to an integrated spectral index determination or, provided  that sufficient angular resolution is available, highlight changes in the spectral index within the sources, usually interpreted as evidence of multi population of accelerated electrons coexisting in the source.

Recently, the Sardinia Radio Telescope~\citep[SRT, ][]{2017A&A...608A..40P}, sensitive in the 0.3--116\,GHz frequency range\footnote{currently limited to the K-band at $\approx 22$\,GHz}, started its regular observations. In particular, SRT observed sources of interest for the ASTRI Mini-Array, such as W~44, IC~433 and Tycho~\citep{2017MNRAS.470.1329E, 2019MNRAS.482.3857L}, making it an excellent observatory for future sinergies in the northern hemisphere.

Prior to the actual construction of the Square Kilometre Array (SKA), a series of demonstration telescopes, the SKA precursors, have been built to develop and test new technologies, as input for the design of SKA, and to anticipate the scientific results of SKA.
Among SKA precursors, MeerKAT, in South Africa, and ASKAP, in Australia, are already fully operational.
Even if located in the Southern emisphere, several synergies using MeerKAT and ASKAP can be anticipated, in particular in the planned observations of the region of the Galactic Centre, including the massive stars clusters, Arch and Quintuplet, see Section~\ref{Subsec:GC}. The recent ASKAP results on the survey of the SCORPIO field~\citep{2021MNRAS.506.2232U} reveal its unique capability to map complex regions at different angular scale, together with its sensitivity and the possibility to perform in-band spectral analysis, which will make it a perfect instrument for: {\it i}) \rf{identifying} radio counterparts; {\it ii}) accurate radio density measurements for structures up to 50\arcmin{}; {\it iii}) spectral analysis within the region, allowing to point out changes in physical parameters.  

In the Northern hemisphere, the Low Frequency Array \citep[LOFAR,][]{2013A&A...556A...2V} is the largest SKA path\-finder, observing at low radio frequencies, in the range (15--240)\,MHz. It reaches a sensitivity more than 100 times better than any previous telescope at low radio frequencies, with a nominal angular resolution of about 6\arcsec{} which can be improved up to 0.1\arcsec{}. This allows to open a new science window in the low-frequency radio band. LOFAR can monitor 2/3 of the sky nightly in Radio Sky Monitor mode, being an excellent radio transient factory. The LOFAR survey programs include the Two Meter Sky Survey~\citep[LoTSS; 120-170\,MHz,][]{2017A&A...598A.104S,2019A&A...622A...1S} and the LBA Sky Survey at very low frequencies~\citep[LoLSS; 42--66\,MHz,][]{2019A&A...622A...5D}. These surveys, in synergy with deeper observations of selected fields~\citep[][e.g.,]{2021A&A...648A...2S}, provide a long-lasting legacy value in numerous areas of astrophysics and cosmology.

\section{The ASTRI Mini-Array Legacy}\label{legacy}
%
The ASTRI Mini-Array will operate for at least eight years. The first period of about four years will be devoted to the ``core science''. At the completion of the goals of the core science, the ASTRI Mini-Array will gradually enter a second period of about four years and will be managed as much as possible as an ``observatory'', open to the scientific community. Extensive multi-wavelength synergies are planned with several international facilities, including both ground- and space-based facilities. Its location is close to several international observing facilities, both in the optical and in the VHE energy range. This will foster scientific synergies and collaborations among different groups. 

In the previous Sections we discussed the scientific breakthroughs that we expect to obtain with the ASTRI Mini-Array. We can anticipate that, while the science topics will remain the ones we described, particular sources and sky regions might vary, according to new results obtained in the near future by both current IACTs and PSAs. The recent LHAASO results~\citep{2021Natur.594...33C} clearly stress the importance of having in the Northern hemisphere an array of Cherenkov telescopes which can reach energies of a few hundred TeV (typical of PSAs) with an angular resolution of a few arcminutes and an energy resolution of few percent, typical of IACTs, in order to provide crucial morphological and spectral information. The ASTRI Mini-Array, therefore, will be extremely important for VHE observation also during the era of the CTA Observatory.
The ASTRI Mini-Array will also provide fundamental results for CTAO observations, eventually allowing a better planning and fine-tuning of their Key Science Projects described in \citet[][]{2019scta.book.....C}.
Last but not least, the ASTRI Mini-Array data constitute a legacy for both current and future VHE facilities and other multi-wavelength observatories, in terms of light-curves, spectra, and high resolution images of extended sources. This will allow the scientific community to use these data in combination with data at other wavelengths and perform, e.g. along the Galactic Plane, population studies.

\newpage

\section*{Acknowledgments}
\noindent
\rf{We thanks the reviewer for their constructive criticisms that helped improving the paper.} This work was conducted in the context of the ASTRI Project. This work is supported by the Italian Ministry of Education, University, and Research (MIUR) with funds specifically assigned to the Italian National Institute of Astrophysics (INAF). We acknowledge support from the Brazilian Funding Agency FAPESP (Grant 2013/10559-5) and from the South African Department of Science and Technology through Funding A\-gree\-ment 0227/2014 for the South African Gamma-Ray Astronomy Programme. This work has been supported by H2020-ASTERICS, a project funded by the European Commission Framework Programme Horizon 2020 Research and Innovation action under grant agreement n. 653477. IAC is supported by the Spanish Ministry of Science and Innovation (MICIU). \rf{The research activity of E.P. has received funding from the European Union’s Horizon 2020 research and innovation program under the Marie Sklodowska-Curie grant agreement No. 847523 ‘INTERACTIONS’ and by Villum Fonden (project n. 18994). G.P. acknowledges funding from the European Research Council (ERC) under the European Union’s Horizon 2020 research and innovation programme (Grant agreement No. [865637]). R.A.B. is funded by the ``la Caixa'' Foundation (ID 10\-00\-10\-434) and the European Union's Horizon 2020 research and innovation program under the Marie Sk\l{}odowska-Curie grant agreement No 847648, fellowship code LCF/\-BQ\-/PI21\-/11830030.}
The ASTRI project is becoming a reality thanks to Giovanni ``Nanni'' Bignami, Nicol\`{o} ``Nichi'' D'Amico two outstanding scientists who, in their capability of INAF Presidents,  provided continuous support and invaluable guidance. While Nanni was instrumental to start the ASTRI telescope, Nichi transformed it into the Mini Array in Tenerife. Now the project is being built owing to the unfaltering support of Marco Tavani, the current INAF President. Paolo Vettolani and Filippo Zerbi, the past and current INAF Science Directors, as well as Massimo Cappi, the Coordinator of the High Energy branch of INAF, have been also very supportive to our work. We are very grateful to all of them. Nanni and Nichi, unfortunately, passed away but their vision is still guiding us.

\newpage

\appendix

\section{On the ASTRI Mini-Array simulation procedure}\label{Sec:Appendix1}
%
Section~\ref{Sec:SciSimSet} describes the scientific simulation setup, while in this Appendix we show an example of the scatter we estimate to have when performing 100 realizations to simulate a spectrum of a source. The procedure described here is based on the {\tt ctools} package, but it can be generalized to the {\tt gammapy} package too and it is based on \citet{ASTRI-MAN-4000-001}. See also~\citet{2018MNRAS.481.5046R,2020MNRAS.494..411R} for a detailed description of the procedure and its application on the study of extra-galactic sources with CTA. 

We define a {\it simulation} as a set of $N$ independent {\it realisations}. 
Each realisation is performed through a script that drives a sequential series of {\tt ctools}  tasks,
 {\tt ctobssim} and  {\tt ctlike}. 
\begin{enumerate}
\item {\bf STEP 1.} 
In our specific case, a realisation includes first running the task {\tt ctobssim} 
to create one event list based on our input model, 
including background events that were randomly drawn from the IRF background model. 
The randomisation is controlled by a seed that is unique to this realisation. 
To overcome the impact of a given statistical realization on the fit results,
for each energy bin, we perform sets of $N=100$ statistically independent realisations 
of the input model, by changing in each of them the  {\tt ctobssim}  seed value.
We note that for each energy bin we use seeds in a natural progression from 1 to 100, 
so that the results can be checked and reproduced after running the simulations,
or at a later time.

\item {\bf STEP 2.} 
Subsequently, the task {\tt ctlike} reads in each event file created in STEP 1 and the input model file and, 
using a maximum likelihood model fitting, determines the best-fit spectral 
parameters from which we derive the flux, as well as the test statistics (TS) value. 
It is reasonable that (if the model is smooth enough and/or the energy bins are small enough) 
we can use as input to  {\tt ctlike}  a power-law model  
$M_{\rm spectral}(E)=k_0 \left( \frac{E}{E_0} \right) ^{-\Gamma}$,  
where $k_0$ is the normalisation, 
$E_0$ is the pivot energy, 
and  $\Gamma$ is the power-law photon index. 
In our method, $k_0$ and $\Gamma$ are free to vary 
while $E_0$ is set to the geometric mean of the boundaries of the energy bin. 
We thus obtain $N=100$ sets of spectral parameters and TS values. 
For each realisation the best fit spectral parameters are used to calculate 
$N$ values of flux in the given energy bin. 
At the end of this procedure we have, for each energy bin,
$N=100$ sets of event files, $N$ sets of best fit
parameters (and TS) from which we calculated $N$ values of the flux. 

\item {\bf STEP 3.} 
Then, in each energy bin, 
the {\it mean TS value} of the $N$ realisations and its uncertainty are calculated as the mean, 
$\overline{TS_{\rm sim}}  = \frac{1}{N}\sum_{k=1}^{N}TS_{\rm sim}(k)$, 
and square root of the standard deviation of the sample of $N$ TS values, 
$s^2_{\rm sim}=\frac{1}{N-1}\sum_{k=1}^{N}(TS_{\rm sim}(k)-\overline{TS_{\rm sim}})^2$.
Similarly a  (simulation) {\it mean flux} and its uncertainty are calculated as the mean, 
and square root of the standard deviation of the sample of $N$ flux values. 

\item {\bf STEP 3B.} 
A special mention is the case when the source is not detected, i.e., when the 
simulation $\overline{TS_{\rm sim}}$ value is below a given threshold.
In that case, a {\it 95\,\% confidence level upper limits} on flux can be calculated 
from the distribution of the simulated fluxes.  
\end{enumerate}

\begin{figure}[]
	\centering
	\includegraphics[width=0.4\textwidth, angle=0]{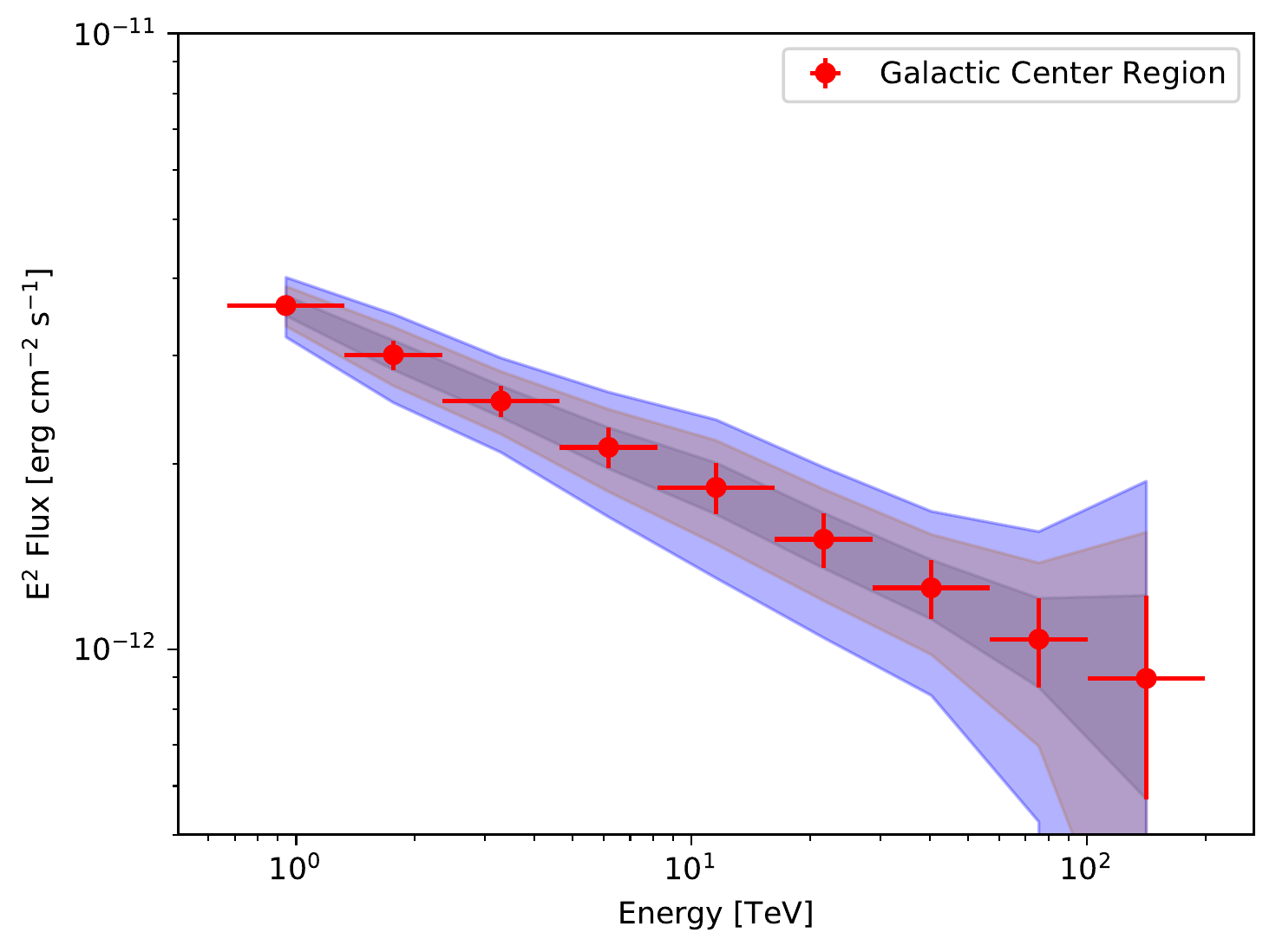}
	\caption{Simulated spectrum of the Galactic Centre. Red dots show the simulated flux as derived following the procedure described in the text and in Section~\ref{Subsec:GC}, Figure~\ref{FIG:GC_bestfit}. The dashed areas represent the 1$\sigma$ (68$\%$), 2$\sigma$ (95$\%$) and 3$\sigma$ (99.8$\%$) regions of uncertainty.}
	\label{FIG:Appendix1_GC}
\end{figure}

We consider the case of the Galactic Center (GC) as an exemplary situation for the approach described above.
In Figure~\ref{FIG:Appendix1_GC}, we show the average spectrum (red points) obtained from the 100 realizations where we superimposed the 1$\sigma$ (68$\%$), 2$\sigma$ (95$\%$) and 3$\sigma$ (99.8$\%$) regions of uncertainty. This implies that every single realization of the global source spectrum will fall into the shaded regions.

\smallskip
\noindent

\bibliographystyle{aa}


\end{document}